\documentclass[%
 aip,pof,
 amsmath,amssymb,
preprint,%
]{revtex4-1}
\usepackage{color}
\usepackage{hyperref}
\usepackage{graphicx}
\usepackage{dcolumn}
\usepackage{bm}
\usepackage[mathlines]{lineno}
\usepackage{mathrsfs}
\usepackage[utf8]{inputenc}
\usepackage[T1]{fontenc}
\usepackage{mathptmx}
\usepackage{subfigure}
\usepackage{tabularx}
\usepackage{booktabs}
\usepackage{bm}
\usepackage{amsmath}
\usepackage{mathrsfs}
\usepackage{float}
\usepackage{multirow}

\begin{document}

\preprint{AIP/POF}
\title{Simulation of two-phase flows at large density ratios and high Reynolds numbers using a discrete unified gas kinetic scheme}

\author{Jun Lai
}
\affiliation{
State Key Laboratory for Turbulence and Complex Systems, College of engineering, Peking University, Beijing 100871, P.R. China
}
\affiliation{
Guangdong Provincial Key Laboratory of Turbulence Research and Applications, Center for Complex Flows and Soft Matter Research and Department of Mechanics and Aerospace Engineering, Southern University of Science and Technology, Shenzhen 518055, P.R. China
}
\author{Zuoli Xiao
}
\affiliation{
	State Key Laboratory for Turbulence and Complex Systems, College of engineering, Peking University, Beijing 100871, P.R. China
}
\author{Lian-Ping Wang
}
\email{wanglp@sustech.edu.cn}
\affiliation{
	Guangdong Provincial Key Laboratory of Turbulence Research and Applications, Center for Complex Flows and Soft Matter Research and Department of Mechanics and Aerospace Engineering, Southern University of Science and Technology, Shenzhen 518055, P.R. China
}
\affiliation{Guangdong-Hong Kong-Macao Joint Laboratory for Data-Driven Fluid Mechanics and
	Engineering Applications, Southern University of Science and Technology, Shenzhen
	518055, China}


%
%

\date{\today}

\begin{abstract}
\centerline{\bf Abstract}
In order to treat immiscible two-phase flows at large density ratios and high Reynolds numbers,
a three-dimensional code based on the discrete unified gas kinetic scheme (DUGKS) 
 is developed, incorporating two major improvements.
First, the particle
distribution functions at cell interfaces are reconstructed using
a weighted essentially non-oscillatory scheme.
Second, the conservative lower-order Allen-Cahn equation is chosen, instead of the higher-order Cahn-Hilliard equation, to evolve
 the free-energy-based phase field governing the dynamics of two-phase interfaces.
Five benchmark problems are simulated to demonstrate the capability of the approach in treating 
two-phase flows at large density ratios and high Reynolds numbers,
including three two-dimensional problems (a stationary droplet, Rayleigh-Taylor instability, and a droplet splashing on a thin liquid film)
and two three-dimensional problems (binary droplets collision and Rayleigh-Taylor instability).
All results agree well with the previous numerical and the experimental results.
In these simulations, the density ratio and Reynolds number can reach a large value of ${\cal O} (1000)$.
Our improved approach sets the stage for the DUGKS scheme to handle realistic two-phase flow problems.
\end{abstract}


\maketitle


\section{Introduction} \label{sec: Introd}
Air-water two-phase flows are widespread phenomena in 
industry (hydroelectric power), agriculture (irrigation), medicine (intravenous injection),
nature (rain) and our daily life (shower).
The density ratio between water and air is about 800.
A deep understanding of this kind of multiphase flows with large-density-ratio can promote the development of science and technology.

There are three approaches to study multiphase flows, including theoretical analysis, experiment observations and numerical simulations.
Theoretical analysis can only deal with the simplest problems.
Experiments rely on instruments, and may have a limited ability to capture 
the space-time evolution at the air-water interface scale, and are affected by the environmental conditions.
In recent years, numerical simulation has developed rapidly and becomes an important approach for multiphase flow research.
One of the common challenges with numerical methods is the numerical instability at large-density-ratios, and
reliable numerical methods at  large-density-ratio remain much desired.
Much attention has been paid to dealing with the multiphase flow simulations at large-density-ratios in recent years.~\cite{WANG2015404,WANG201541,PhysRevE.98.063314,PhysRevE.97.033309,PhysRevE.99.063306,2017Improved,2019Phase,Yangzeren2019,doi:10.1063/5.0086723}

In so-called interface-resolved multiphase flow simulations, the tracking of two-phase interfaces is essential.
The traditional approaches based on Navier-Stokes solvers include the front-tracking method,~\cite{1996A} volume-of-fluid (VOF) method,~\cite{Scardovelli1999DIRECT,2021A} level-set (LS) method,~\cite{SUSSMAN1998663,2021A} {\it etc.}
It remains challenging for the front-tracking method to model interface breakup and coalescence, because the interface needs to be artificially  ruptured.~\cite{1996A,PhysRevE.85.046309} For the VOF and LS methods, an interface reconstruction step is required, which would be 
a complex task to implement.~\cite{Scardovelli1999DIRECT,PhysRevE.85.046309}
Furthermore, the VOF method relies on the refinement of the grid when dealing with the flow interface, 
the mass conservation is well observed, but the accuracy in dealing with complex sharp interfaces is limited. The LS method has 
a superior capability in treating complex interfaces, and its tracking accuracy of free surface is typically higher than the VOF method. However, frequent reinitializations of the LS function are needed, which may cause errors in mass conservation. 

The physical width of the fluid-fluid interface is much smaller than the macroscopic scale.
It is beneficial to deal with the interface from a multiscale perspective.
The mesoscopic simulation approaches for fluid-fluid two-phase flows have been developed based on the model Boltzmann equations over the past three decades.
The two-phase mesoscopic models contain
color-gradient model,~\cite{1991Lattice} pseudo-potential model,~\cite{shan1993lattice,1994Simulation} free-energy model,~\cite{swift1996lattice} {\it etc.}
Gunstensen~{\it et al.}~\cite{1991Lattice} proposed the color-gradient model,
which is based on the cellular automata model presented
by Rothmann and Keller.~\cite{rothman1988immiscible} 
The surface tension is generated by a perturbation operator, causing some unphysical velocity near the interface.~\cite{chen1998lattice,PhysRevE.85.046309,2016Multiple}
Shan and Chen~\cite{shan1993lattice} developed the pseudo-potential model (Shan-Chen model), 
which features simple formulation and high computational efficiency. 
Furthermore, Shan-Chen model can separate fluid phases or components naturally, due to microscopic particle interactions.~\cite{chen1998lattice,PhysRevE.87.053301}
However, the surface tension is related to the density ratio, and spurious currents are non-negligible around the interface.~\cite{ZHANG20191128,PhysRevE.85.046309}
Later, Swift~{\it et al.}~\cite{swift1996lattice} proposed the free-energy model. 
The free-energy-based phase-field model is thermodynamically more consistent than the above models,~\cite{chen1998lattice,PhysRevE.85.046309} and is expected to be physically more capable.

As noted before, large-density-ratio multiphase flows are not easy to simulate.
If Reynolds number is also considered, maintaining numerical stability becomes more challenging.
Coupled Shan-Chen model with multiple-relaxation-time (MRT) collision operator, Li~{\it et al.}~\cite{PhysRevE.87.053301}
simulated two-phase flows (droplet splashing on a thin film) with
density ratio and Reynolds number up to $\rho^*\sim 700, Re\sim 1000$.
The lattice velocity model is D2Q9, hence they only performed two-dimensional (2D) simulations.
The main contribution to improved numerical stability is the MRT collision operator, which is a common and effective tool to calculate large-density-ratio multiphase flows in the mesoscopic approaches.
They pointed out that
the MRT collision operators are generally more stable than the standard Bhatnagar-Gross-Krook (BGK) collision operator.~\cite{qian1992lattice}
We note that there are two possible reasons. 
From a mathematical point of view, the MRT model has more tunable parameters and more degrees of freedom.
From a physical point of view, the relaxation time corresponding to different physical quantities may be different, and the MRT model can be more consistent with the physical principle.  
The MRT model was also incorporated into the color-gradient method for the large-density-ratio simulations, which was studied by
Ba~{\it et al.}~\cite{2016Multiple} with the D2Q9 lattice velocity model.
Their results showed that $\rho^*\sim 1000$ for the steady flows (a static droplet and the layered channel flow), and $\rho^*\sim 100$, $Re\sim 500$ can be reached for the unsteady flow (a droplet splashing on a thin film).



In the phase-field model, the Allen-Cahn (AC) equation~\cite{ALLEN1976425,SUN2007626,CHIU2011185} and Cahn-Hilliard (CH)~\cite{cahn1958free,cahn1959free} equation are the two common models for the evolution of the phase field $\phi$.
It is noted that the CH equation is a fourth-order partial differential
equation (PDE), while the AC equation is a second-order PDE which is easier to implement numerically and tends to be 
more stable.
Wang {\it et al.}~\cite{H2016Comparative}
compared AC equation and CH equation in the framework of LB models, and demonstrated that AC equation has a better 
numerical stability.
Hence the AC equation is more suitable for the large-density-ratio two-phase-flow simulations~\cite{2017Improved,PhysRevE.97.033309,PhysRevE.99.063306,2019Phase,Yangzeren2019,doi:10.1063/5.0086723}.

The CH equation can also treat large-density-ratio flows under some specific formulations.
Wang~{\it et al.}~\cite{WANG2015404,WANG201541} proposed multiphase lattice Boltzmann flux solvers (MLBFS) to simulate flows with $\rho^*\sim 1000$, and the Reynolds number could be greater than $1000$.
In MLBFS, the hydrodynamic equations are the combination of Navier-Stokes equations (NSE) and lattice Boltzmann equations (LBE).
The LBE is utilized for the computation of the flux terms in NSE, while the NSE is discretized by the finite volume  scheme.
The Poisson equation for pressure does not need to be solved, hence this method combines the advantages of traditional and mesoscopic algorithms.
Furthermore, a finite difference scheme is used for the CH equation, in which the convective term is solved by the weighted essentially non-oscillatory (WENO) scheme.
We note that the WENO scheme~\cite{jiang1996efficient} is based on the essentially non-oscillatory (ENO) scheme.~\cite{HARTEN1987231}
In ENO scheme, the smoothest stencil is selected to approximate the flux at cell boundary, which can reduce the numerical oscillations.
In WENO scheme, the convex combination of all stencils is used and the weights of the stencils are properly chosen to improve
the local accuracy.
It is expected that the WENO scheme can also reduce the oscillation and enhance the numerical stability in the multiphase flow simulations.
Chen~{\it et al.}~\cite{PhysRevE.98.063314} proposed a simplified multiphase lattice Boltzmann method (SMLBM) for the 2D (D2Q9) large-density-ratio ($\rho^*\sim 1000$) simulations.
In SMLBM, they directly updated the macroscopic variables resolved in a predictor-corrector scheme, instead of the mesoscopic distribution functions.

A few studies coupled the LB method with the more stable phase-field equation (AC equation) to simulate the large-density-ratio problems.
We denote this method as PF(AC)-LBM.
Liang~{\it et al.}~\cite{PhysRevE.97.033309,PhysRevE.99.063306} simulated large-density-ratio multiphase flows by PF(AC)-LBM with BGK collision operator.
The parameters can reach $\rho^*\sim 1000, Re\sim 500$ for the case of a 2D droplet splashing on a thin liquid film,~\cite{PhysRevE.97.033309}
and $\rho^*> 800, Re\sim 500$ for the case of a 3D droplet impact on a wetting solid.~\cite{PhysRevE.99.063306}
Furthermore, the numerical stability performance can be improved when combined with other techniques.
Incorporating the MRT collision operator, 
Fakhari~{\it et al.}~\cite{2017Improved}
used PF(AC)-LBM to perform the 2D Rayleigh-Taylor instability simulations with
density ratio and Reynolds number up to $\rho^*\sim 1000, Re\sim 3000$.
Based on the open-source LB framework,~\cite{LANIEWSKIWOLLK2016833}
Kumar~{\it et al.}~\cite{2019Phase} simulated 2D Rayleigh-Taylor instability with $\rho^*\sim 1000, Re\sim 3000$ when combining PF(AC)-LBM with the MRT collision operator.

As a relatively new mesoscopic approach, the discrete unified gas kinetic scheme (DUGKS)~\cite{Guo2013,Guo2015}
 combines the advantages of the LBM~\cite{chen1998lattice,aidun2010lattice} and unified gas kinetic scheme (UGKS).~\cite{XU20107747} In DUGKS, the model Boltzmann equation is solved using an accurate finite-volume formulation coupling tightly the kinetic particle transport and particle collisions. 
Compared to LBM, DUGKS can more easily incorporate irregular meshes and different kinetic particle velocity models.
Hence we wish to explore the capability for DUGKS to simulate large-density-ratio multiphase flows.
Based on the PF(AC)-DUGKS method, 
Yang~{\it et al.}~\cite{Yangzeren2019} could simulate 2D Rayleigh-Taylor instability with large-density-ratio ($\rho^*\sim 1000$) but the Reynolds number is small ($Re\sim 50$).
Combined with the adaptive mesh refinement technique, a multilevel PF(AC)-DUGKS was proposed by Yang~{\it et al.}~\cite{doi:10.1063/5.0086723} to simulate a 2D droplet splashing on a thin film with $\rho^*\sim 1000,~Re\sim 500$.

To summarize, there are several ways to help improve the numerical stability and increase the density ratio in the multiphase flow simulations:
\begin{enumerate}
\item Incorporating the WENO scheme when treating the convective term;~\cite{WANG2015404,WANG201541}
\item Combining the macroscopic equations /  variables and mesoscopic approaches;~\cite{WANG2015404,WANG201541,PhysRevE.98.063314}
\item Adopting the MRT collision operator instead of the BGK collision operator;~\cite{PhysRevE.87.053301,2016Multiple,2017Improved,2019Phase}
\item Using the AC equation instead of the CH equation in the phase-field approach.~\cite{PhysRevE.97.033309,PhysRevE.99.063306,2017Improved,2019Phase,Yangzeren2019,doi:10.1063/5.0086723}
\end{enumerate}

In the present work, we aim to extend the DUGKS scheme to simulate 3D multiphase flows at large density ratios
and higher Reynolds numbers.
For this purpose, we couple Yang~{\it et al.}~\cite{Yangzeren2019}'s PF(AC)-DUGKS algorithm with the WENO scheme at the cell interface to improve the numerical stability.
The D3Q19 lattice velocity model is used,  with a parallel implementation strategy to  address the 
 computing resources requirement.

The rest of the paper is organized as follows. In Section~\ref{sec: Meth}, the governing equations and the numerical methods for the large-density-ratio two-phase flows are presented. 
In Section~\ref{sec: 2D} and~\ref{sec: 3D}, 2D benchmark problems (a stationary droplet, Rayleigh-Taylor instability, and a droplet splashing on a thin liquid film) and 3D benchmark problems (binary droplets collision and Rayleigh-Taylor instability) are validated and analyzed.
Finally, conclusions are given in Section~\ref{sec: Concl}.

\section{Methodology} \label{sec: Meth}


\subsection{The phase-field theory and governing equations for two-phase flows}
In the phase-field theory, the free energy functional
of fluid-fluid two-phase flow contains the bulk free-energy and the interfacial free-energy, and can be expressed
as~\cite{2002Molecular,liu2003phase,jacqmin1996energy,yue2004diffuse,zhang2018discrete,chen2019simulation,liang2014phase}
\begin{equation}
{\cal F}(\phi,\nabla \phi)=\int_{V}\left[\psi(\phi)+\frac{\kappa}{2}|\nabla \phi|^{2}\right] d V,\label{Fphi}
\end{equation}
where $ \phi $ is the order parameter corresponding to the two phases. $\psi(\phi)=\beta\left(\phi-\phi_{A}\right)^{2}\left(\phi-\phi_{B}\right)^{2}$ is the 
free-energy density of the bulk fluids,
representing separation of the two phases into the bulk region.
$\frac{\kappa}{2}|\nabla \phi|^{2}$ is the free-energy density of the interfacial region, representing mixing of the two phases.
$V$ is the volume of the considered domain.
$\phi_{A}=1$ and $\phi_{B}=0$ are parameters corresponding to the two phases.
$\beta$ and $ \kappa $ are positive coefficients that determine the relative magnitudes of bulk free-energy density and interfacial free-energy density, respectively.
They also relate to the surface tension $\sigma$ and the interfacial thickness parameter $W$, {\it i.e.},
\begin{equation}
\sigma=\frac{\left|\phi_{A}-\phi_{B}\right|^{3}}{6} \sqrt{2 \kappa \beta}, \quad
W=\frac{1}{\phi_{A}-\phi_{B}} \sqrt{\frac{8 \kappa}{\beta}}.
\end{equation}

The chemical potential $\mu_{\phi}$ is
the variation of the free energy functional ${\cal F}(\phi,\nabla \phi)$ with respect to $\phi$,~\cite{zhang2018discrete,chen2019simulation}
\begin{equation}
\begin{aligned}
\mu_{\phi}=\frac{\delta {\cal F}}{\delta \phi}
= 4 \beta\left(\phi-\phi_{A}\right)\left(\phi-\phi_{B}\right)\left(\phi-\frac{\phi_{A}+\phi_{B}}{2}\right)-\kappa \nabla^{2} \phi.\label{muphi}
\end{aligned}
\end{equation}
For a flat interface at equilibrium, $\mu_{\phi}^{eq}=0$. Then $\phi$ can be obtained as~\cite{chen2019simulation,jacqmin1996energy}
\begin{equation}\label{flatphi}
\phi^{eq}(\zeta)=\frac{\phi_{A}+\phi_{B}}{2}+\frac{\phi_{A}-\phi_{B}}{2} \tanh \left(\frac{2 \zeta}{W}\right),
\end{equation}
where $\zeta$ is the signed distance normal to the interface.

Next, we discuss the time evolution of the order parameter $\phi$.
The governing equation of $\phi$ can be written in terms of the flux densities as a conservative form~\cite{PhysRevE.91.063309}
\begin{equation}\label{Eqphi}
\frac{\partial \phi}{\partial t}=-\nabla\cdot \left( \boldsymbol{j}_A+\boldsymbol{j}_D+ \boldsymbol{j}_S\right) ,
\end{equation}
where $t$ is the time. $\boldsymbol{j}_A$, $\boldsymbol{j}_D$ and $\boldsymbol{j}_S$ are the advective flux density, the
diffusive flux density, and the phase separation flux density.
 The advective flux is $\boldsymbol{j}_A=\phi \boldsymbol{u}$, where $\boldsymbol{u}$ is the flow velocity.


The AC equation is utilized here to help treat  large density ratios. 
For the conservative AC equation, the diffusive flux is expressed as~\cite{PhysRevE.91.063309}
\begin{equation}\label{diffusiveflux}
\boldsymbol{j}_D=-M_{AC}\nabla \phi,
\end{equation}
where $M_{AC}$ is the mobility.

The phase separation flux density can be determined by the equilibrium profile of $\phi$, Eq.~\eqref{flatphi}.
Considering a flat interface located on $x=0$ at the equilibrium state, Eq.~\eqref{Eqphi} should be satisfied.
In this case, $\partial_t \phi^{eq}=0$ and $\boldsymbol{j}_A^{eq}= \boldsymbol{0}$, leading to $\nabla\cdot \left( \boldsymbol{j}_D^{eq}+ \boldsymbol{j}_S^{eq}\right)=0$.
A simple choice is
\begin{equation}\label{fluxrelation}
\boldsymbol{j}_S^{eq}=-\boldsymbol{j}_D^{eq},
\end{equation}
where
\begin{equation}\label{diffusivefluxeq}
\begin{aligned}
\boldsymbol{j}_D^{eq}=&-M_{AC}
\boldsymbol{e}_x
\frac{d}{dx} \phi^{eq}\left( x\right) 
\\=&
-M_{AC}
\boldsymbol{e}_x
\frac{\phi_{A}-\phi_{B}}{W} \left[1-\tanh^2 \left(\frac{2 x}{W}\right) \right] 
\\=&
-M_{AC}
\boldsymbol{e}_x
\frac{4\left[\phi_{A}-\phi^{eq}\left( x\right) \right]\left[\phi^{eq}\left( x\right)-\phi_{B} \right]}{W\left(\phi_{A}-\phi_{B} \right) }
.
\end{aligned}
\end{equation}
Therefore, for a curved interface, the phase separation flux density can be designed as
\begin{equation}\label{separationflux}
\boldsymbol{j}_S=
M_{AC}
\frac{4\left(\phi_{A}-\phi \right)\left(\phi-\phi_{B} \right)}{W\left(\phi_{A}-\phi_{B} \right) }
\frac{\nabla\phi}{\left|\nabla\phi \right| }.
\end{equation}

Combining with the mass equation and the momentum equation,
then the macroscopic governing equations (ACNS system) of immiscible two-phase flows are~\cite{Yangzeren2019}
\begin{subequations}\label{macroEqs}
	\begin{equation}\label{EqAC}
	\frac{\partial \phi}{\partial t}+\nabla \cdot(\phi \boldsymbol{u})=\nabla\cdot \left[  M_{AC}\left( \nabla \phi-\theta \boldsymbol{n}\right) \right] ,
	\end{equation}     	
	\begin{equation}\label{EqMa}
	\nabla \cdot \boldsymbol{u}=0,
	\end{equation}
	\begin{equation}\label{EqMo}
	\frac{\partial(\rho \boldsymbol{u})}{\partial t}+\nabla \cdot(\rho \boldsymbol{u} \boldsymbol{u})=-\nabla p+\nabla \cdot\left[\mu\left(\nabla \boldsymbol{u}+ \boldsymbol{u}\nabla\right)\right]+\boldsymbol{F},
	\end{equation} 	
\end{subequations}
where 
$\theta=\frac{-4\left(\phi-\phi_{A}\right)\left(\phi-\phi_{B}\right)}{W\left(\phi_{A}-\phi_{B}\right)}$ is a parameter related to $\phi$,
$\boldsymbol{n}=\frac{\nabla\phi}{|\nabla\phi|}$ is the normal unit vector of the $\phi$ field.
$p$ is the pressure. $\boldsymbol{F}=\boldsymbol{F}_s+\boldsymbol{F}_b$ is the total body force, where $\boldsymbol{F}_{s}=\mu_{\phi}\nabla\phi$ is the interfacial force and $\boldsymbol{F}_b$ contains other body forces such as gravity. 
It is noted that  $\boldsymbol{F}_{s}=\mu_{\phi}\nabla\phi$ instead of the alternative form  $\boldsymbol{F}_{s}= - \phi  \nabla \mu_{\phi}$
to reduce the order of spatial derivatives in the simulations.
The fluid density $\rho$ and dynamic viscosity $\mu$ are given by the linear models~\cite{HE1999642} 
	as
\begin{subequations} 	
	\begin{equation}\label{Eqrho}
	\rho=\frac{\phi-\phi_{B}}{\phi_{A}-\phi_{B}} \rho_{A}+\frac{\phi-\phi_{A}}{\phi_{B}-\phi_{A}} \rho_{B},
	\end{equation}
	\begin{equation}
	{\mu}= \frac{\phi-\phi_{B}}{\phi_{A}-\phi_{B}}{\mu_A}+ \frac{\phi-\phi_{A}}{\phi_{B}-\phi_{A}}{\mu_B},    
	\end{equation}  
\end{subequations}  
where $\rho_{A}$, $\rho_{B}$ and $\mu_{A}$, $\mu_{B}$ are
the densities and
the dynamic viscosities of the two phases. 

It is noted that the density $\rho$ is
determined by the order parameter $\phi$,
and is independent of the pressure $p$.
Substituting Eq.~\eqref{Eqrho} and Eq.~\eqref{EqMa} into Eq.~\eqref{EqAC}, leading to the equation of density as
\begin{equation}\label{EqMass}
\frac{\partial \rho}{\partial t}+\nabla \cdot(\rho \boldsymbol{u})=\frac{\rho_{A}-\rho_{B}}{\phi_{A}-\phi_{B}} \nabla \cdot\left[M_{AC}(\nabla \phi-\theta \boldsymbol{n})\right].
\end{equation}
Therefore, the mass is not locally conserved in this model.
In the single-phase region, $\nabla \phi=\boldsymbol{0}$ and $\theta=0$, then the right hand side (RHS) of Eq.~\eqref{EqMass} is zero.
If the densities of the two phases are the same, then $\rho_{A}=\rho_{B}$ and the RHS of Eq.~\eqref{EqMass} is also zero.
As a result, the nonconservative mass problem only happens near the two-phase interface when the densities of the two fluids are different.
The mass conservation could be restored by
a quasi-incompressible model,~\cite{PhysRevE.93.043303,zhang2018discrete,chen2019simulation} where
the velocity-divergence is modified. Hence in the quasi-incompressible model, the flow field is not incompressible precisely near the two-phase interface if the density-ratio is not one. 

\par 

\subsection{The mesoscopic model for the ACNS system}

The double-distribution function model is used to reproduce the conservative AC equation and the hydrodynamic equations. 
The following two model Boltzmann equations with a 
Bhatnager-Gross-Krook (BGK) collision model~\cite{PhysRev.94.511} are employed~\cite{Yangzeren2019}
\begin{subequations}\label{Boltzmann}   
	\begin{equation}\label{Boltzmannf}
	\frac{\partial f_{\alpha}}{\partial t}+\boldsymbol{\xi}_{\alpha} \cdot \nabla f_{\alpha}=-\frac{f_{\alpha}-f_{\alpha}^{e q}}{\tau_{f}}+S_{\alpha}^{f},
	\end{equation}
	\begin{equation}\label{Boltzmanng}
	\frac{\partial g_{\alpha}}{\partial t}+\boldsymbol{\xi}_{\alpha} \cdot \nabla g_{\alpha}=-\frac{g_{\alpha}-g_{\alpha}^{e q}}{\tau_{g}}+S_{\alpha}^{g},
	\end{equation}
\end{subequations}
where the distribution functions $f_{\alpha} = f_{\alpha}\left(\boldsymbol{x}, t\right)$ and $g_{\alpha} = g_{\alpha}\left(\boldsymbol{x}, t\right)$   corresponding to a discrete particle velocity $\xi_{\alpha}$ 
are functions of position $\boldsymbol{x}$ and time $t$.  
$\tau_{f}$ and $\tau_{g}$ are the relaxation times. 
The key to recover the macroscopic governing equations is to properly design the two equilibrium distribution functions ($f_{\alpha}^{eq}$, $g_{\alpha}^{eq}$) and the two source terms ($S_{\alpha}^{f}$, $S_{\alpha}^{g}$). 

Applying the Chapman-Enskog analysis,~\cite{Chapman1970}
we can derive the moment-integral constraints for the equilibrium distribution functions and the source terms,
to recover the ACNS system, as shown in Appendix~\ref{ap:InverseDesign}.
There are many possible designs, and one specific design is presented here. For the phase-field, we adopt the following design~\cite{H2016Comparative,PhysRevE.97.033309,Yangzeren2019,doi:10.1063/5.0086723}
\begin{subequations}\label{feqSf}
	\begin{equation}\label{feq}
	f_{\alpha}^{e q}=\omega_{\alpha} \phi\left(1+\frac{\boldsymbol{\xi}_{\alpha} \cdot \boldsymbol{u}}{R T}\right),
	\end{equation}
	\begin{equation}\label{Sf}
	S_{\alpha}^{f}=\omega_{\alpha} \boldsymbol{\xi}_{\alpha}\cdot \theta\boldsymbol{n} +\frac{\omega_{\alpha}\boldsymbol{\xi}_{\alpha}}{R T}\cdot \frac{\partial\left(\phi \boldsymbol{u}\right)}{\partial t}. 
	\end{equation}
\end{subequations}
The equilibrium distribution function and the source terms for the hydrodynamic variables can be specified as~\cite{zu2013phase,liang2014phase,PhysRevE.97.033309,Yangzeren2019}
\begin{subequations}\label{geqSg}
	\begin{equation}\label{geq}
	g_{\alpha}^{e q} =\left\{\begin{aligned}
	&{s_{0}\rho  +\left(\omega_{0}-1\right)\frac{p}{R T},} & {\alpha=0,} \\ &{s_{\alpha}\rho  +\omega_{\alpha}\frac{p}{R T}}, & {\alpha \neq 0,}
	\end{aligned}\right.
	\end{equation}
	\begin{equation}\label{Sg}
	S_{\alpha}^{g}
	=\frac{\boldsymbol{\xi}_{\alpha}-\boldsymbol{u}}{RT}\cdot\left[\left(\omega_{\alpha}+s_{\alpha}\right)\boldsymbol{F}+s_{\alpha} RT \nabla \rho\right],
	\end{equation}
\end{subequations}
where 
\begin{equation}
s_{\alpha}=\omega_{\alpha}\left[\frac{\boldsymbol{\xi}_{\alpha} \cdot \boldsymbol{u}}{R T}+\frac{\left(\boldsymbol{\xi}_{\alpha} \cdot \boldsymbol{u}\right)^{2}}{2 (R T)^{2}}-\frac{u^{2}}{2 R T}\right].
\end{equation}

For the discrete-velocity model, the D3Q19 model (the number of the discrete velocities $Q=19$) is used, with
\begin{equation}
\boldsymbol{\xi}_{\alpha}=\left\{\begin{array}{cc}{(0,0,0)c,}  & {\alpha=0,} \\ {(\pm 1,0,0)c,(0,\pm 1,0)c,(0,0,\pm 1)c,} & {\alpha=1-6,} \\ {(0,\pm 1,\pm 1)c,(\pm 1,0,\pm 1)c,(\pm 1,\pm 1,0)c,}  & {\alpha=7-18,}\end{array}\right.
\end{equation}
where $c=\sqrt{3RT}=1$ in lattice unit, $R$ is the model gas constant, $T$ is the reference temperature. The weighting coefficients are $\omega_{0}=1/3$, $\omega_{1-6}=1/18$, $\omega_{7-18}=1/36$, respectively. 

\subsection{Discrete unified gas kinetic scheme (DUGKS)}

The DUGKS approach~\cite{Guo2013,Guo2015} is a finite-volume scheme to solve the model Boltzmann equation.
For the sake of discussion, Eqs.~(\ref{Boltzmann}) are written into the following unified form~\cite{zhang2018discrete,Yangzeren2019}
\begin{equation}\label{Boltzmannvarphi}
\frac{\partial \varphi_{\alpha}}{\partial t}+\boldsymbol{\xi}_{\alpha} \cdot \nabla \varphi_{\alpha}=\Omega_{\alpha}^{\varphi}+S_{\alpha}^{\varphi},
\end{equation} 
where $\varphi$ denotes the distribution function $f$ or $g$, and $\Omega_{\alpha}^{\varphi}=-\left(\varphi_{\alpha}-\varphi_{\alpha}^{e q}\right) / \tau_{\varphi}$ is the corresponding collision term.

First, we integrate Eq.~(\ref{Boltzmannvarphi}) in space (control volume $V_j$) and time (from $t_n$ to $t_{n+1}$), obtaining
\begin{equation}\label{intvarphi}
\begin{aligned}
&\varphi_{\alpha}^{n+1}-\varphi_{\alpha}^{n}+\frac{\Delta t}{\left|V_{j}\right|} J_{\alpha}^{n+1 / 2}\\
=& \frac{\Delta t}{2}\left(\Omega_{\alpha}^{\varphi, n+1}+\Omega_{\alpha}^{\varphi, n}\right)+\frac{\Delta t}{2}\left(S_{\alpha}^{\varphi, n+1}+S_{\alpha}^{\varphi, n}\right),
\end{aligned}
\end{equation}
where
\begin{subequations}
	the volume average of the distribution function at $t_n$ is
	\begin{equation}
	\varphi_{\alpha}^{n} = \frac{1}{\left|V_{j}\right|} \int_{V_{j}} \varphi_{\alpha}\left(\boldsymbol{x}, t_{n}\right) d V,
	\end{equation}
	the flux across the cell interface at $t_{n+1 / 2}$ is
	\begin{equation}
	J_{\alpha}^{n+1 / 2}=\int_{A_{j}}(\boldsymbol{\xi}_{\alpha} \cdot \boldsymbol{n}) \varphi_{\alpha}\left(\boldsymbol{x}, t_{n+1 / 2}\right) d A,\label{flux}
	\end{equation}
	the volume average of the collision term and the source term at $t_n$ are, respectively,
	\begin{equation}
	\Omega_{\alpha}^{\varphi, n} = \frac{1}{|V_{j}|} \int_{V_{j}} \Omega_{\alpha}^{\varphi}\left(\boldsymbol{x}_{j}, t_{n}\right) d V,
	\end{equation}
	\begin{equation}
	S_{\alpha}^{\varphi, n} = \frac{1}{|V_{j}|} \int_{V_{j}} S_{\alpha}^{\varphi}\left(\boldsymbol{x}_{j}, t_{n}\right) d V.
	\end{equation}
\end{subequations}
$\Delta t=t_{n+1}-t_n$. $|V_{j}|$ and $A_{j}$ are the volume and surface of $V_{j}$, respectively.
It is noted that the midpoint rule is utilized for the advection term and trapezoidal rule for the collision term and the source term in Eq.~\eqref{intvarphi}.

Separating the terms at the same time step, Eq.~(\ref{intvarphi}) can be written in an explicit form as
\begin{equation}
\widetilde{\varphi}_{\alpha}^{n+1}=\widetilde{\varphi}_{\alpha}^{+, n}-\frac{\Delta t}{\left|V_{j}\right|} J_{\alpha}^{n+1 / 2}, \label{plusone}
\end{equation}
in which the auxiliary distribution functions are
\begin{subequations}
	\begin{equation}
	\begin{aligned}
	\widetilde{\varphi}_{\alpha}=&\varphi_{\alpha}-\frac{\Delta t}{2}\left(\Omega_{\alpha}^{\varphi}+S_{\alpha}^{\varphi}\right) \\
	=&\frac{2 \tau_{\varphi}+\Delta t}{2 \tau_{\varphi}} \varphi_{\alpha}-\frac{\Delta t}{2 \tau_{\varphi}} \varphi_{\alpha}^{e q}-\frac{\Delta t}{2} S_{\alpha}^{\varphi} ,
	\end{aligned}
	\end{equation}
	\begin{equation}
	\begin{aligned}
	\widetilde{\varphi}_{\alpha}^{+} =&\varphi_{\alpha}+\frac{\Delta t}{2}\left(\Omega_{\alpha}^{\varphi}+S_{\alpha}^{\varphi}\right) \\
	=&\frac{2 \tau_{\varphi}-\Delta t}{2 \tau_{\varphi}+\Delta t} \widetilde{\varphi}_{\alpha}+\frac{2 \Delta t}{2 \tau_{\varphi}+\Delta t} \varphi_{\alpha}^{e q}+\frac{2 \tau_{\varphi} \Delta t}{2 \tau_{\varphi}+\Delta t} S_{\alpha}^{\varphi}.
	\end{aligned}
	\end{equation}
\end{subequations}

To update $\widetilde{\varphi}_{\alpha}^{n+1}$, the key is to evaluate the flux across the cell interface at half time step $t_{n+1/2}$. 
Integrating Eq.~(\ref{Boltzmannvarphi}) from $t_n$ to $t_{n+1/2}$ along the characteristic line, yields
\begin{equation}
\begin{aligned}
&\varphi_{\alpha}\left(\boldsymbol{x}_{j+1 / 2}, t_{n}+h\right)-\varphi_{\alpha}\left(\boldsymbol{x}_{j+1 / 2}-\boldsymbol{\xi}_{\alpha} h, t_{n}\right)
\\=&\frac{h}{2}\left[\Omega_{\alpha}^{\varphi}\left(\boldsymbol{x}_{j+1 / 2}, t_{n}+h\right)+\Omega_{\alpha}^{\varphi}\left(\boldsymbol{x}_{j+1 / 2}-\boldsymbol{\xi}_{\alpha} h, t_{n}\right)\right]
\\&+\frac{h}{2}\left[S_{\alpha}^{\varphi}\left(\boldsymbol{x}_{j+1 / 2}, t_{n}+h\right)+S_{\alpha}^{\varphi}\left(\boldsymbol{x}_{j+1 / 2}-\boldsymbol{\xi}_{\alpha} h, t_{n}\right)\right] 
,
\end{aligned}\label{characteristicline}
\end{equation}
where $h=\Delta t / 2$ denotes the half time step size, the interface location $\boldsymbol{x}_{j+1 / 2}=\left(\boldsymbol{x}_{j}+\boldsymbol{x}_{j+1}\right) / 2$ for the uniform grid.

Similarly, Eq.~(\ref{characteristicline}) can be reduced in an explicit form as
\begin{equation}\label{varphibar}
\overline{\varphi}_{\alpha}\left(\boldsymbol{x}_{j+1 / 2}, t_{n}+h\right)=\overline{\varphi}_{\alpha}^{+}\left(\boldsymbol{x}_{j+1 / 2}-\boldsymbol{\xi}_{\alpha} h, t_{n}\right),
\end{equation}
where another two auxiliary distribution functions are introduced as
\begin{subequations}
	\begin{equation}
	\overline{\varphi}_{\alpha}=\frac{2 \tau_{\varphi}+h}{2 \tau_{\varphi}} \varphi_{\alpha}-\frac{h}{2 \tau_{\varphi}} \varphi_{\alpha}^{e q}-\frac{h}{2} S_{\alpha}^{\varphi}, \label{transform3}
	\end{equation}
	\begin{equation}
	\overline{\varphi}_{\alpha}^{+}=\frac{2 \tau_{\varphi}-h}{2 \tau_{\varphi}+h} \overline{\varphi}_{\alpha}+\frac{2 h}{2 \tau_{\varphi}+h} \varphi_{\alpha}^{e q}+\frac{2 \tau_{\varphi} h}{2 \tau_{\varphi}+h} S_{\alpha}^{\varphi}.
	\end{equation}
\end{subequations}

The RHS of Eq.~(\ref{varphibar}) is evaluated using the first-order Taylor expansion,
\begin{equation}
\overline{\varphi}_{\alpha}^{+}\left(\boldsymbol{x}_{j+1 / 2}-\boldsymbol{\xi} h, t_{n}\right) \approx \overline{\varphi}_{\alpha}^{+}\left(\boldsymbol{x}_{j+1 / 2}, t_{n}\right)-\boldsymbol{\xi}_{\alpha} h \cdot \boldsymbol{\sigma}_{j+1 / 2},
\end{equation}                                         
where $\boldsymbol{\sigma}_{j+1 / 2} = \nabla \overline{\varphi}_{\alpha}^{+}\left(\boldsymbol{x}_{j+1 / 2}, t_{n}\right)$.
To enhance the numerical stability of the PF(AC)-DUGKS approach, 
the distribution function $\overline{\varphi}_{\alpha}^{+}\left(\boldsymbol{x}_{j+1 / 2}, t_{n}\right)$ is reconstructed using the WENO scheme, which is discussed in Section~\ref{subsec: WENO}.

The original distribution function at the half time step $\varphi_{\alpha}\left(\boldsymbol{x}_{j+1 / 2}, t_{n}+h\right)$ can be obtained by Eq.~(\ref{transform3}) as
\begin{equation}
\varphi_{\alpha}=\frac{2 \tau_{\varphi}}{2 \tau_{\varphi}+h} \overline{\varphi}_{\alpha}+\frac{h}{2 \tau_{\varphi}+h} \varphi_{\alpha}^{e q}+\frac{\tau_{\varphi} h}{2 \tau_{\varphi}+h}S_{\alpha}^{\varphi}.
\end{equation}
It is used to evaluate $\boldsymbol{J}_{\alpha}^{n+1 / 2}$ in Eq.~(\ref{flux}).

Furthermore, we can show the following relationship of the auxiliary distribution functions,
\begin{subequations}
\begin{equation}
\widetilde{\varphi}_{\alpha}^{+}=\frac{4}{3} \overline{\varphi}_{\alpha}^{+}-\frac{1}{3} \widetilde{\varphi}_{\alpha},
\end{equation}
\begin{equation}
\overline{\varphi}_{\alpha}^{+}=\frac{2 \tau_{\varphi}-h}{2 \tau_{\varphi}+\delta t} \widetilde{\varphi}_{\alpha}+\frac{3 h}{2 \tau_{\varphi}+\delta t} \varphi_{\alpha}^{e q}+\frac{3 \tau_{\varphi} h}{2 \tau_{\varphi}+\delta t} S_{\alpha}^{\varphi}.
\end{equation}
\end{subequations}

In the end, from Eq.~(\ref{plusone}) we can obtain $\widetilde{\varphi}_{\alpha}$, which is the distribution function tracked in the DUGKS approach.

For the ACNS system, the macroscopic variables are evaluated as~\cite{PhysRevE.97.033309,Yangzeren2019}
\begin{subequations}
	\begin{equation}
	\phi\left(\boldsymbol{x}_{j}, t_{n}+\Delta t\right)=\sum_{\alpha=0}^{Q-1} \widetilde{f}_{\alpha},
	\end{equation}
	\begin{equation}
	\boldsymbol{u}\left(\boldsymbol{x}_{j}, t_{n}+\Delta t\right)=\frac{1}{\rho}\left(\sum_{\alpha=0}^{Q-1} \boldsymbol{\xi}_{\alpha} \widetilde{g}_{\alpha} +\frac{\Delta t}{2} \boldsymbol{F}\right)
	,
	\end{equation}
	\begin{equation}
	\begin{aligned}
	p\left(\boldsymbol{x}_{j}, t_{n}+\Delta t\right)=
	\frac{RT}{1-\omega_{0}}
	\left( \sum_{\alpha=1}^{Q-1} \widetilde{g}_{\alpha}+\frac{\Delta t}{2} \boldsymbol{u} \cdot \nabla \rho+\rho s_0 \right) . \label{pressuremoment}
	\end{aligned}
	\end{equation}
\end{subequations}

The time step size is related to the Courant-Friedrichs-Lewy (CFL) condition~\cite{XU20107747,Guo2013,zhang2018discrete,2020Simulation}
\begin{equation}
\Delta t= CFL  \frac{\Delta x_{\min }} {\sqrt{3RT} },
\end{equation}
where $\Delta x_{\min}$ is the minimal grid spacing, $CFL$ is the CFL number.

It should be noted that the two source terms and the computation of macroscopic variables involve first-order and second-order spatial derivatives, they are evaluated by second-order spatial central finite difference schemes.
The first-order time derivative in the computation the macroscopic variables is approximated by a backward Euler scheme.

\subsection{The WENO  treatment for the distribution functions at the cell interface}\label{subsec: WENO}
The key idea in the WENO scheme is that
a combination of all neighbor stencils is used to approximate the flux at the cell boundary, and the weights of the stencils are chosen appropriately.
The smoother stencil has a larger weight.
Hence it is expected to reduce the numerical oscillation and improve the stability of the multiphase flow simulations.~\cite{jiang1996efficient,WANG2015404,WANG201541}
Chen~{\it et al.}~\cite{2020Simulation} has coupled WENO with DUGKS scheme to improve the spatial accuracy and the numerical stability, which can capture shocklet in single-phase compressible decaying isotropic turbulence. Their results showed that the incorporation of WENO allows DUGKS to obtain better numerical results.

Here the 3rd-order classical WENO scheme~\cite{jiang1996efficient,2020Simulation} is employed to improve
 the numerical stability of the PF(AC)-DUGKS two-phase scheme.
We take a cell interface located at $x_{j+1/2}$ and normal to $x$-direction at time $t_n$ as an example. 
If the $x$ component of the discrete particle velocity is positive (${\xi}_{\alpha x}>0$) at the position $x_{j+1/2}$, then the approximate cell interface results based on different stencils are
\begin{subequations}
	\begin{equation}
	\bar{\varphi}_{\alpha}^{+(0)}\left(x_{j+1 / 2}, t_{n}\right)=-\frac{1}{2} \bar{\varphi}_{\alpha}^{+}\left(x_{j-1}, t_{n}\right)+\frac{3}{2} \bar{\varphi}_{\alpha}^{+}\left(x_{j}, t_{n}\right),
	\end{equation}
	\begin{equation}
	\bar{\varphi}_{\alpha}^{+(1)}\left(x_{j+1 / 2}, t_{n}\right)=\frac{1}{2} \bar{\varphi}_{\alpha}^{+}\left(x_{j}, t_{n}\right)+\frac{1}{2} \bar{\varphi}_{\alpha}^{+}\left(x_{j+1}, t_{n}\right).
	\end{equation}
\end{subequations}

The indicators of the smoothness are first defined by Liu~{\it et al.},~\cite{LIU1994200}
which can be expressed as
\begin{subequations}
	\begin{equation} 
	\beta_{0}= \left[\bar{\varphi}_{\alpha}^{+}\left(x_{j-1}, t_{n}\right)-\bar{\varphi}_{\alpha}^{+ }\left(x_{j}, t_{n}\right)\right]^{2}  ,
	\end{equation}
	\begin{equation}
	\beta_{1} =\left[\bar{\varphi}_{\alpha}^{+}\left(x_{j}, t_{n}\right)-\bar{\varphi}_{\alpha}^{+ }\left(x_{j+1}, t_{n}\right)\right]^{2} .
	\end{equation} 
\end{subequations}
It is noted that $\beta_{0},\beta_{1}$ would be larger/smaller when $\bar{\varphi}_{\alpha}^{+}$ is discontinuous/continuous on the stencils considered.
Hence they represent the smoothness of $\bar{\varphi}_{\alpha}^{+}$.~\cite{LIU1994200}
Then the nonlinear weights are
\begin{equation}
\omega_{0}=\frac{\tilde{\omega}_{0}}{\tilde{\omega}_{0}+ \tilde{\omega}_{1}}, 
\quad
\omega_{1}=\frac{\tilde{\omega}_{1}}{\tilde{\omega}_{0}+ \tilde{\omega}_{1}},
\end{equation}
with
\begin{equation}
\tilde{\omega}_{0}=\frac{\gamma_{0}}{\left(\varepsilon+\beta_{0}\right)^{2}}, \quad
\tilde{\omega}_{1}=\frac{\gamma_{1}}{\left(\varepsilon+\beta_{1}\right)^{2}},
\end{equation}
where $\gamma_0=1/3$, $\gamma_1=2/3$ are the linear weights. A small constant $\varepsilon=10^{-6}$ is needed in order to make the denominator non-zero.~\cite{LIU1994200,jiang1996efficient,2020Simulation}

In the end, the distribution function at the cell interface $x_{j+1/2}$ can be reconstruction as the following convex combination 
\begin{equation}
\begin{aligned} \bar{\varphi}_{\alpha}^{+}\left(x_{j+1 / 2}, t_{n}\right)= \omega_{0} \bar{\varphi}_{\alpha}^{+(0)}\left(x_{j+1 / 2}, t_{n}\right)+\omega_{1} \bar{\varphi}_{\alpha}^{+(1)}\left(x_{j+1 / 2}, t_{n}\right) . \end{aligned}
\end{equation}

For ${\xi}_{\alpha x}<0$, the results are similar based on the upwind scheme. For ${\xi}_{\alpha x}=0$,  an average of the
 results from ${\xi}_{\alpha x}>0$ and ${\xi}_{\alpha x}<0$ is taken.
For the cell interfaces in the $y$ direction
and $z$ direction, the procedures are the same as those in the $x$ direction.

\section{Two dimensional simulations and discussions} \label{sec: 2D}
First, we utilize the 3D code to model 2D problems in this section, to demonstrate that this code can simulate dimensionality reduction problems very well. The computational domain is $[0,N_x]\times[0,N_y]\times[0,2]$ in the 2D simulations.
Since the variables of the two grid nodes in $z$ direction are the same, we only mention $[0,N_x]\times[0,N_y]$ or $[-N_x/2,N_x/2]\times[-N_y/2,N_y/2]$ for simplicity.

\subsection{A stationary droplet and the Young-Laplace law}

A stationary liquid droplet is first simulated to validate our PF(AC)-DUGKS-WENO code.
All the parameters are in lattice units in this paper. Initially, a circular droplet is placed at the center of the computational domain of size $[-L,L]\times[-L,L]=128^2$. 
The periodic boundary conditions are applied in
all spatial directions.
To model the two phases similar as water and air,
the density ratio is $\rho^*=\rho_{A}/\rho_{B}=1000$
and kinematic viscosity ratio is $\nu^*=\nu_{A}/\nu_{B}=0.06$,
these same ratios were used by Lycett-Brown and Luo.~\cite{lycett2016cascaded} 
$M_{AC}=1.0\times 10^{-6}$ is chosen and $W=5.0$.
There is no gravity in this simulation.
The CFL number is $CFL=0.25$. 
The order parameter is initialized as~\cite{zhang2018discrete,zu2013phase,chen2019simulation,ZHANG20191128} 
\begin{equation}
\phi\left(r_2\right)=\frac{\phi_{A}+\phi_{B}}{2}+\frac{\phi_{A}-\phi_{B}}{2} \tanh \left(2 \frac{R_0-r_2}{W}\right),\label{phini}
\end{equation}
where $r_2=\sqrt{\left(x-x_{c}\right)^{2}+\left(y-y_{c}\right)^{2}}$ is the distance between any point $(x,y)$ in the computational domain and the droplet center $(x_{c},y_{c})$.
The initial velocity is zero in the whole domain. 
The pressure is initialized as $p=\phi\sigma/R_0$, in order for the simulation to reach the steady state faster. 
In this case, the dimensionless time is defined as $t^*=\frac{t}{R_0}\sqrt{\frac{\sigma}{\rho R_0}}$.
For testing the Young-Laplace law and comparing with the previous studies,~\cite{zhang2018discrete,Yangzeren2019} 18 cases are simulated. The radii of the droplets are selected as $R_0=20,24,28,32,36,40$, and the surface tensions are $\sigma=0.001,0.005,0.01$.

\begin{figure}[]
	\centering
	\includegraphics[width=0.5\columnwidth,trim={0cm 0cm 0cm 0cm},clip]{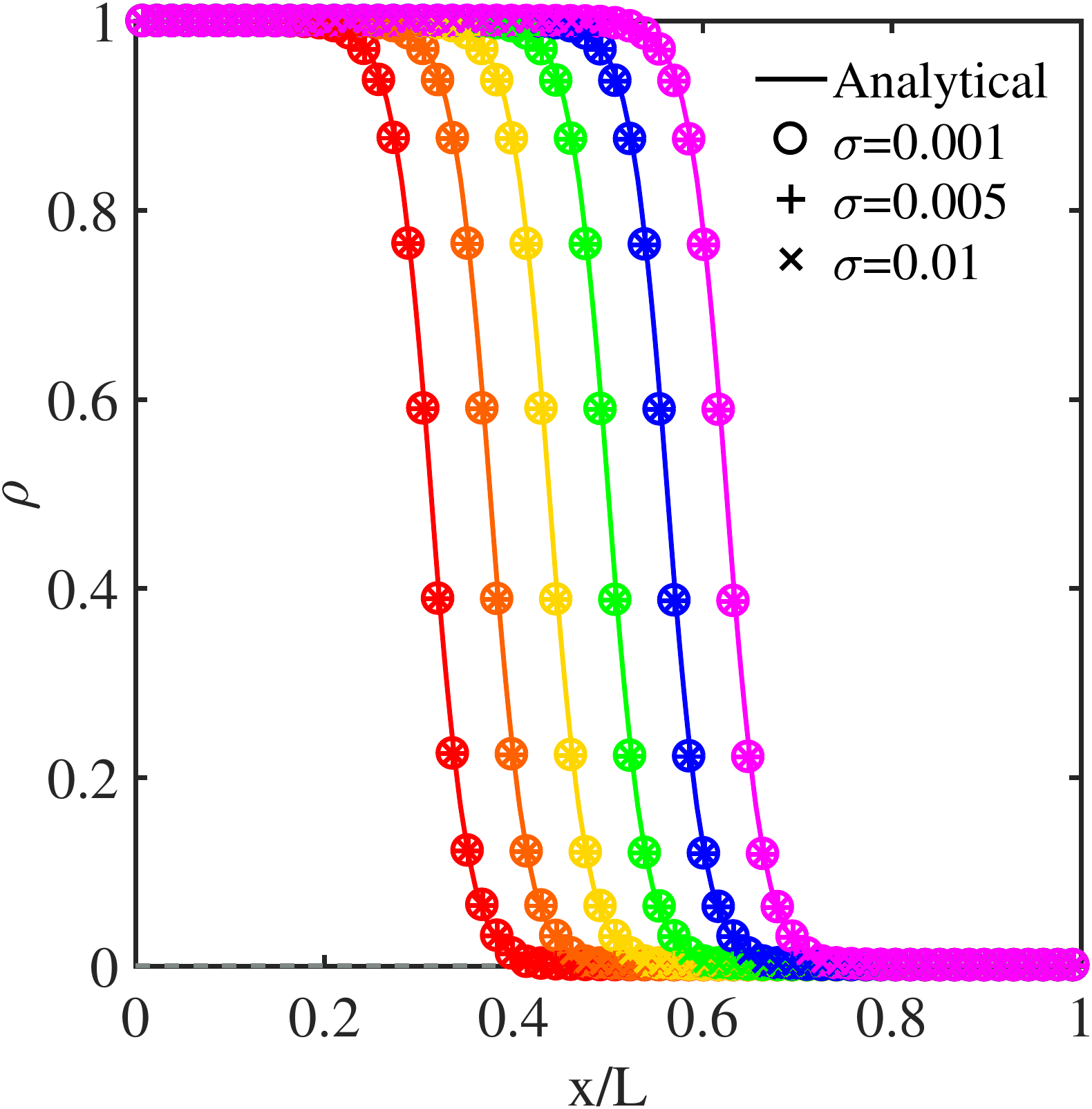}
	\centering
	\caption{The density profile of each stationary droplet at $\rho^*=1000$ and $\nu^*=0.06$. $R_0=20,24,28,32,36,40$ from left to right. The gray dashed line represents the constant $\rho_{B}=0.001$. The "Analytical" results are based on Eq.~(\ref{phini}) and Eq.~(\ref{Eqrho}) with different radii.}
	\label{fig:RhoProfileof2DSD}
\end{figure}

\begin{figure}[]
	\centering
	\includegraphics[width=0.5\columnwidth,trim={0cm 0cm 0cm 0cm},clip]{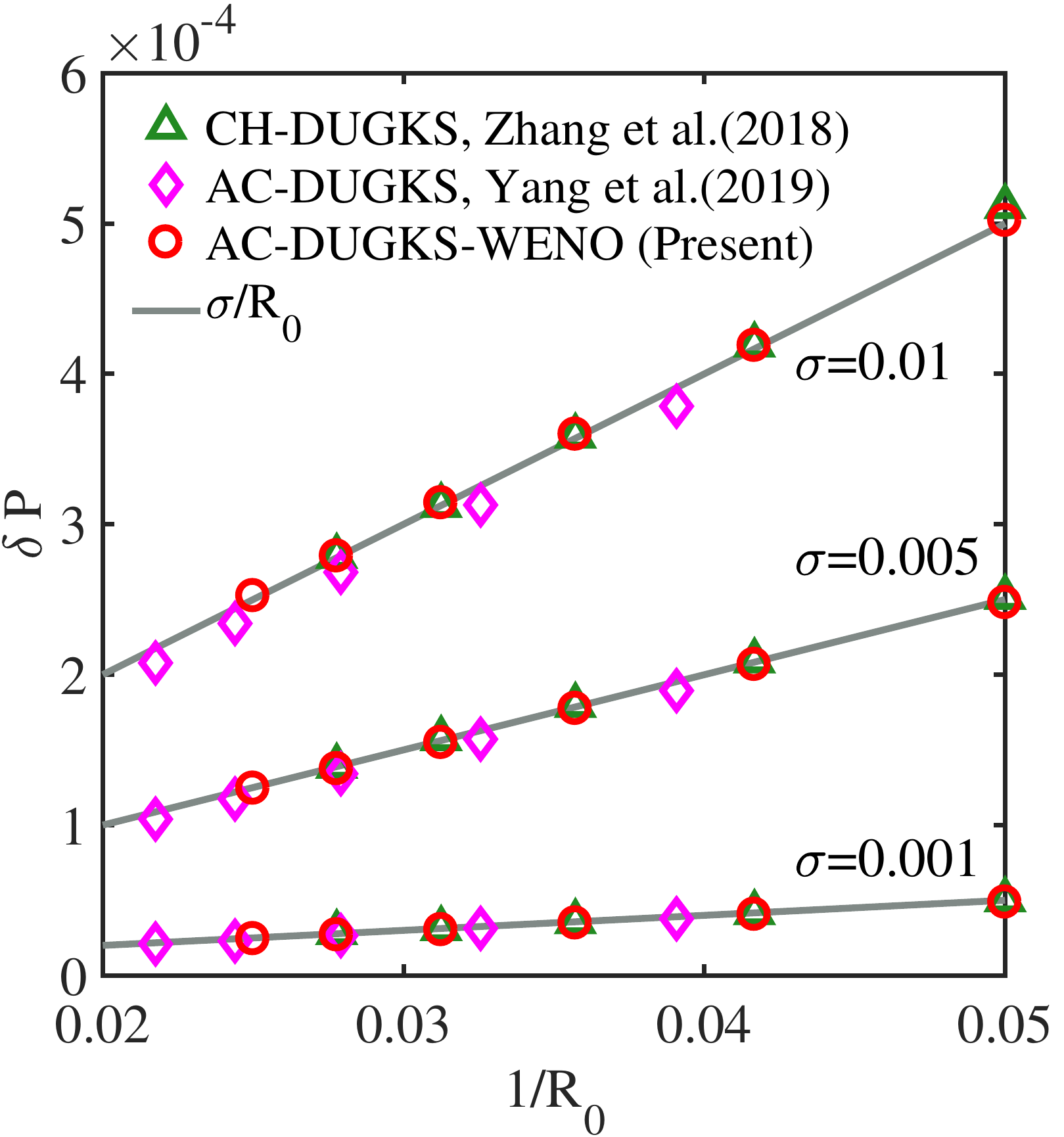}
	\centering
	\caption{Examination of the Laplace law for a stationary droplet.  
	$(\rho^*,\nu^*)=(5,1),(1000,1),(1000,0.06)$ in Zhang {\it et al.}~\cite{zhang2018discrete}'s paper, Yang {\it et al.}~\cite{Yangzeren2019}'s paper, and our present simulation, respectively.
		$\delta P=\sigma/R_0$ is the analytical result of the 2D Laplace law, where $P=p+\phi d\psi/d\phi-\psi-\kappa \phi \nabla^{2} \phi-\kappa|\nabla \phi|^{2} / 2-\phi\mu_{\phi}$ in our simulation.}
	\label{fig:LaplaceLawof2DSD}
\end{figure}

The convergence criterion for the static flows in Yang {\it et al.}~\cite{Yangzeren2019} is
\begin{equation}
\delta\Phi=\frac{\sum_{i,j}\left|\phi(i,j,n)- \phi(i,j,n-1000)\right| ^2}{\sum_{i,j}\left|\phi(i,j,n)\right| ^2} < 1.0\times 10^{-8}.
\end{equation}
In our simulation, we iterate $n=4\times 10^{5}$ time steps for each case, and $\delta\Phi< 1.0\times 10^{-9}$ for all the cases, which are more than one order of magnitude smaller than those in the literature.~\cite{Yangzeren2019} Therefore, the droplet is stable at this time. We plot the density profiles (Fig.~\ref{fig:RhoProfileof2DSD}) and the Young-Laplace law (Fig.~\ref{fig:LaplaceLawof2DSD}) at this stable time.
Our numerical results show good agreement with the analytical results.
Fig.~\ref{fig:LaplaceLawof2DSD} also shows that our results are better than Yang {\it et al.}~\cite{Yangzeren2019}'s large-density-ratio results without the WENO scheme. 
Yang {\it et al.}~\cite{Yangzeren2019} mentioned that the ratio of numerical surface tensions to analytical ones is about 96.5\% in their results, while our results are around 99\%.
We note that $\nu^*=1$ in Yang {\it et al.}~\cite{Yangzeren2019}'s paper, not the same as the ratio between water and air.
Zhang {\it et al.}~\cite{zhang2018discrete}'s Laplace law results utilizing Cahn-Hilliard model are also better then Yang {\it et al.}~\cite{Yangzeren2019}'s comparing to the analytical result, but with a small density ratio $\rho^*=5$.

\subsection{2D Rayleigh-Taylor instability}\label{subsec: 2DRTI}
The Rayleigh–Taylor instability (RTI) occurs when a heavy fluid $A$ is on top of a light fluid $B$, and is driven by the gravity. The heavy fluid falls into the light fluid and creates the instability between the fluid-fluid interface. The RTI is a common phenomenon, and it is also a classical benchmark problem widely simulated in the previous studies.~\cite{zu2013phase,2017Improved,zhang2018discrete,Yangzeren2019,chen2019simulation,2019Phase,2021Direct} We use RTI to test the ability to capture complex interface evolution at high Reynolds number and large density ratio for our DUGKS-PF(AC)-WENO approach.

The computational domain of 2D RTI in our simulation is $[0,4L]\times[0,L]$ with $L=256$. The reference velocity is $U_0=\sqrt{gL}$. The order parameter, density, dynamic viscosity of the heavy fluid and light fluid are $\phi_{A}, \rho_{A}, \mu_{A}$ and $\phi_{B}, \rho_{B}, \mu_{B}$, respectively. The main dimensionless numbers of this physical problem are the Atwood number $At=\left(\rho_A-\rho_B \right)/\left(\rho_A+\rho_B \right)$ and the Reynolds number $Re=\rho_A U_0 L/\mu_A$. Other dimensionless parameters are the capillary number $Ca=\mu_A U_0/\sigma$, Peclet number $Pe=U_0 L/M_{AC}$, and the viscosity ratio. The reference time is defined as $T=\sqrt{\frac{L}{g\cdot At}}$, then the normalized time is $t^*=t/T$.
The periodic boundary condition is applied to the lateral boundaries and the no-slip bounce-back
boundary condition is applied to the top and bottom walls.
Initially, the two-phase interface is located at 
\begin{equation}
	x(y)=2L+0.1L\cos\left(\frac{2\pi y}{L} \right).
\end{equation}

\begin{figure}[]
	\centering
	\includegraphics[width=1\columnwidth,trim={1.5cm 2cm 1cm 1.5cm},clip]{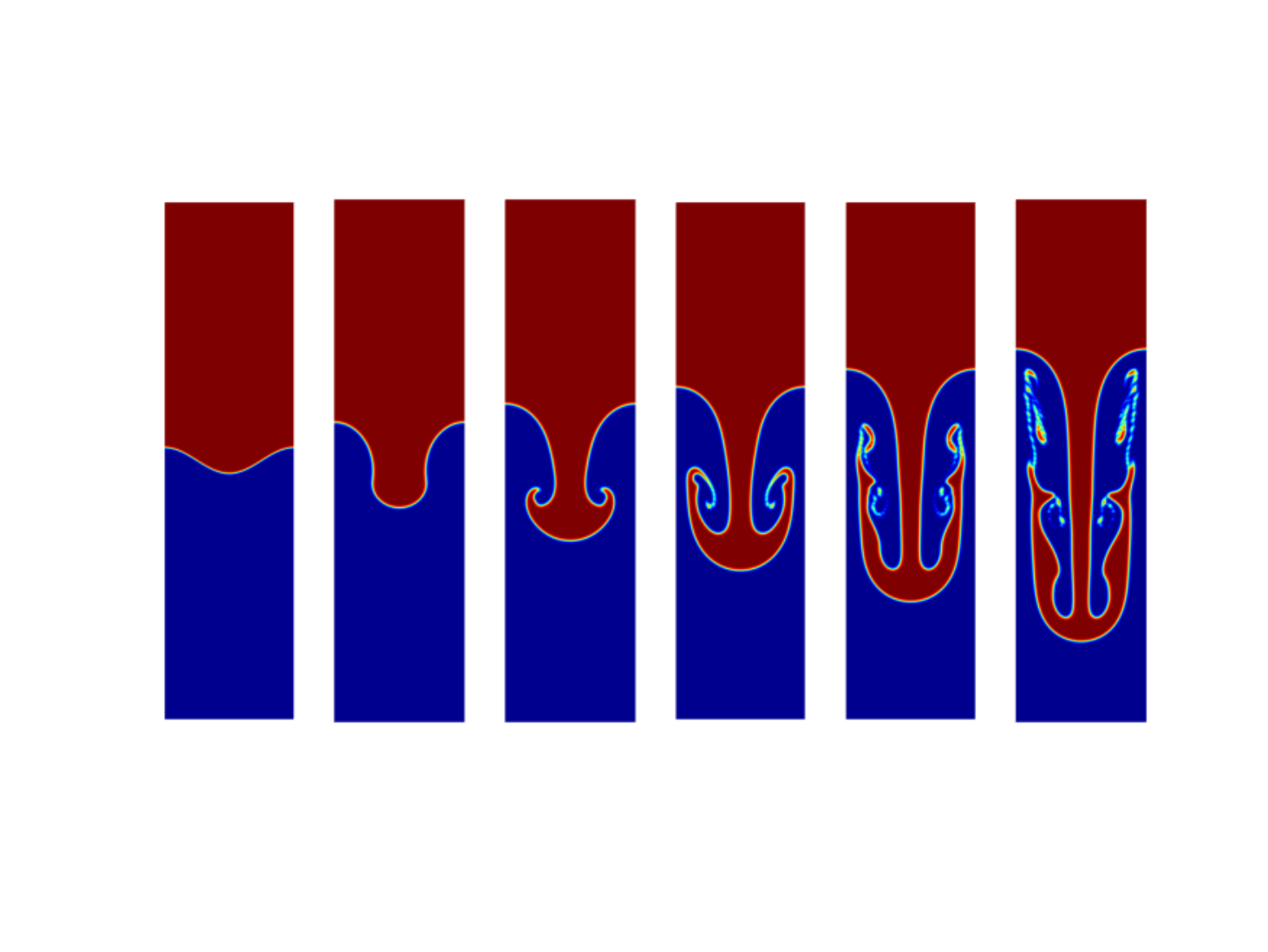}
	\centering
	\caption{Time evolution of $\phi$ for 2D RTI at $At=0.5$ and $Re=3000$. $t^*=0,1,1.5,2,2.5,3$.}
	\label{fig:phi2DRT05}
\end{figure}

\begin{figure}[]
	\centering
	\includegraphics[width=0.5\columnwidth,trim={0cm 0cm 0cm 0cm},clip]{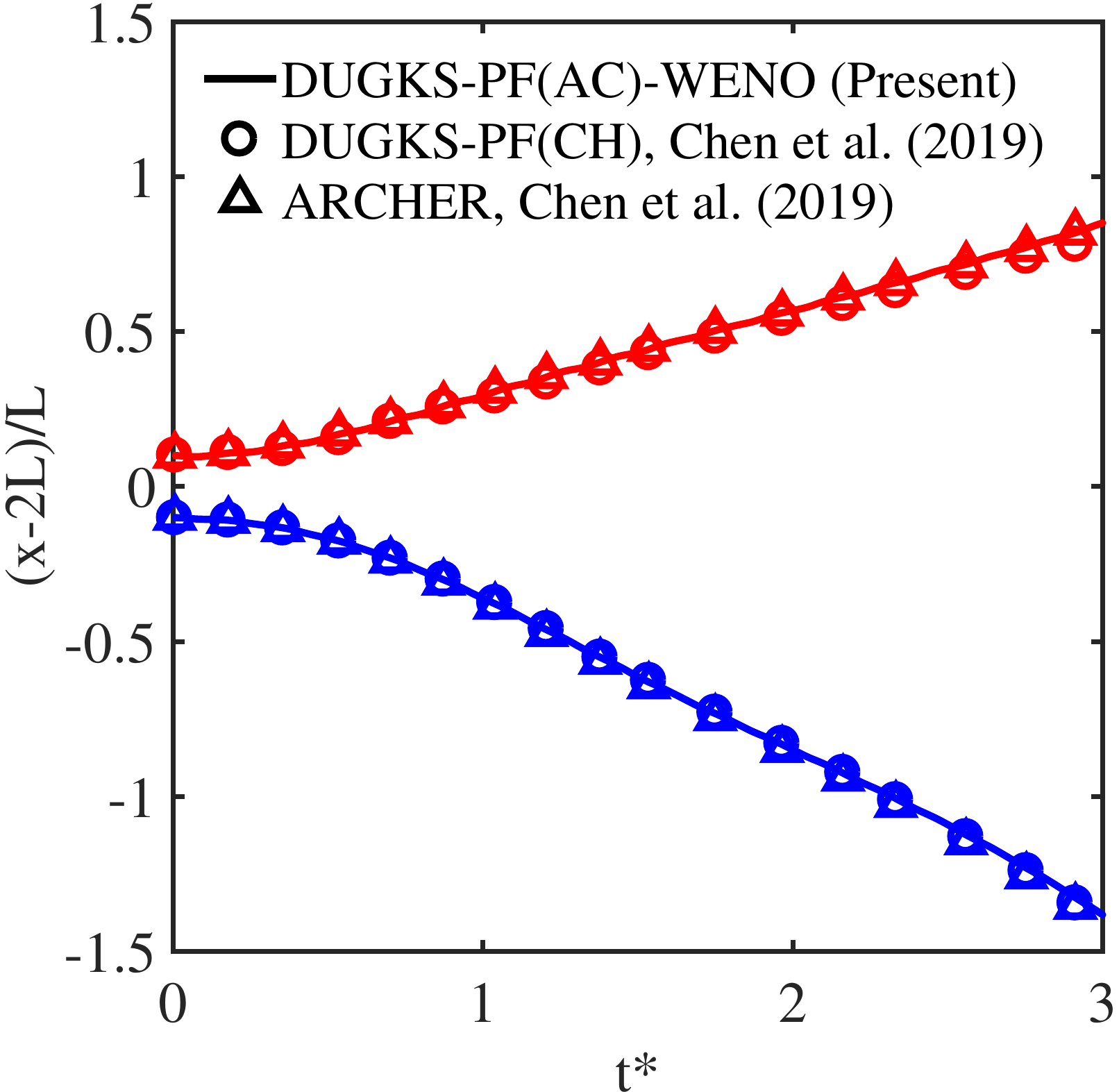}
	\centering
	\caption{Time evolution of the positions of bubble front (red) and spike tip (blue) for 2D RTI at $At=0.5$ and $Re=3000$.}
	\label{fig:Timeof2DRT05}
\end{figure}

First, we simulate a common high Reynolds number case, $At=0.5$ and $Re=3000$, which has been widely tested in the literatures.~\cite{zu2013phase,2017Improved,zhang2018discrete,Yangzeren2019,chen2019simulation,2019Phase} The other parameters in our simulation are $g$ = 2e-6,
$CFL=0.25$, $\mu_A/\mu_B = 1.0$, $W=5.0$, $Ca=0.44$, $Pe=1000$. The contours of the order parameter vary with the normalized time are shown in Fig.~\ref{fig:phi2DRT05}. It can be seen that the heavy fluid accelerates into the light fluid with a symmetric rolling-up process due to the gravitational field, and many small structures are generated at the later times. 
The evolution contours are in good agreement with the previous studies~\cite{zu2013phase,2017Improved,zhang2018discrete,Yangzeren2019,chen2019simulation,2019Phase}
 at the early times, but are not precisely the same at the later times because the small structures are
 sensitive to the minor differences in different methods.
We compare the evolution of the positions of bubble front and spike tip with Chen {\it et al.}~(2019),~\cite{chen2019simulation} as shown in Fig.~\ref{fig:Timeof2DRT05}. The results agree well with their results with two different approaches (DUGKS-PF(CH) and ARCHER). Since Chen {\it et al.}~(2019)~\cite{chen2019simulation} have compared their results with the previous studies.~\cite{DING20072078,li2012additional,zu2013phase,ren2016improved}
Therefore, our results are also consistent with these previous results.


\begin{figure}[]
	\centering
	\includegraphics[width=0.9\columnwidth,trim={0cm 0cm 0cm 0cm},clip]{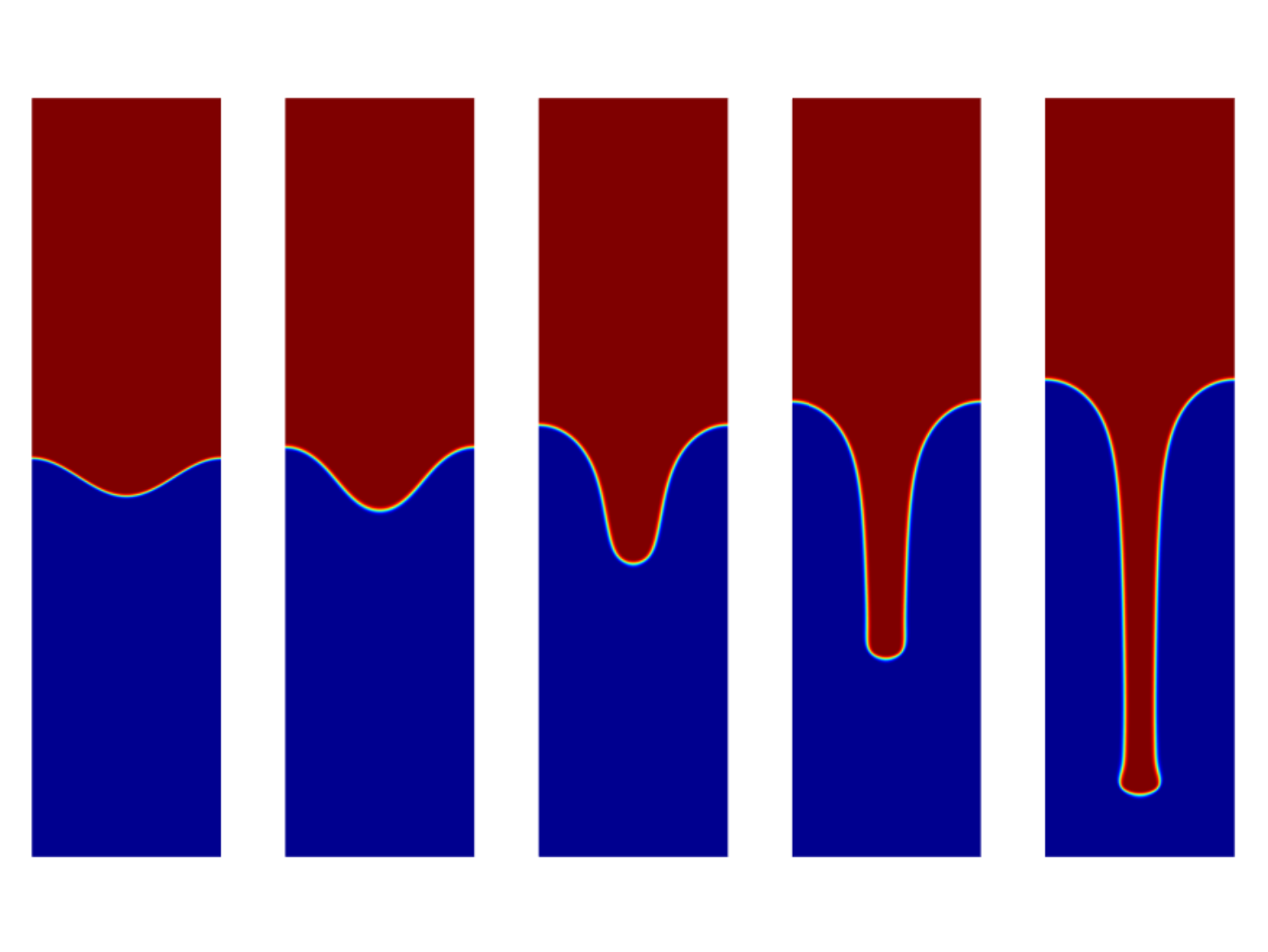}
	\centering
	\caption{Time evolution of $\phi$ for 2D RTI at $At=0.998$ and $Re=3000$. $t^*=0,0.5,1,1.5,2$.}
	\label{fig:phi2DRT0998}
\end{figure}

\begin{figure}[]
	\centering
	\includegraphics[width=0.5\columnwidth,trim={0cm 0cm 0cm 0cm},clip]{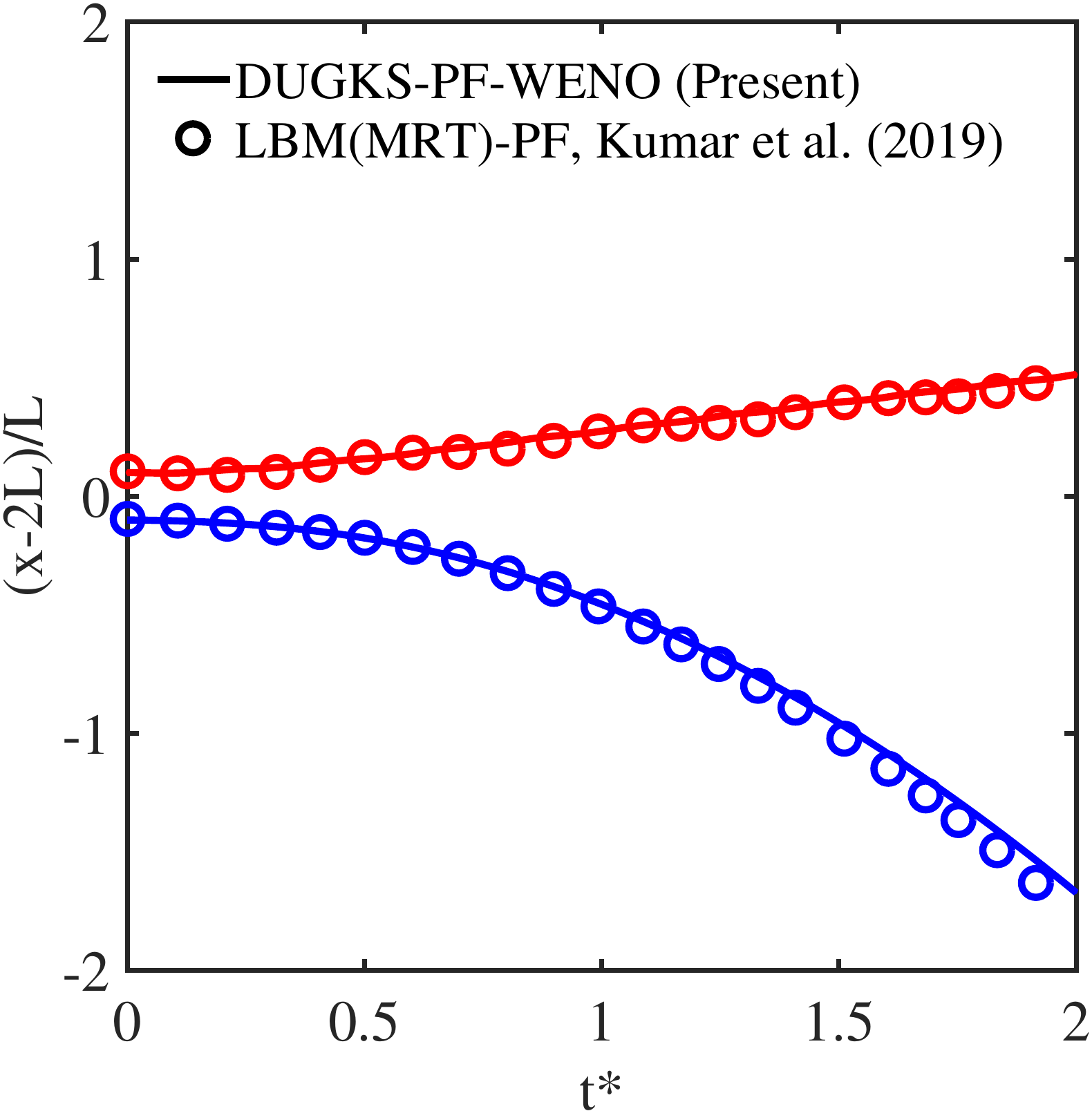}
	\centering
	\caption{Time evolution of the positions of bubble front (red) and spike tip (blue) for 2D RTI at $At=0.998$ and $Re=3000$.}
	\label{fig:Timeof2DRT0998}
\end{figure}

Next, a large density ratio of 1000 (or $At=0.998$) and high Reynolds number ($Re=3000$) case is carried out, which is a difficult 
case for the phase-field model.~\cite{2017Improved,2019Phase} In our simulation, the other parameters are $U_0=0.02$, $CFL=0.25$, $\mu_B/\mu_A = 2.0$, $W=5.0$, $Ca= 0.44$, $Pe=2000$.
Fig.~\ref{fig:phi2DRT0998} shows the contours of $\phi$ at several normalized times. 
Since the density ratio is large, the influence of light fluid is small, hence the interface is smooth during the evolution process, and there is no rolling-up process and no small droplet structures.
These qualitative features are in good agreement with the previous studies.~\cite{2017Improved,2019Phase} For the positions of bubble front and spike tip at $At=0.998$ and $Re=3000$, we can only find one related paper (Kumar {\it et al.}~\cite{2019Phase}) and compare with them in Fig.~\ref{fig:Timeof2DRT0998}. The results are also in reasonable agreement.


Yang {\it et al.}~\cite{Yangzeren2019} reported that they could not simulate a case with $At=0.98$ and $Re=600$ by the DUGKS-PF(AC) approach.
Therefore, our approach with the WENO scheme can better model the larger density ratio and higher Reynolds number than theirs. Furthermore, they need a greater interfacial thickness $W=8$ for the large density ratio RTI case, while we can apply the common value $W=5$.

\subsection{A droplet splashing on a thin liquid film}

Droplet splashing occurs when the droplets fall down onto a wet ground, such as rain and shower.
This physical problem is more challenge to simulate than RTI, because the initial droplet velocity is nonzero.
The previous studies focus on the early stage of droplet impact on the liquid film.~\cite{josserand2003droplet,lee2005stable,wang2015multiphase}
They found that there are two possible phenomena resulting from this impact process, deposition and splashing, mainly depending on the Reynolds number.
Furthermore, when droplet splashing appears, the impact radius obeys a power law at this early stage.

The computational domain is $1024\times 512$.
The periodic boundary condition is applied to the lateral boundaries and the no-slip bounce-back boundary condition is applied to the top and bottom walls.
Initially, a circular liquid droplet with radius $R_0=64$ is on top of a thin liquid film with height $h=32$. The droplet, with velocity magnitude $U_0=0.0102$, would fall down to the film.
Gravity is not considered in this simulation.
The interfacial thickness is $W=5$.
The mobility is $M_{AC}=0.001$.
The CFL number is still fixed at $CFL=0.25$.
The main dimensionless parameters are the Reynolds number $Re=2R_0U_0/\nu_A$, Weber number $We=2R_0\rho_A U_0^2/\sigma$, the density ratio $\rho^*=\rho_{A}/\rho_{B}$ and the kinematic viscosity ratio $\nu^*=\nu_{A}/\nu_{B}$.
Here $A$ represents the liquid phase (including the droplet and the thin film) and $B$ is the background phase.

\begin{figure}[]
	\centering
	\includegraphics[width=0.5\columnwidth,trim={0cm 0cm 0cm 0cm},clip]{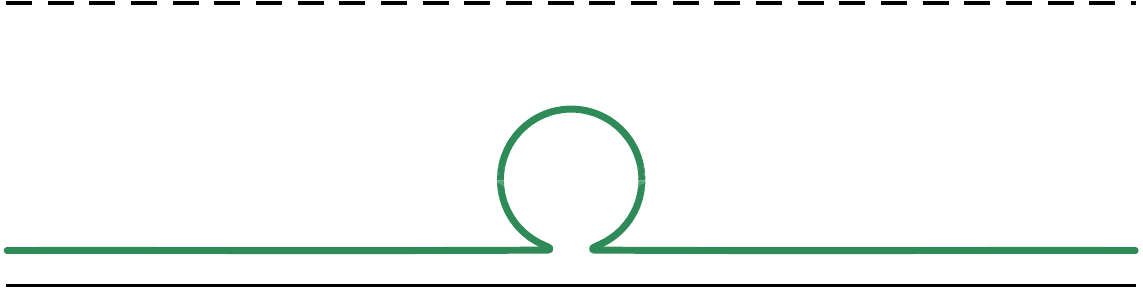}
	\includegraphics[width=0.5\columnwidth,trim={0cm 0cm 0cm 0cm},clip]{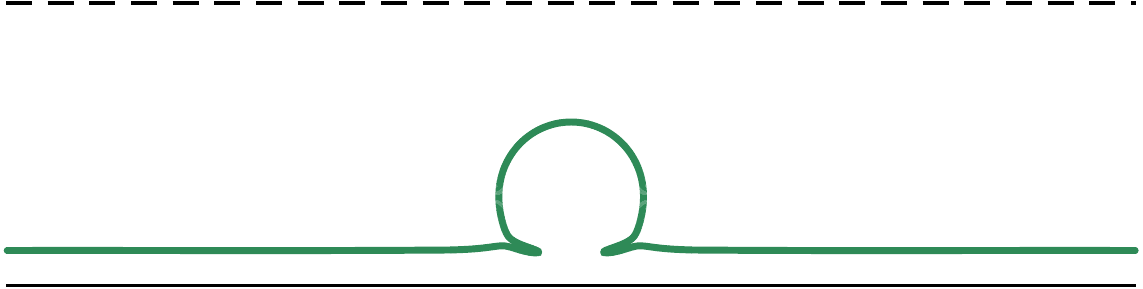}\\
	\includegraphics[width=0.5\columnwidth,trim={0cm 0cm 0cm 0cm},clip]{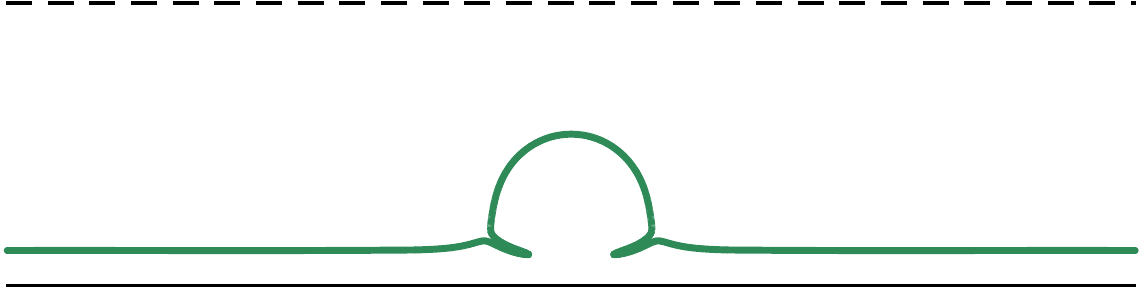}
	\includegraphics[width=0.5\columnwidth,trim={0cm 0cm 0cm 0cm},clip]{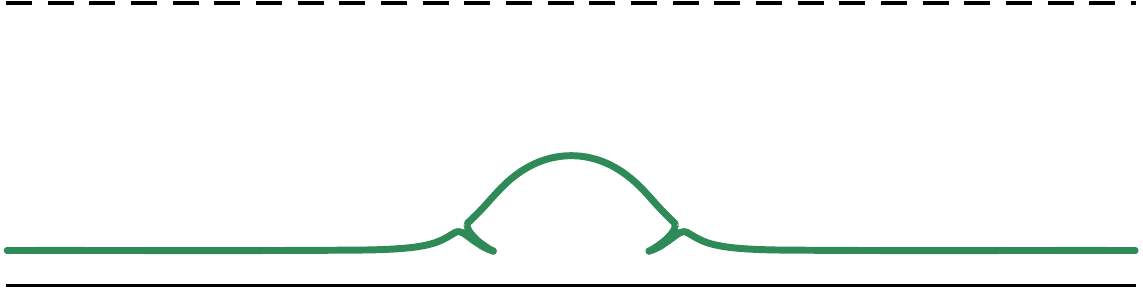}\\
	\includegraphics[width=0.5\columnwidth,trim={0cm 0cm 0cm 0cm},clip]{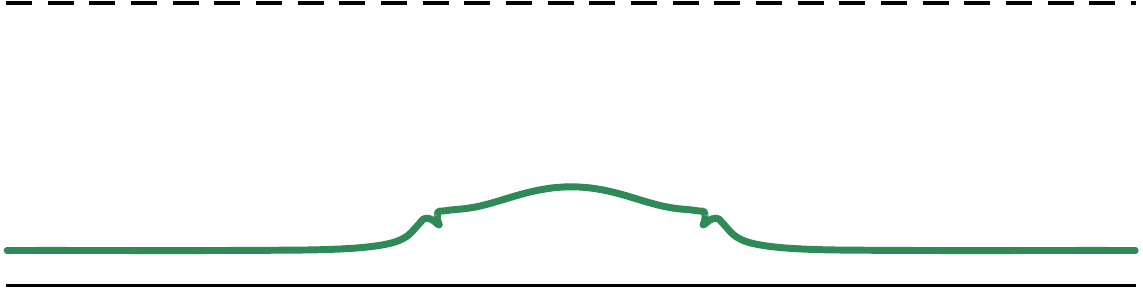}
	\includegraphics[width=0.5\columnwidth,trim={0cm 0cm 0cm 0cm},clip]{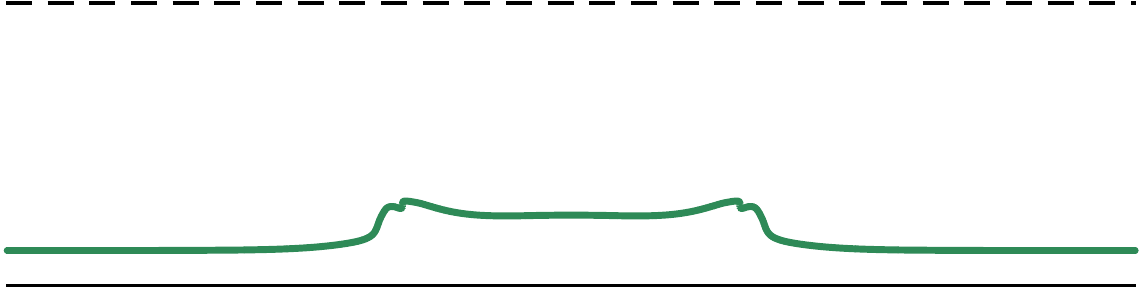}
	\centering
	\caption{Evolution of the fluid-fluid interface for a droplet splashing on a thin liquid film at $\rho^*=1000$, $Re=20$, and $We=8000$. $U_0 t/D_0=0, 0.1, 0.2, 0.4, 0.8, 1.6$. The dashed line marks the centerline of the computational domain.}
	\label{fig:DSRe20new}
\end{figure}

First, we model the deposition phenomenon. A small Reynolds number, $Re=20$, is selected. Other dimensionless parameters are $We=8000$, $\rho^*=1000$, $\nu^*=0.06$. The impact process is shown in Fig.~\ref{fig:DSRe20new}. The droplet spreads gently on the film, and an outward moving surface wave is observed in this case, which is typical for the deposition process.~\cite{josserand2003droplet,lee2005stable,wang2015multiphase}

\begin{figure}[]
	\centering
	\includegraphics[width=0.5\columnwidth,trim={0cm 0cm 0cm 0cm},clip]{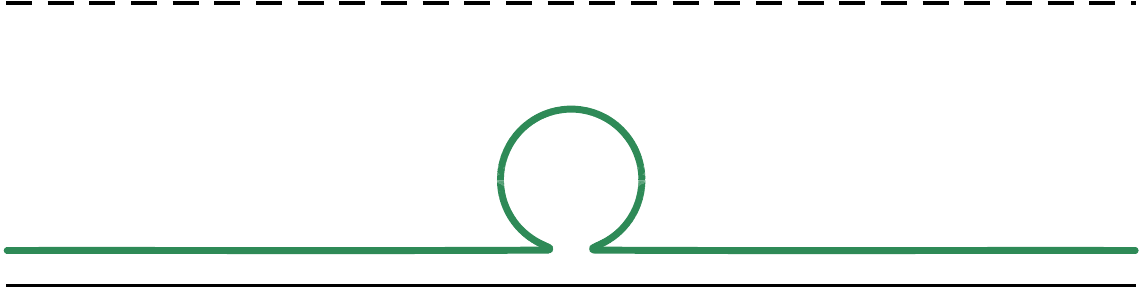}\\
	\includegraphics[width=0.5\columnwidth,trim={0cm 0cm 0cm 0cm},clip]{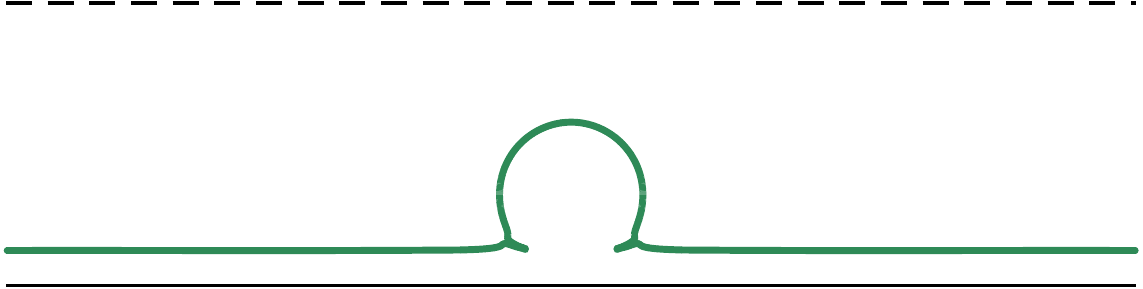}\\
	\includegraphics[width=0.5\columnwidth,trim={0cm 0cm 0cm 0cm},clip]{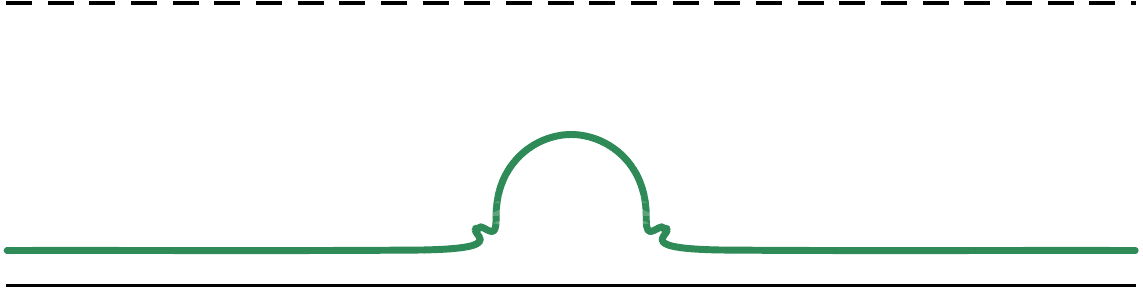}\\
	\includegraphics[width=0.5\columnwidth,trim={0cm 0cm 0cm 0cm},clip]{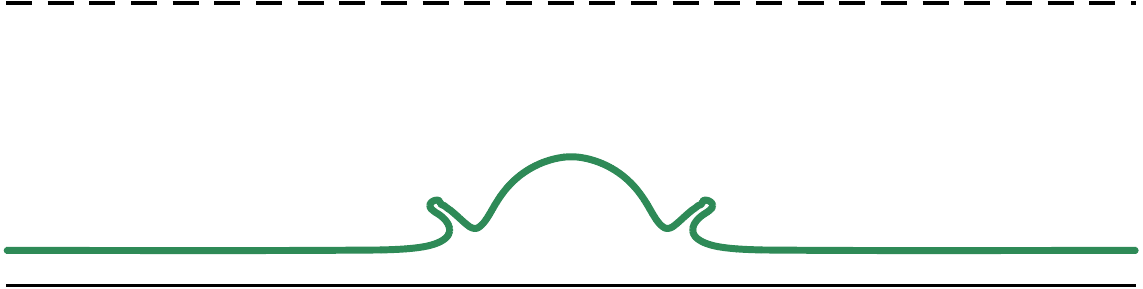}\\
	\includegraphics[width=0.5\columnwidth,trim={0cm 0cm 0cm 0cm},clip]{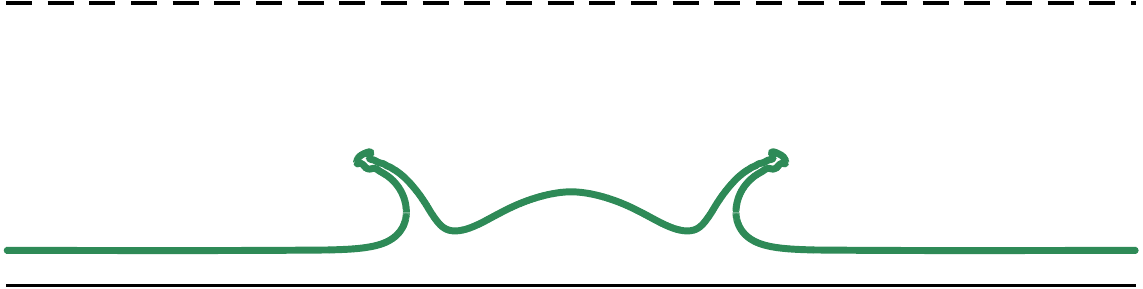}\\
	\includegraphics[width=0.5\columnwidth,trim={0cm 0cm 0cm 0cm},clip]{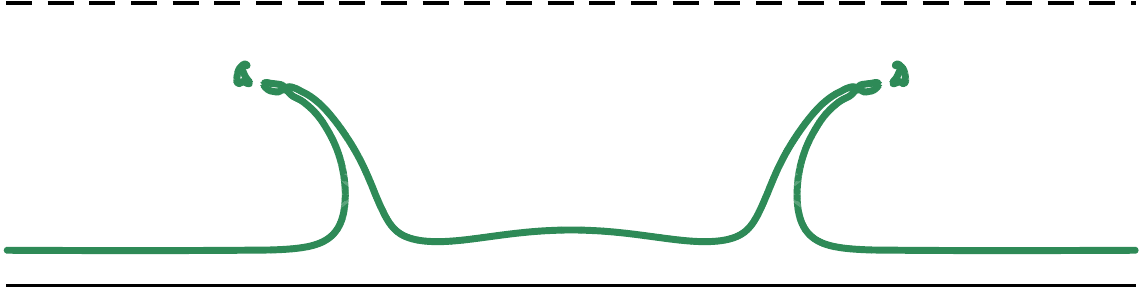}
	\centering
	\caption{Evolution of the fluid-fluid interface for a droplet splashing on a thin liquid film at $\rho^*=1000$, $Re=500$, and $We=8000$. $U_0 t/D_0=0, 0.1, 0.2, 0.4, 0.8, 1.6$. The dashed line marks the centerline of the computational domain.}
	\label{fig:DSRe500new}
\end{figure}

Then we simulate the splashing phenomenon. A large Reynolds number, $Re=500$, is assumed in this case. Other dimensionless parameters are $We=8000$, $\rho^*=1000$, $\nu^*=0.001$. The evolution of droplet splashing process is shown in Fig.~\ref{fig:DSRe500new}.
We can observe that two liquid fingers are generated after the impact of the droplet onto the thin film.
In the end, the fingers may break up into small droplets, due to the Rayleigh–Plateau instability.~\cite{josserand2003droplet,lee2005stable,wang2015multiphase}

\begin{figure}[]
	\centering
	\includegraphics[width=0.5\columnwidth,trim={0cm 0cm 0cm 0cm},clip]{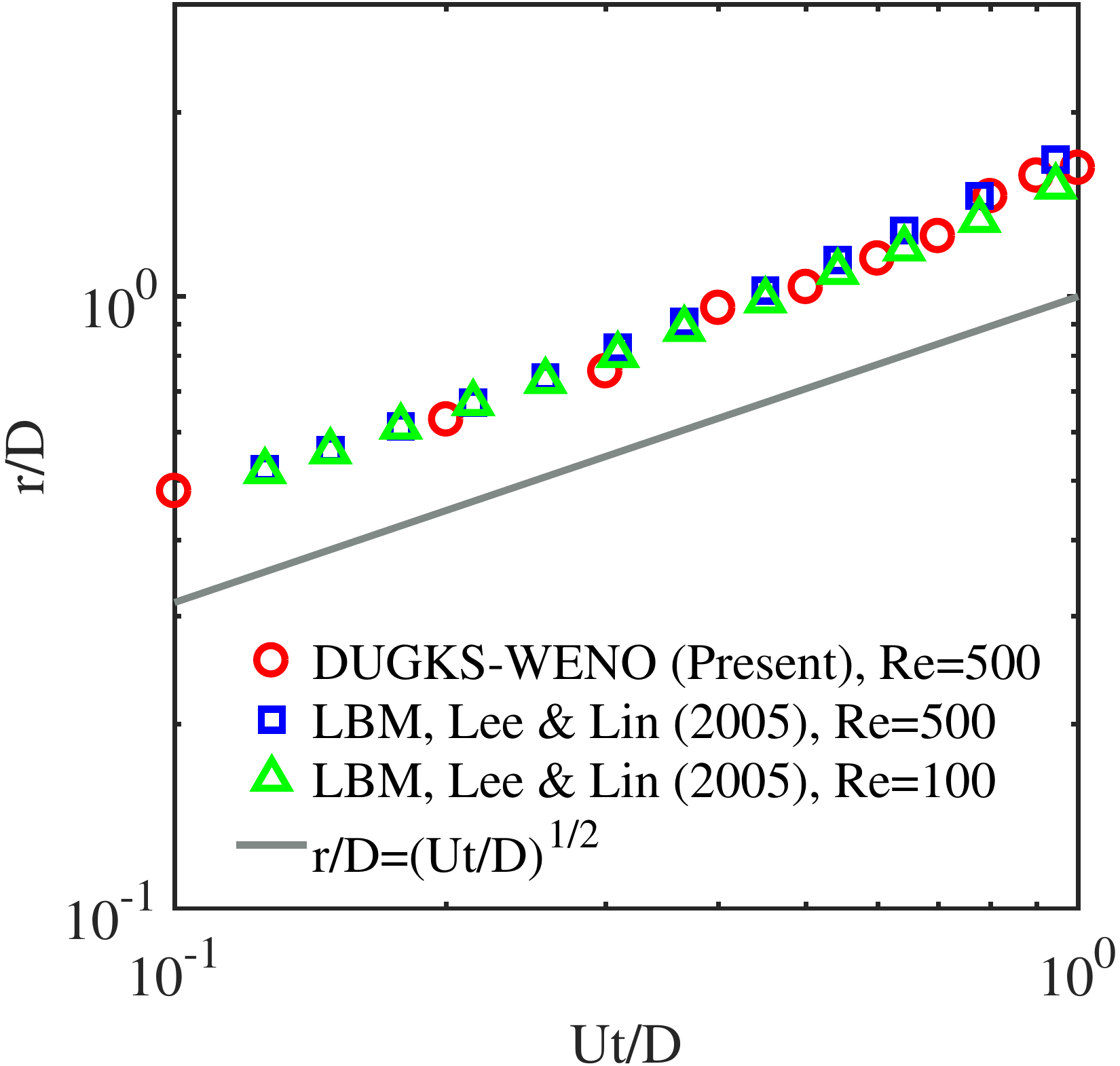}
	\centering
	\caption{Time evolution of the spread factor $r/D_0$ for 2D droplet splashing.}
	\label{fig:TimeofDSRe500}
\end{figure}

The evolution of spread factor $r/D_0$ is compared with Lee and Lin~\cite{lee2005stable} in Fig.~\ref{fig:TimeofDSRe500}.
Here the spreading radius $r$ is defined as the radius of the position where the velocity magnitude has a maximum value in the whole domain.~\cite{josserand2003droplet}
Our results agrees well with the previous studies, and the power law evolution of the radius is reproduced, which demonstrates the ability of our approach.

%
%
%
%
%
%
%
%
%
%


\section{Three dimensional simulations and discussions} \label{sec: 3D}
Practical problems are mostly three dimensional, thus accurate simulation of 3D physical problems are needed.
In this section, numerical simulations are carried out to model the 3D two-phase flow problems.

\subsection{Collision of two droplets}
The phenomenon of droplets collision appears widely in the nature and industry, such as raining, printing, and spray combustion.
In this subsection, the collision of binary equal-sized droplets is modeled to demonstrate that our approach can capture well the dynamics of 3D droplets.
Initially, two spherical droplets with radii $R_0=16$ are 
placed at the center line of the computational domain of size $128^3$. 
The distance of the centers of the two droplets is 48.
The periodic boundary conditions are applied in
all spatial directions.
The main dimensionless parameters of this problem are the Reynolds number $Re=2R_0U_0/\nu_A=1720$, Weber number $We=2R_0\rho_A U_0^2/\sigma=58$, the density ratio $\rho^*=\rho_{A}/\rho_{B}$ and the kinematic viscosity ratio $\nu^*=\nu_{A}/\nu_{B}=0.01$.
Here $A$ is the droplet phase and $B$ is the background phase.
The other parameters are $W=5$, $M_{AC}=0.01$, the acceleration of gravity $\boldsymbol{g}=\boldsymbol{0}$, and the initial relative velocity $U_0=0.02$.
The pressure is initialized as $p=2\phi\sigma/R_0$.
The CFL number is still fixed at $CFL=0.25$. 

\begin{figure}[]
	\centering
	\includegraphics[width=0.8\columnwidth,trim={0cm 0cm 0cm 0cm},clip]{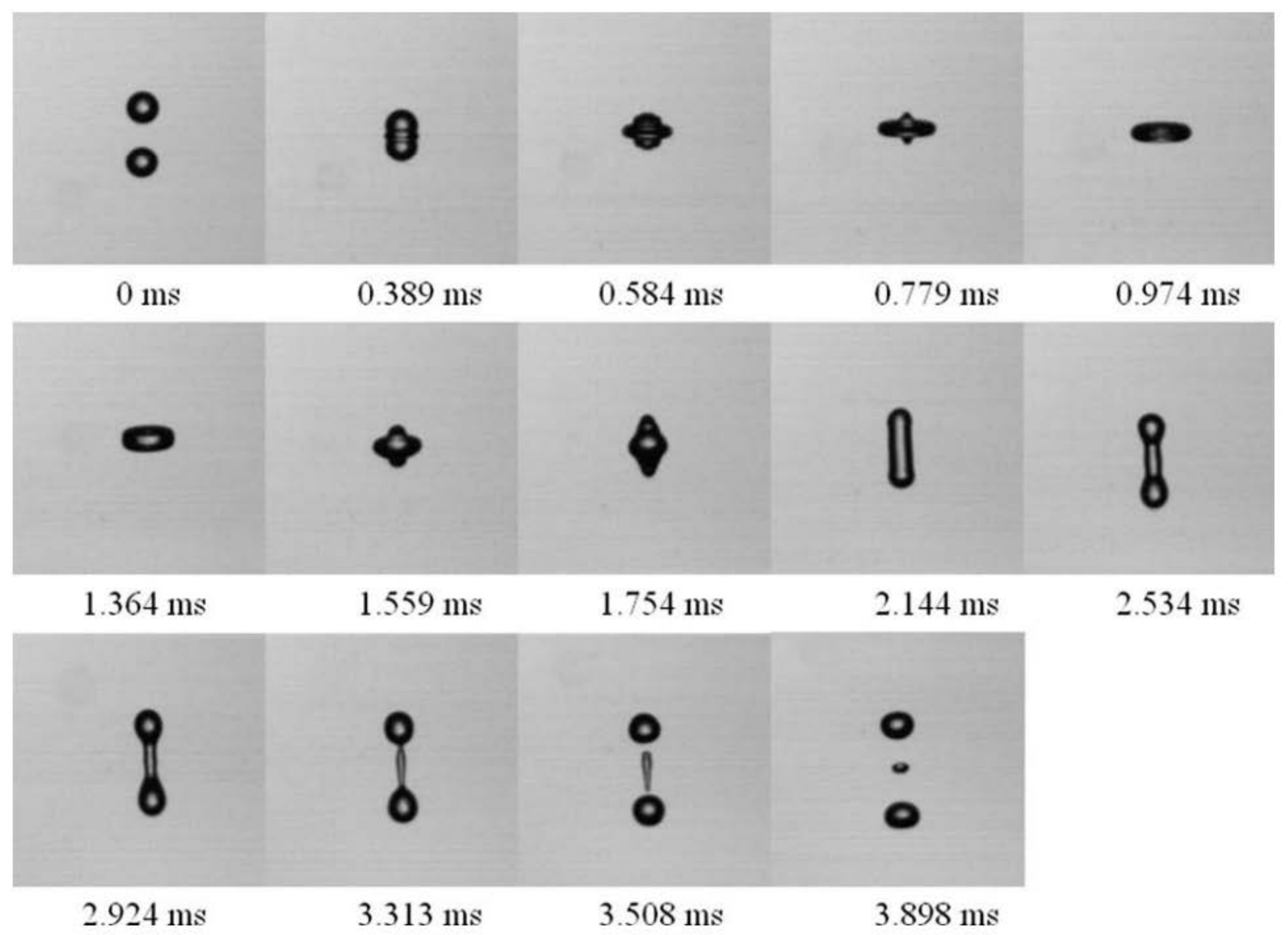}
	\centering
	\caption{Experimental results by Pan {\it et al.}~\cite{pan2009binary} for the head-on collisions of binary droplets, 
	showing that a satellite droplet is formed between two large droplets. $Re=1720$ and $We=58$.}
	\label{fig:Pan2009Fig2}
\end{figure}

\begin{figure}[]
	\centering
	\includegraphics[width=0.19\columnwidth,trim={2cm 0cm 2cm 0cm},clip]{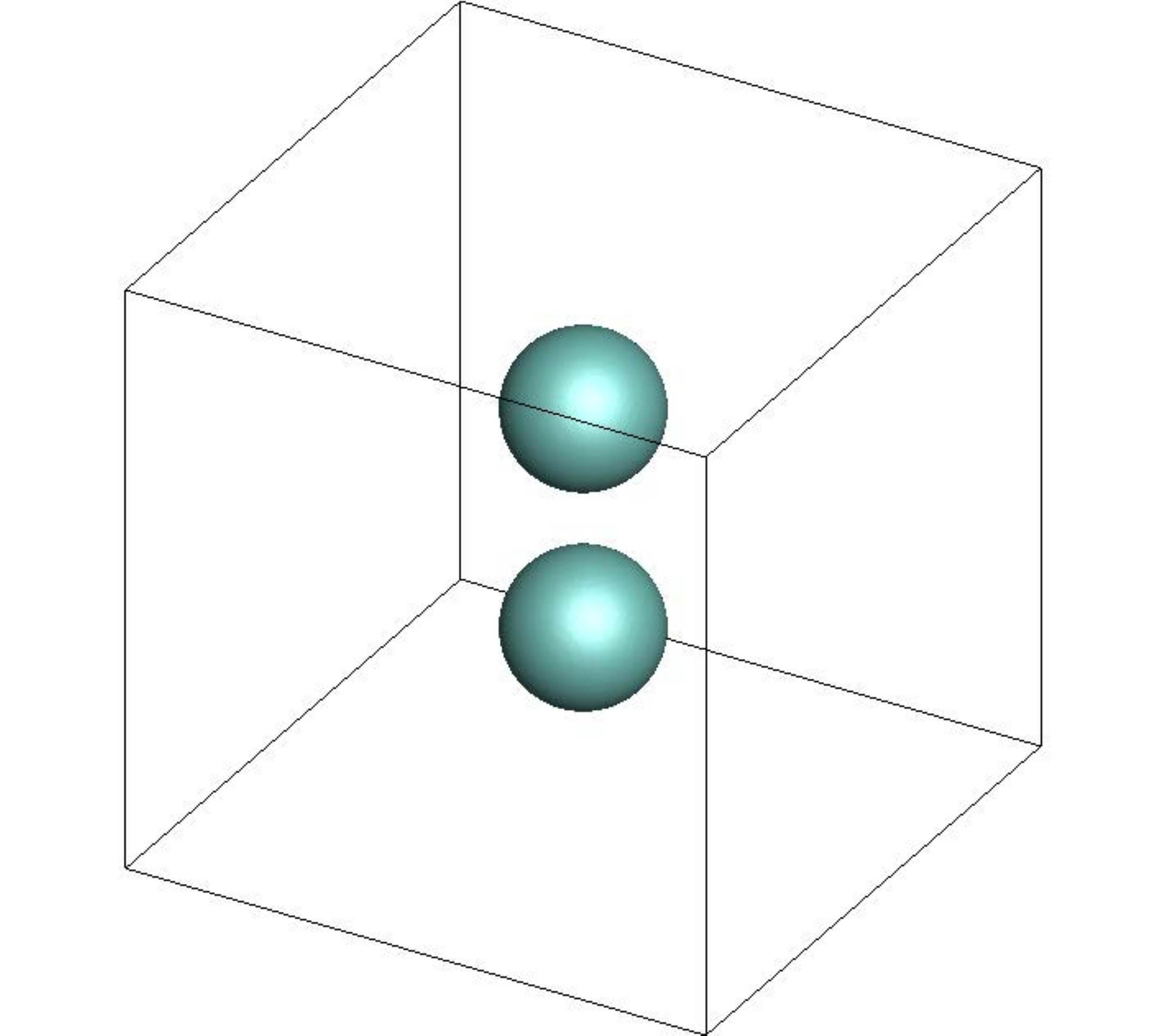}
	\includegraphics[width=0.19\columnwidth,trim={2cm 0cm 2cm 0cm},clip]{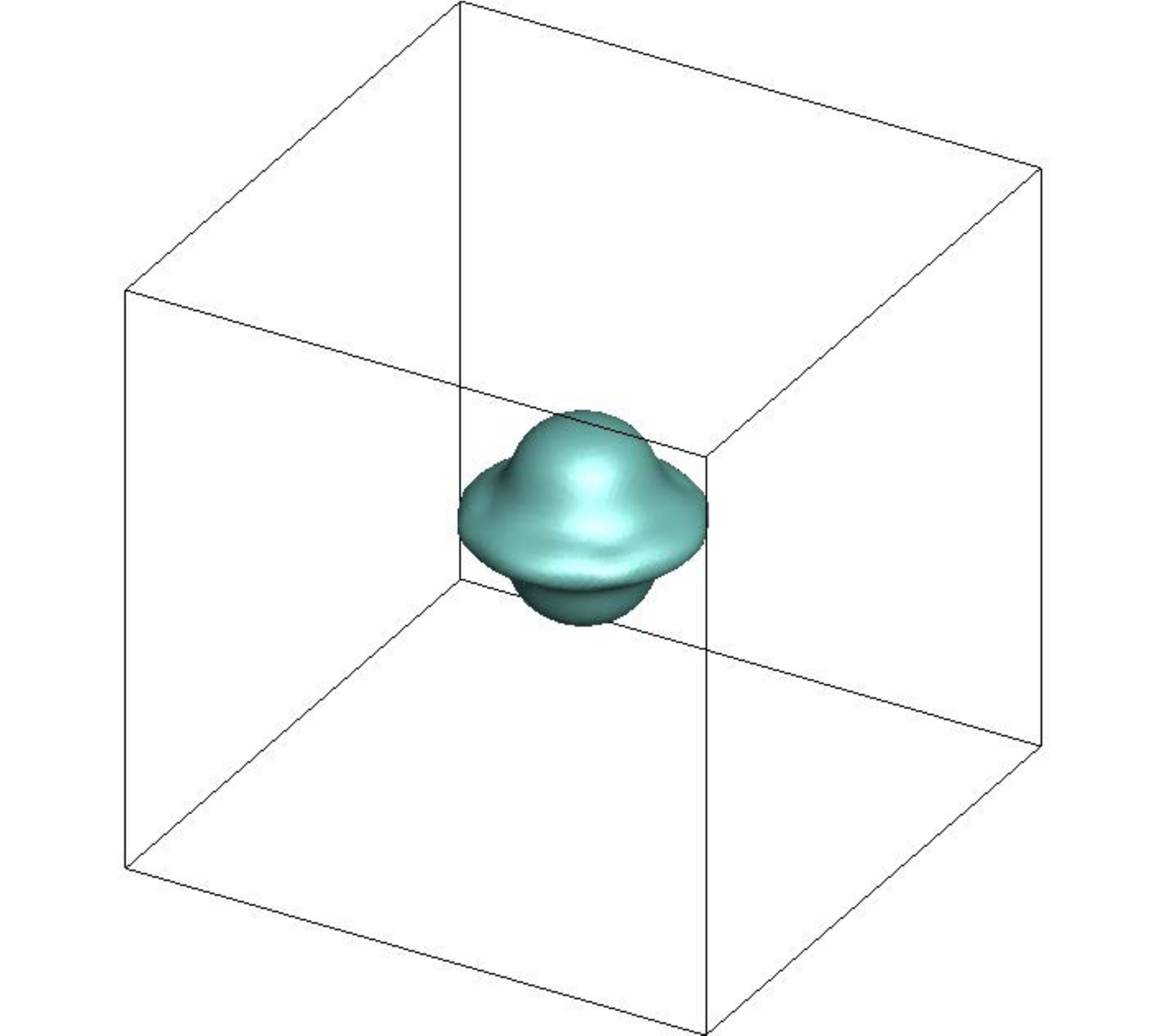}
	\includegraphics[width=0.19\columnwidth,trim={2cm 0cm 2cm 0cm},clip]{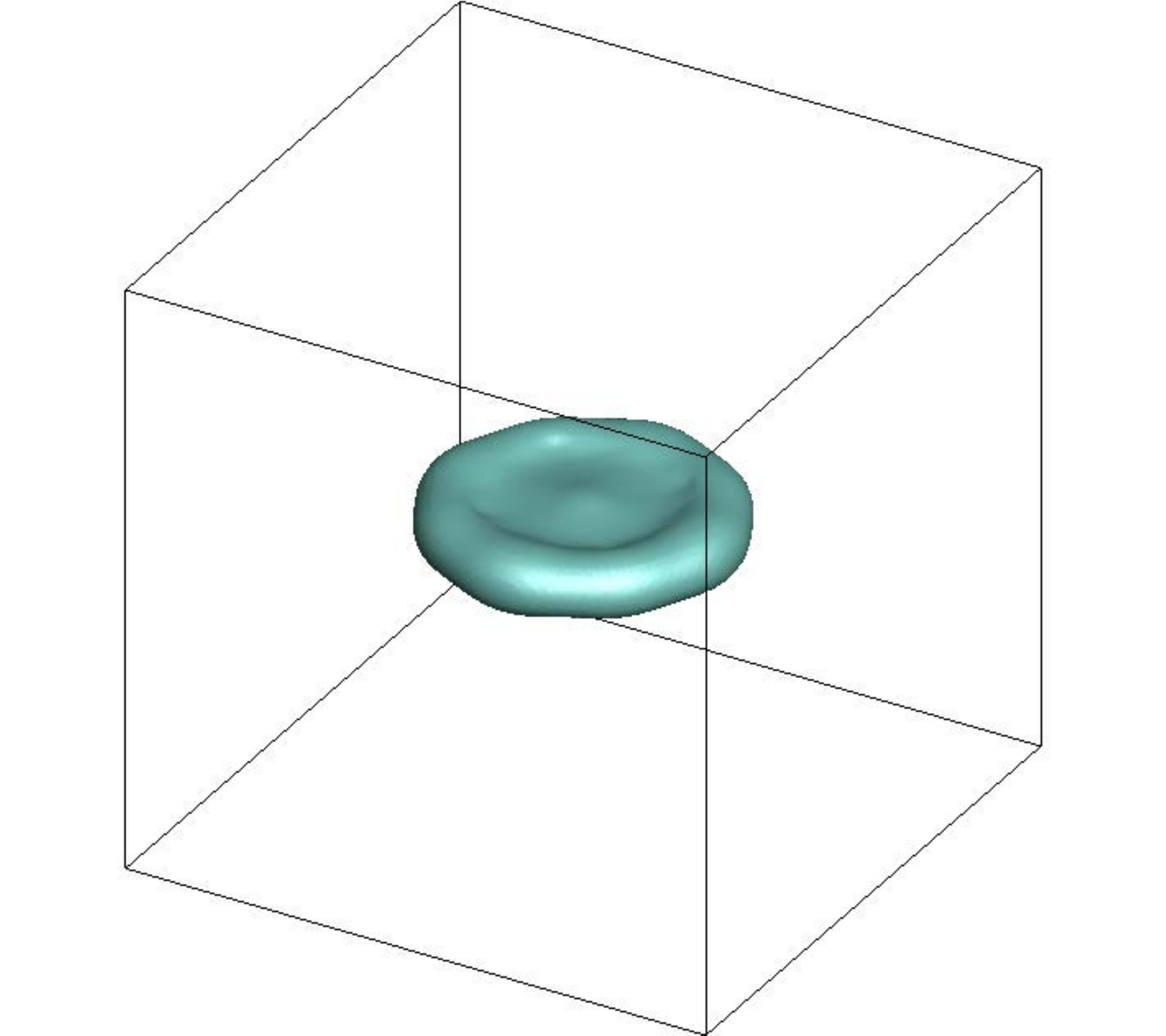}
	\includegraphics[width=0.19\columnwidth,trim={2cm 0cm 2cm 0cm},clip]{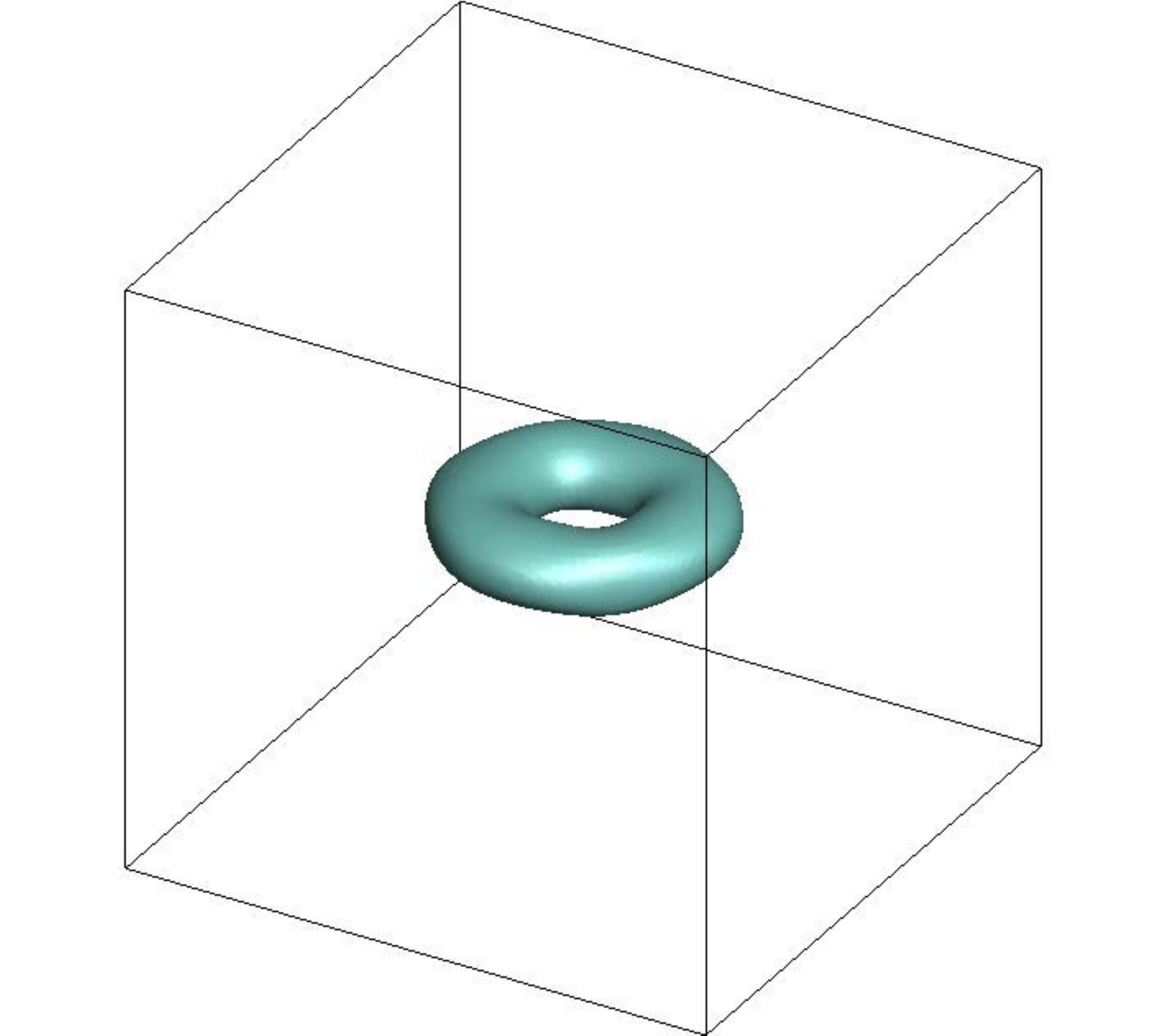}
	\includegraphics[width=0.19\columnwidth,trim={2cm 0cm 2cm 0cm},clip]{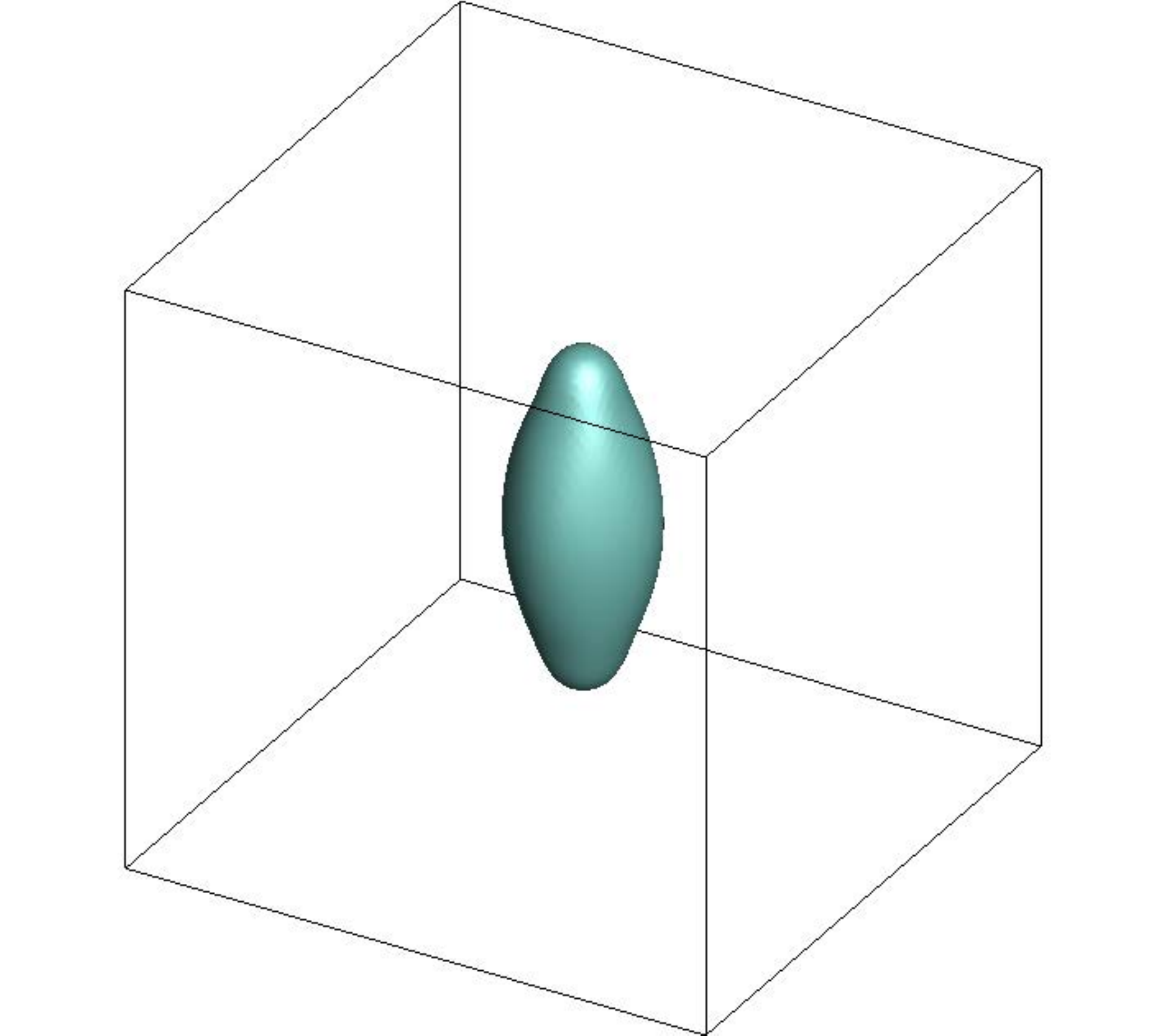}\\
	\includegraphics[width=0.19\columnwidth,trim={2cm 0cm 2cm 0cm},clip]{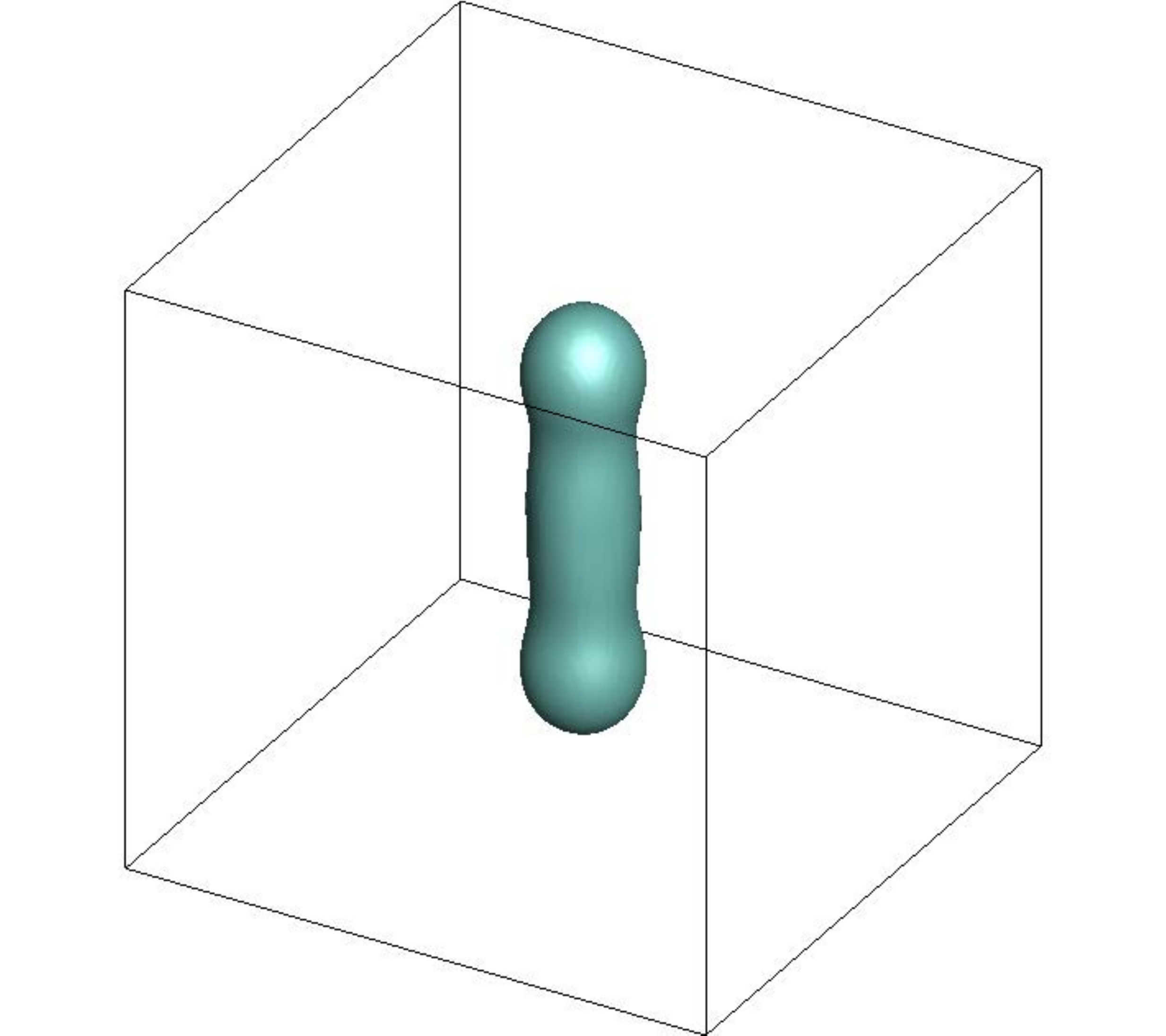}
	\includegraphics[width=0.19\columnwidth,trim={2cm 0cm 2cm 0cm},clip]{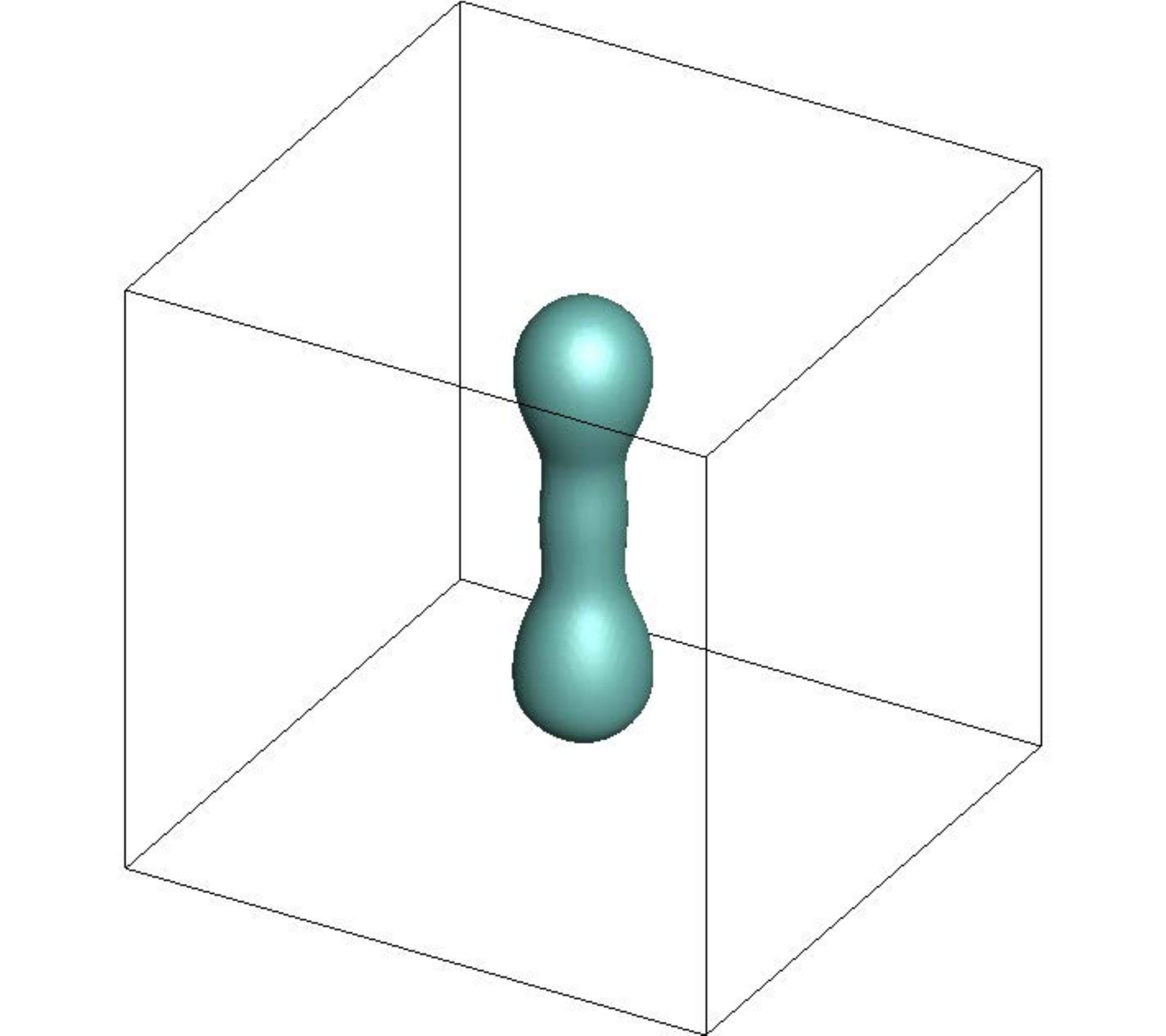}
	\includegraphics[width=0.19\columnwidth,trim={2cm 0cm 2cm 0cm},clip]{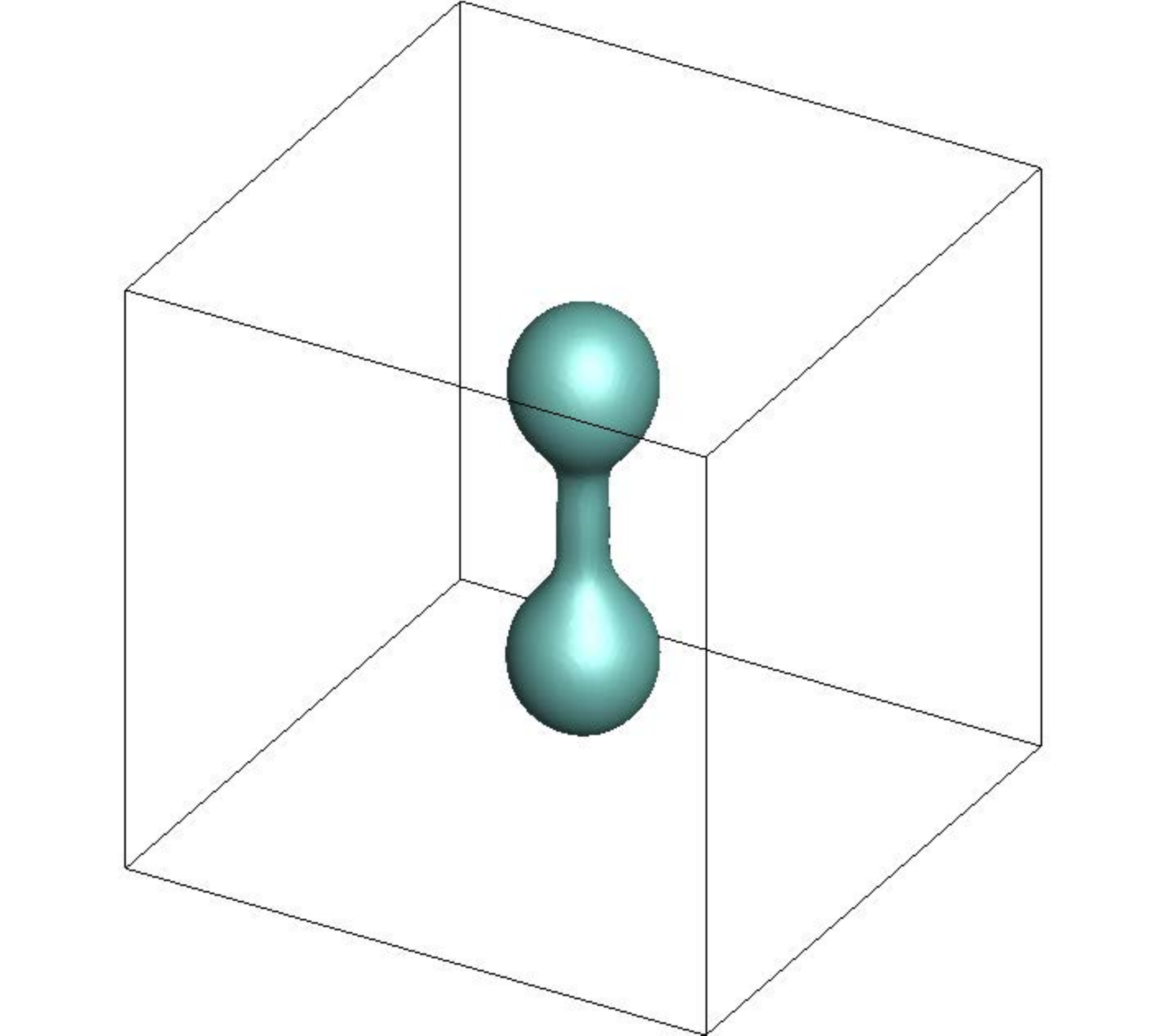}
	\includegraphics[width=0.19\columnwidth,trim={2cm 0cm 2cm 0cm},clip]{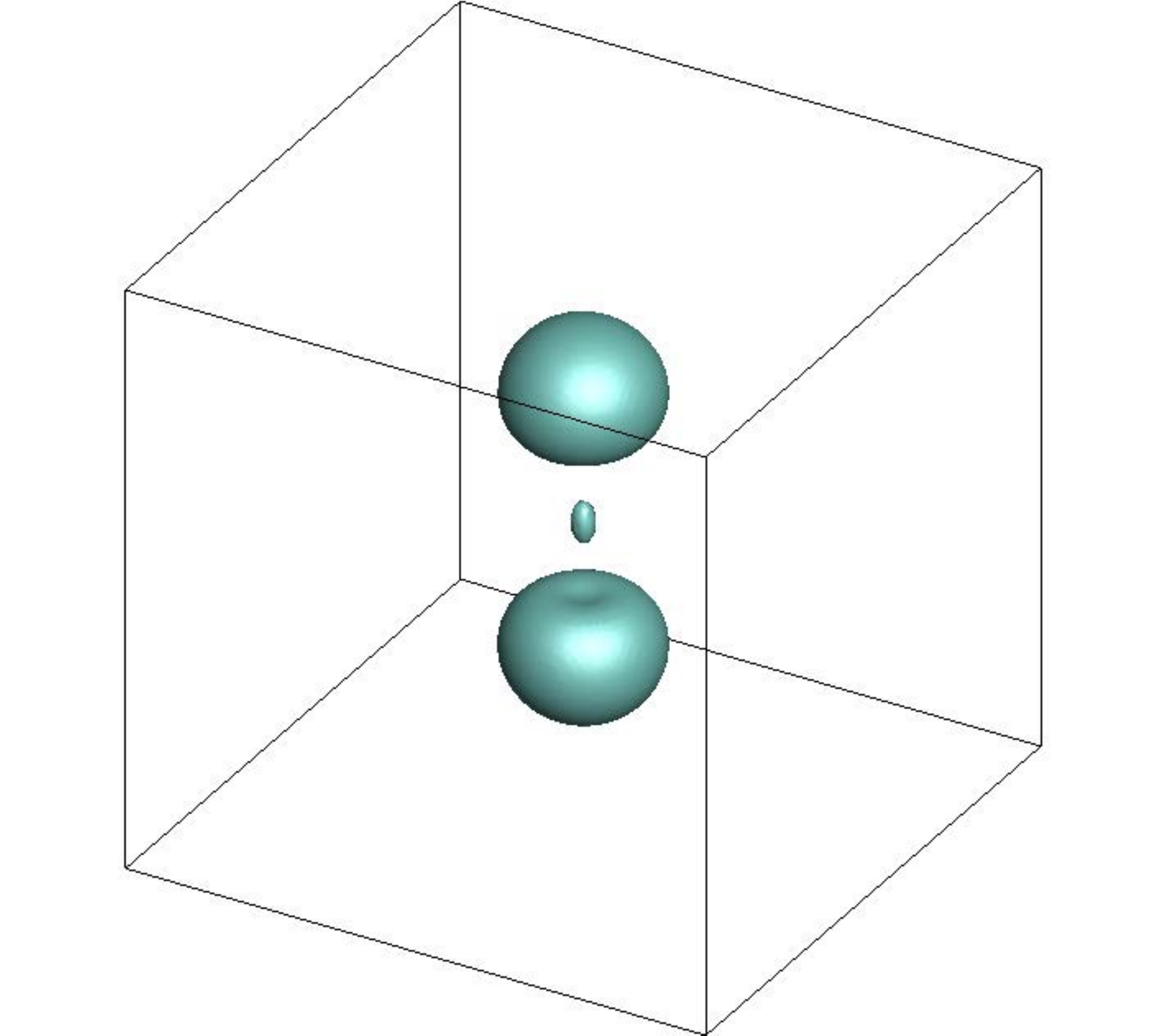}
	\includegraphics[width=0.19\columnwidth,trim={2cm 0cm 2cm 0cm},clip]{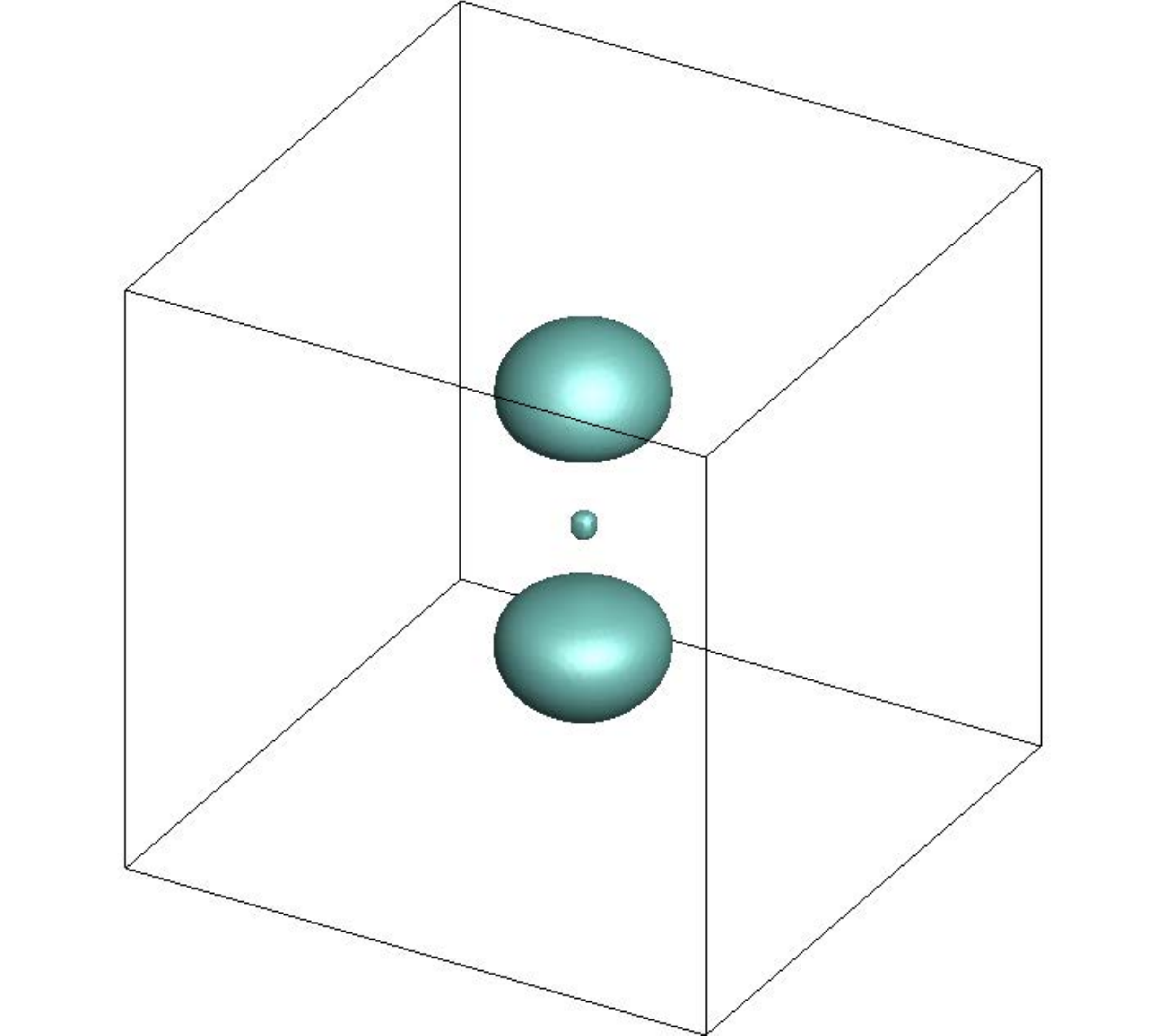}
	\centering
	\caption{Evolution of fluid-fluid interface for head-on collisions of binary droplets, showing that 
	a satellite droplet is formed between the two droplets. $\rho^*=1000$, $Re=1720$, $We=58$. $U_0 t/(2R_0)=0, 1.56, 3.13, 7.03, 10.2, 12.5, 14.8, 16.4, 18.0, 18.8$.}
	\label{fig:DC1em6rhor1000}
\end{figure}

First, the density ratio is selected as $\rho^*=1000$ to compare with the previous experimental (Fig.~\ref{fig:Pan2009Fig2}) and numerical results.~\cite{pan2009binary,lycett2016cascaded} The droplets head-on collision dynamic for this case is shown in Fig.~\ref{fig:DC1em6rhor1000}.
At the beginning, the two droplets move towards each other and merge, like two hats that fit together.
Then the hats flatten gradually by inertia.
As mass is conserved, the brim thickens gradually.
The center region thinned out so that a hole opens, the hats become a doughnut.
The doughnut contracts to the center and the hole disappears due to surface tension.
Again, because of inertia force and conserved mass, the large droplet elongates to the two sides in the initial droplets direction, and becomes a stick.
As time proceeds, the center of the stick is stretched and eventually broken.
One small satellite droplet is created in the end,
which is also reported by Pan {\it et al.}~\cite{pan2009binary} experimentally (Fig.~\ref{fig:Pan2009Fig2}) and Lycett-Brown and Luo~\cite{lycett2016cascaded} numerically.

\begin{figure}[]
	\centering
	\includegraphics[width=0.19\columnwidth,trim={2cm 0cm 2cm 0cm},clip]{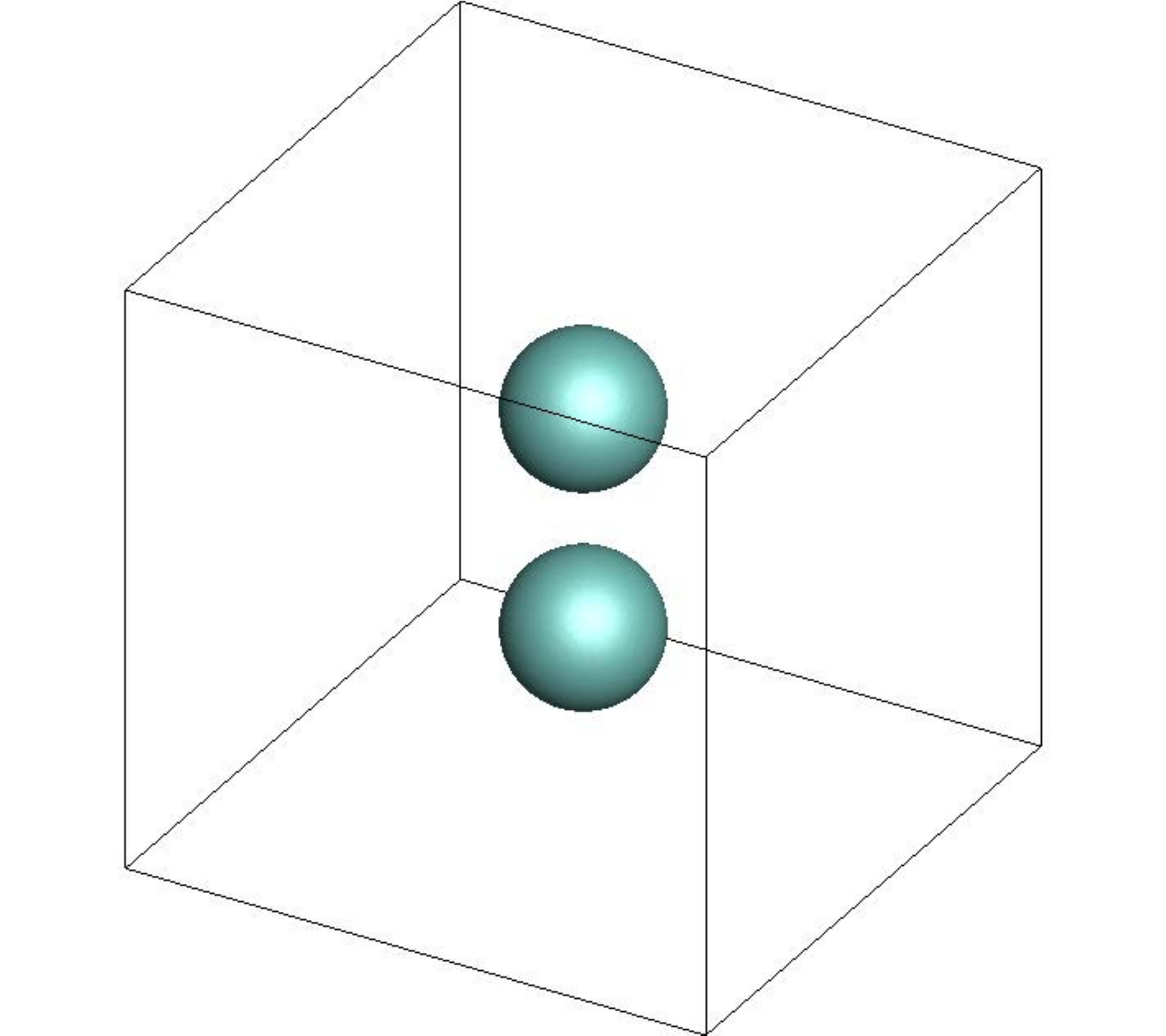}
	\includegraphics[width=0.19\columnwidth,trim={2cm 0cm 2cm 0cm},clip]{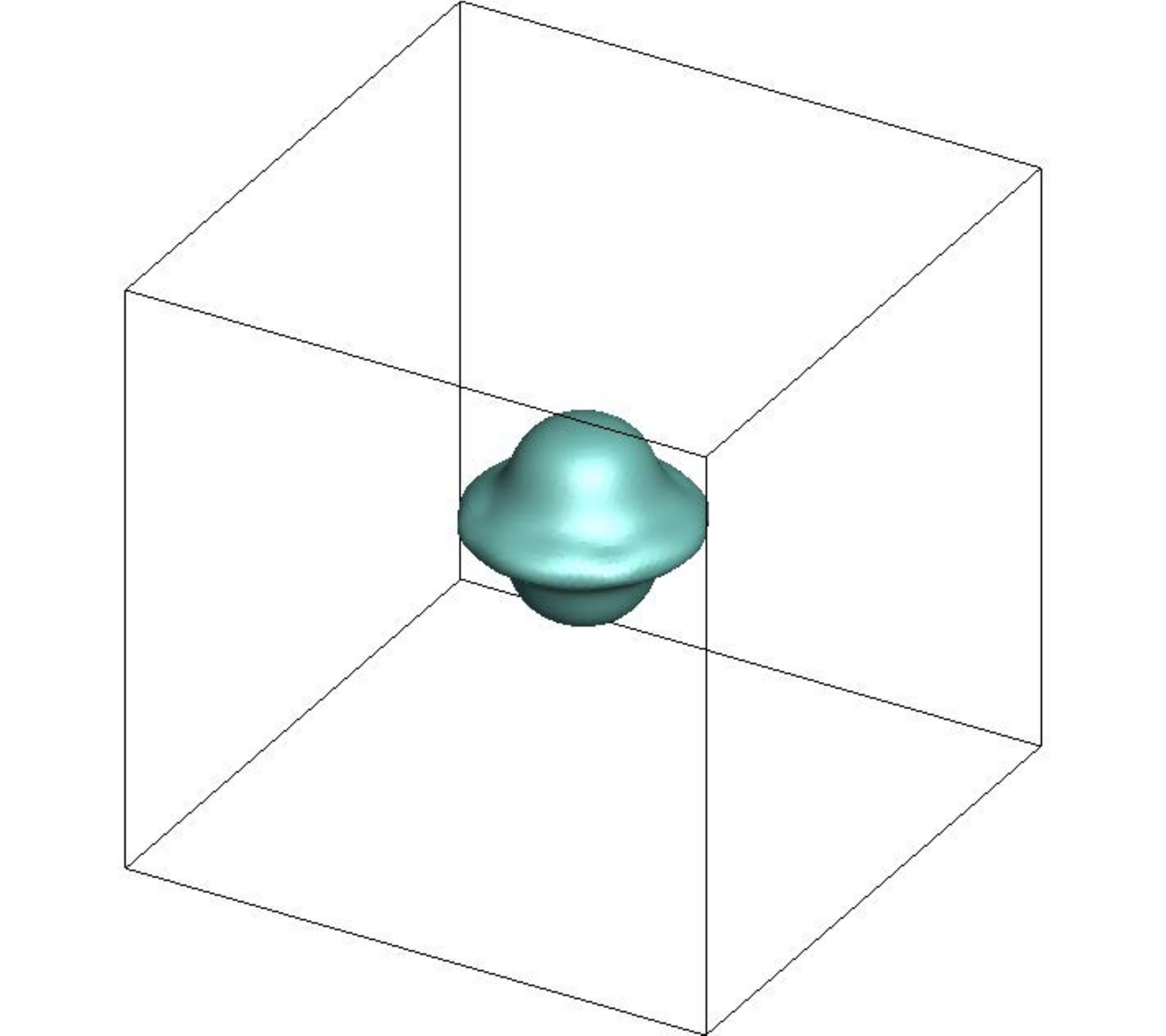}
	\includegraphics[width=0.19\columnwidth,trim={2cm 0cm 2cm 0cm},clip]{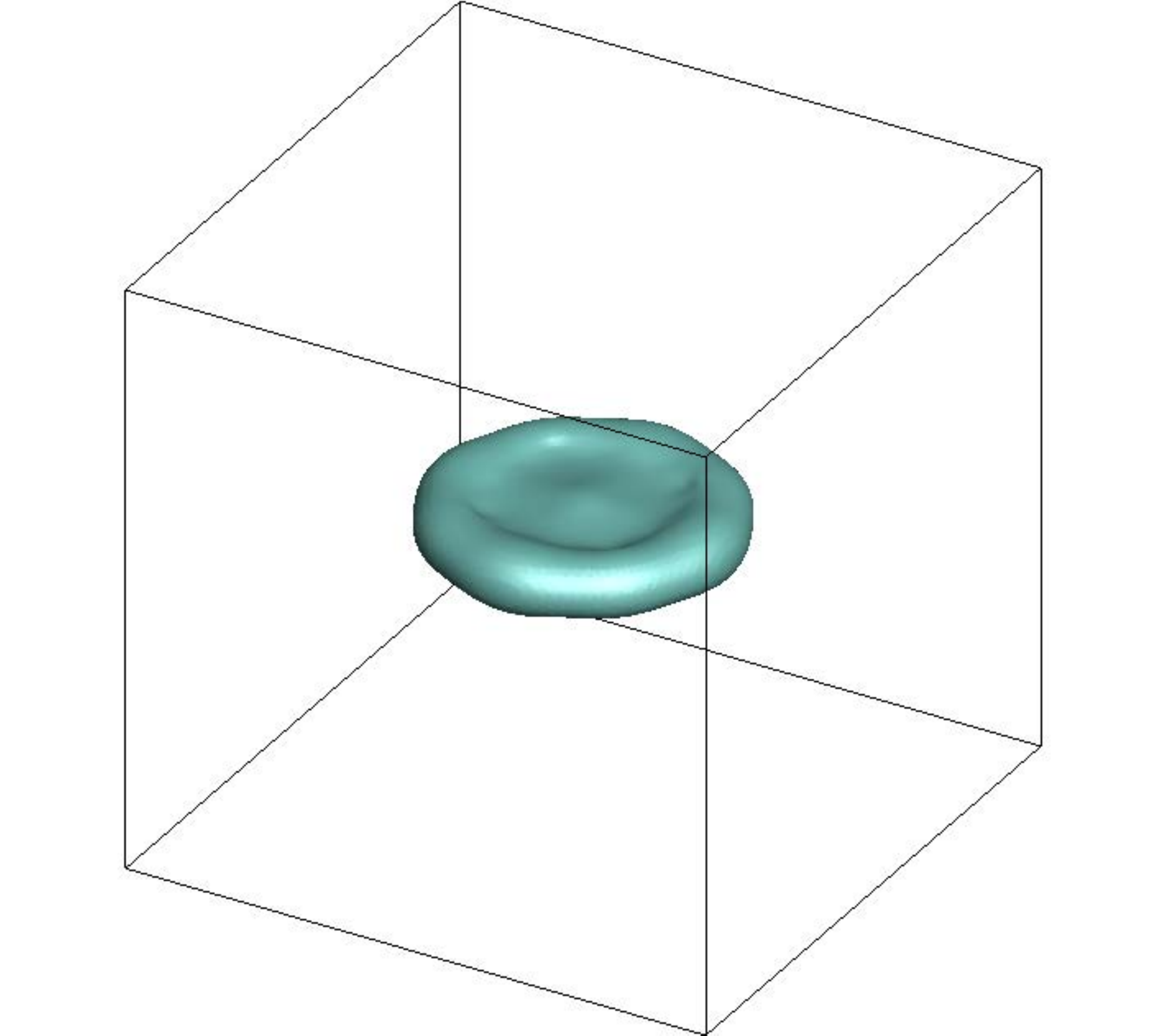}
	\includegraphics[width=0.19\columnwidth,trim={2cm 0cm 2cm 0cm},clip]{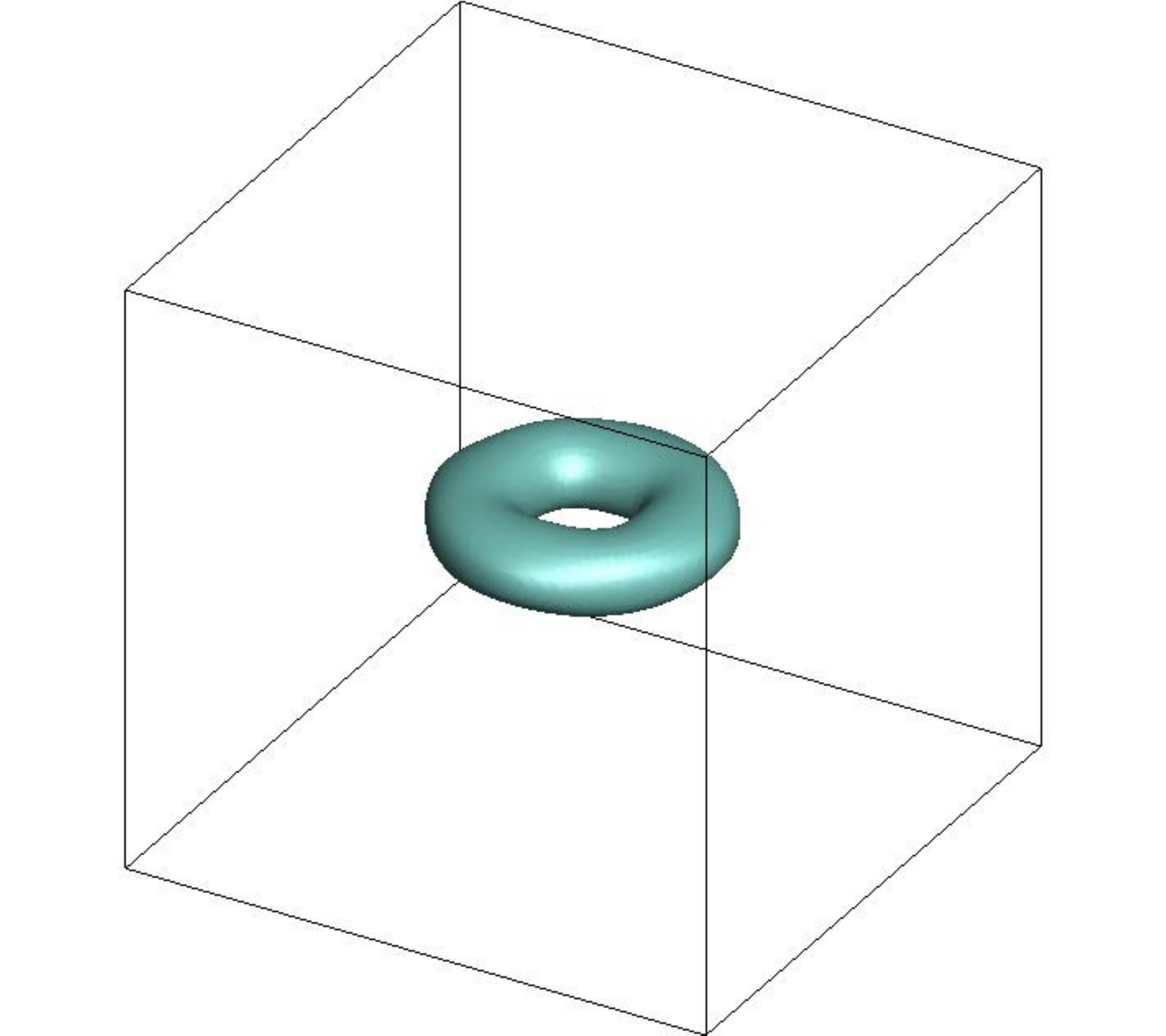}
	\includegraphics[width=0.19\columnwidth,trim={2cm 0cm 2cm 0cm},clip]{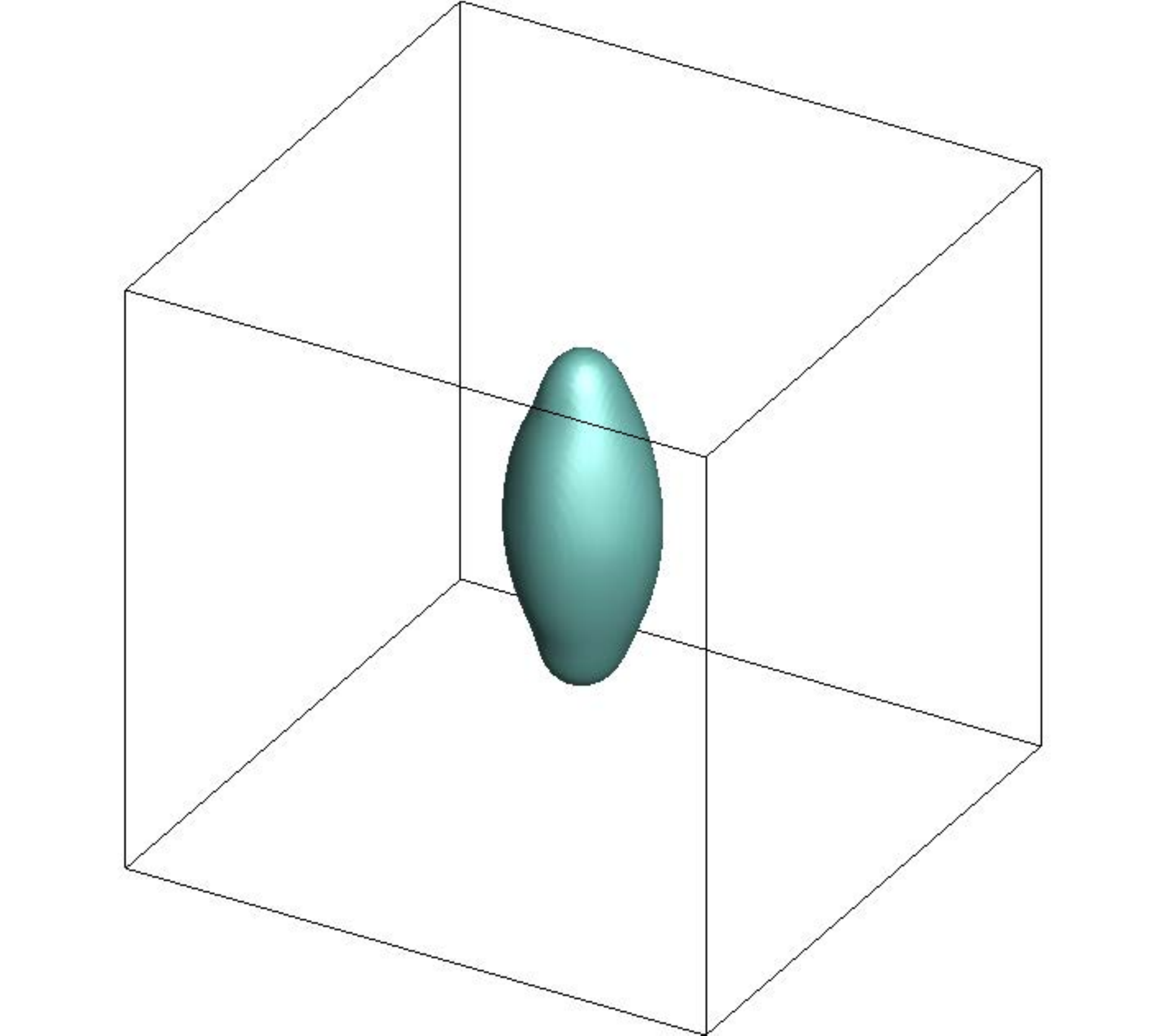}\\
	\includegraphics[width=0.19\columnwidth,trim={2cm 0cm 2cm 0cm},clip]{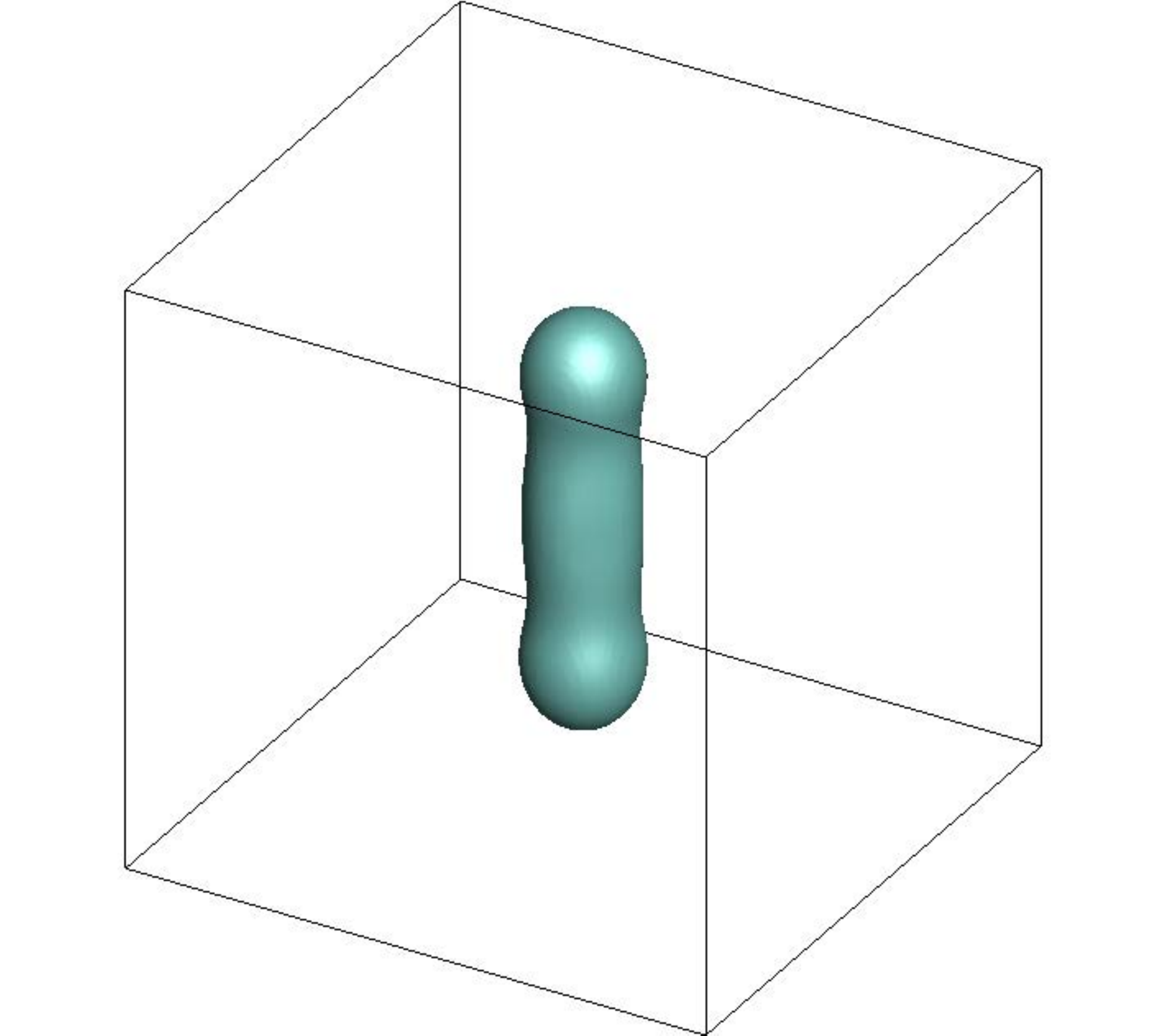}
	\includegraphics[width=0.19\columnwidth,trim={2cm 0cm 2cm 0cm},clip]{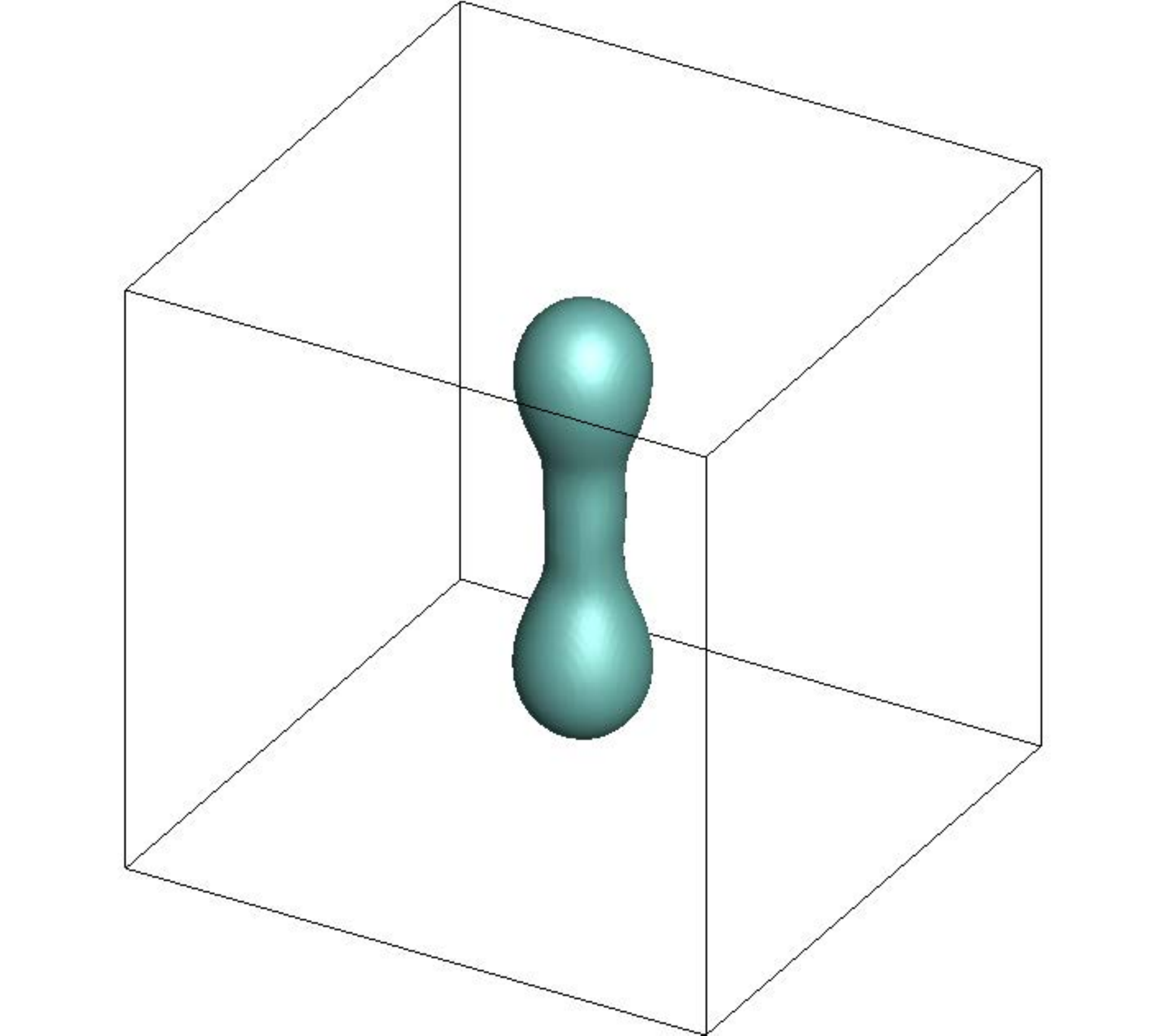}
	\includegraphics[width=0.19\columnwidth,trim={2cm 0cm 2cm 0cm},clip]{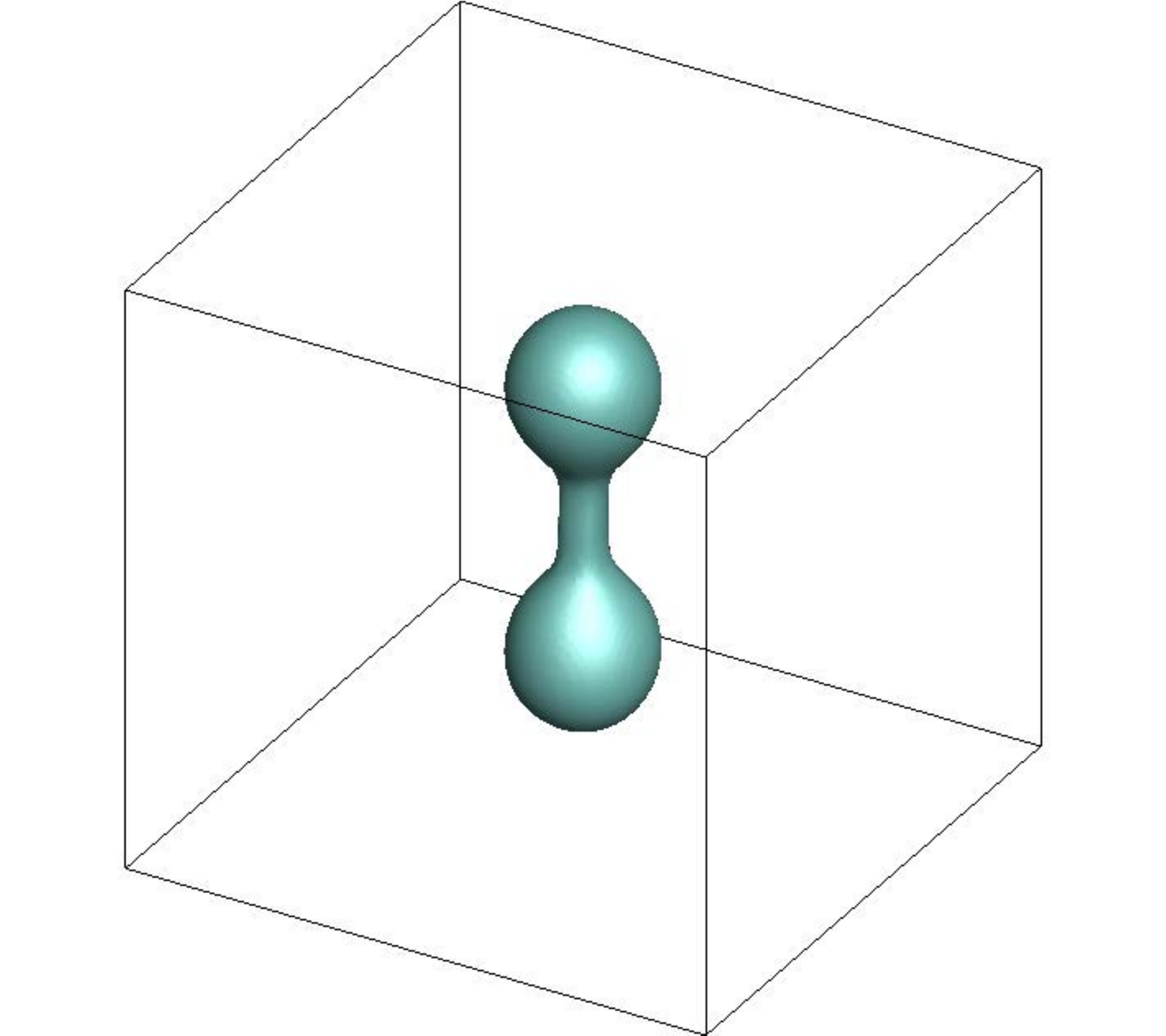}
	\includegraphics[width=0.19\columnwidth,trim={2cm 0cm 2cm 0cm},clip]{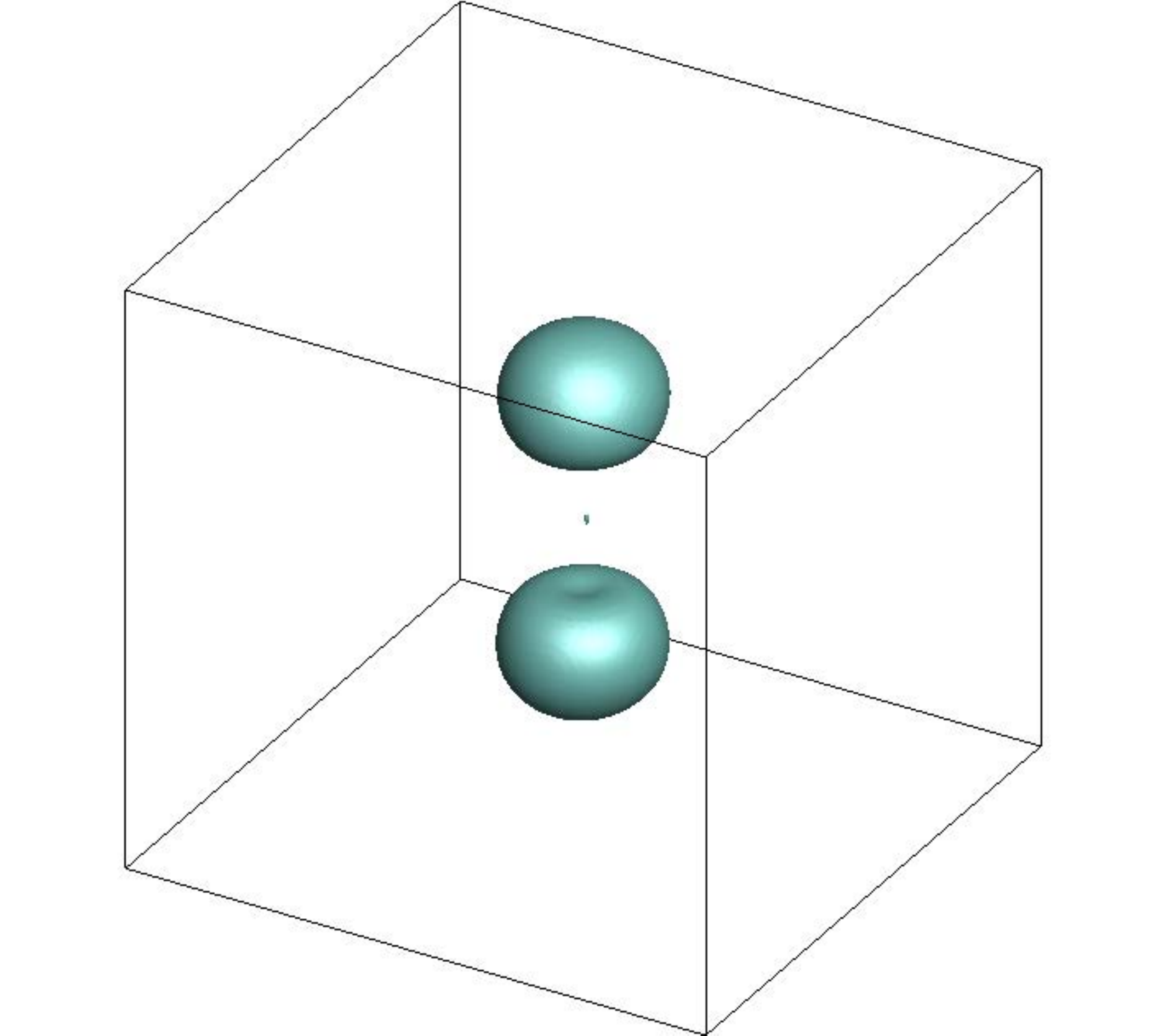}
	\includegraphics[width=0.19\columnwidth,trim={2cm 0cm 2cm 0cm},clip]{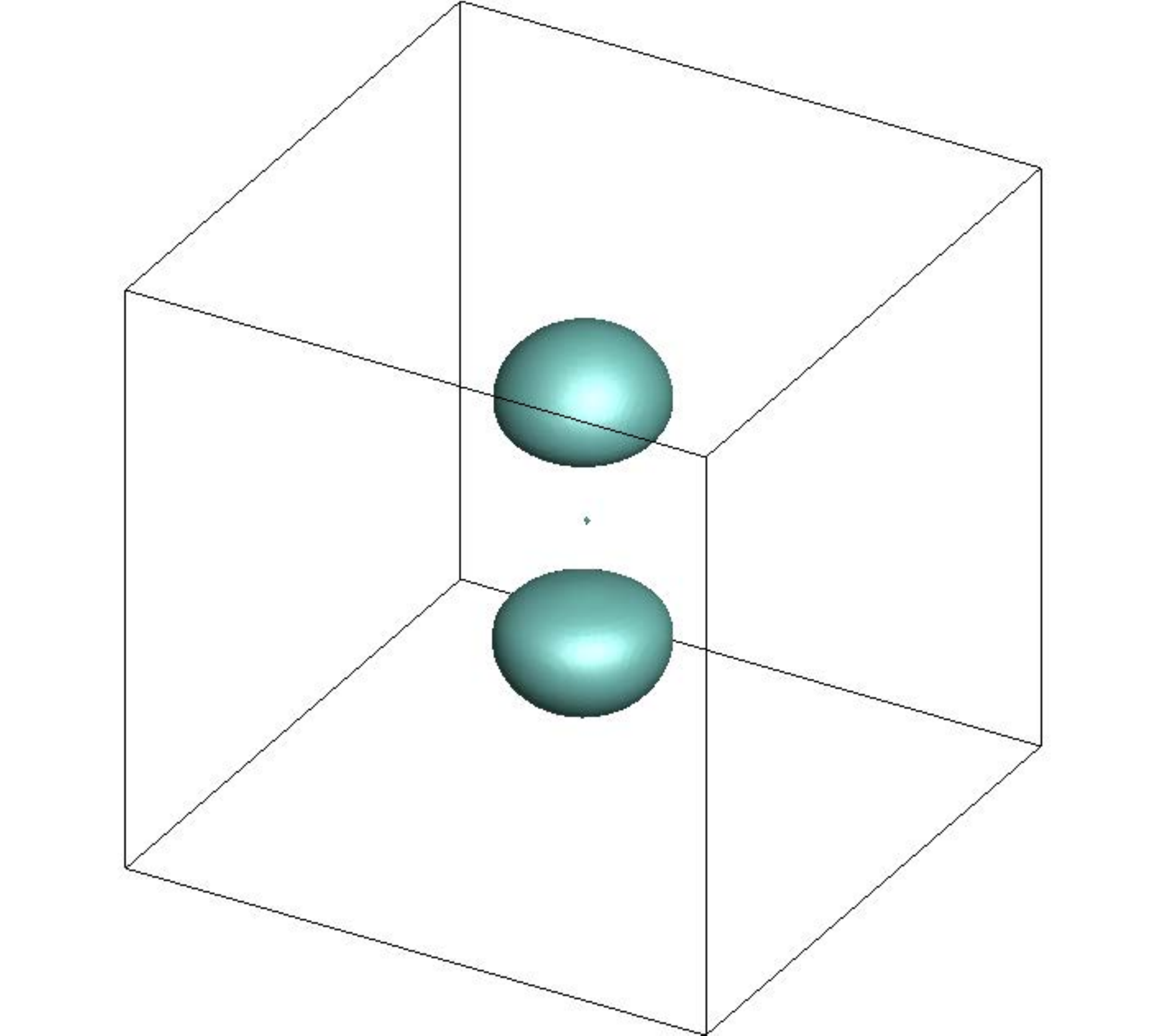}
	\centering
	\caption{Evolution of fluid-fluid interface for head-on collisions of binary droplets. A small satellite droplet is still
	formed in this case. $\rho^*=100$, $Re=1720$, $We=58$. $U_0 t/(2R_0)=0, 1.56, 3.13, 7.03, 10.2, 12.5, 14.8, 16.4, 18.0, 18.8$.}
	\label{fig:DC1em6rhor100}
\end{figure}

\begin{figure}[]
	\centering
	\includegraphics[width=0.19\columnwidth,trim={2cm 0cm 2cm 0cm},clip]{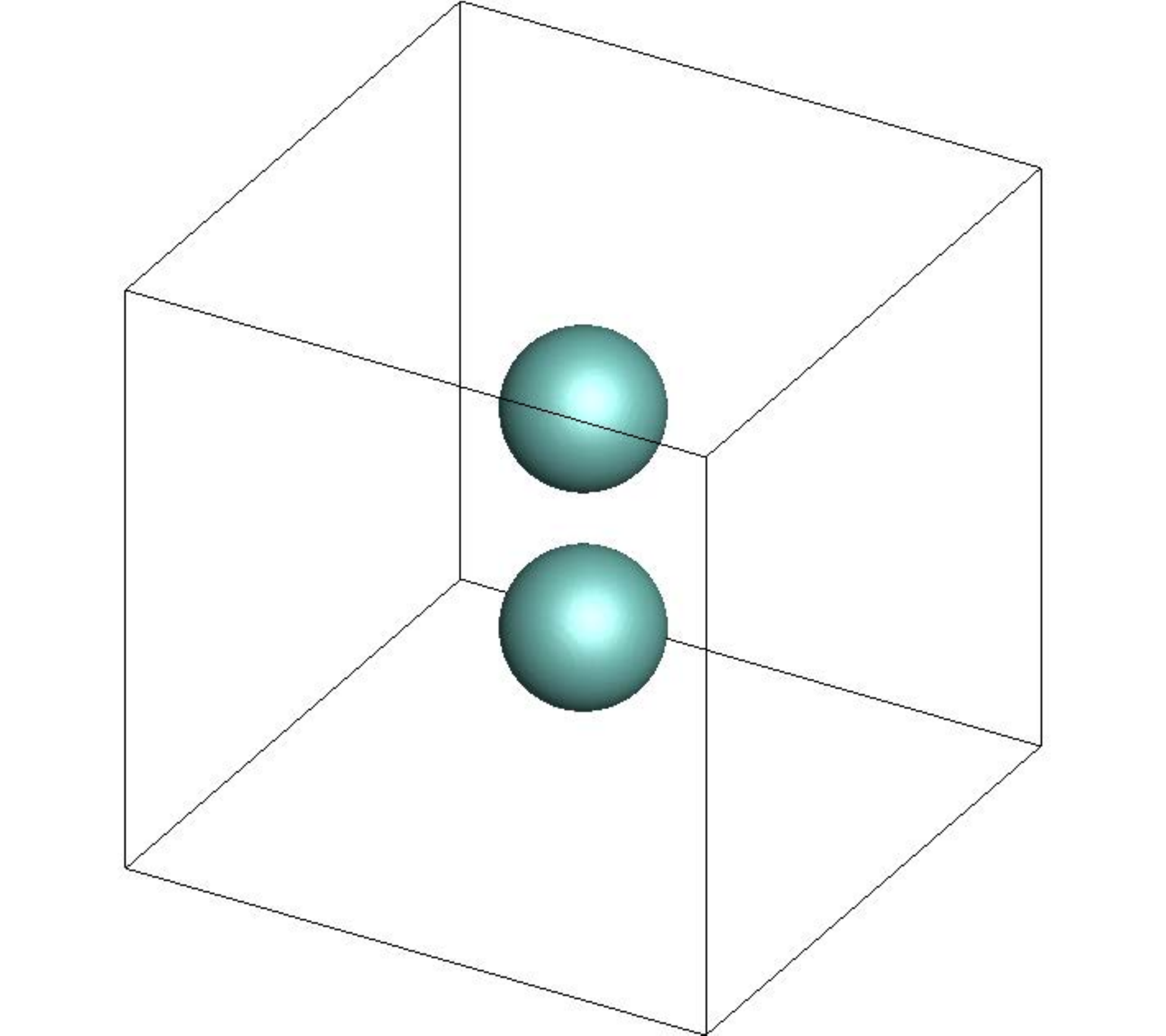}
	\includegraphics[width=0.19\columnwidth,trim={2cm 0cm 2cm 0cm},clip]{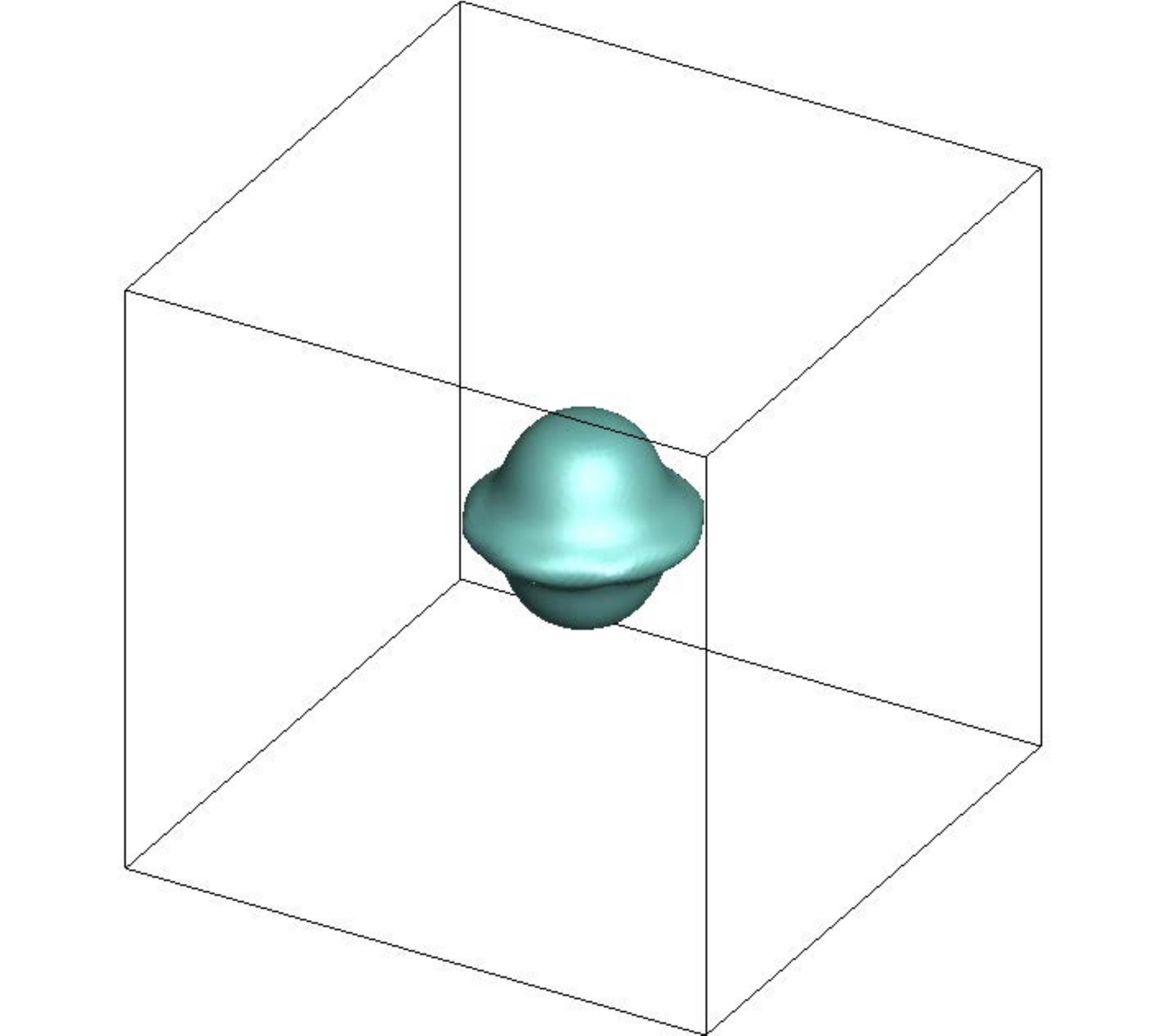}
	\includegraphics[width=0.19\columnwidth,trim={2cm 0cm 2cm 0cm},clip]{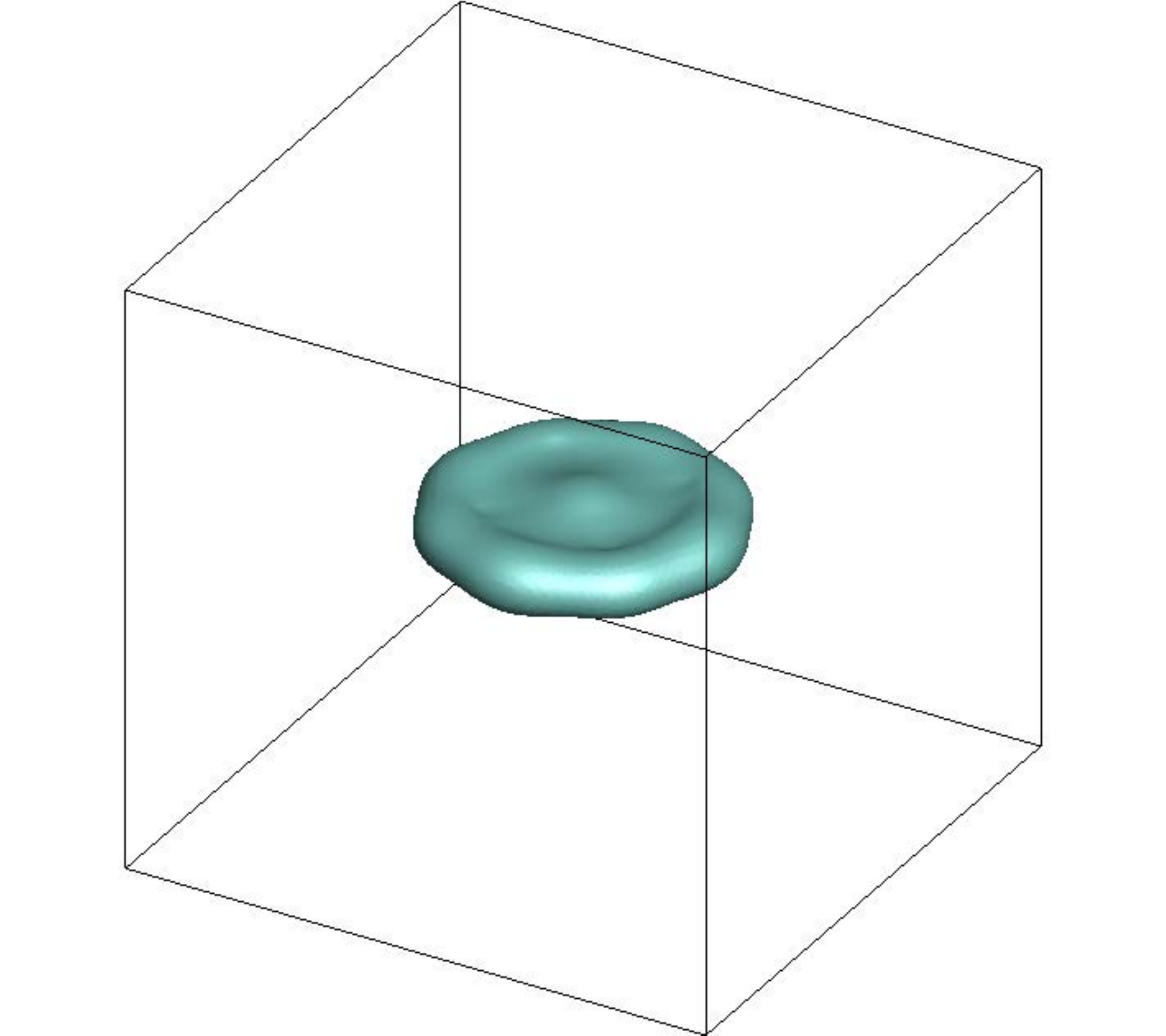}
	\includegraphics[width=0.19\columnwidth,trim={2cm 0cm 2cm 0cm},clip]{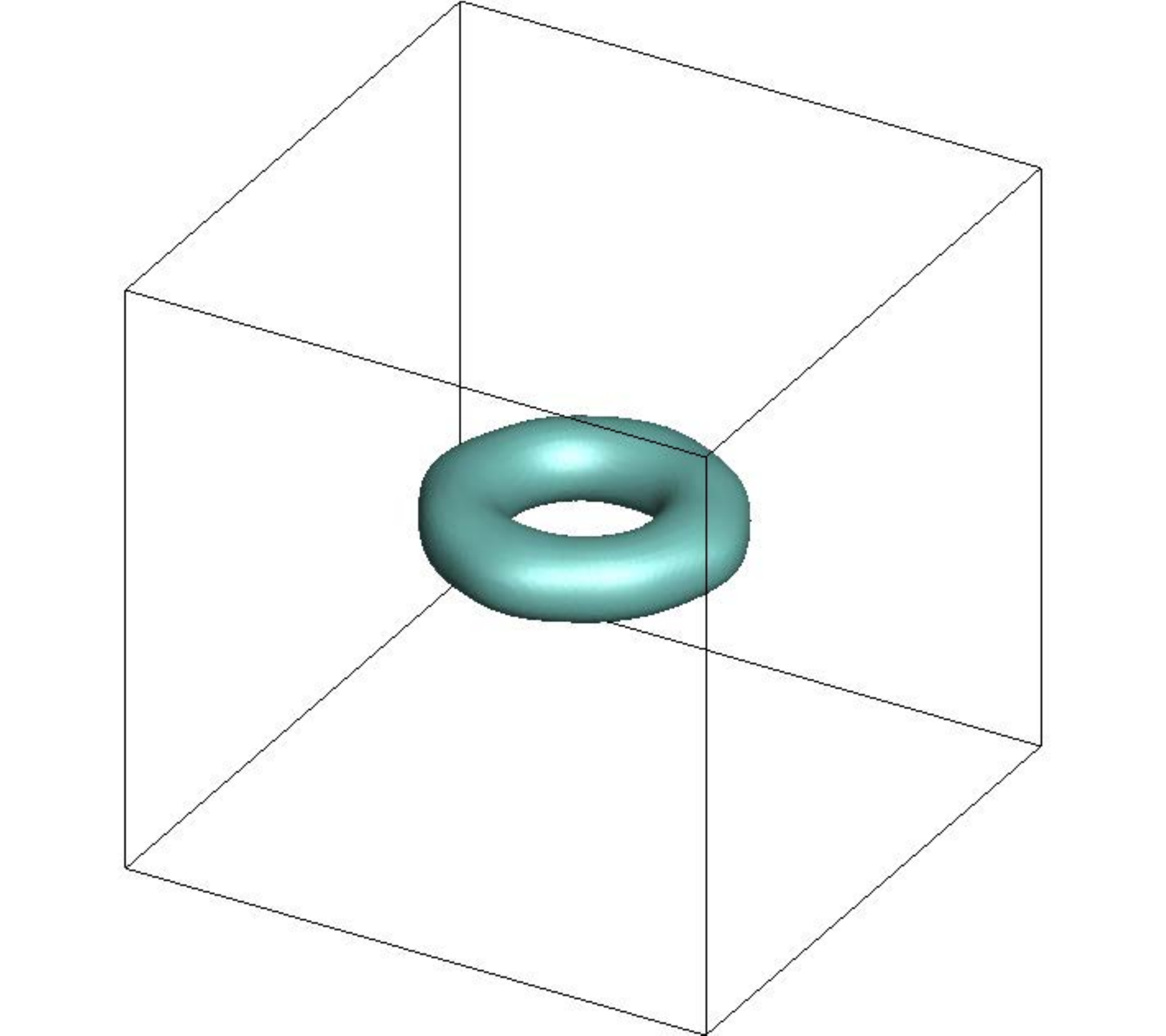}
	\includegraphics[width=0.19\columnwidth,trim={2cm 0cm 2cm 0cm},clip]{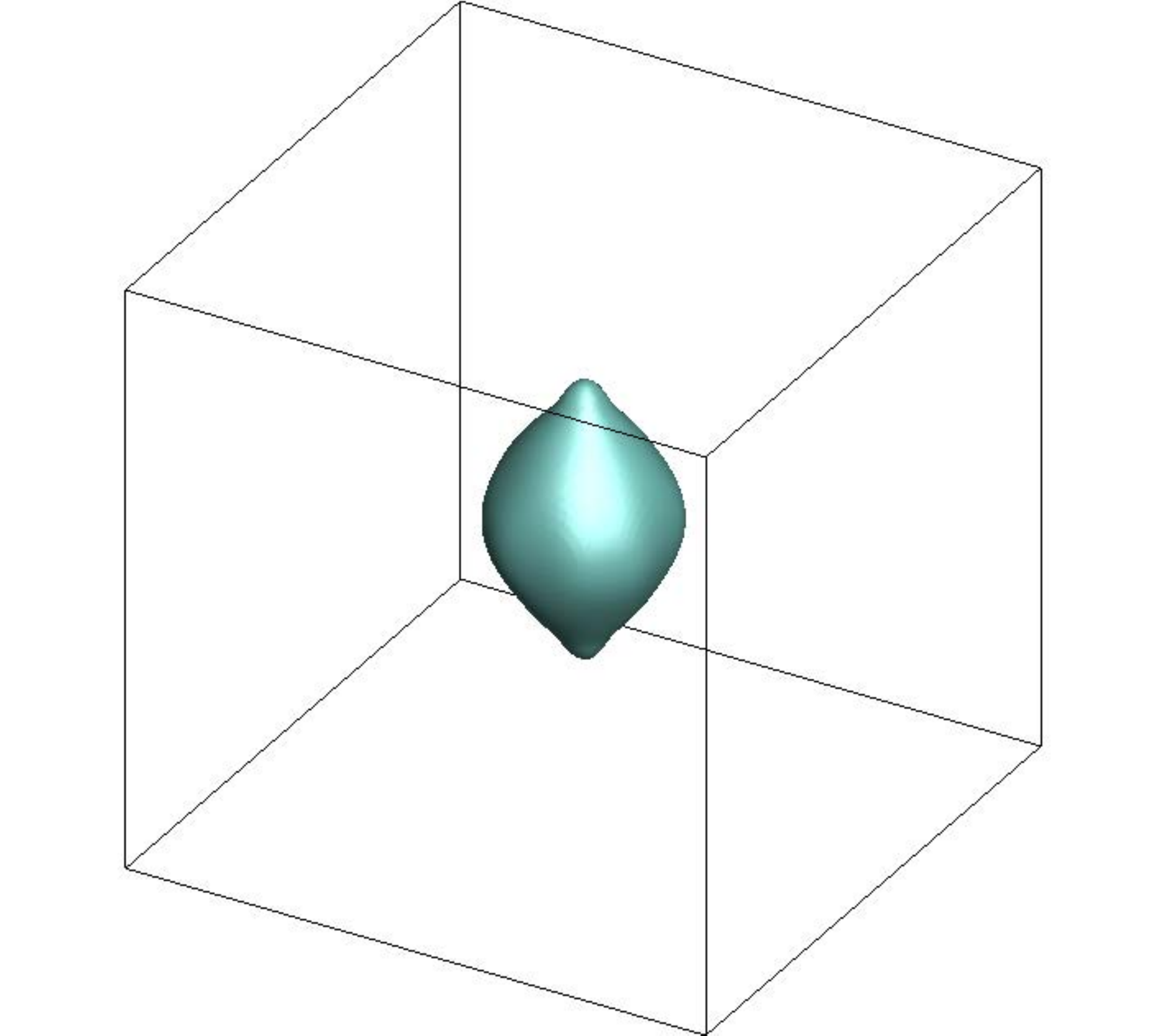}\\
	\includegraphics[width=0.19\columnwidth,trim={2cm 0cm 2cm 0cm},clip]{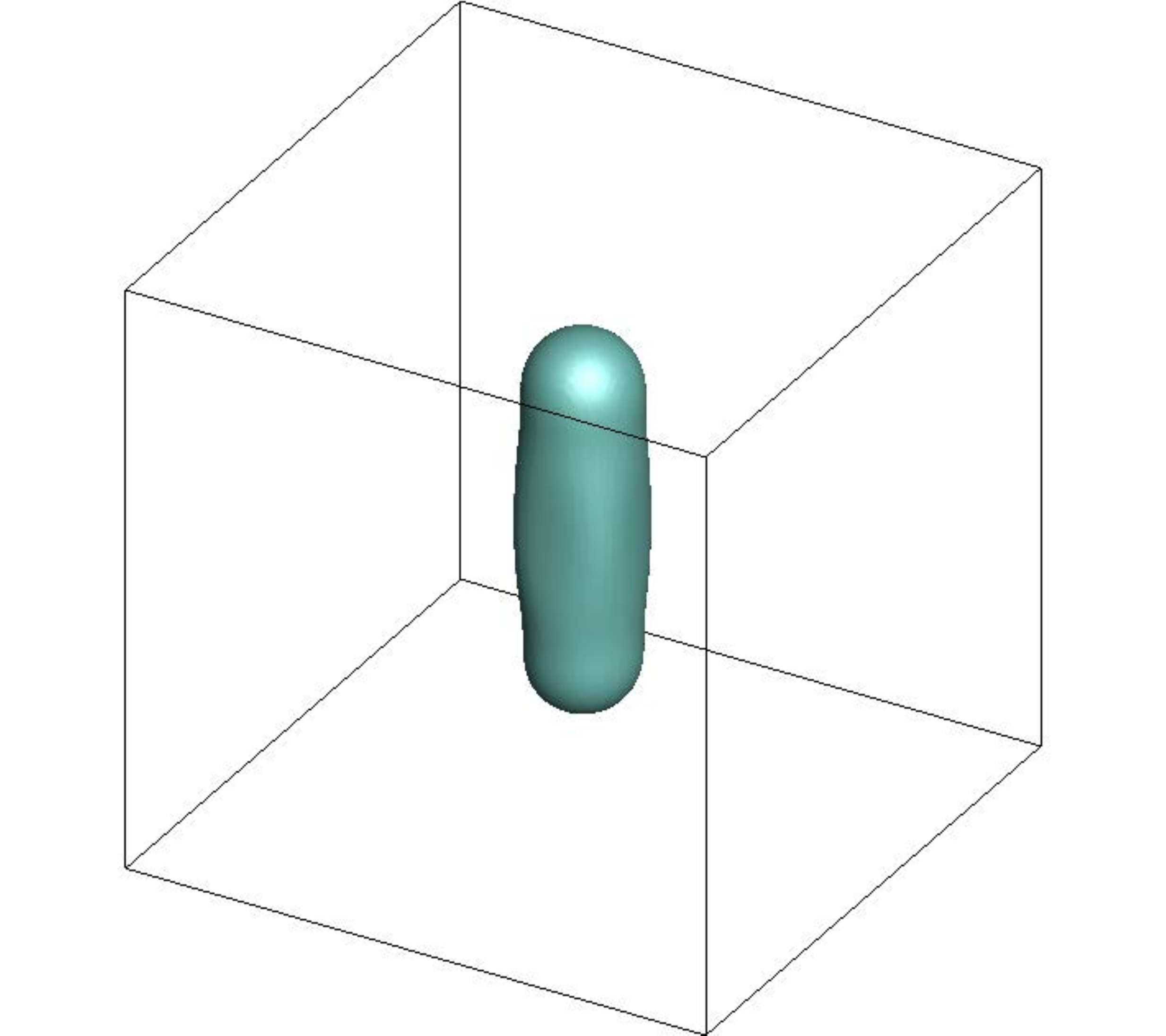}
	\includegraphics[width=0.19\columnwidth,trim={2cm 0cm 2cm 0cm},clip]{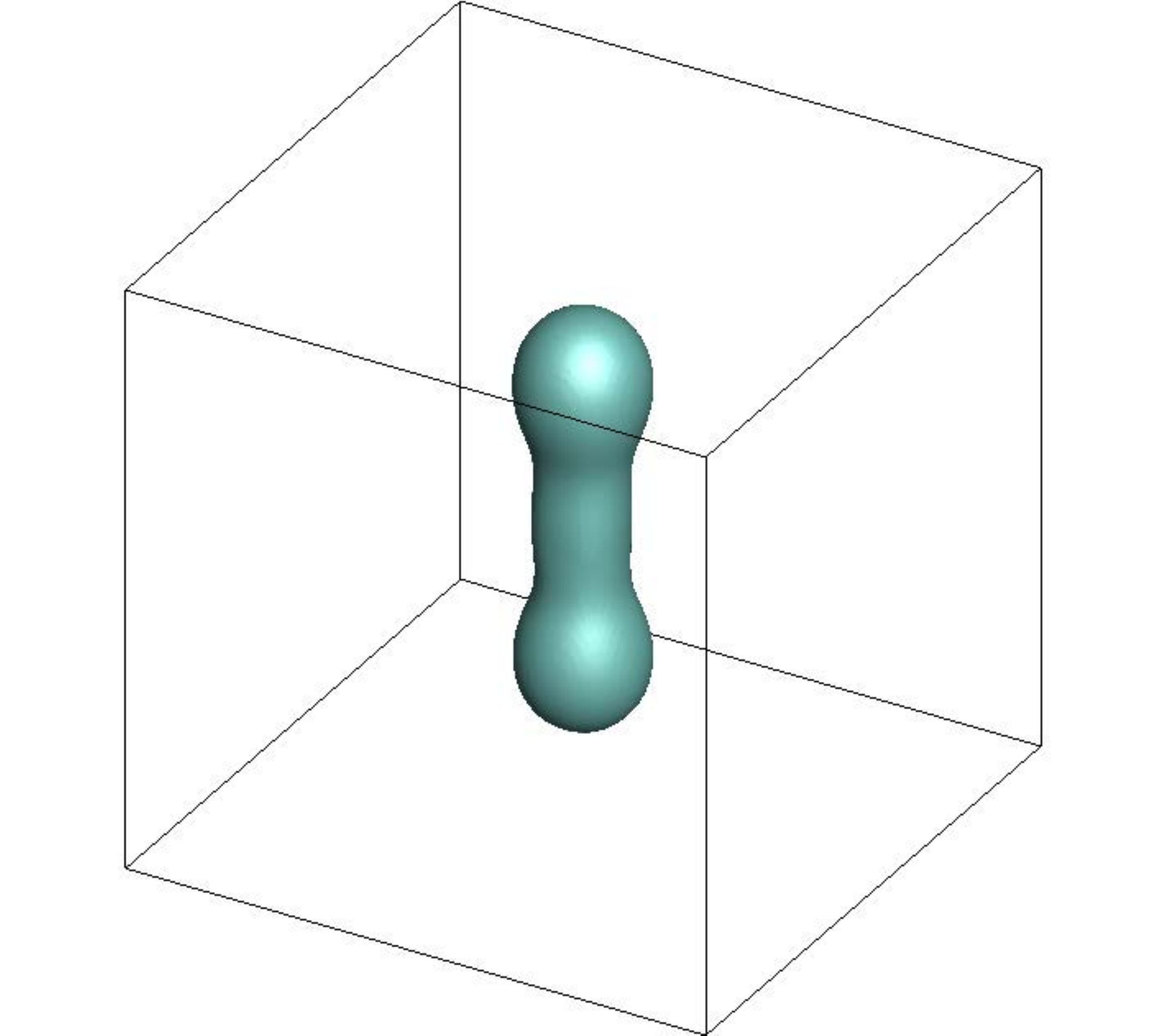}
	\includegraphics[width=0.19\columnwidth,trim={2cm 0cm 2cm 0cm},clip]{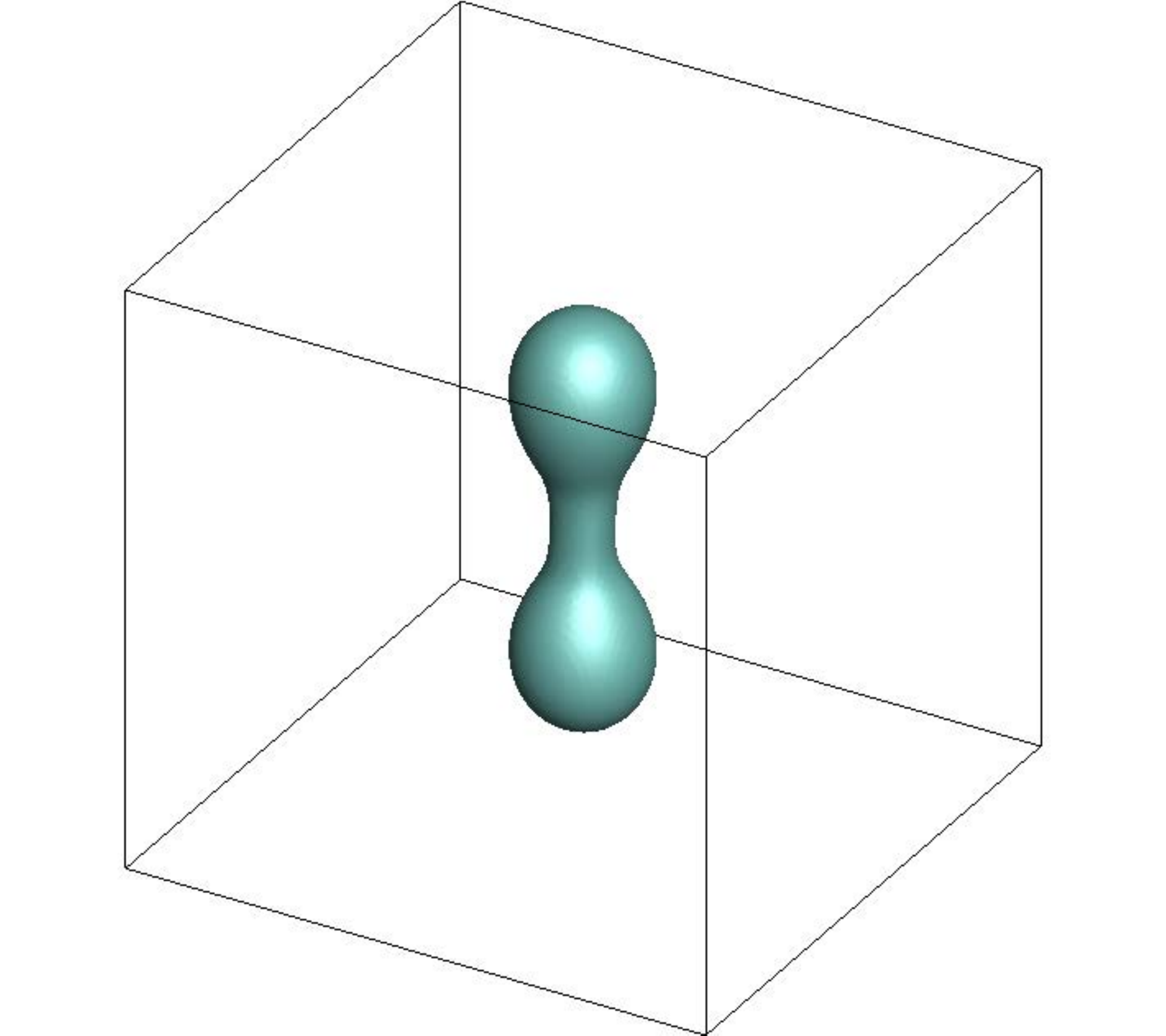}
	\includegraphics[width=0.19\columnwidth,trim={2cm 0cm 2cm 0cm},clip]{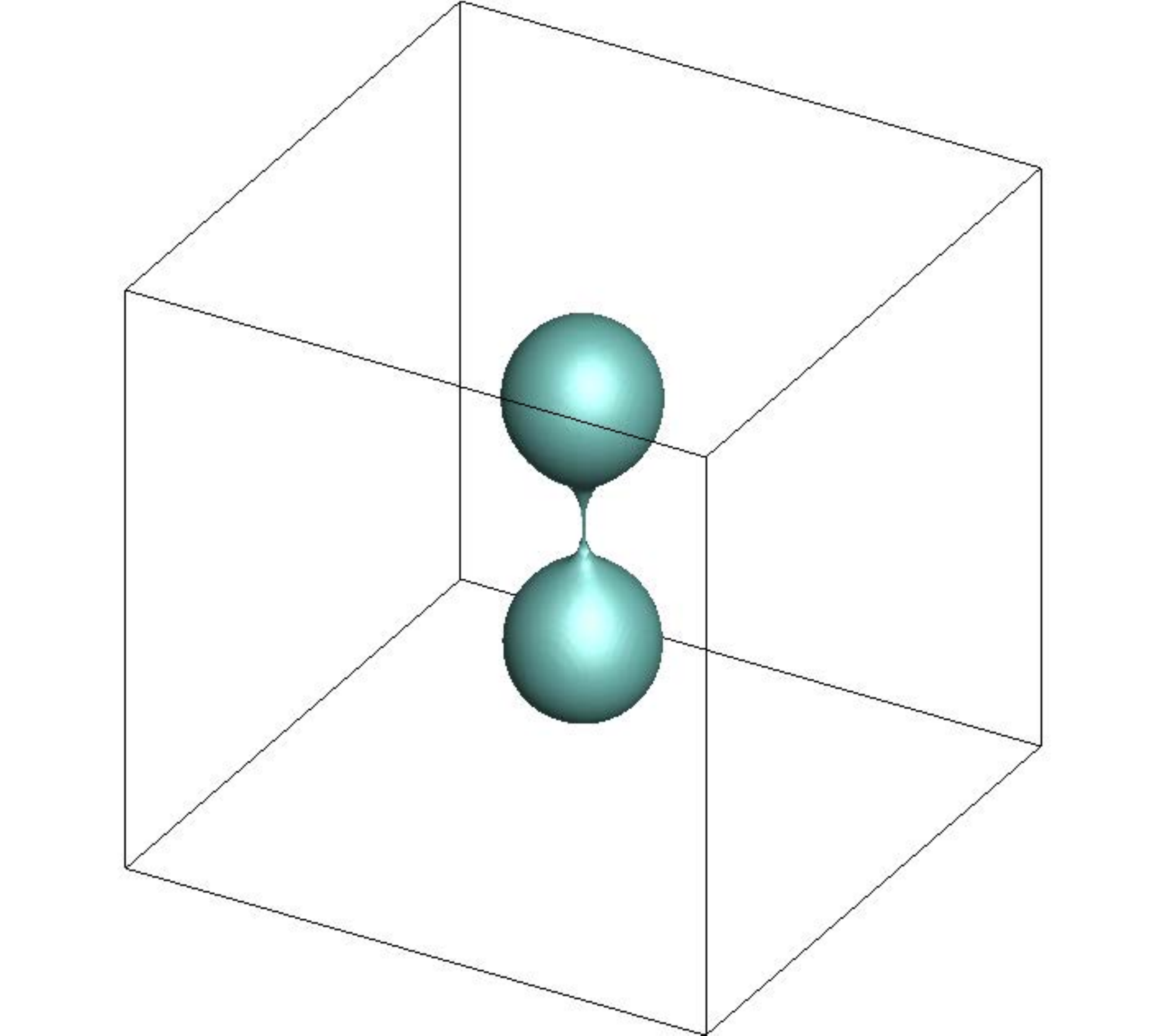}
	\includegraphics[width=0.19\columnwidth,trim={2cm 0cm 2cm 0cm},clip]{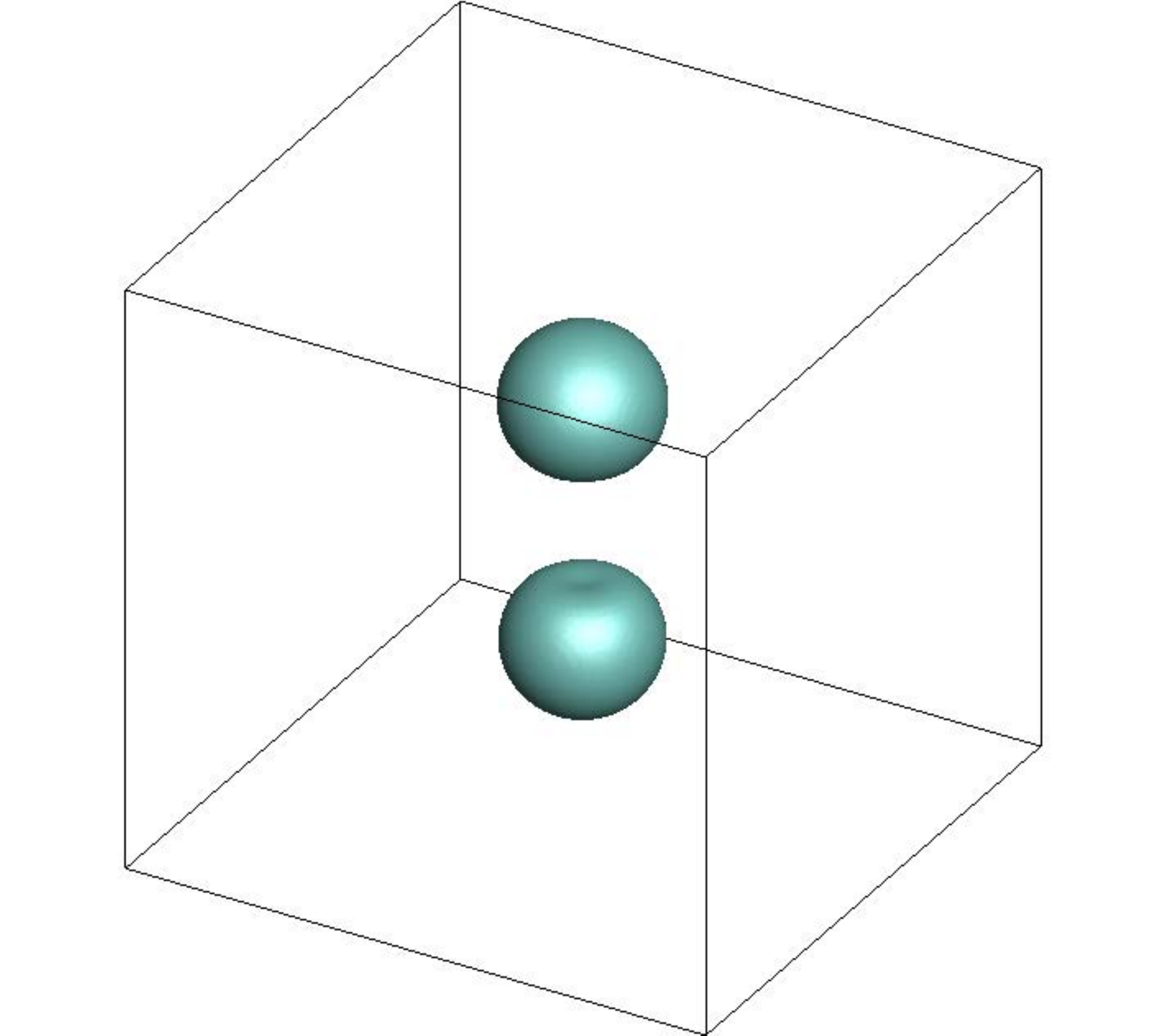}
	\centering
	\caption{Evolution of fluid-fluid interface for head-on collisions of binary droplets. $\rho^*=10$, $Re=1720$, $We=58$. $U_0 t/(2R_0)=0, 1.56, 3.13, 7.03, 10.2, 12.5, 14.8, 16.4, 18.0, 18.8$.}
	\label{fig:DC1em6rhor10}
\end{figure}

\begin{figure}[]
	\centering
	\includegraphics[width=0.19\columnwidth,trim={2cm 0cm 2cm 0cm},clip]{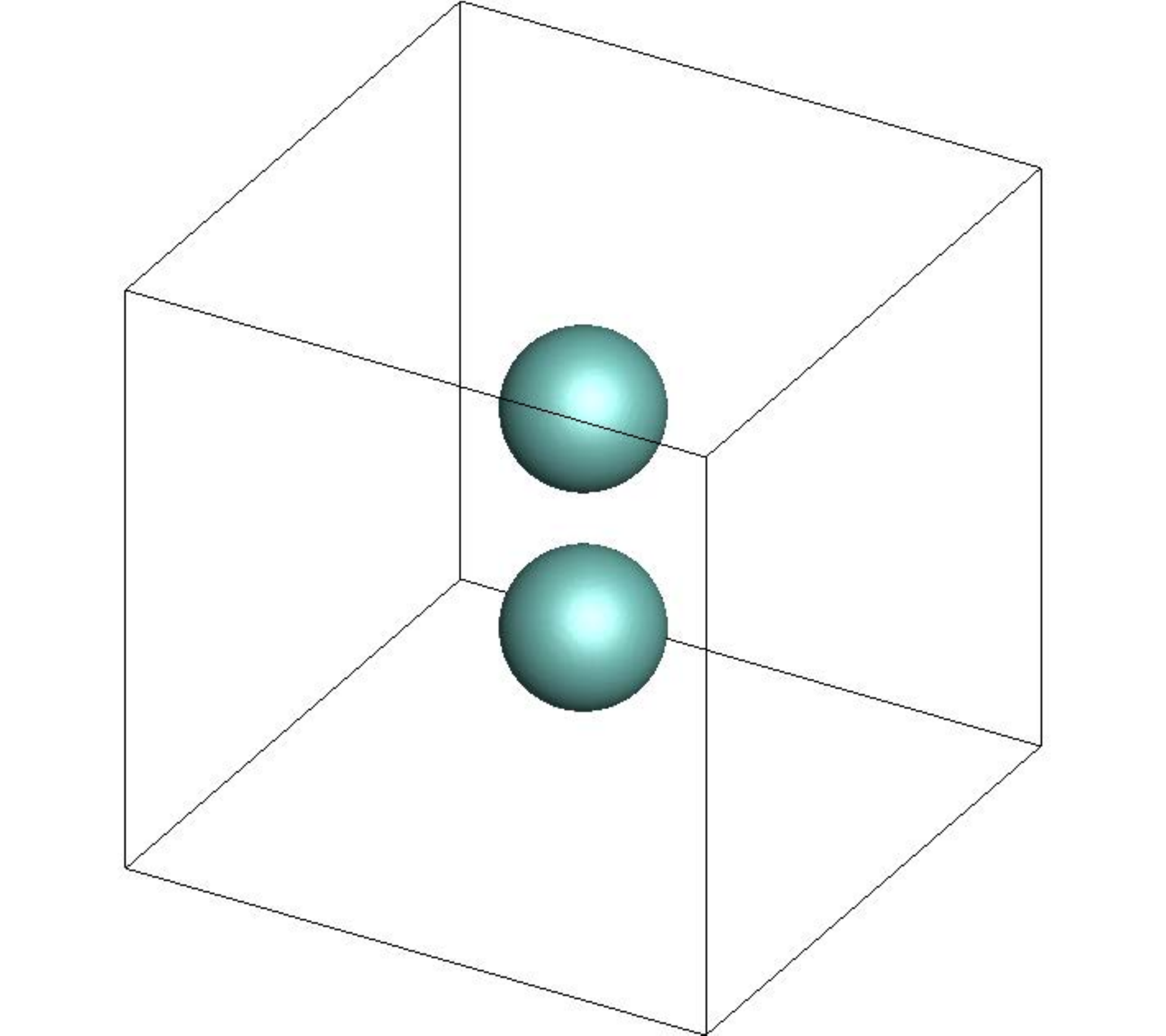}
	\includegraphics[width=0.19\columnwidth,trim={2cm 0cm 2cm 0cm},clip]{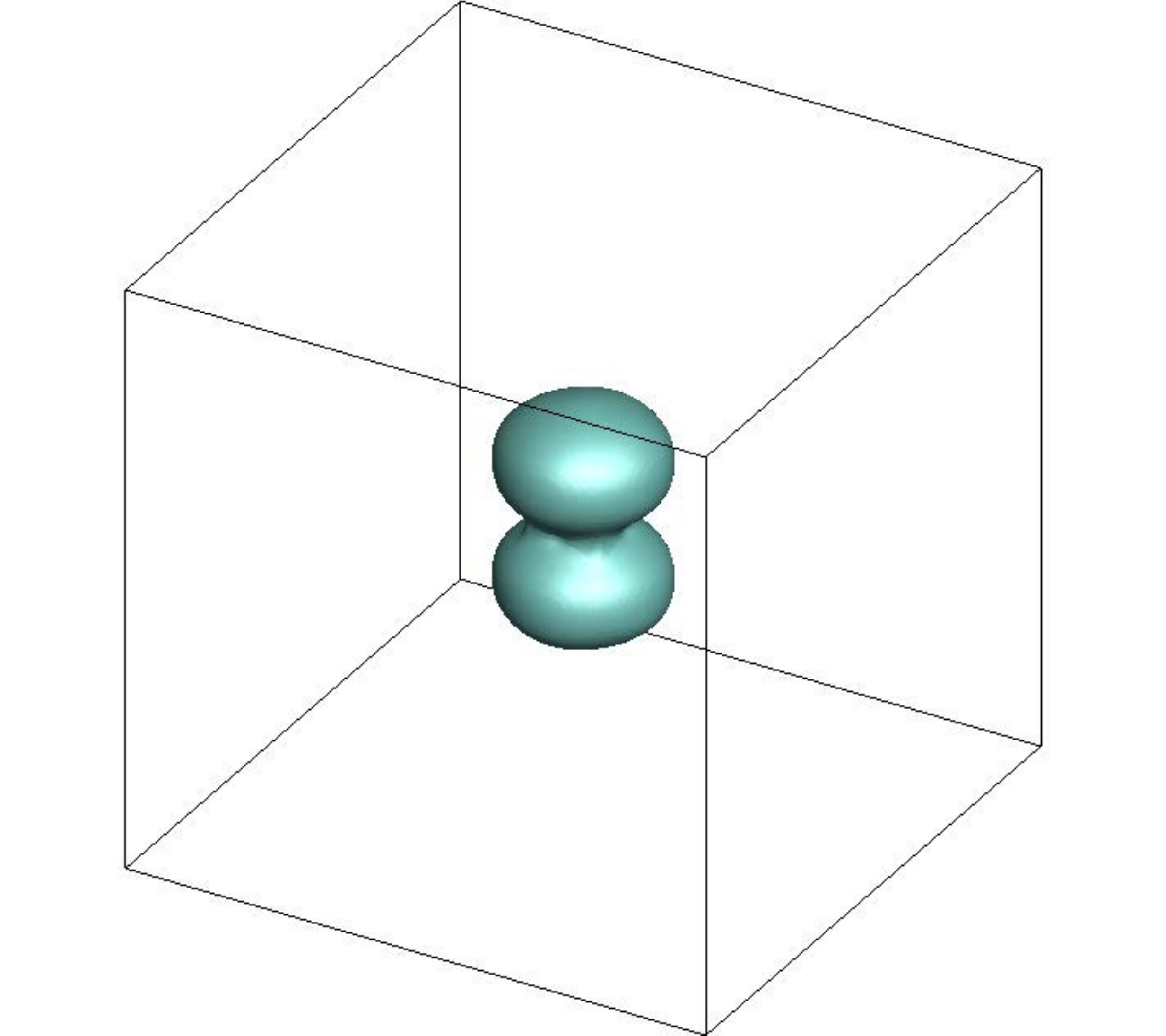}
	\includegraphics[width=0.19\columnwidth,trim={2cm 0cm 2cm 0cm},clip]{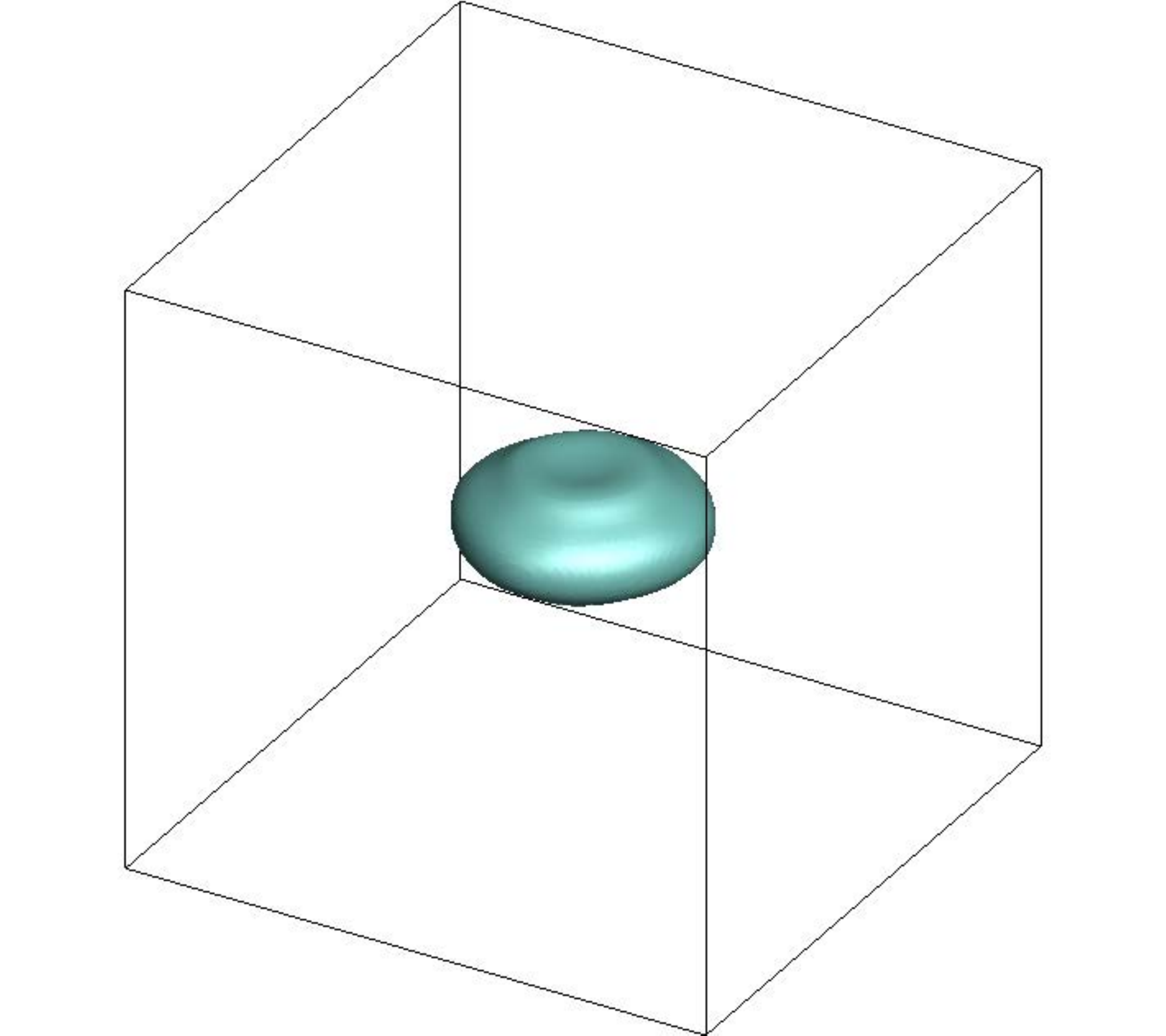}
	\includegraphics[width=0.19\columnwidth,trim={2cm 0cm 2cm 0cm},clip]{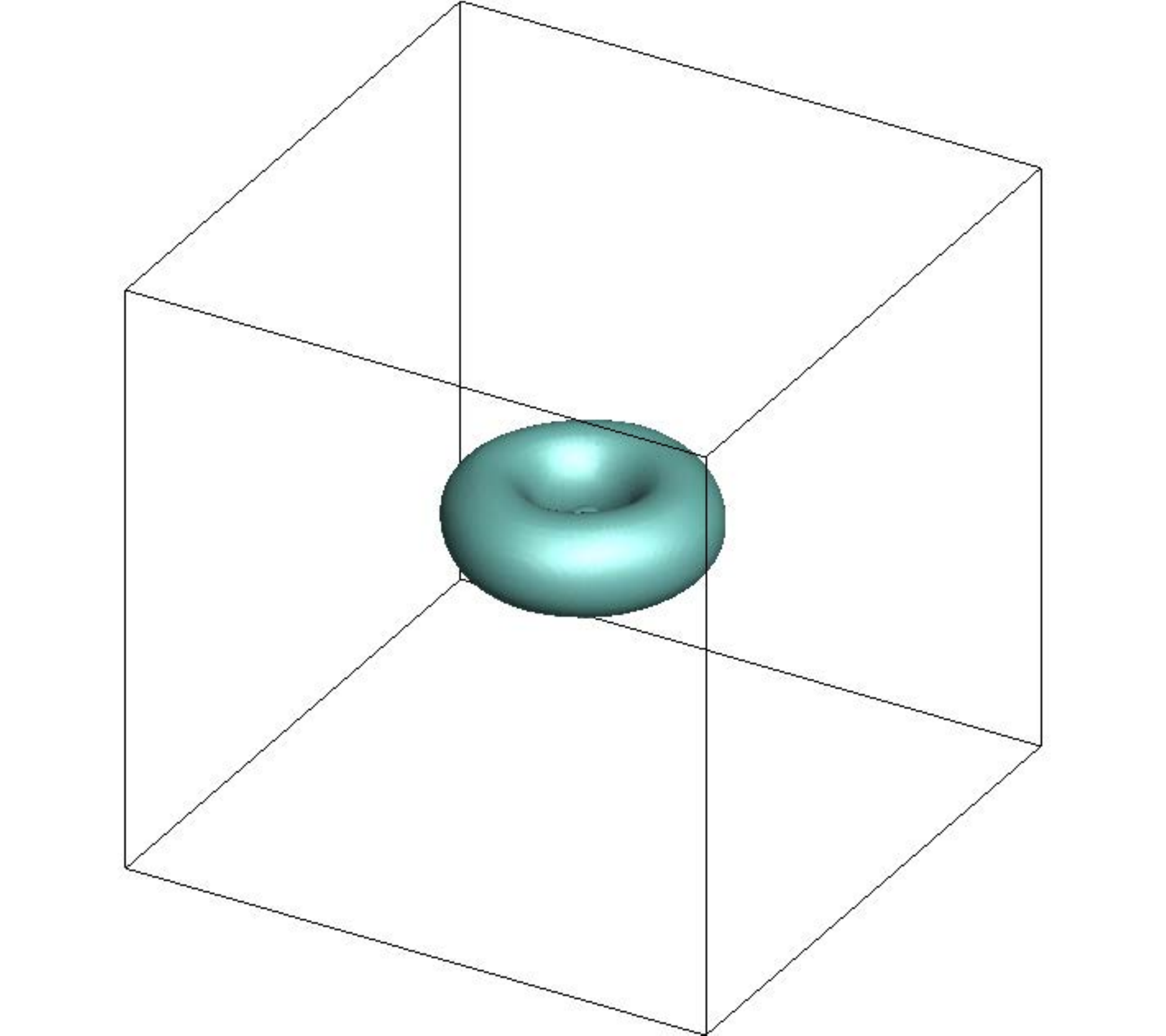}
	\includegraphics[width=0.19\columnwidth,trim={2cm 0cm 2cm 0cm},clip]{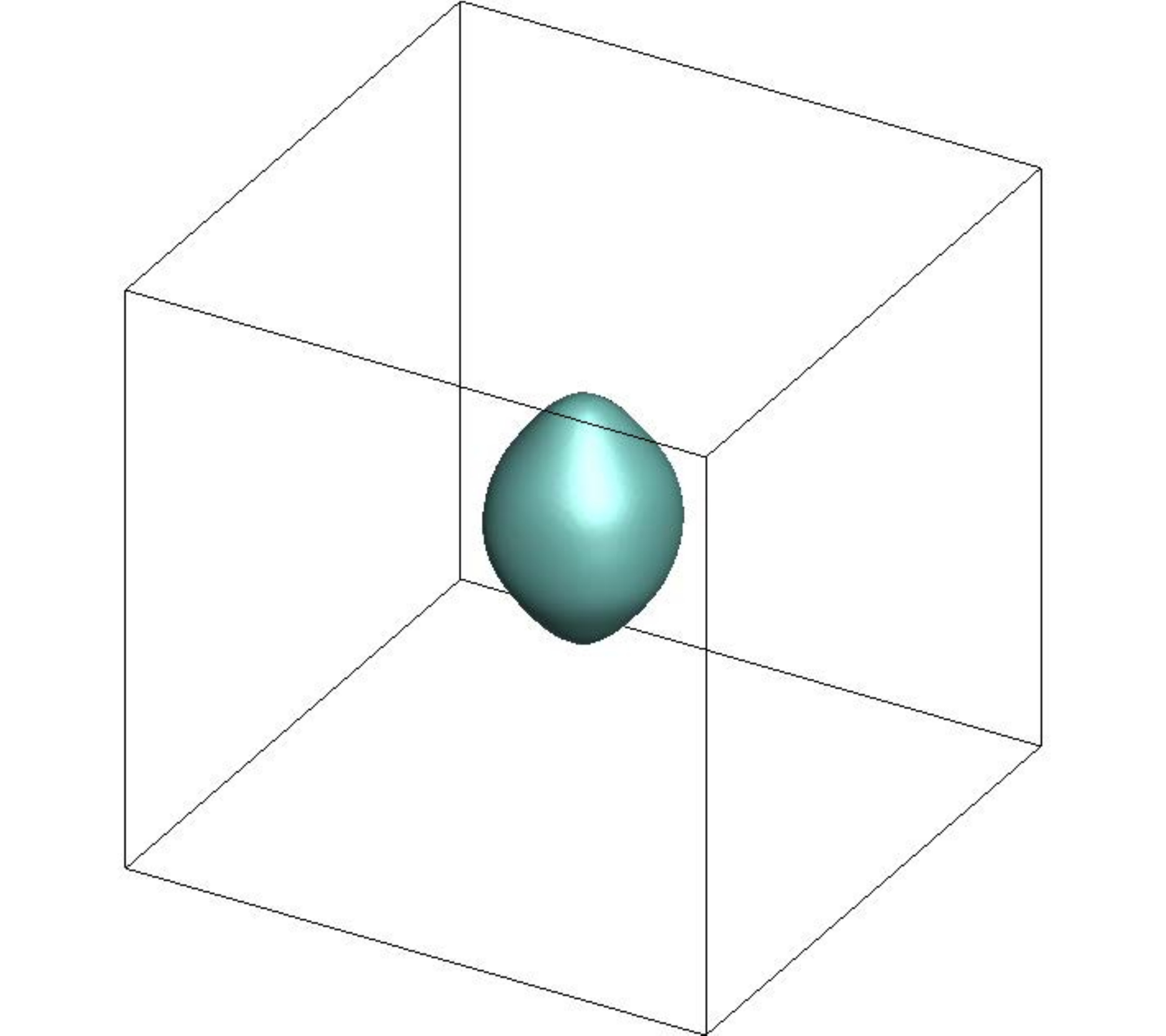}\\
	\includegraphics[width=0.19\columnwidth,trim={2cm 0cm 2cm 0cm},clip]{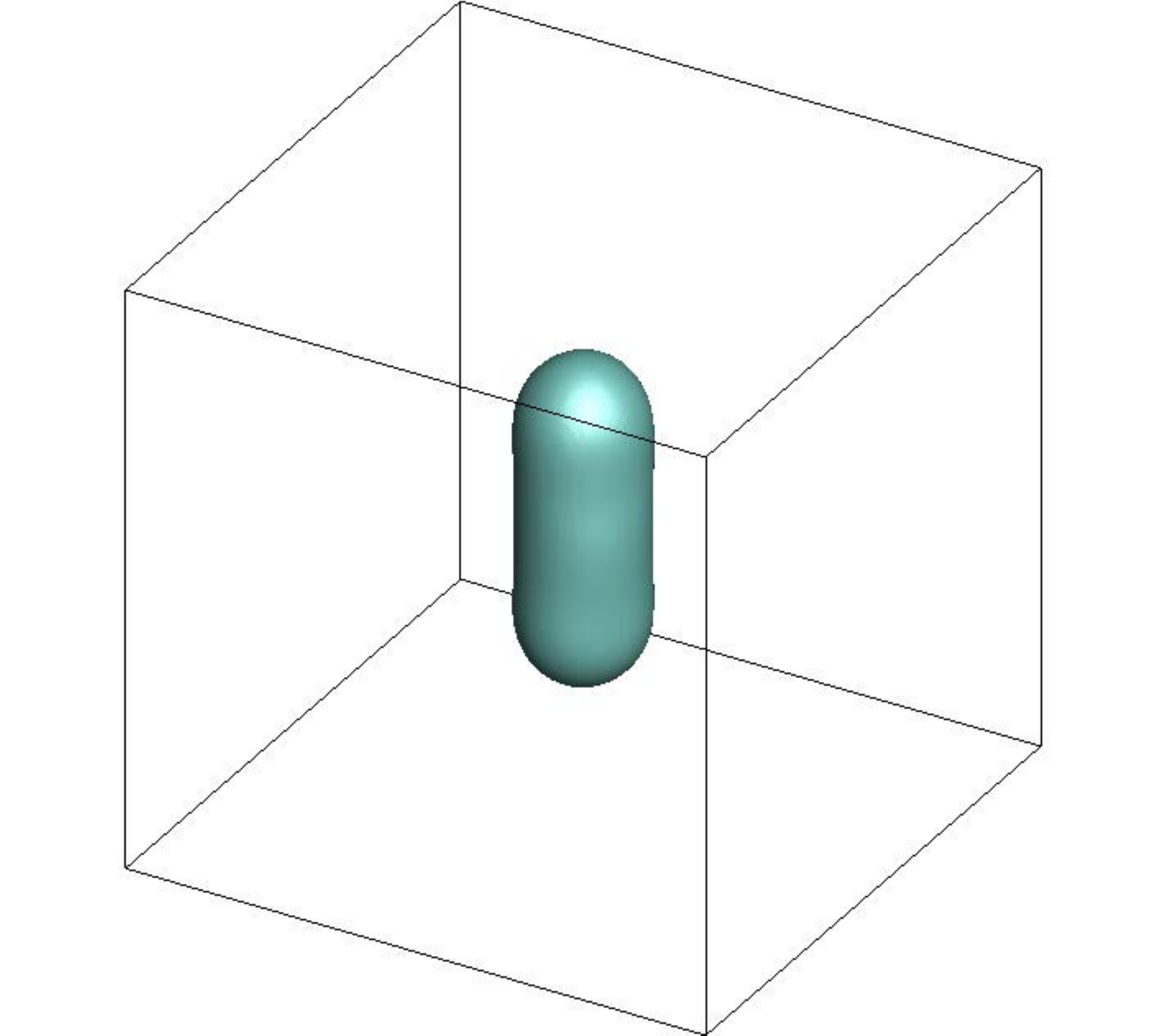}
	\includegraphics[width=0.19\columnwidth,trim={2cm 0cm 2cm 0cm},clip]{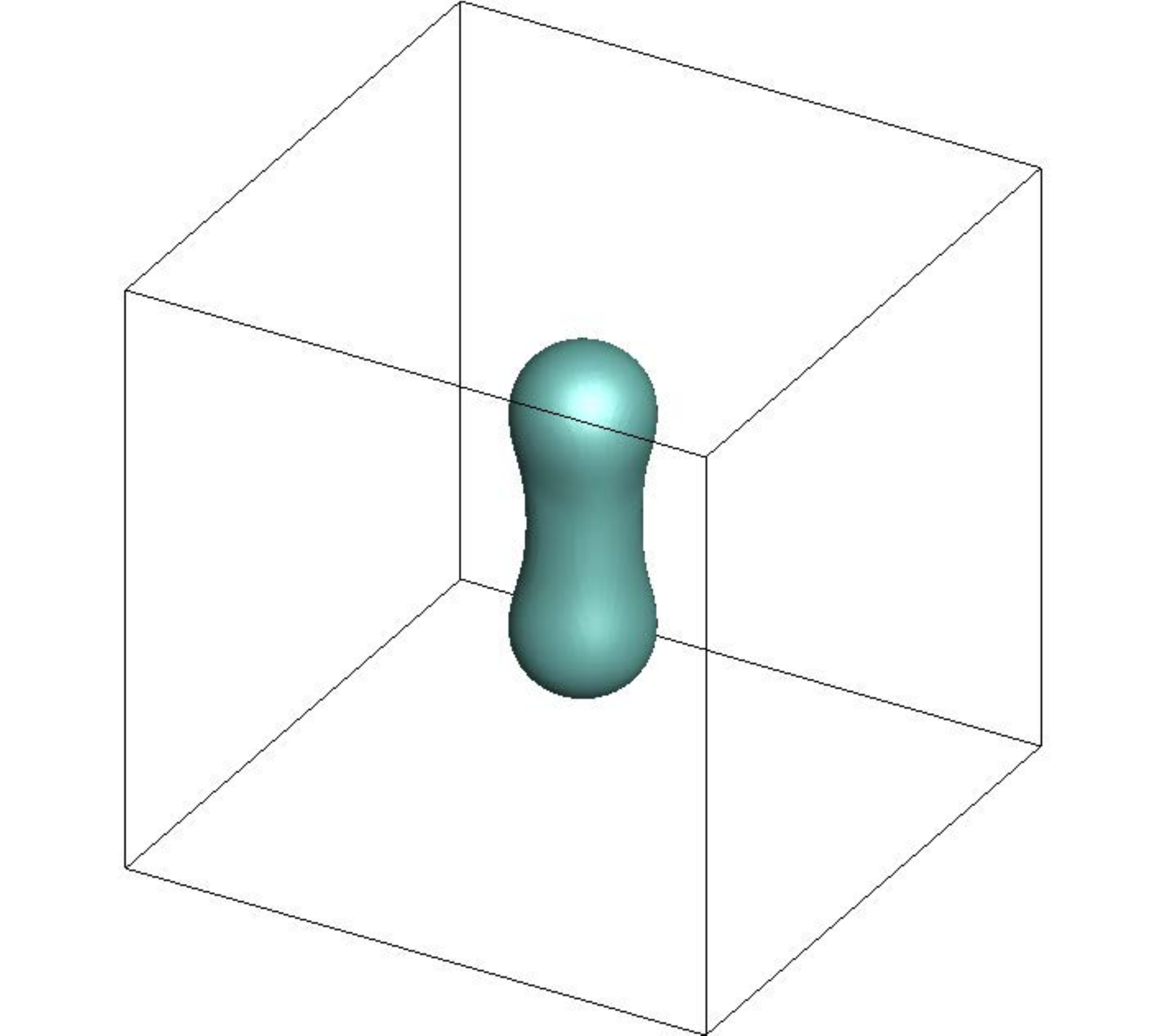}
	\includegraphics[width=0.19\columnwidth,trim={2cm 0cm 2cm 0cm},clip]{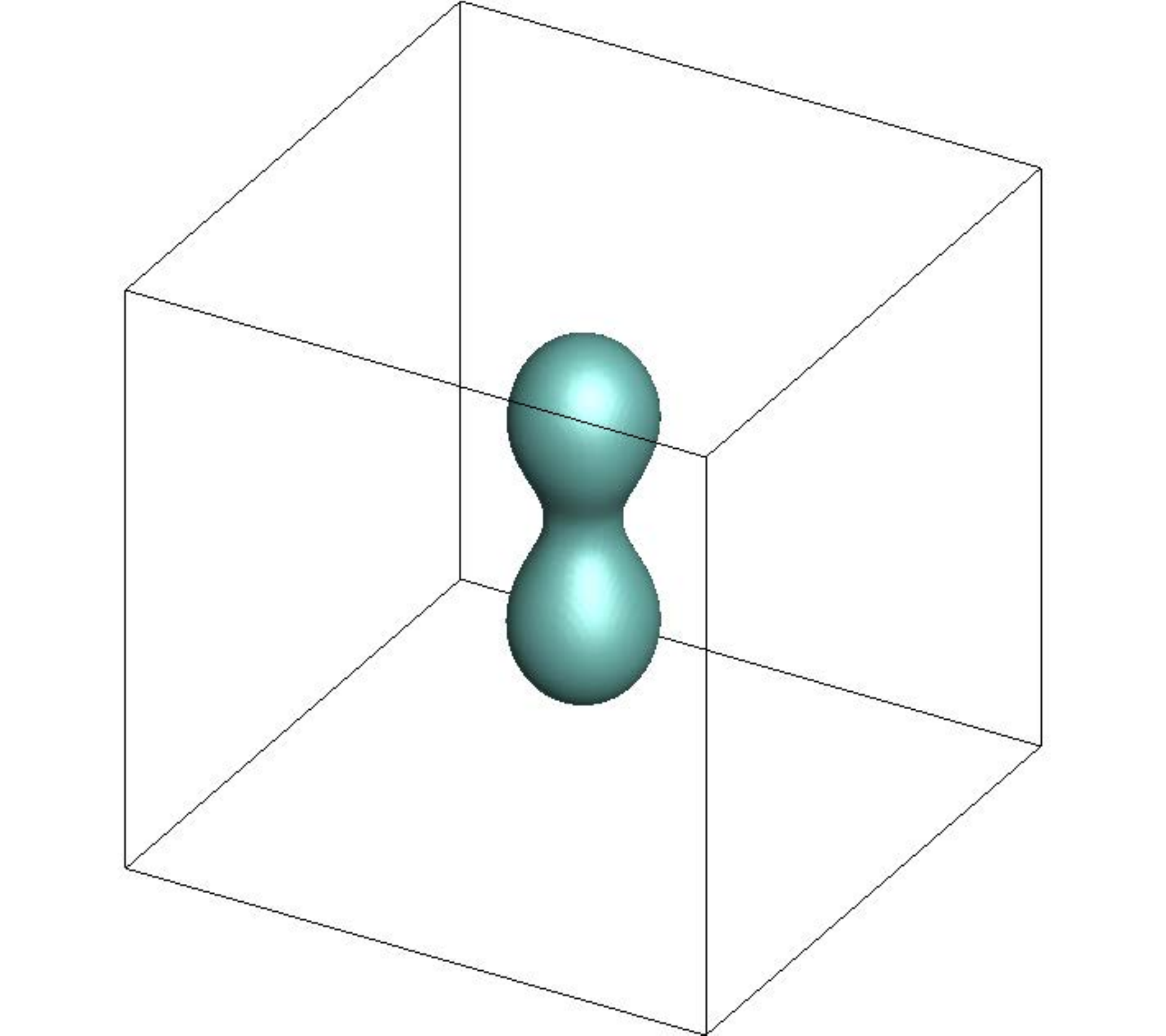}
	\includegraphics[width=0.19\columnwidth,trim={2cm 0cm 2cm 0cm},clip]{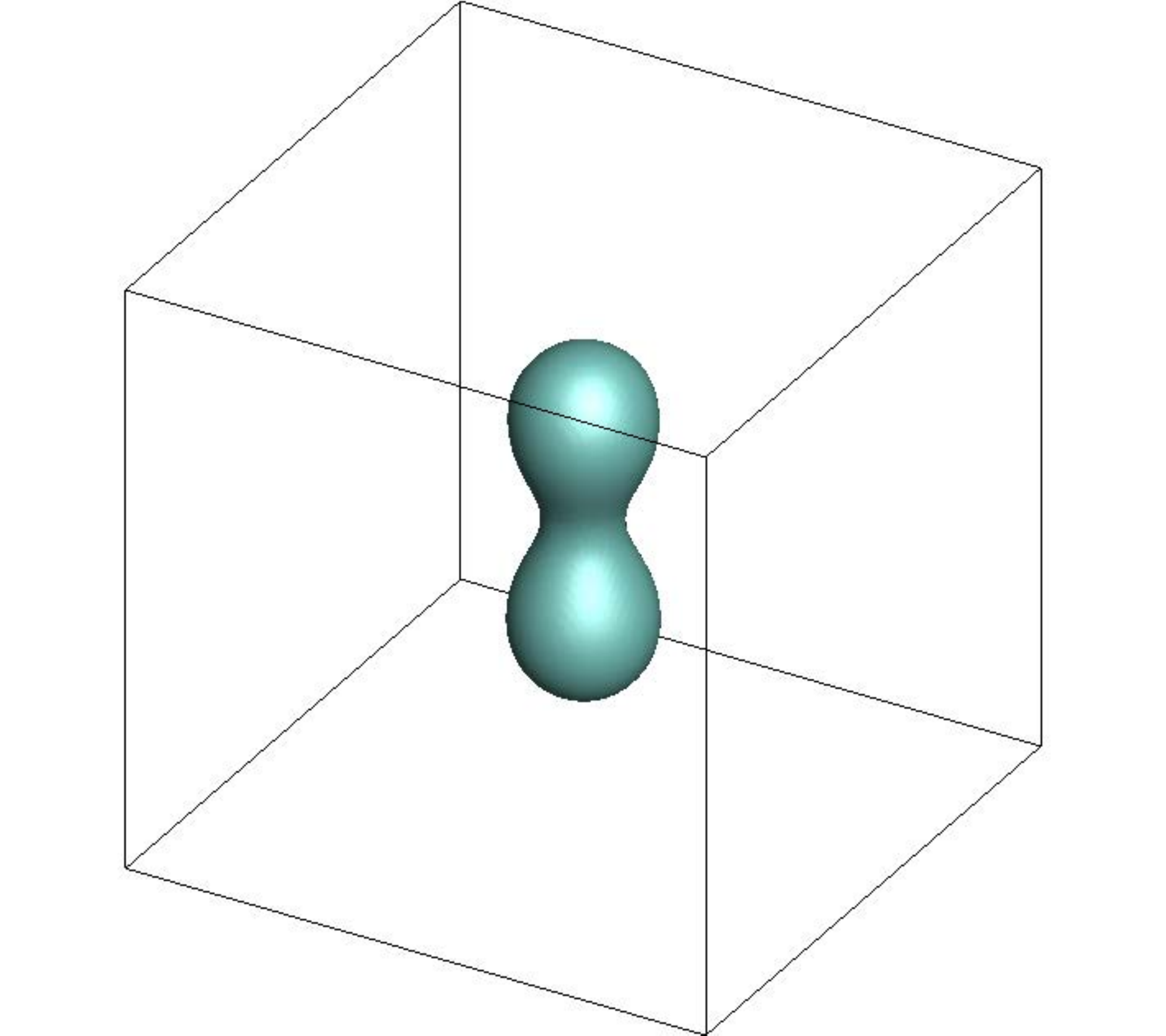}
	\includegraphics[width=0.19\columnwidth,trim={2cm 0cm 2cm 0cm},clip]{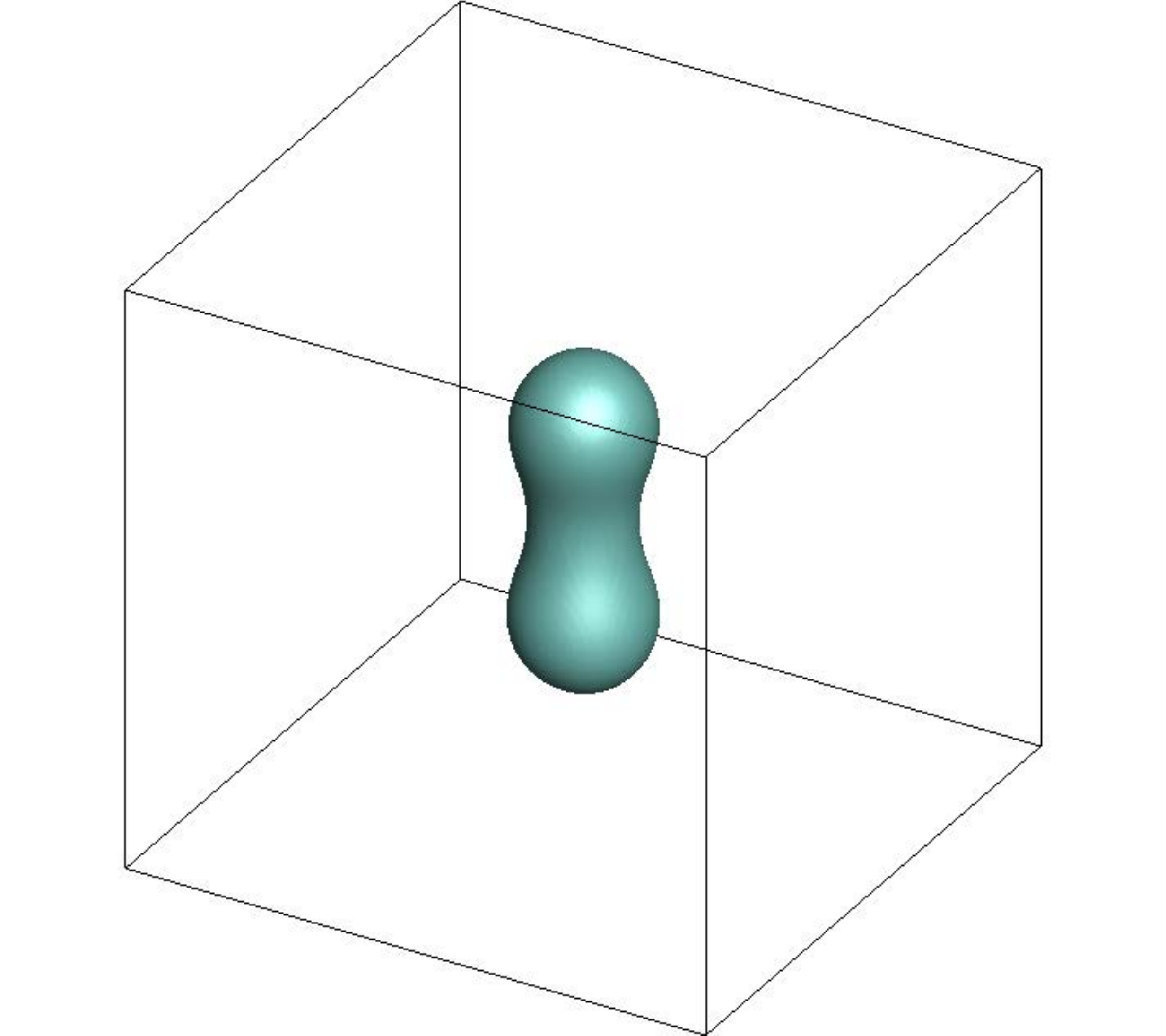}
	\centering
	\caption{Evolution of fluid-fluid interface for head-on collisions of binary droplets. $\rho^*=1$, $Re=1720$, $We=58$. $U_0 t/(2R_0)=0, 1.56, 3.13, 10.2, 13.3, 15.6, 17.2, 20.3, 21.9, 23.4$.}
	\label{fig:DC1em6rhor1}
\end{figure}

Second, we decrease the density ratio to see how it affects the droplets collision process.
The interface dynamics of $\rho^*=100, 10, 1$ are shown in
Figs.~\ref{fig:DC1em6rhor100}-\ref{fig:DC1em6rhor1}, respectively.
For the $\rho^*=100$ case, the general evolution process is almost the same as the $\rho^*=1000$ case, because the density of the background flow is so small that it can be neglected for these two cases. The main difference is that the satellite droplet is much smaller.
For the $\rho^*=10$ case, the density of the background flow cannot be neglected anymore. The impeding effect of background flow field on the droplets appears.  Hence the droplets evolve slightly more slowly, as shown in Fig.~\ref{fig:DC1em6rhor10}, compared to Fig.~\ref{fig:DC1em6rhor1000} or Fig.~\ref{fig:DC1em6rhor100}. For instance, at $U_0 t/(2R_0)=7.03$, the doughnut of the $\rho^*=10$ case is larger than that of the $\rho^*=1000$ or $\rho^*=100$ case. This doughnut will become smaller due to the surface tension force. At $U_0 t/(2R_0)=18.0$, the droplets are still connected with each other for the $\rho^*=10$ case, while they have already separated for the larger-density-ratio cases.
For the $\rho^*=1$ case, the background flow field has obvious influence on the droplets dynamics. The droplets evolve more slowly and we show the large times of this same-density-ratio case (Figs.~\ref{fig:DC1em6rhor1}), compared to the above large density-ratio cases. One major difference for this case is that the droplet would not break up after collision. Should the simulation continues, only one
large droplet would be present.


\subsection{3D Rayleigh-Taylor instability}
In this subsection, 3D RTI is performed to further demonstrate the accuracy and stability of our DUGKS-PF(AC)-WENO approach in solving more complex interfacial flows.
The computational domain is $[0,4L]\times[0,L]\times[0,L]$ with $L=128$. 
The boundary condition is periodic on the lateral
boundaries, while the bounce-back scheme is implemented on the top and bottom walls.
Initially, the two-phase interface is located at 
\begin{equation}
	x(y,z)=2L+0.05L\left[ \cos\left(\frac{2\pi y}{L} \right)+\cos\left(\frac{2\pi z}{L} \right) \right].
\end{equation}
The reference time is defined as $T=\sqrt{L/g}$ in the 3D RTI simulations, for a comparison with the results of He {\it et al.}~\cite{1999On} and Wang {\it et al.}.~\cite{WANG201541}
The definitions of other parameters are the same as those in subsection~\ref{subsec: 2DRTI}.

\begin{figure}[]
	\centering
	\includegraphics[width=0.18\columnwidth,trim={2cm 0cm 2cm 0cm},clip]{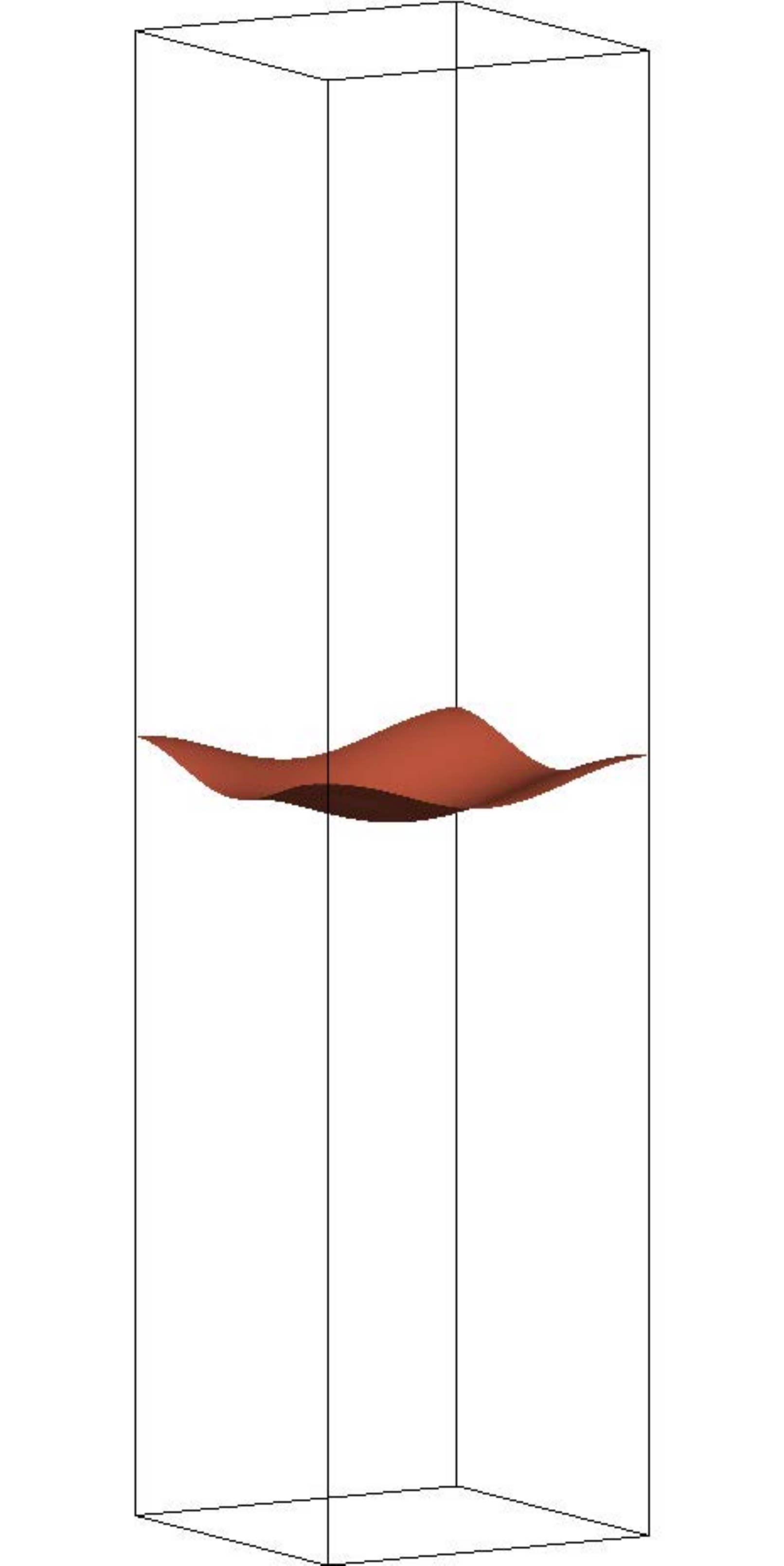}
	\includegraphics[width=0.18\columnwidth,trim={2cm 0cm 2cm 0cm},clip]{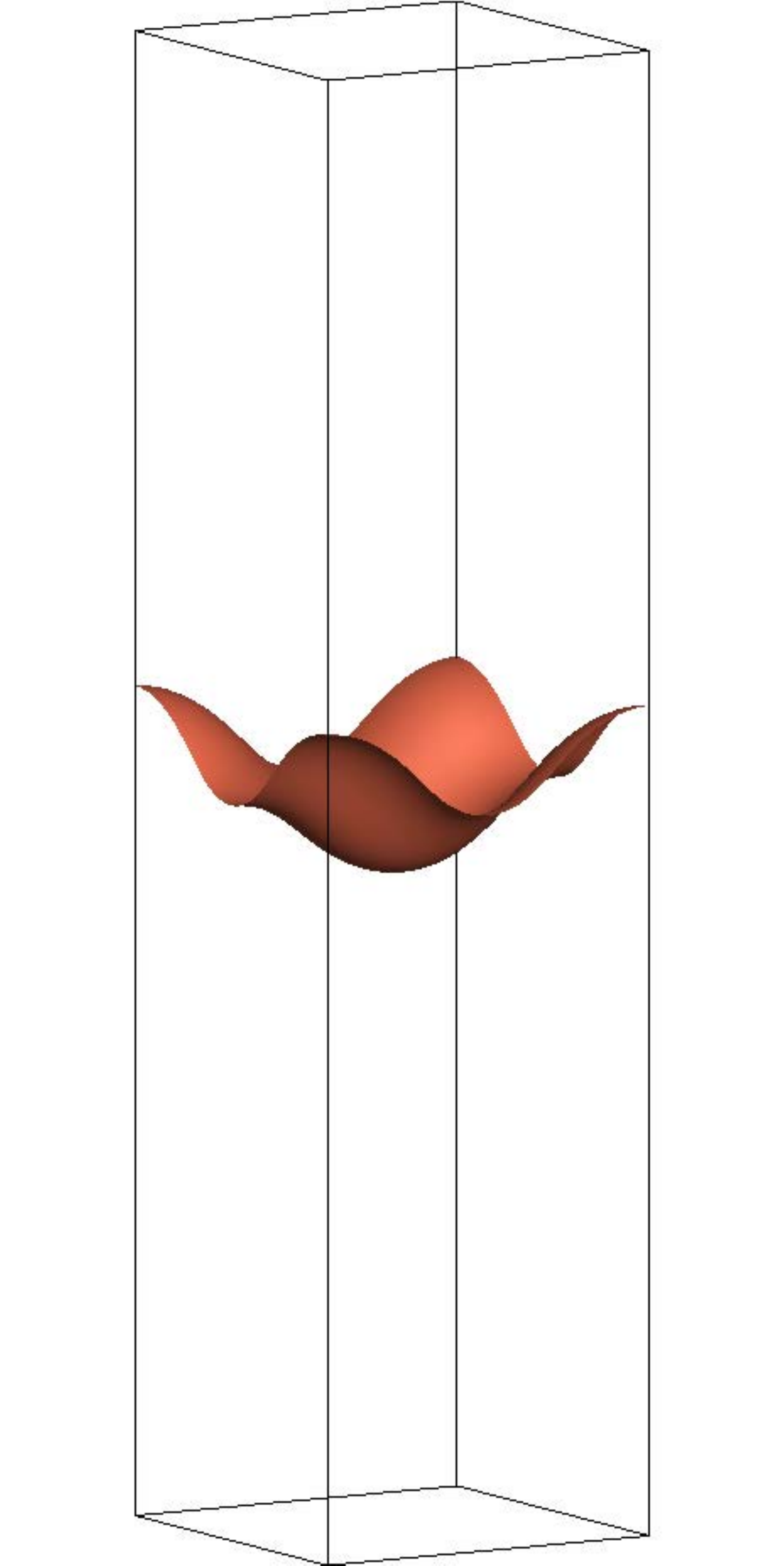}
	\includegraphics[width=0.18\columnwidth,trim={2cm 0cm 2cm 0cm},clip]{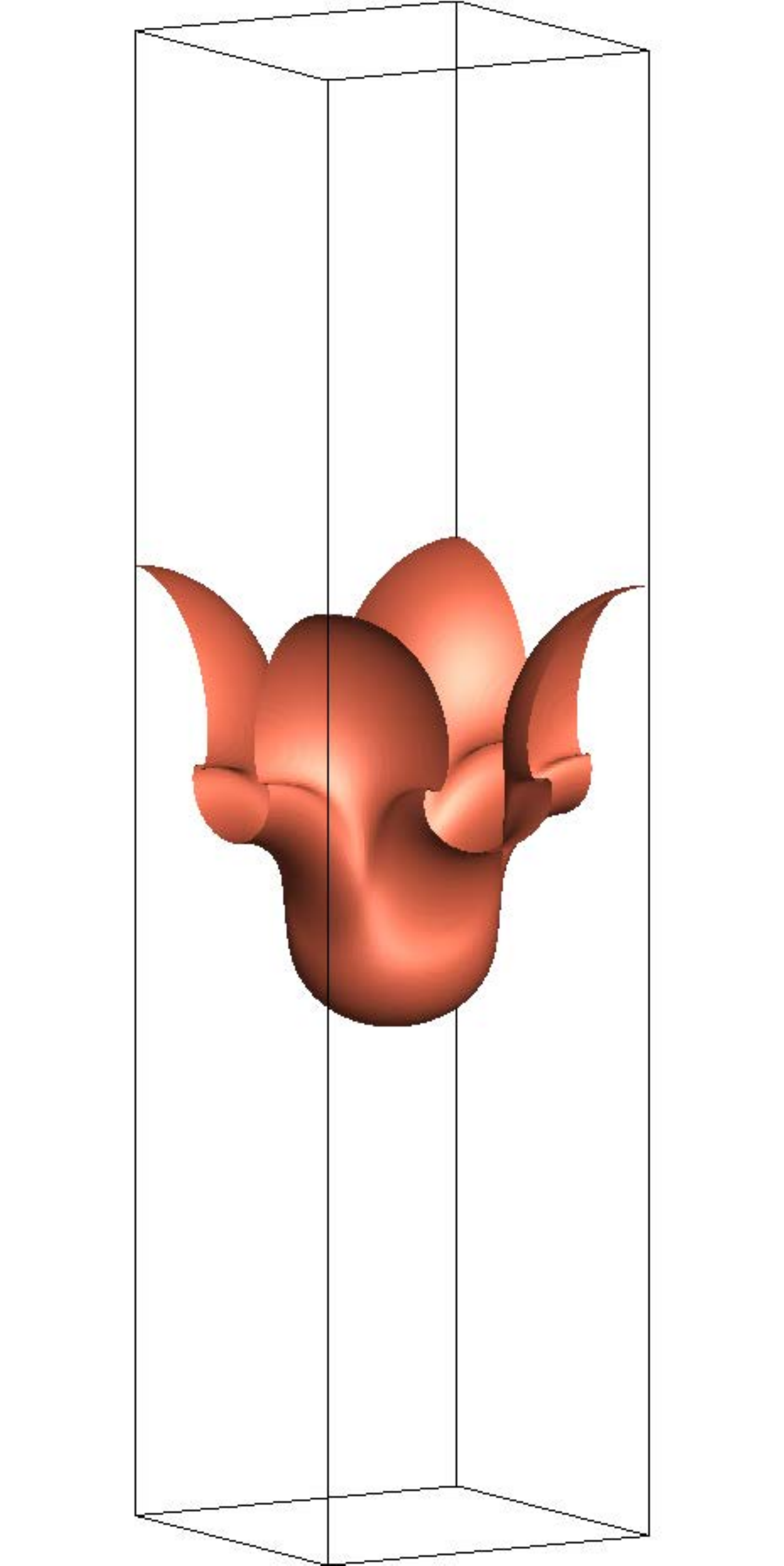}
	\includegraphics[width=0.18\columnwidth,trim={2cm 0cm 2cm 0cm},clip]{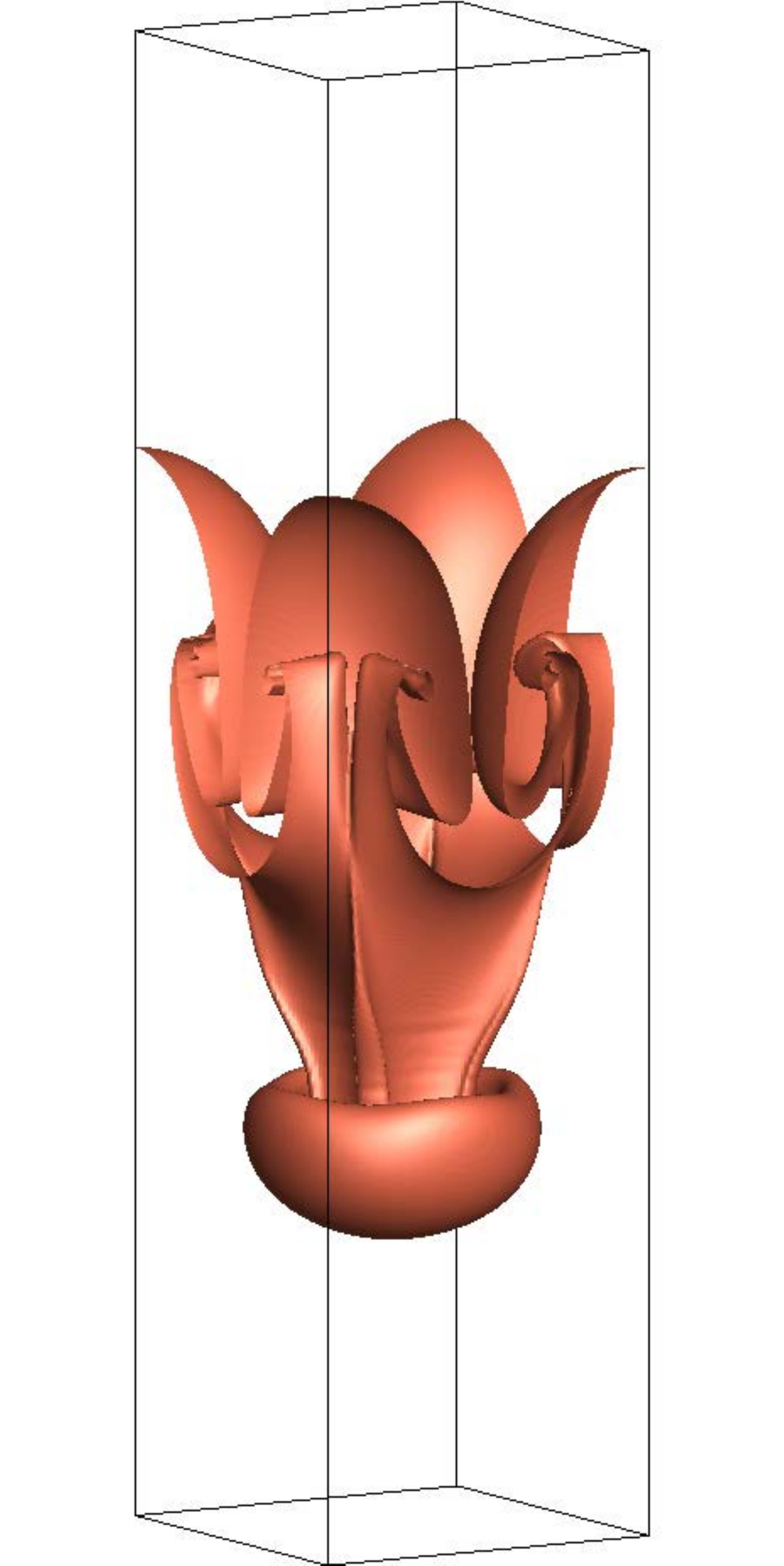}
	\includegraphics[width=0.18\columnwidth,trim={2cm 0cm 2cm 0cm},clip]{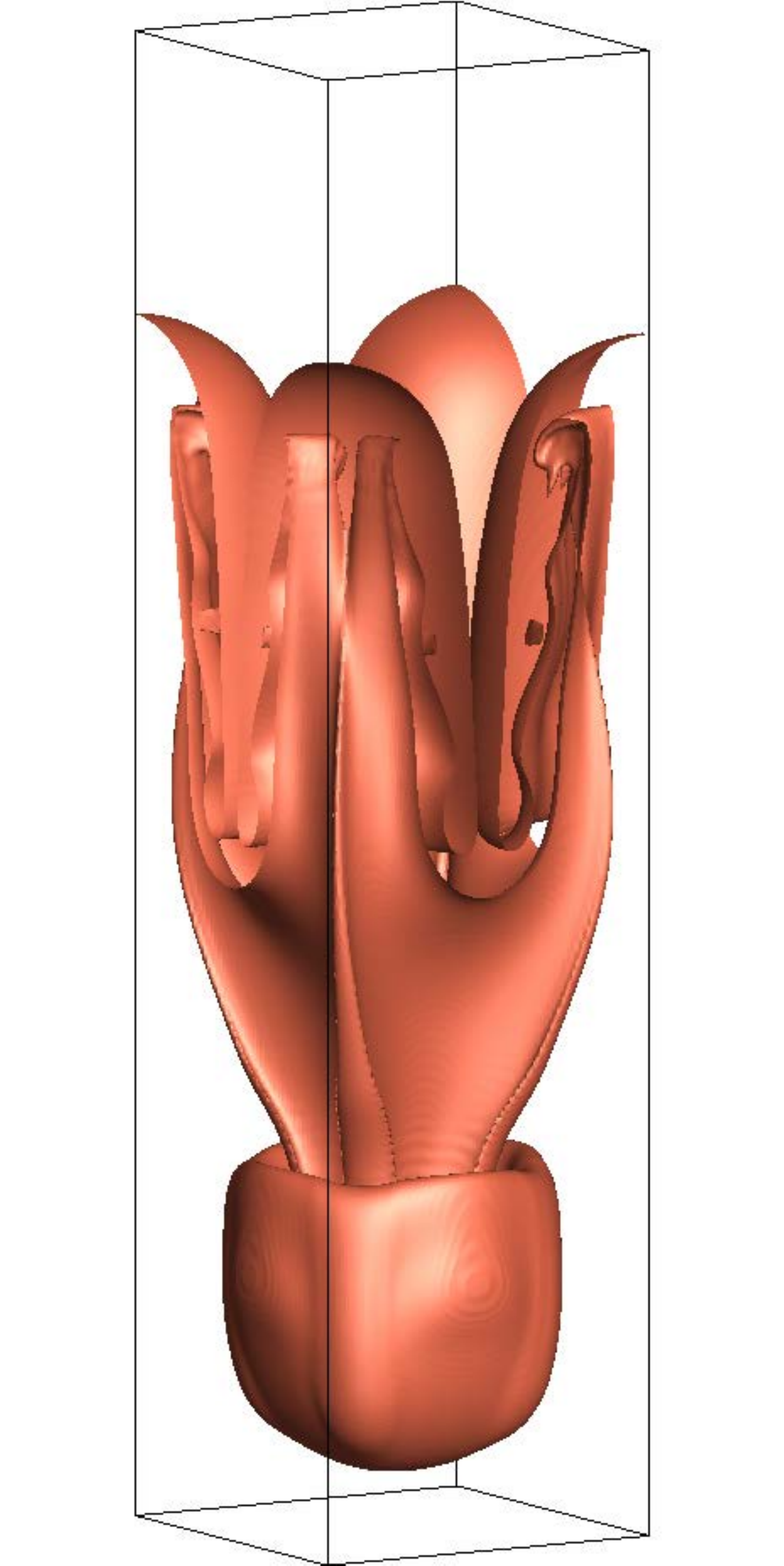}\\	
	\includegraphics[width=0.12\columnwidth,trim={0cm 4cm 0cm 0cm},clip]{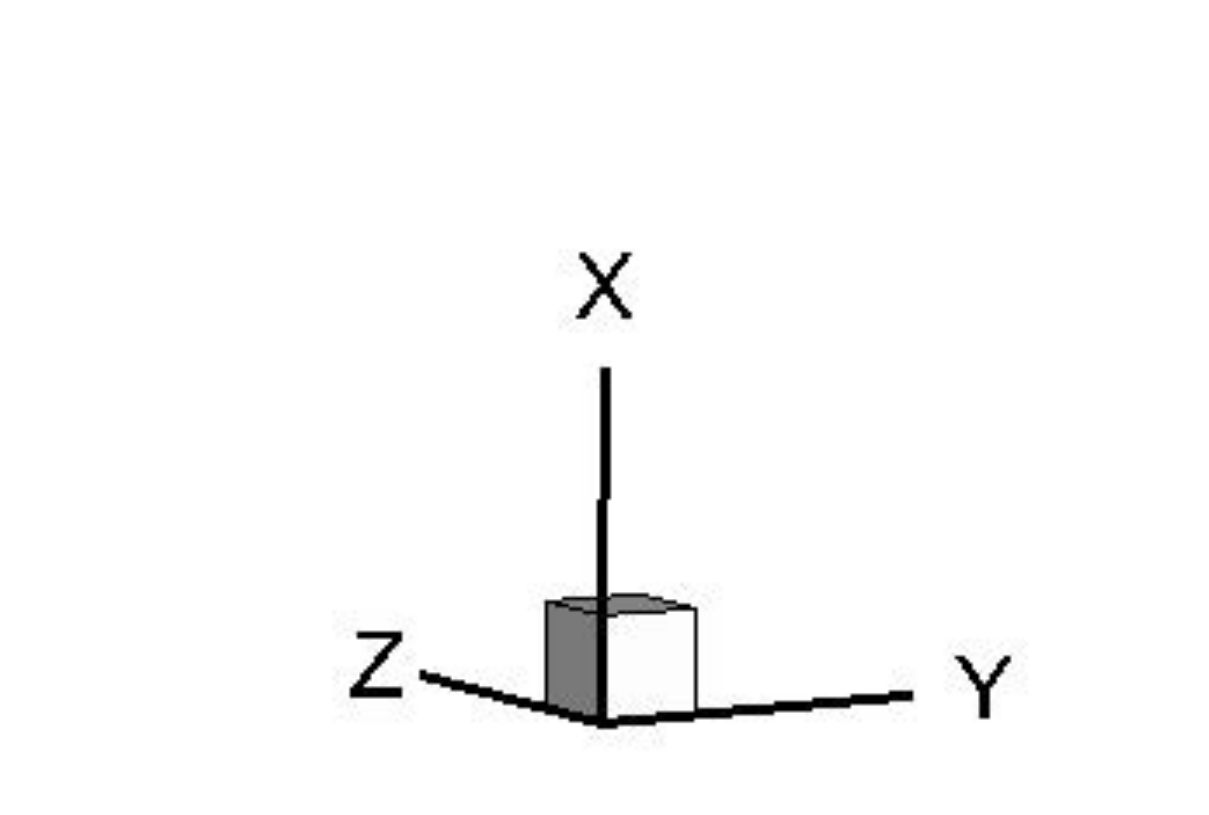}
	\centering
	\caption{Time evolution of fluid-fluid interface for 3D RTI at $At=0.5$ and $Re=1024$. $t^*=0,1,2,3,4$.}
	\label{fig:RT3Dinterface05}
\end{figure}

\begin{figure}[]
	\centering
	\includegraphics[width=0.5\columnwidth,trim={0cm 0cm 0cm 0cm},clip]{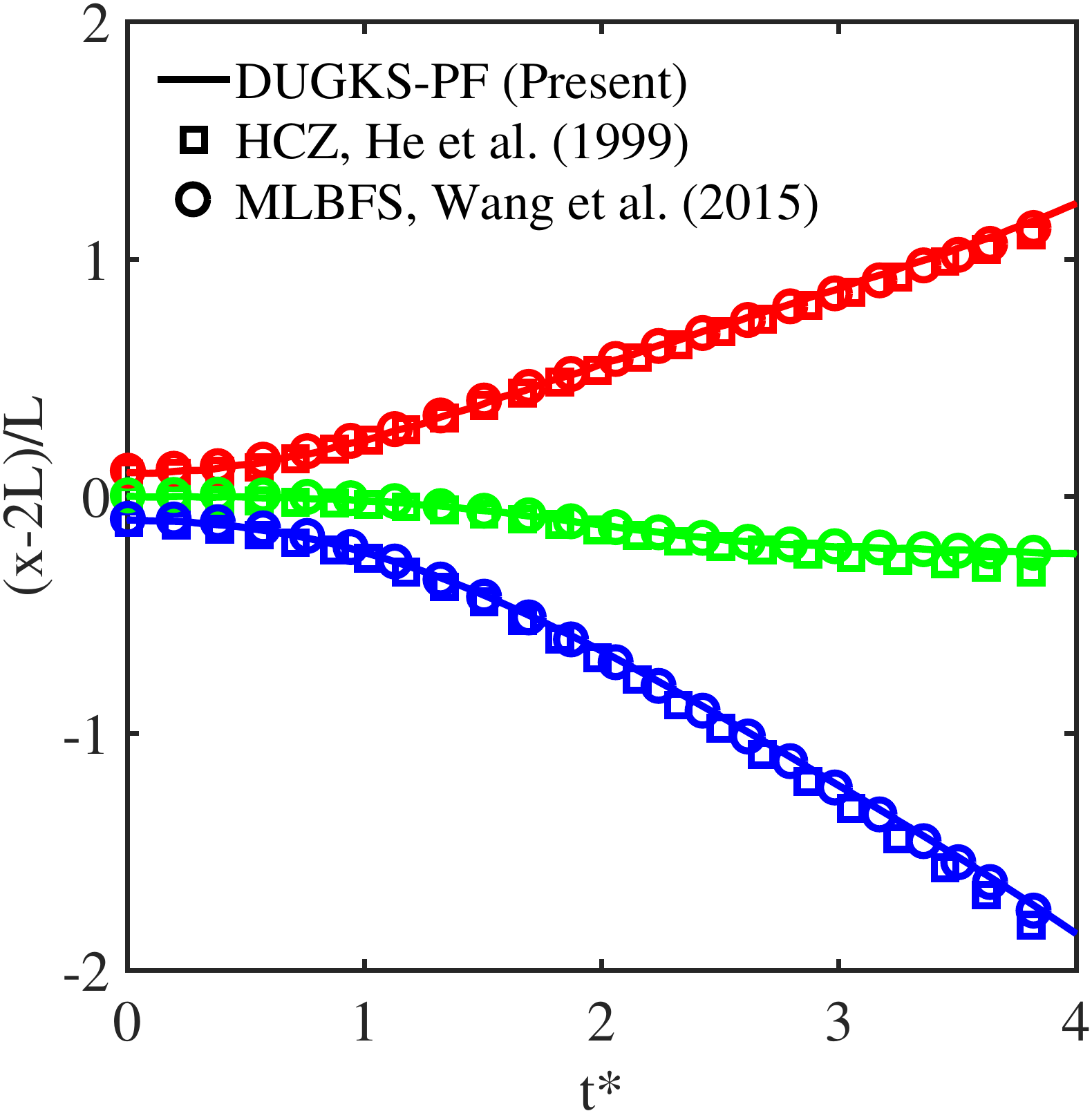}
	\centering
	\caption{Time evolution of the positions of bubble front (red), saddle point (green), and spike tip (blue) for 3D RTI at $At=0.5$ and $Re=1024$.}
	\label{fig:Timeof3DRT05}
\end{figure}

\begin{figure}[]
	\centering
	\includegraphics[width=0.18\columnwidth,trim={0cm 0cm 0cm 0cm},clip]{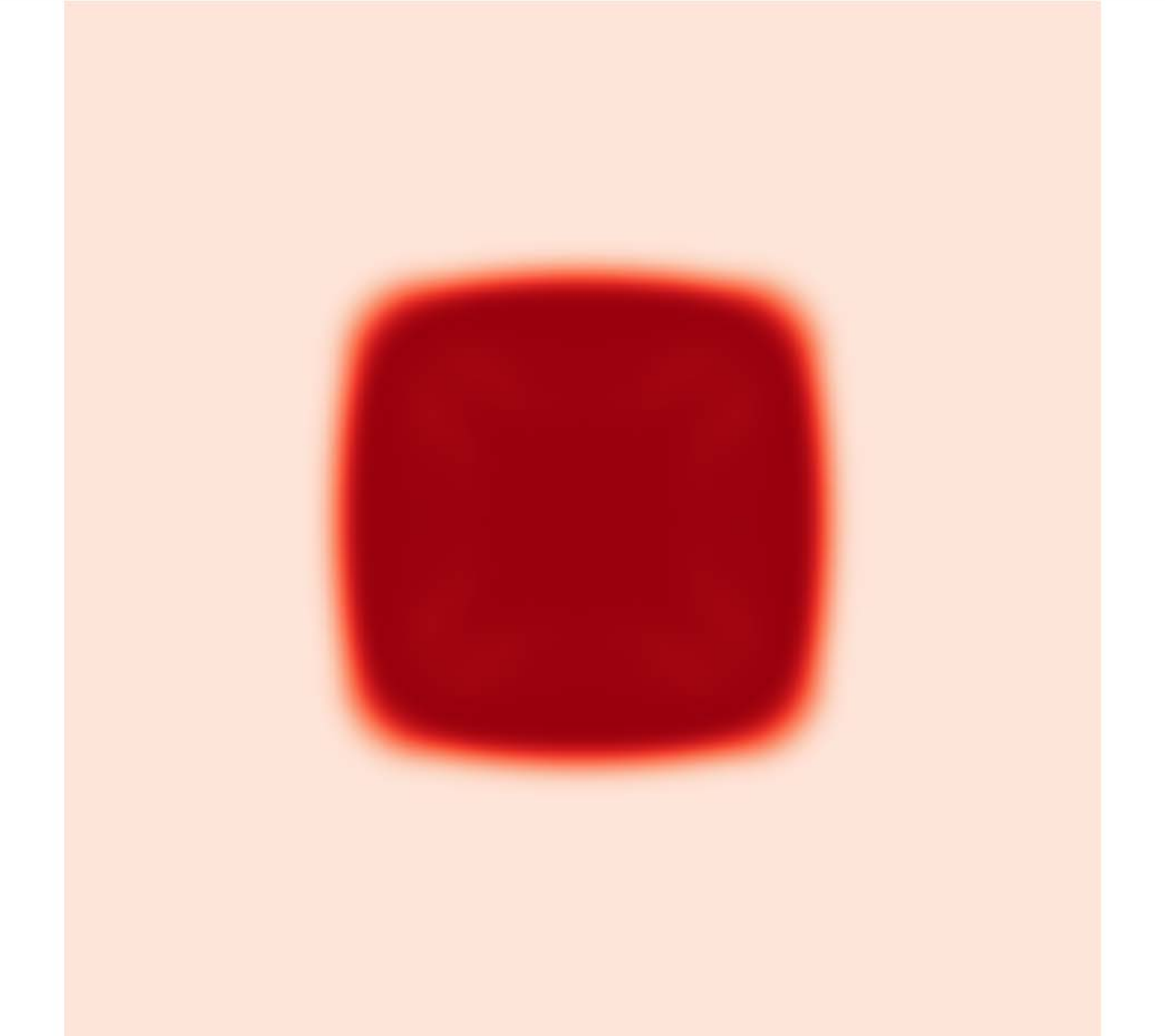}
	\includegraphics[width=0.18\columnwidth,trim={0cm 0cm 0cm 0cm},clip]{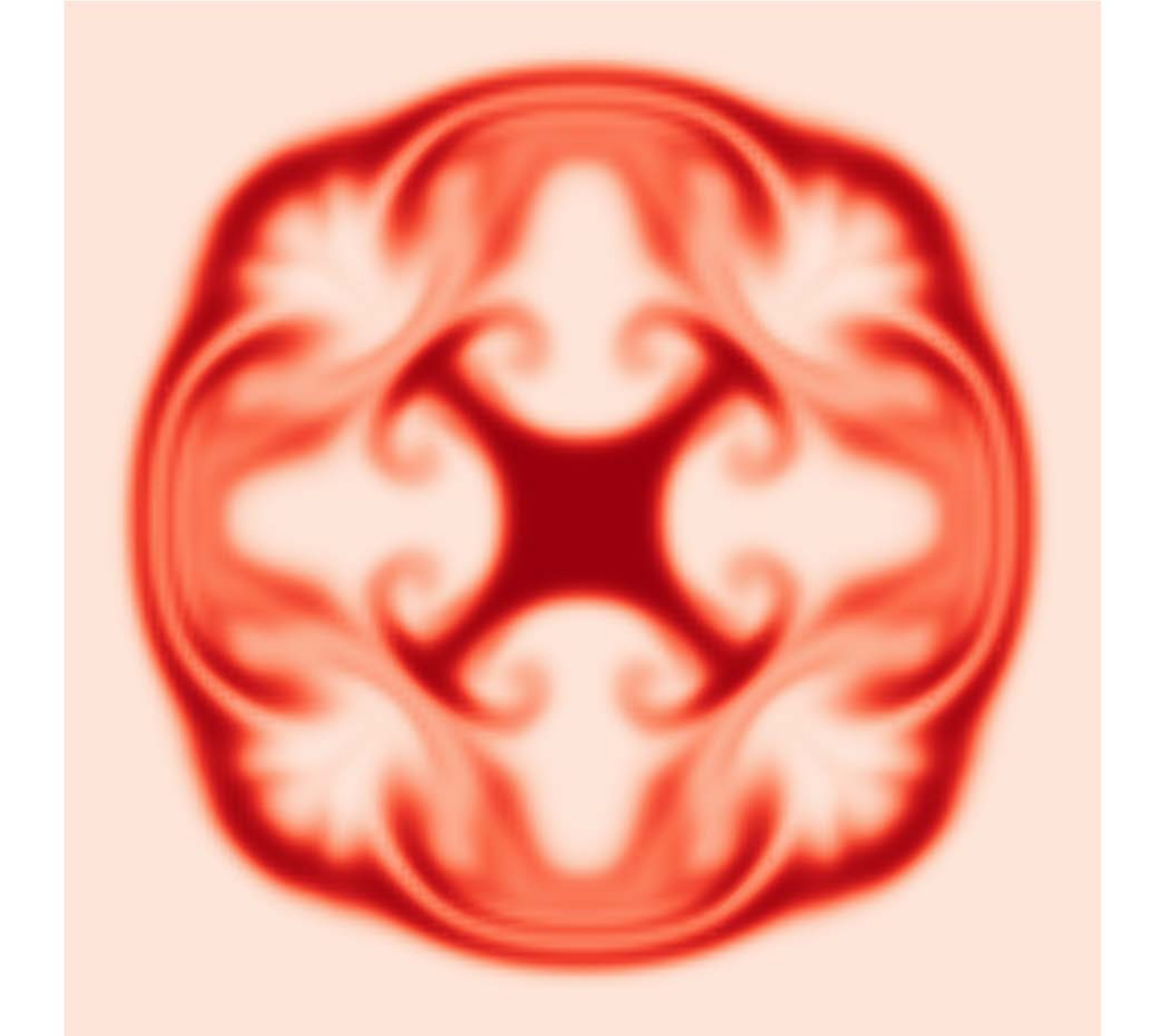}
	\includegraphics[width=0.18\columnwidth,trim={0cm 0cm 0cm 0cm},clip]{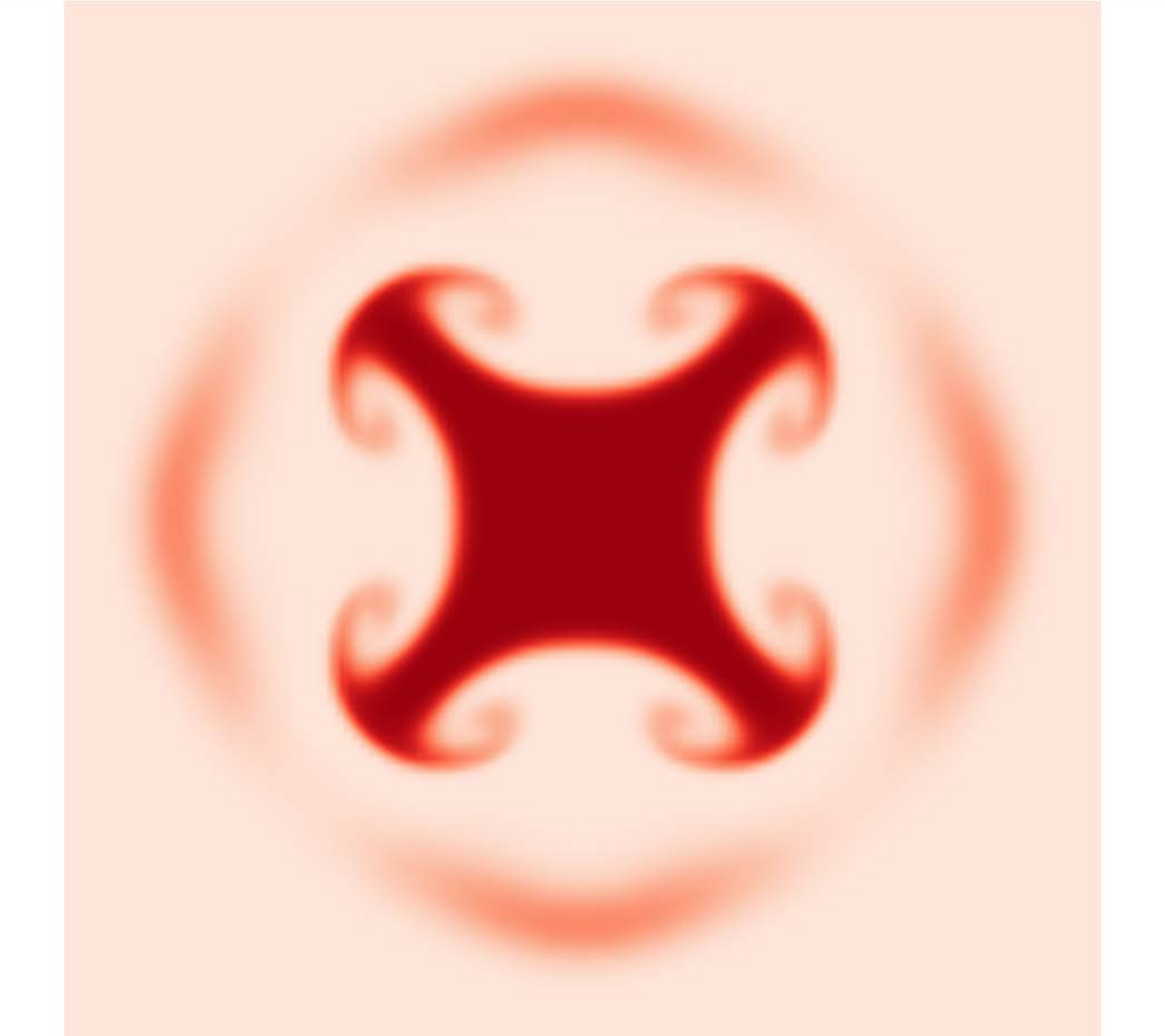}
	\includegraphics[width=0.18\columnwidth,trim={0cm 0cm 0cm 0cm},clip]{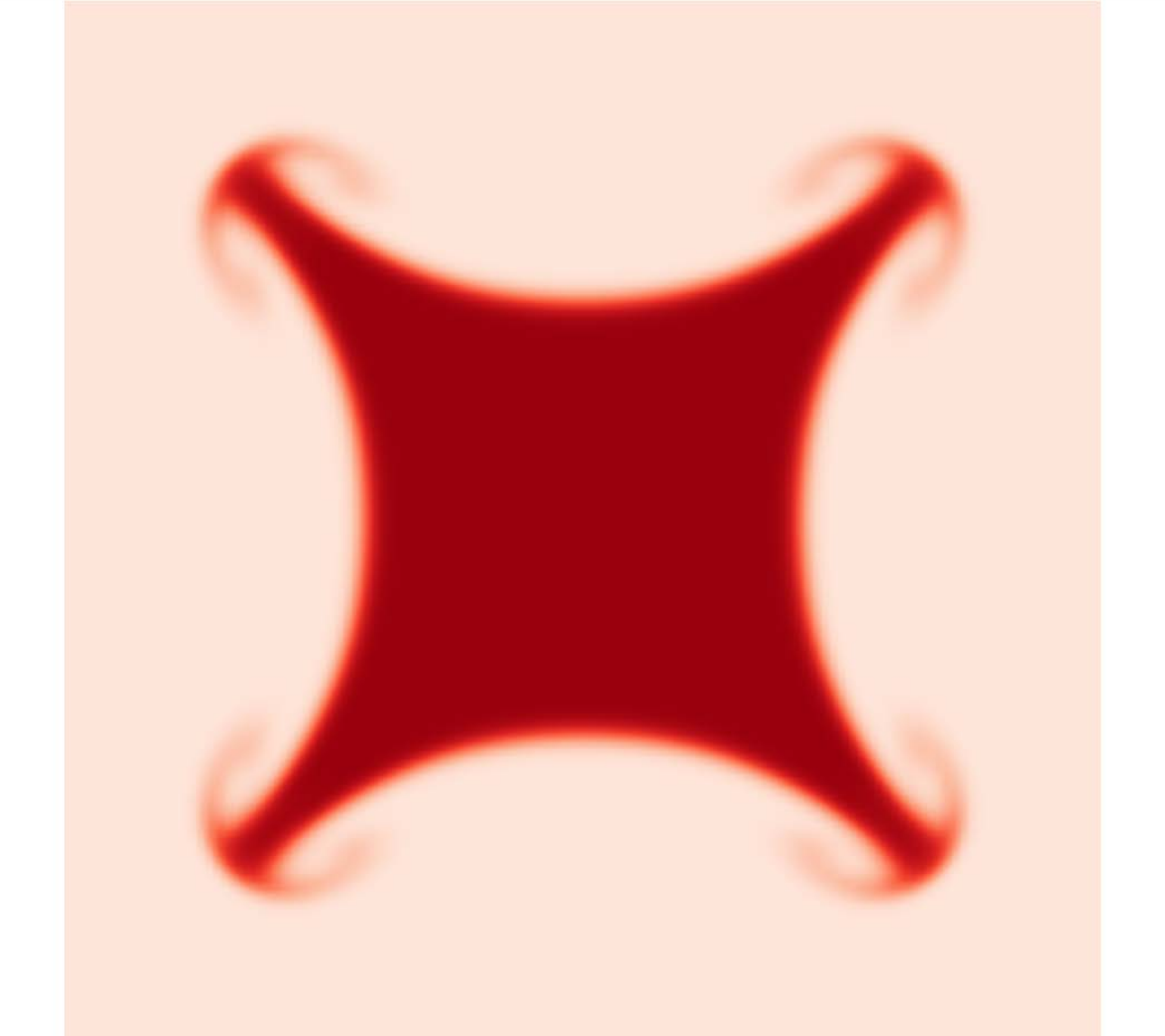}\\
	\includegraphics[width=0.18\columnwidth,trim={0cm 0cm 0cm 0cm},clip]{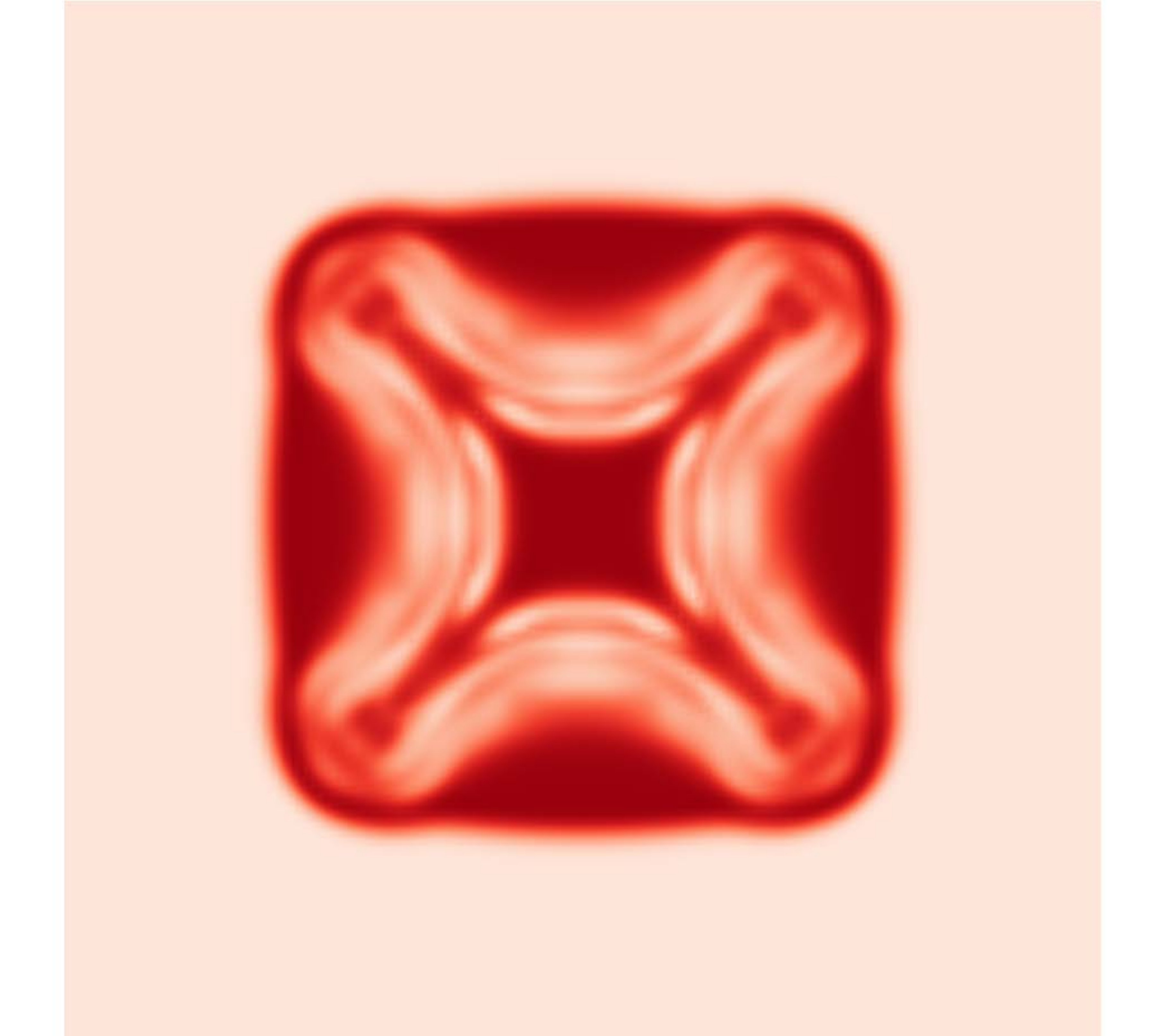}
	\includegraphics[width=0.18\columnwidth,trim={0cm 0cm 0cm 0cm},clip]{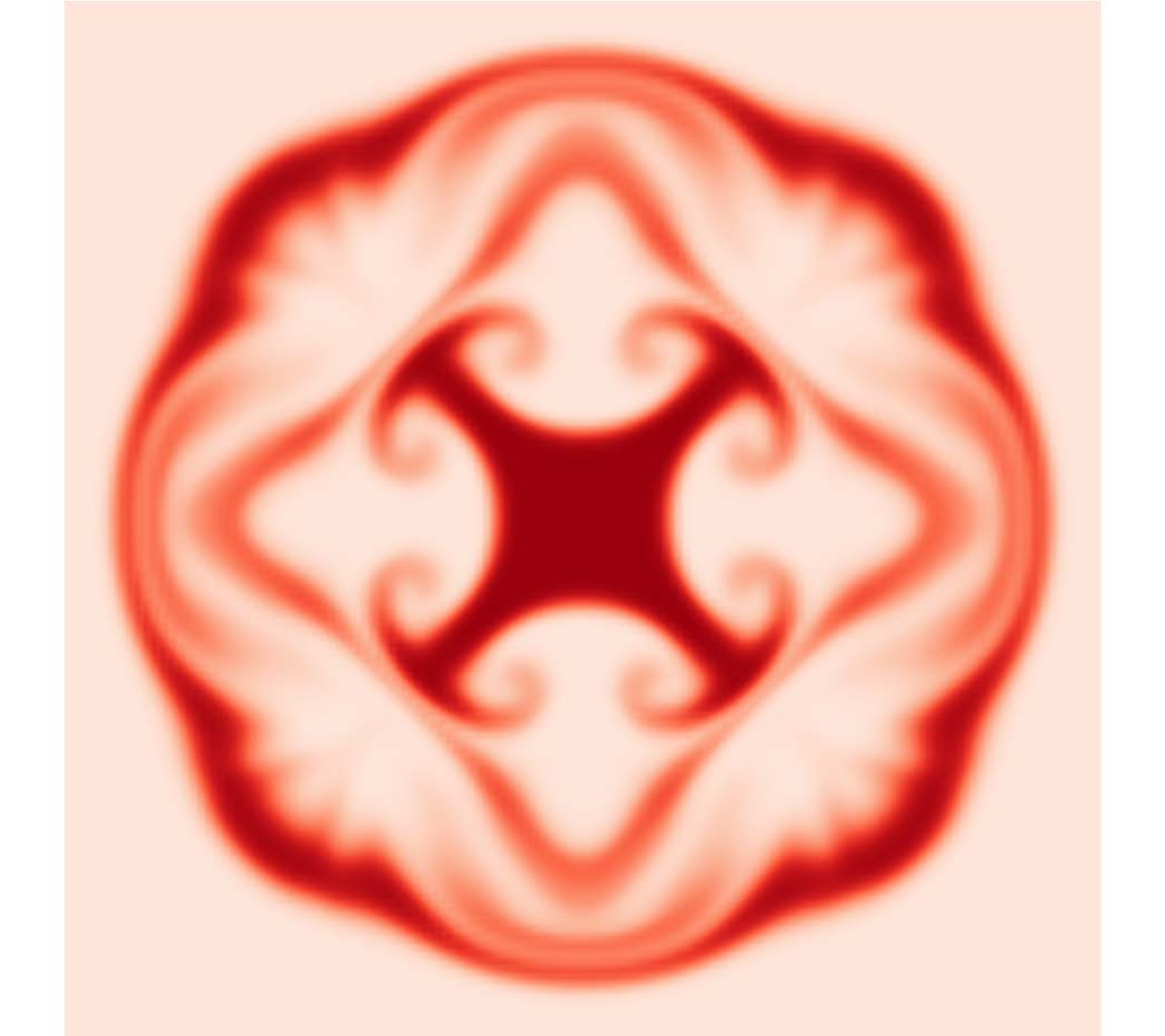}
	\includegraphics[width=0.18\columnwidth,trim={0cm 0cm 0cm 0cm},clip]{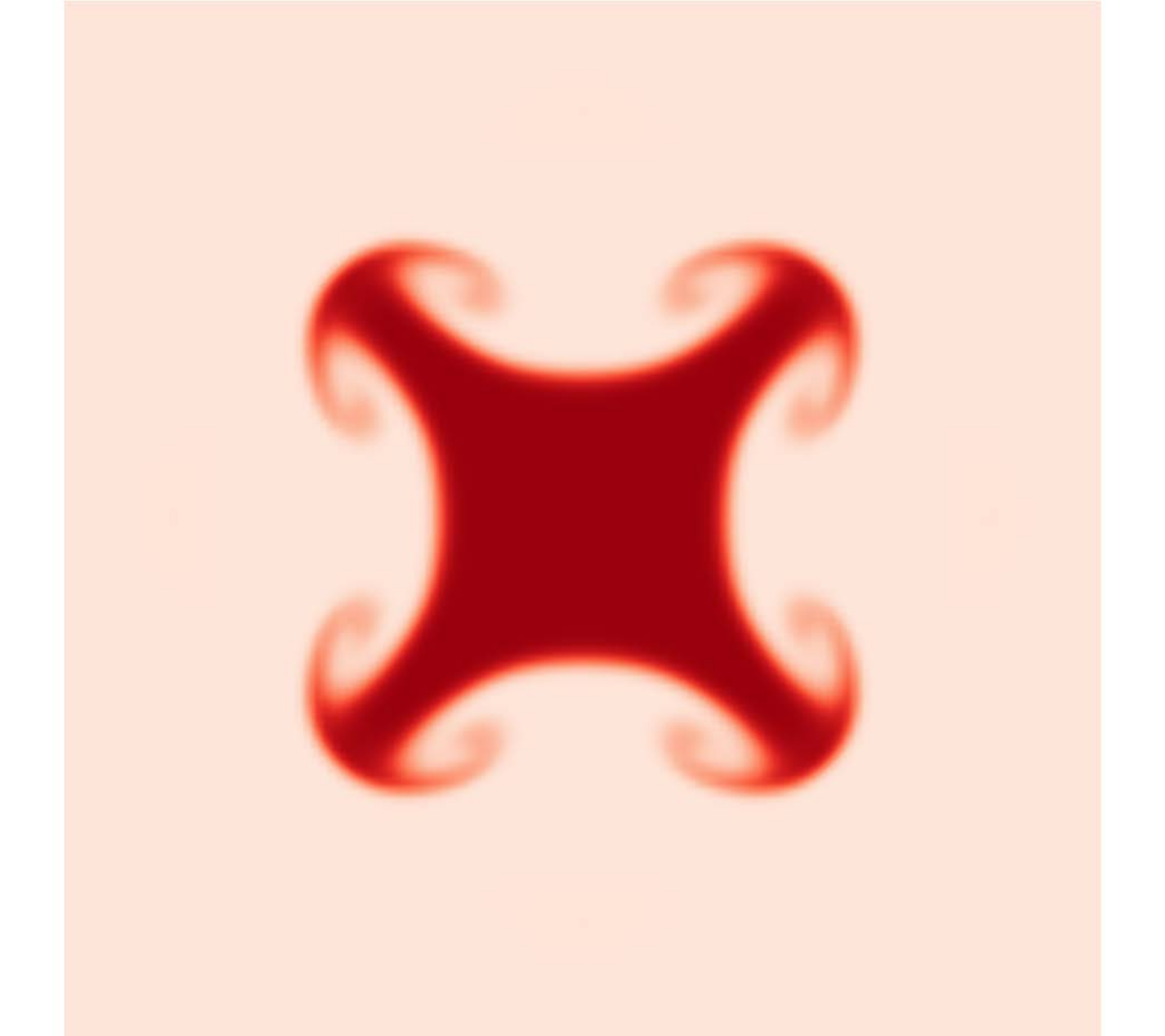}
	\includegraphics[width=0.18\columnwidth,trim={0cm 0cm 0cm 0cm},clip]{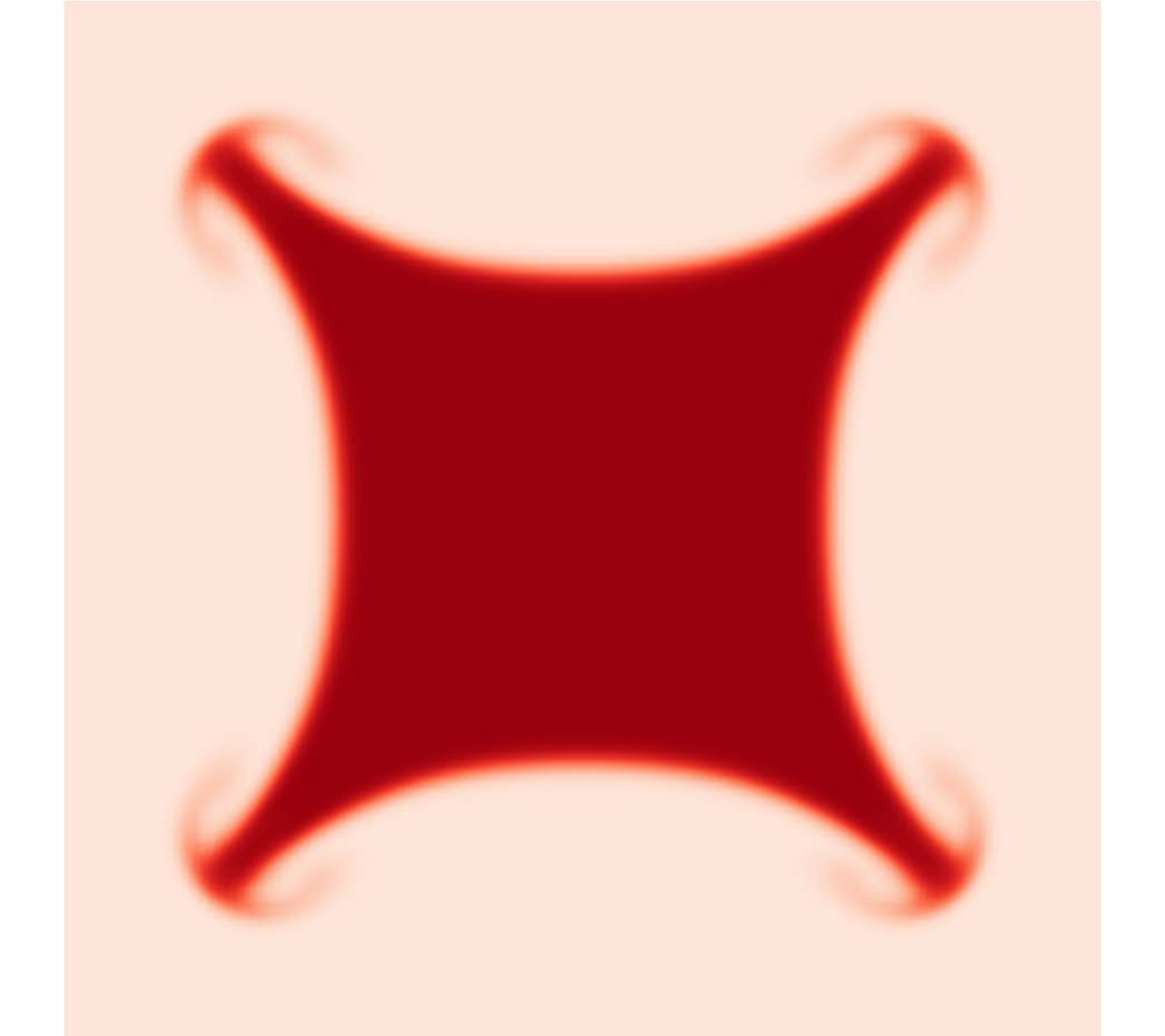}\\
	\includegraphics[width=0.18\columnwidth,trim={0cm 0cm 0cm 0cm},clip]{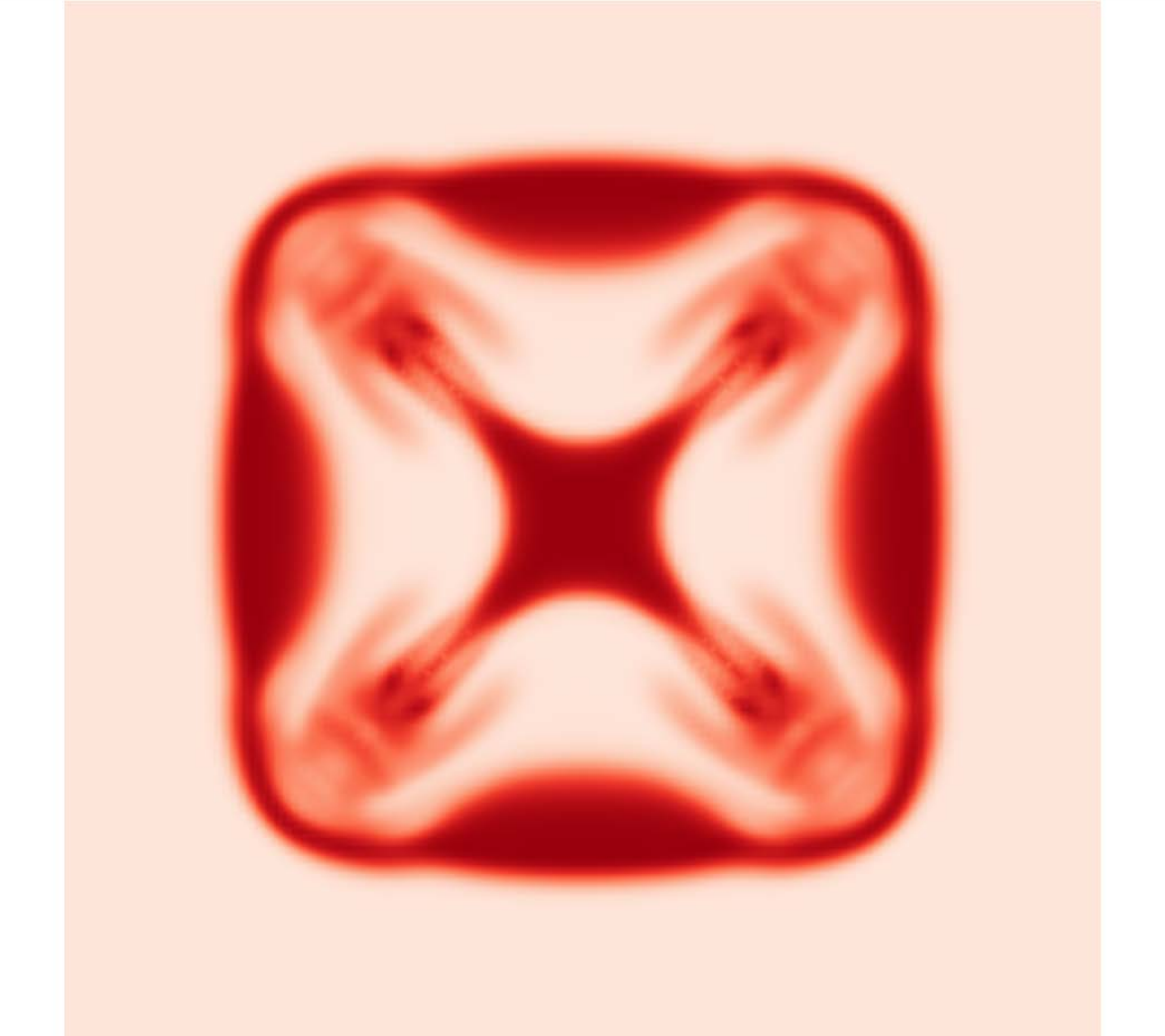}
	\includegraphics[width=0.18\columnwidth,trim={0cm 0cm 0cm 0cm},clip]{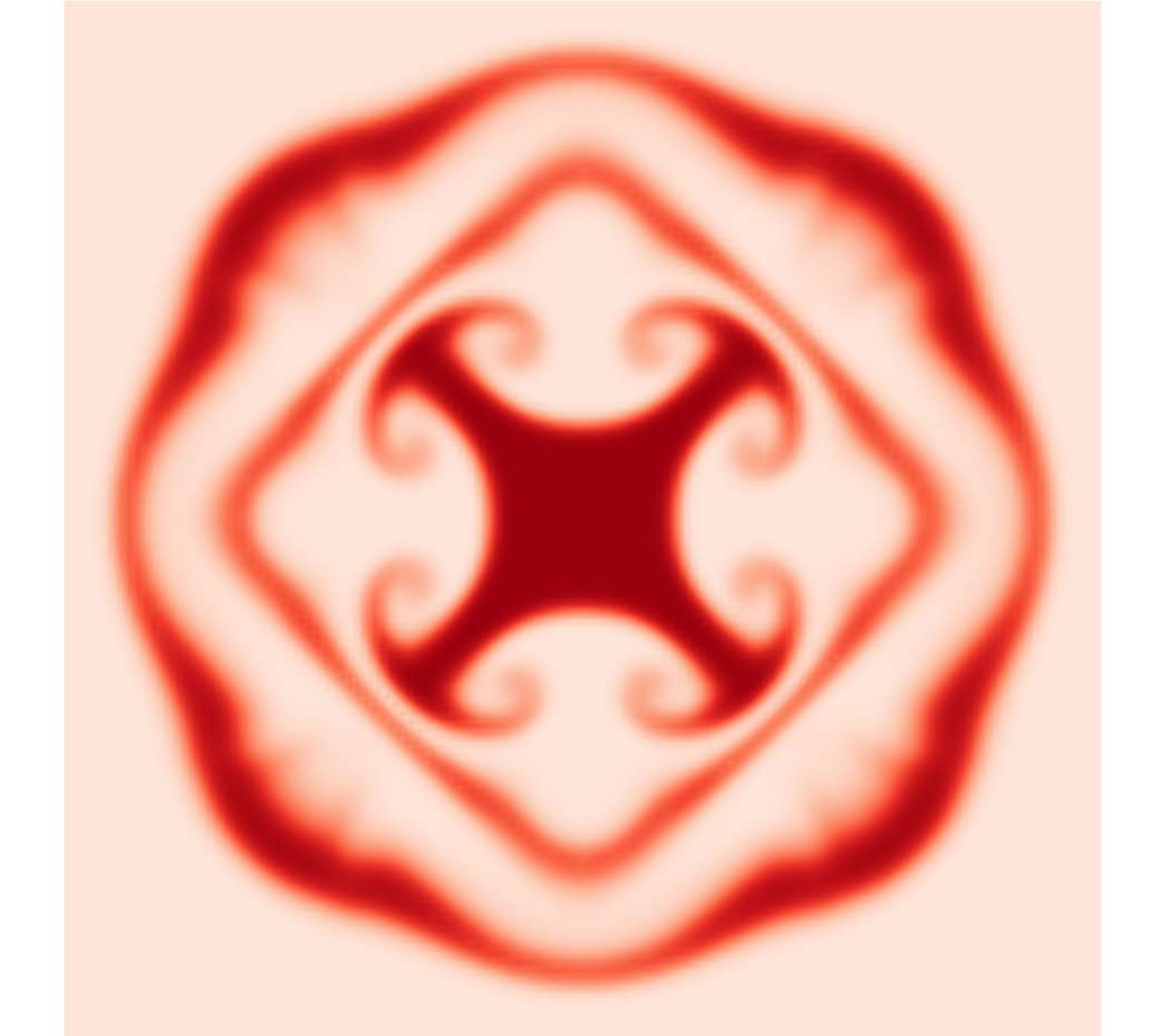}
	\includegraphics[width=0.18\columnwidth,trim={0cm 0cm 0cm 0cm},clip]{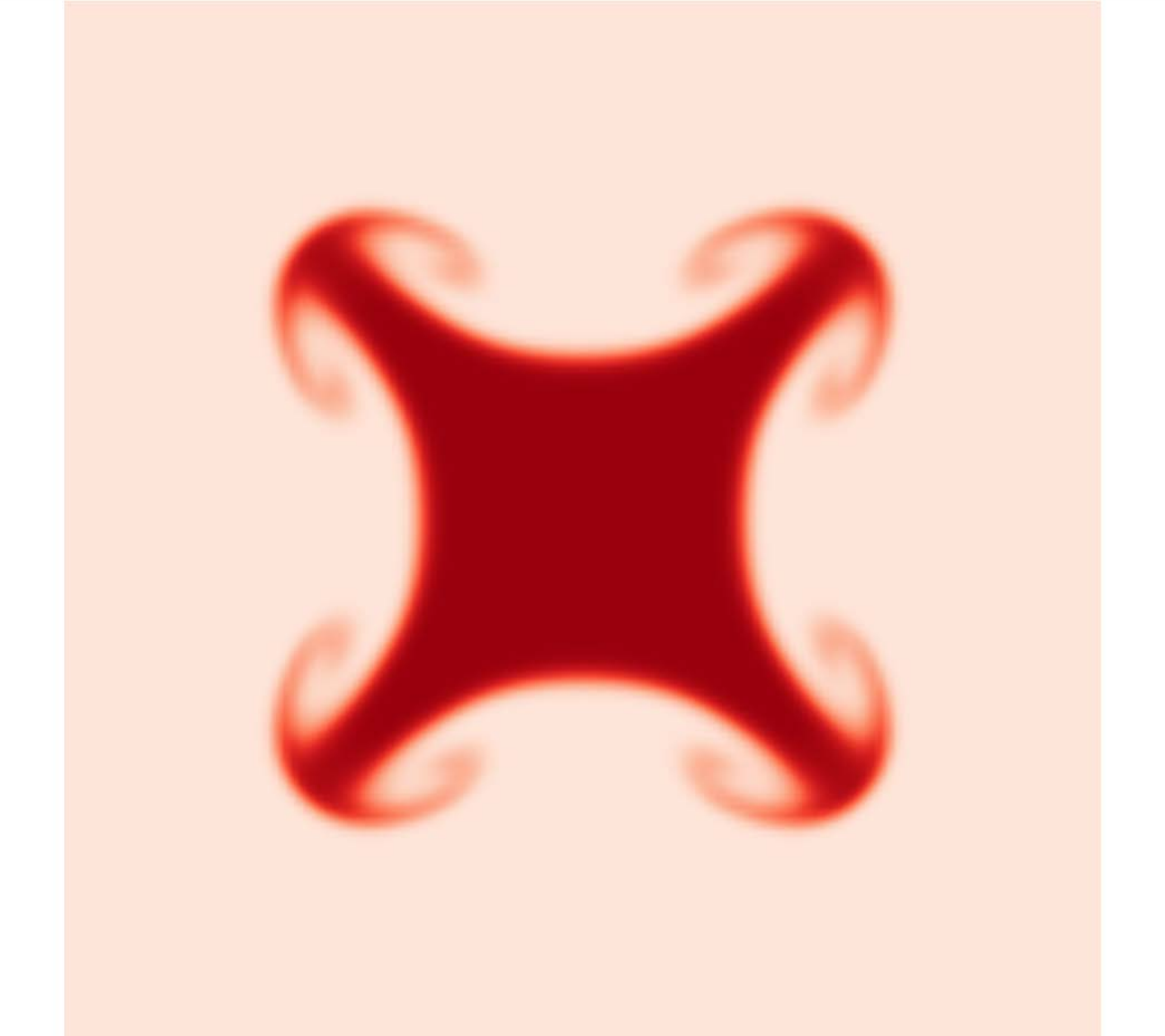}
	\includegraphics[width=0.18\columnwidth,trim={0cm 0cm 0cm 0cm},clip]{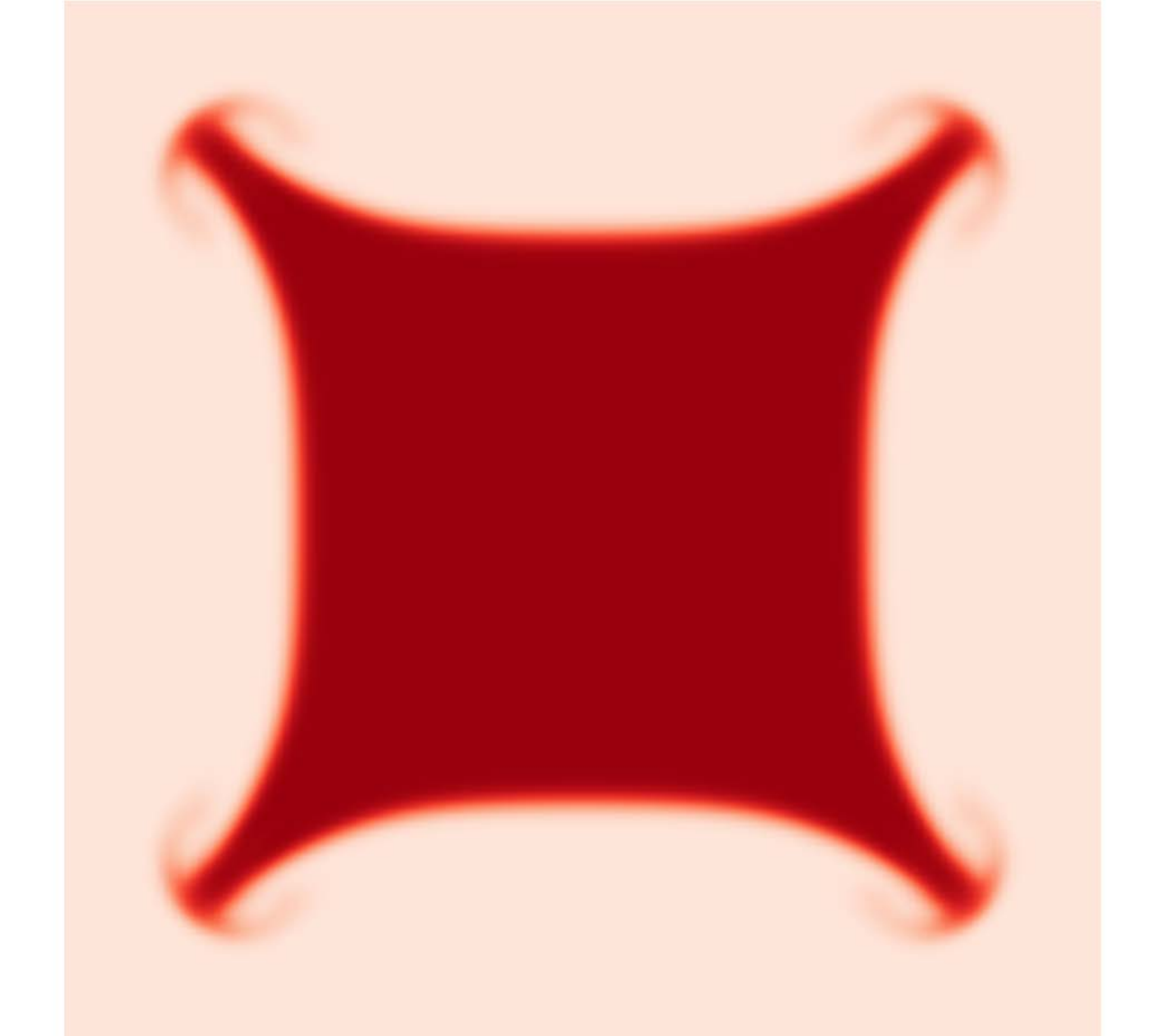}\\
	\includegraphics[width=0.18\columnwidth,trim={0cm 0cm 0cm 0cm},clip]{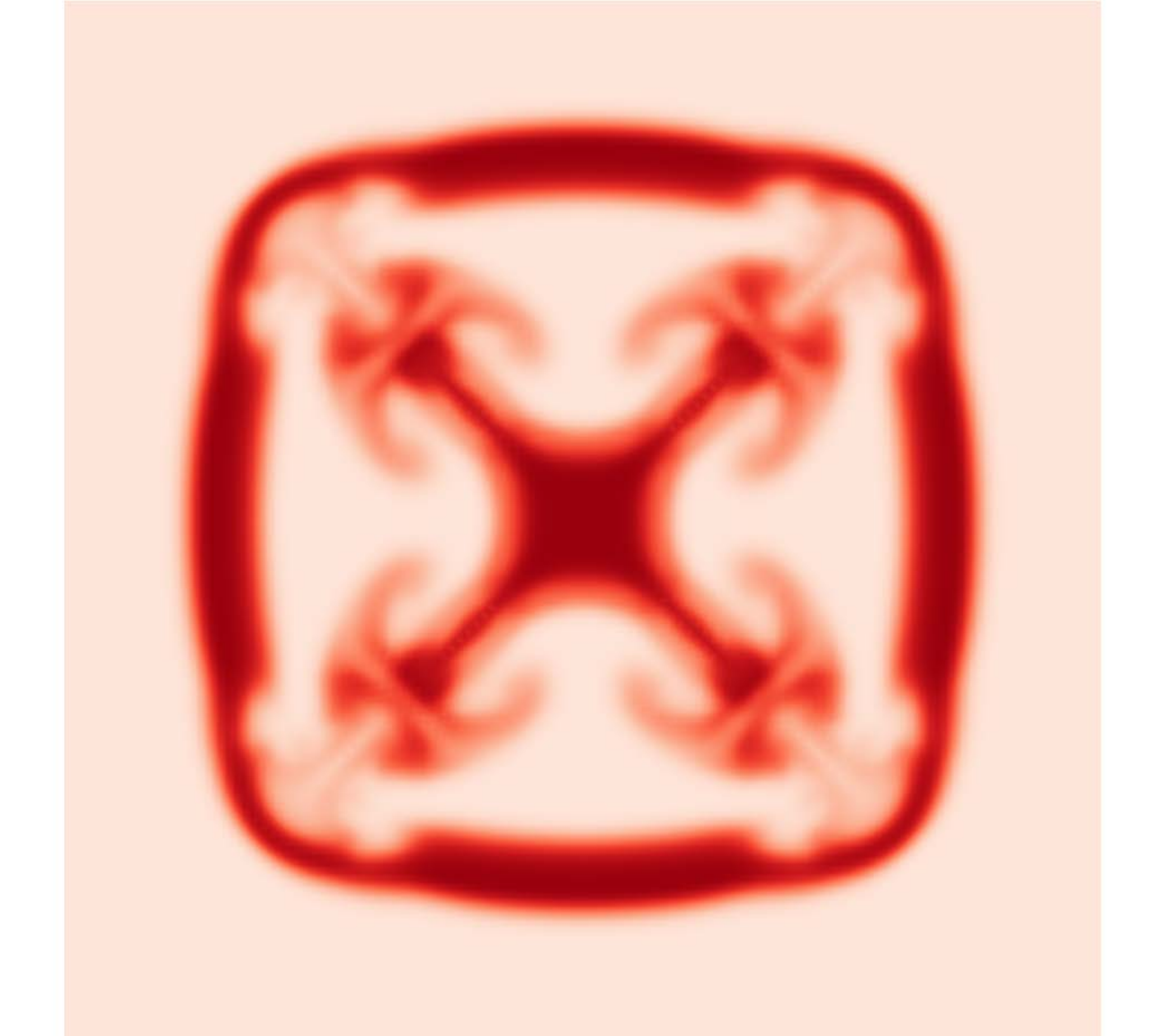}
	\includegraphics[width=0.18\columnwidth,trim={0cm 0cm 0cm 0cm},clip]{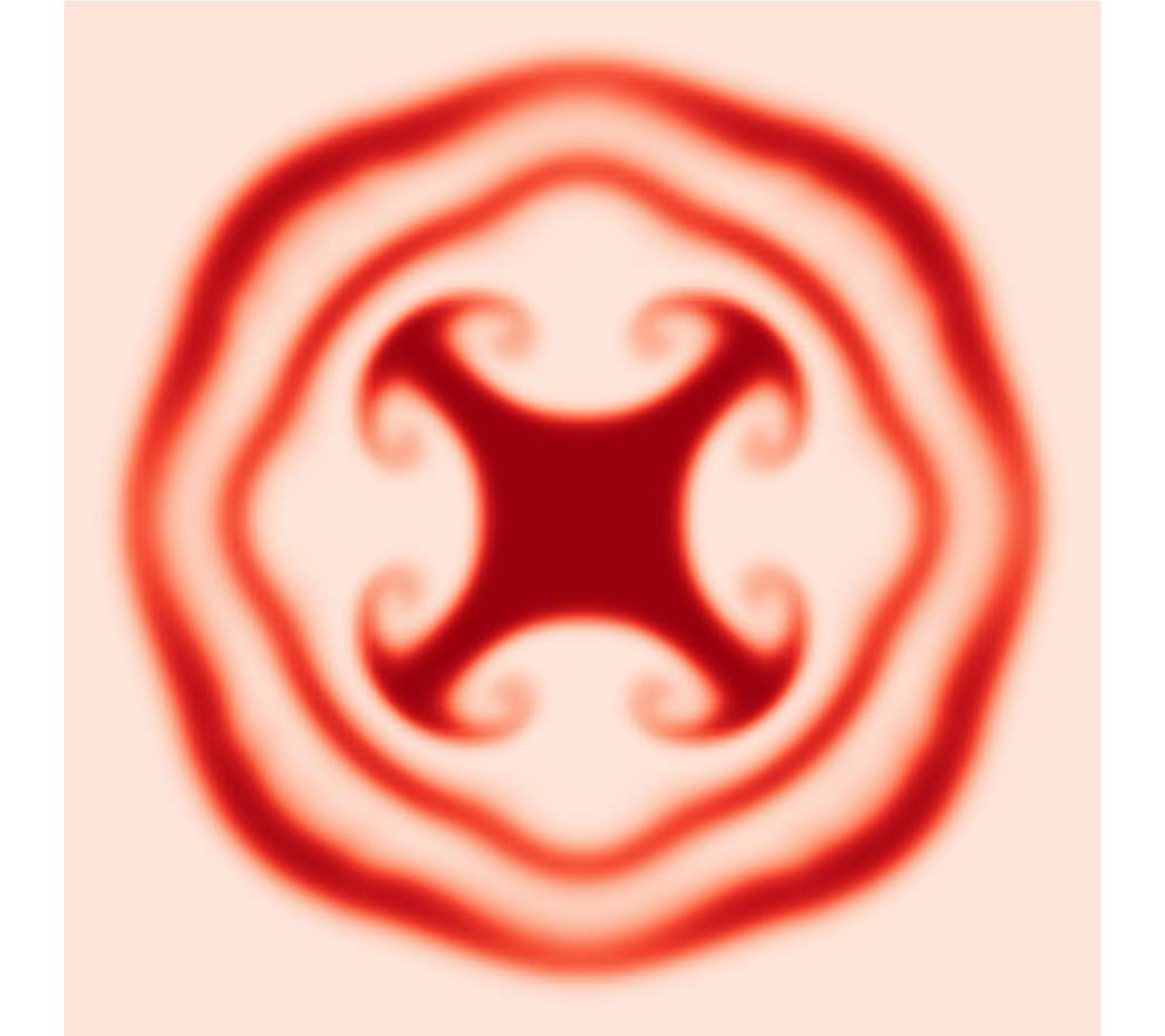}
	\includegraphics[width=0.18\columnwidth,trim={0cm 0cm 0cm 0cm},clip]{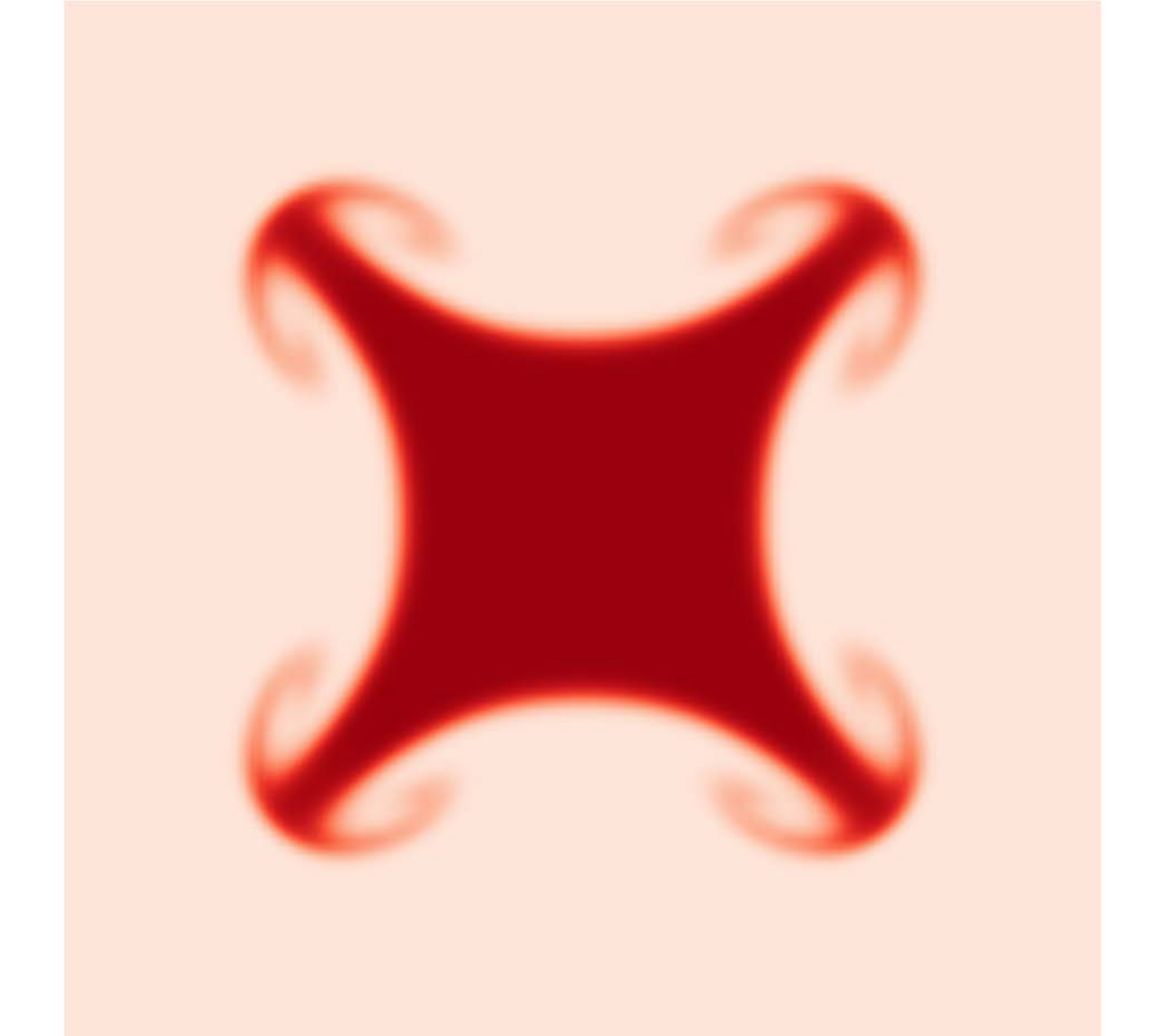}
	\includegraphics[width=0.18\columnwidth,trim={0cm 0cm 0cm 0cm},clip]{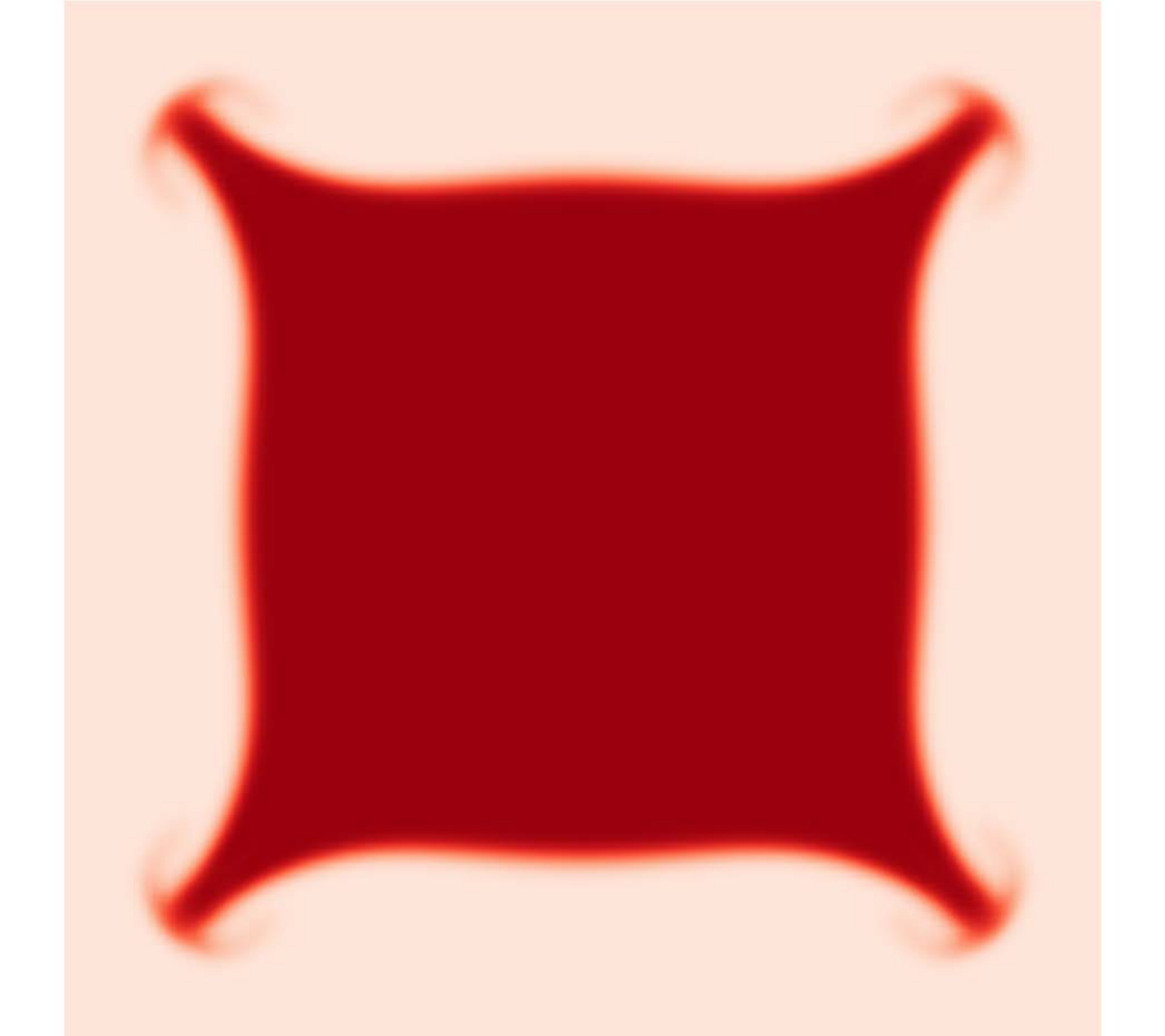}\\
	\includegraphics[width=0.18\columnwidth,trim={0cm 0cm 0cm 0cm},clip]{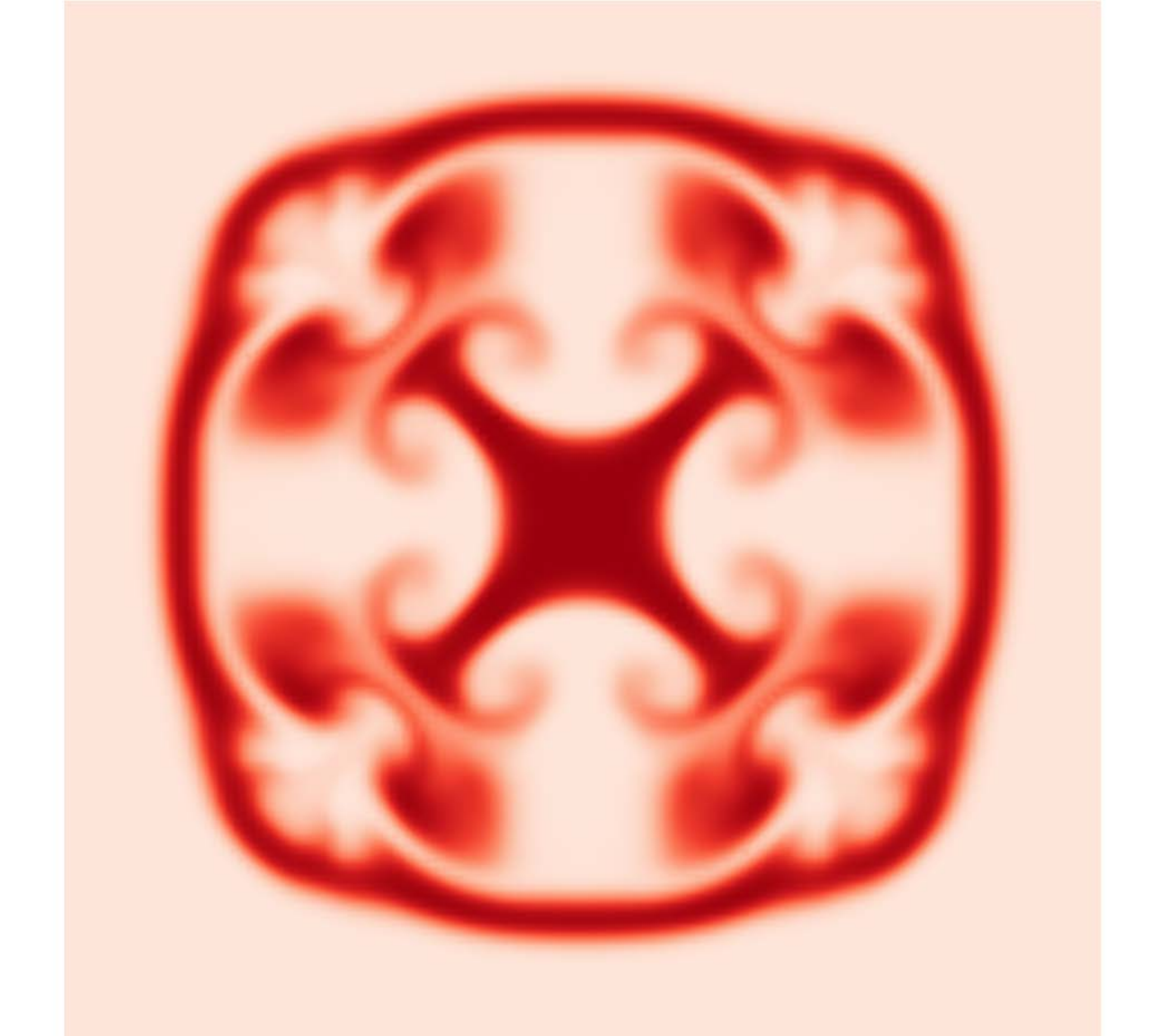}
	\includegraphics[width=0.18\columnwidth,trim={0cm 0cm 0cm 0cm},clip]{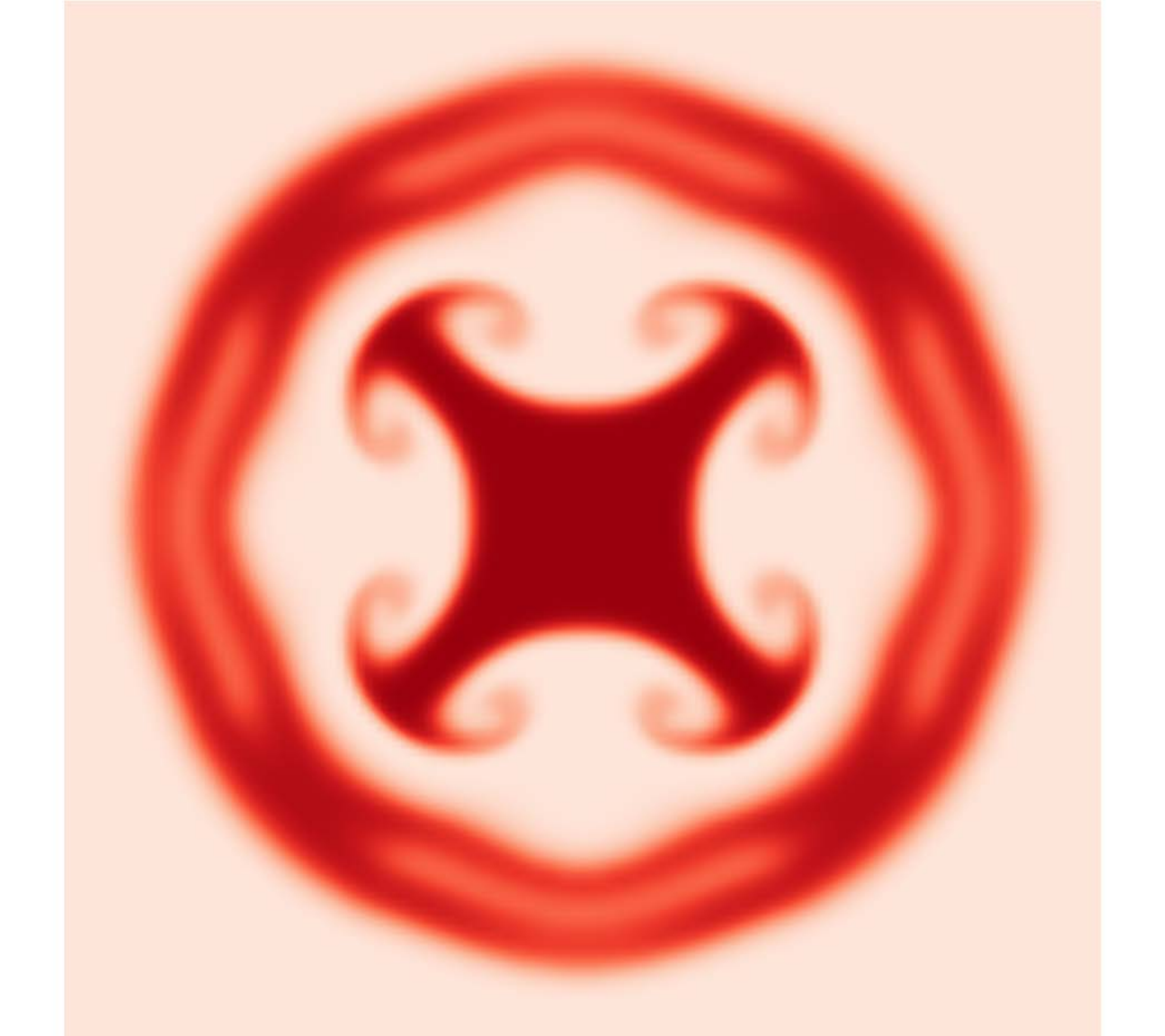}
	\includegraphics[width=0.18\columnwidth,trim={0cm 0cm 0cm 0cm},clip]{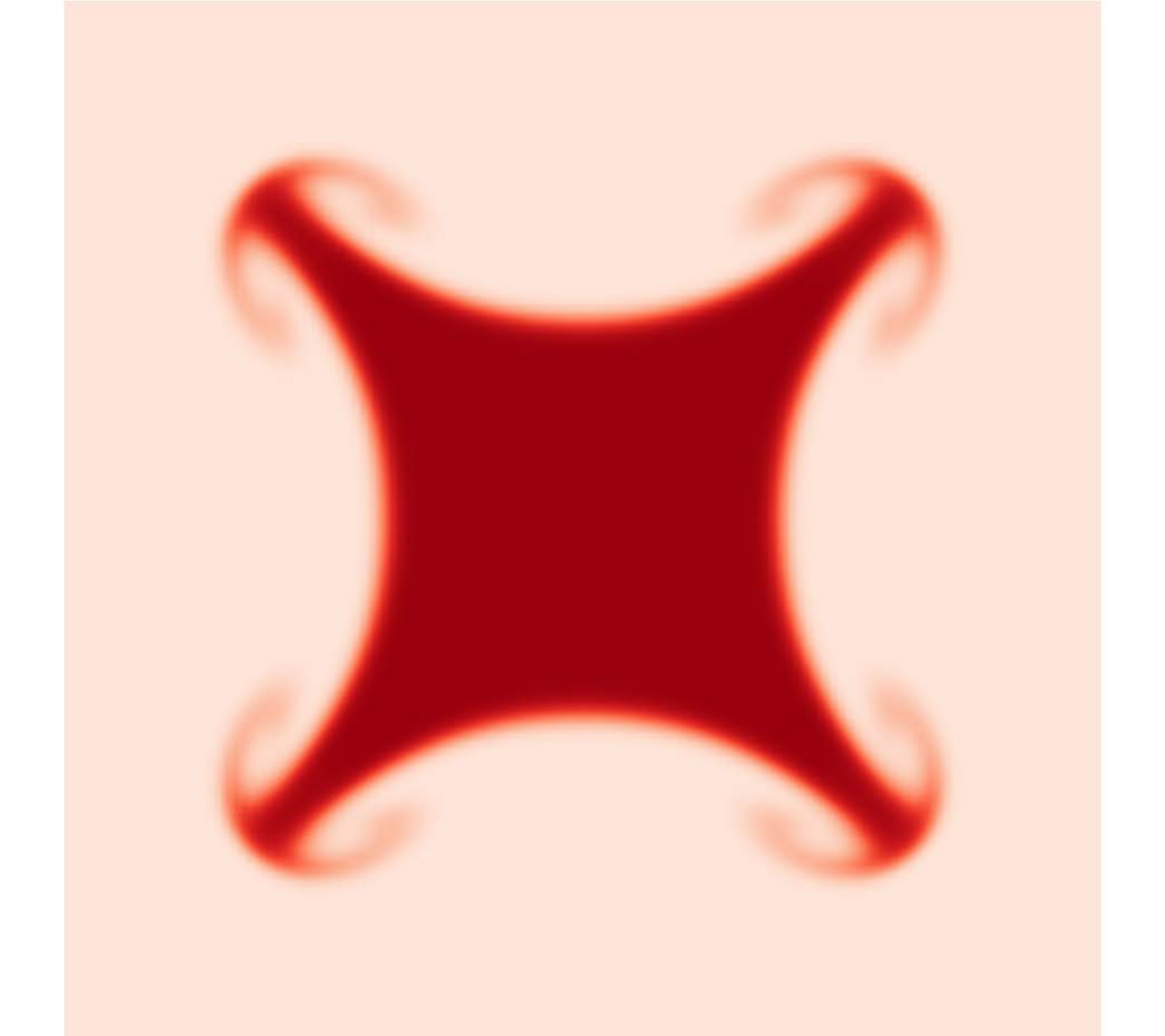}
	\includegraphics[width=0.18\columnwidth,trim={0cm 0cm 0cm 0cm},clip]{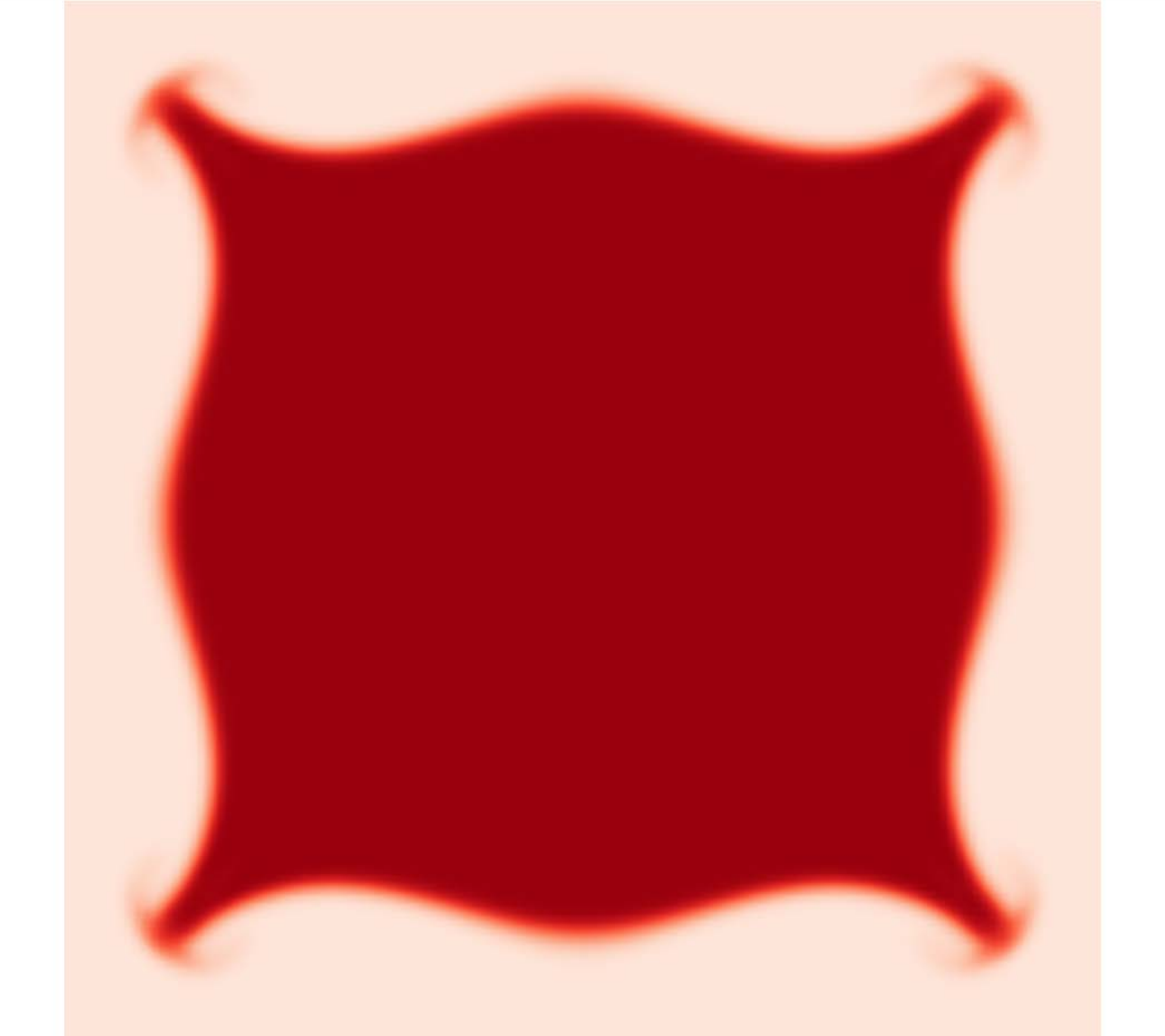}
	\centering
	\caption{Density contours in horizontal $(y,z)$ planes for 3D RTI at $At=0.5$ and $Re=1024$. $t^*=4$.
		$k=30,40,50,60,70$ for the first column,
		$k=80,90,100,110,120$ for the second column,
		$k=130,140,150,160,170$ for the third column,
		$k=180,190,200,210,220$ for the fourth column.
		The plane altitude is $x=k/128$.}
	\label{fig:Densitycontours05}
\end{figure}

To demonstrate the accuracy of our approach, a high Reynolds number case ($At=0.5$ and $Re=1024$) is simulated to compare with the results from the literatures.~\cite{1999On,WANG201541} Since He {\it et al.}~\cite{1999On} did not consider surface tension and Wang {\it et al.}~\cite{WANG201541} did not mention surface tension in the 3D RTI simulation, we also choose a negligible surface tension such that the capillary number is $Ca=1.5\times 10^{15}$. The other parameters in our simulation are
$U_0=0.02$, $CFL=0.25$, $\mu_A/\mu_B = 3.0$, $W=5.0$, $Pe=20000$. The 3D interface evolution is shown in Fig.~\ref{fig:RT3Dinterface05}. 
Due to the local Kelvin–Helmholtz instability, the rolling-up processes and mushroom structures are observed.
The profiles are almost the same as those reported by the previous studies using He-Chen-Zhang (HCZ) model (He {\it et al.}~\cite{1999On}) and multiphase lattice Boltzmann flux solver (MLBFS) method (Wang {\it et al.}~\cite{WANG201541}).
We also compare the evolution of three main positions (bubble front, saddle point, and spike tip) with their data in Fig.~\ref{fig:Timeof3DRT05} and the results show a very good agreement.
For more details of the interface evolution, we plot the density contours in different horizontal planes at $t^*=4$ (Fig.~\ref{fig:Densitycontours05}) to compare with He {\it et al.}.~\cite{1999On} The planes are labeled by $k$ and the altitudes are $x=k/L$, respectively. The results are very similar to those shown in He {\it et al.}~\cite{1999On} except for the small structures where $k<100$, since the minor details in the numerical methods will affect the small structures of the interface.


\begin{figure}[]
	\centering
	\includegraphics[width=0.18\columnwidth,trim={2cm 0cm 2cm 0cm},clip]{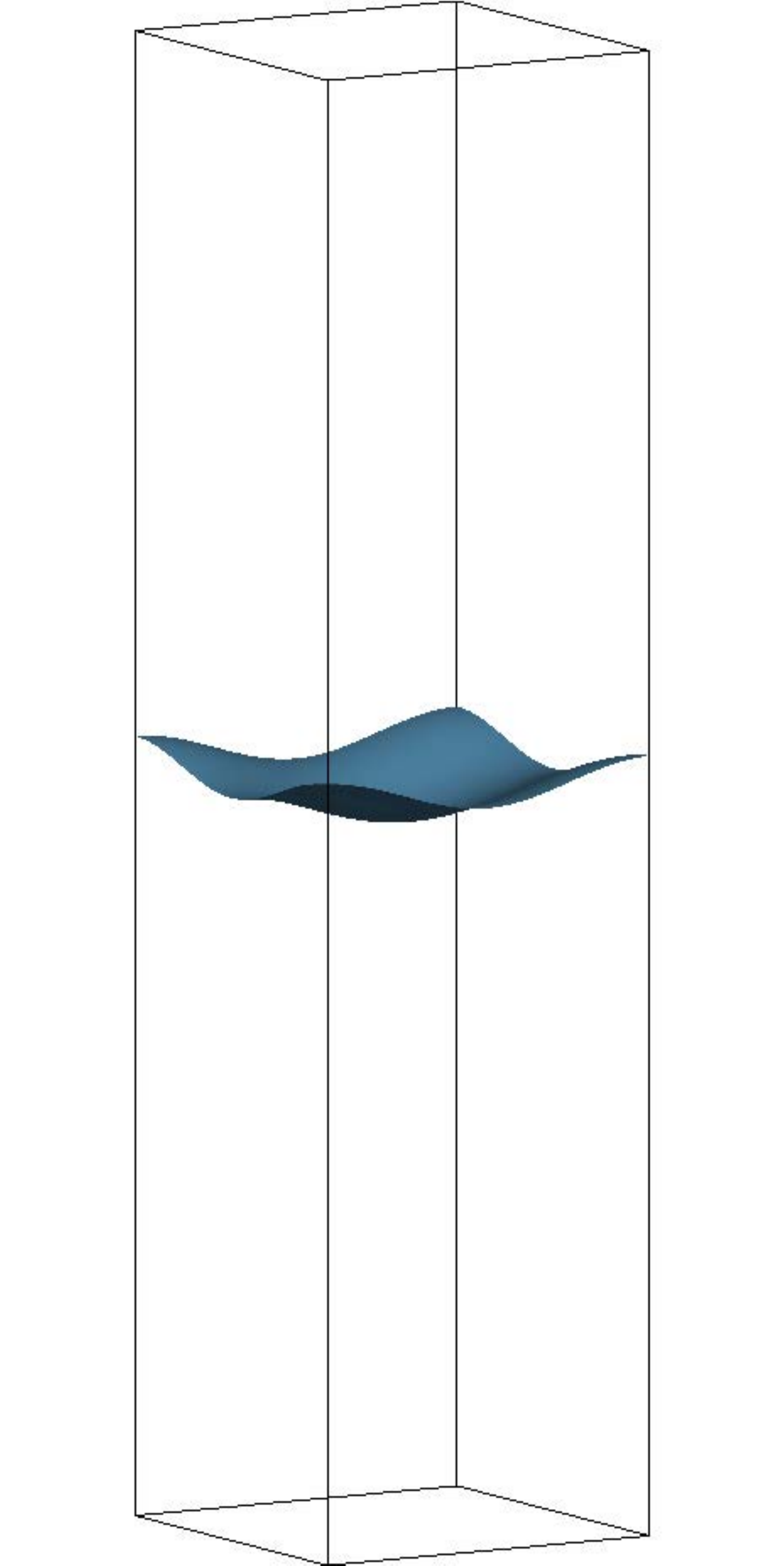}
	\includegraphics[width=0.18\columnwidth,trim={2cm 0cm 2cm 0cm},clip]{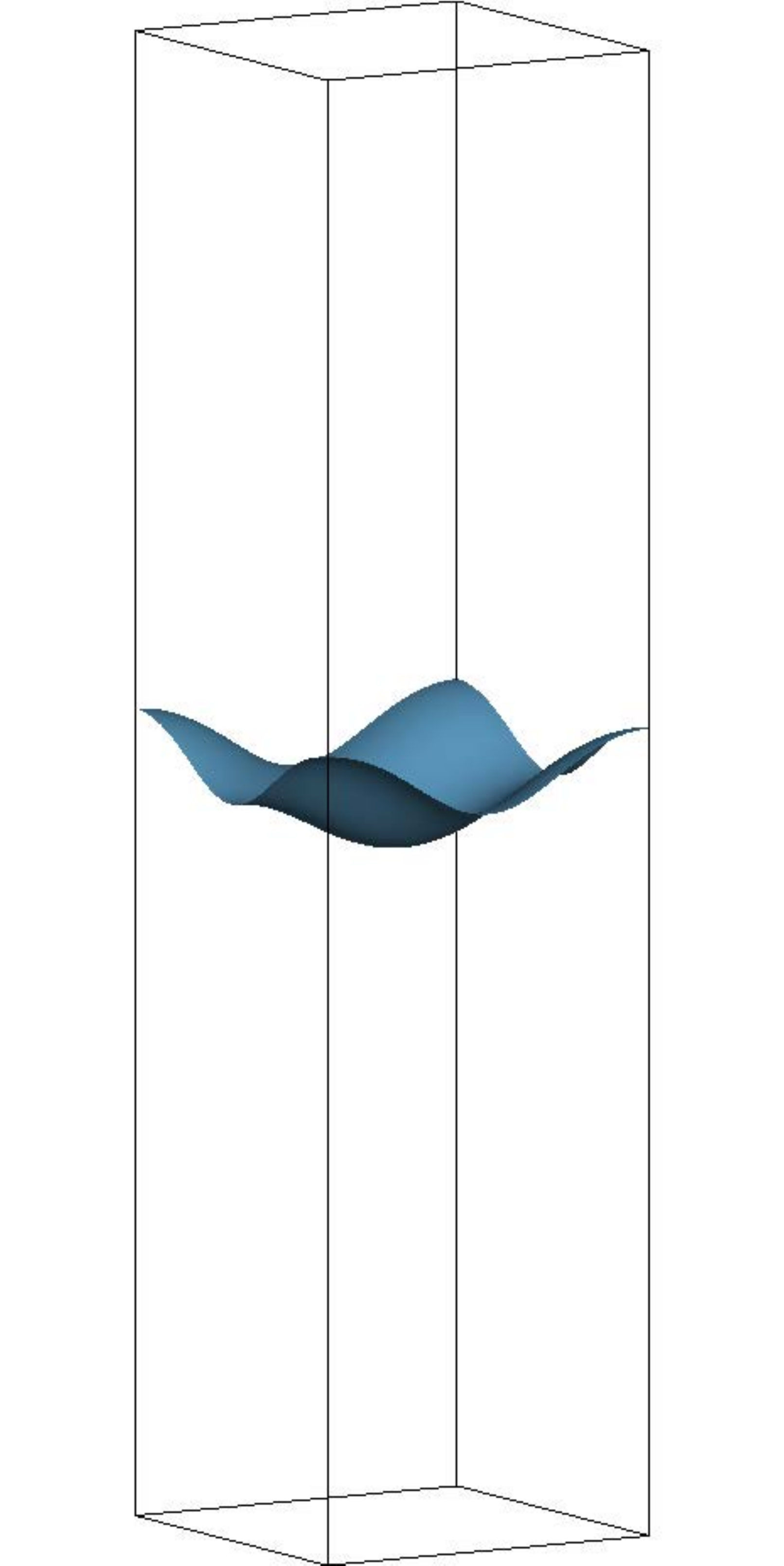}
	\includegraphics[width=0.18\columnwidth,trim={2cm 0cm 2cm 0cm},clip]{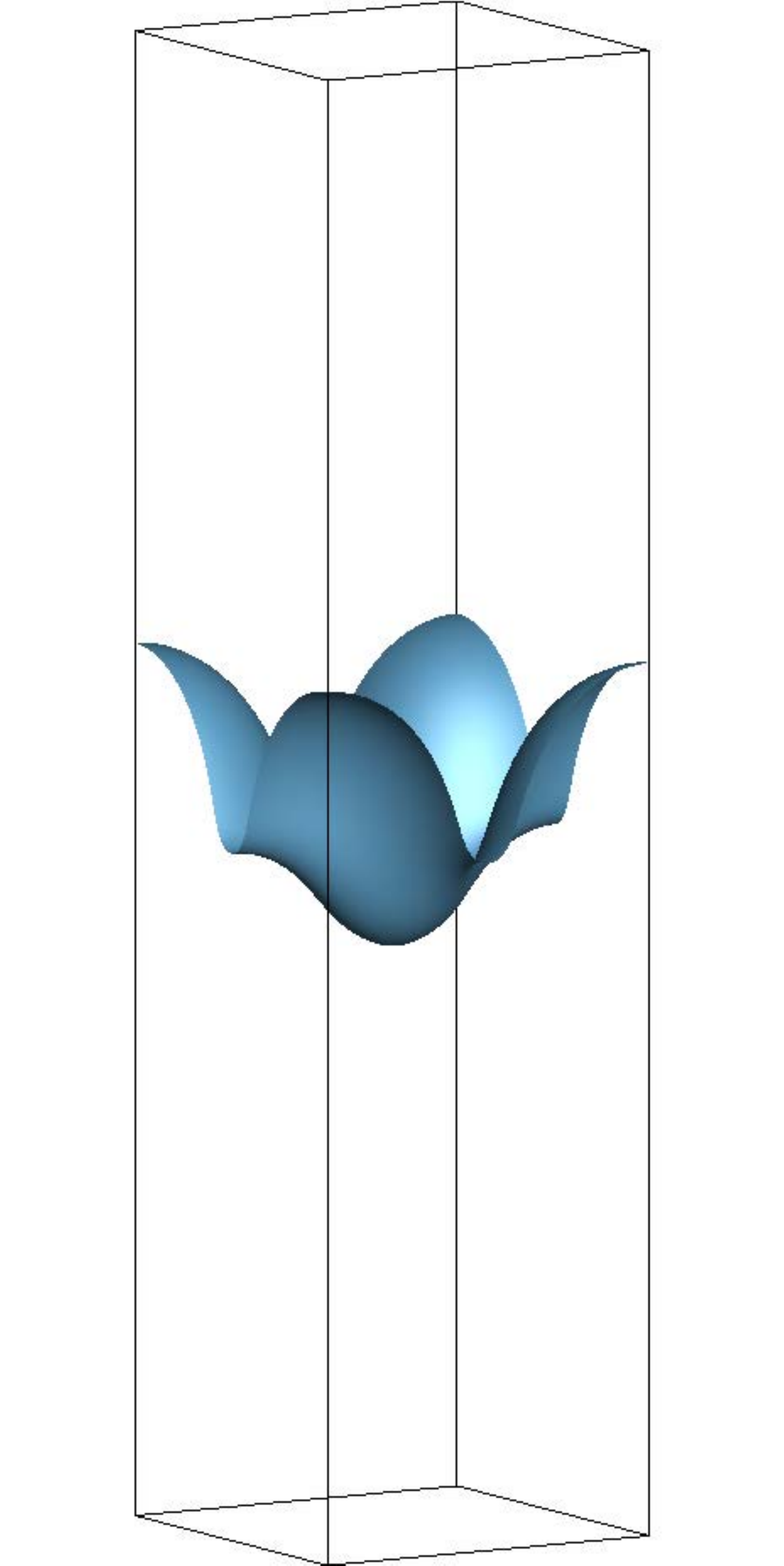}
	\includegraphics[width=0.18\columnwidth,trim={2cm 0cm 2cm 0cm},clip]{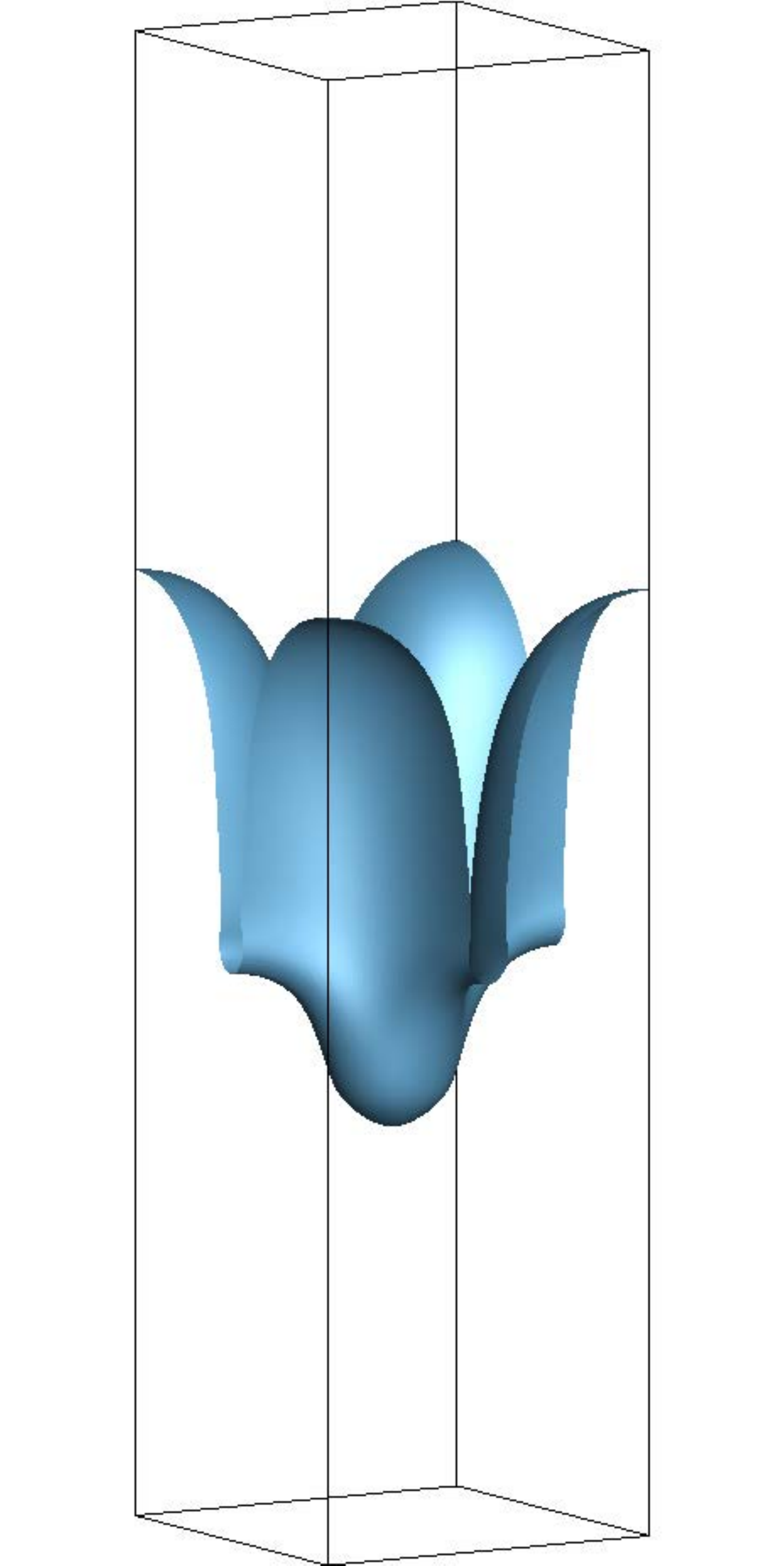}
	\includegraphics[width=0.18\columnwidth,trim={2cm 0cm 2cm 0cm},clip]{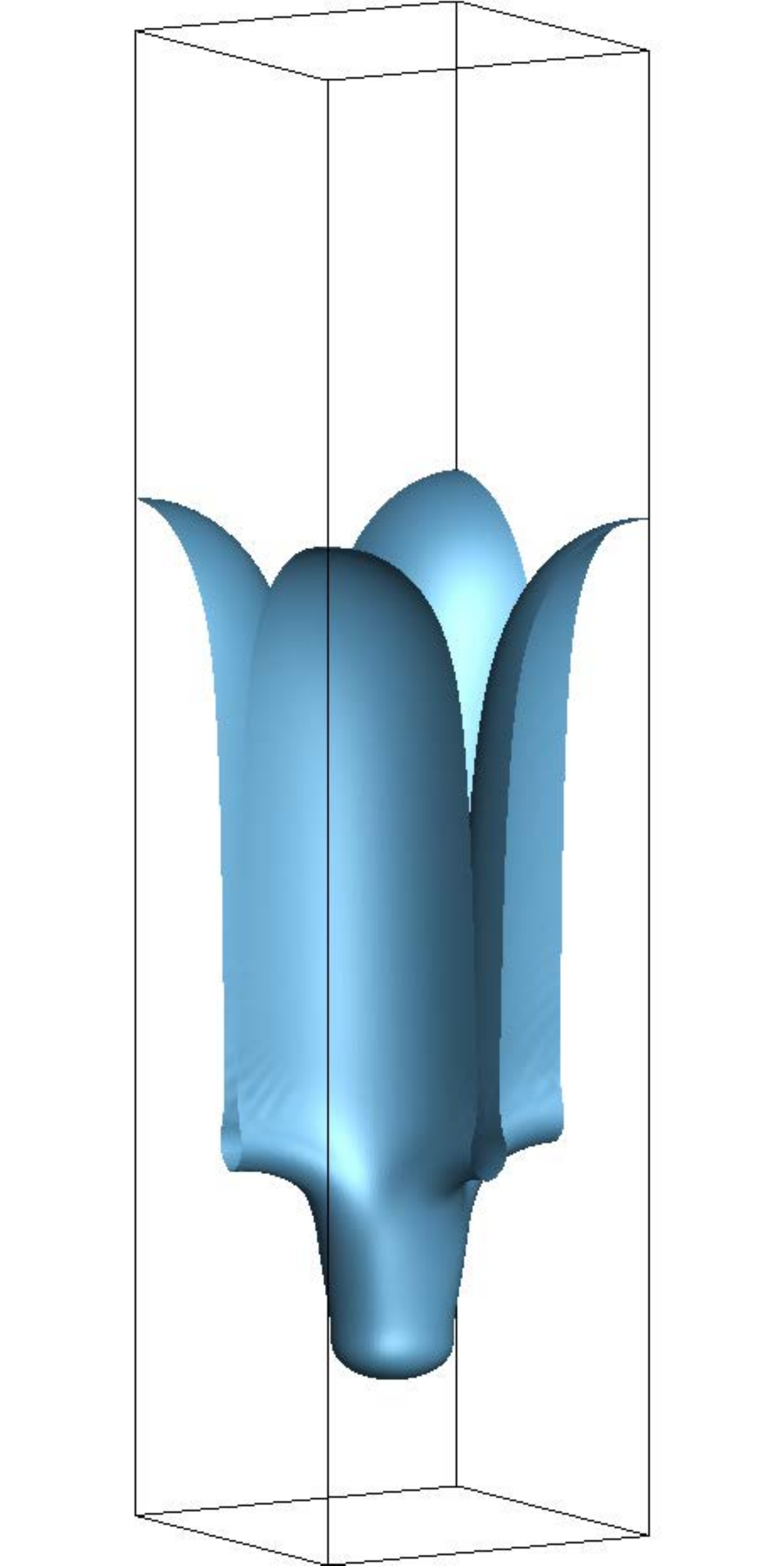}\\	
	\includegraphics[width=0.12\columnwidth,trim={0cm 4cm 0cm 0cm},clip]{RT3D_pngpdf/RTxyz.pdf}
	\centering
	\caption{Time evolution of fluid-fluid interface for 3D RTI at $At=0.998$ and $Re=3000$. $t^*=0,0.5,1,1.5,2$.}
	\label{fig:RT3Dinterface0998}
\end{figure}

\begin{figure}[]
	\centering
	\includegraphics[width=0.5\columnwidth,trim={0cm 0cm 0cm 0cm},clip]{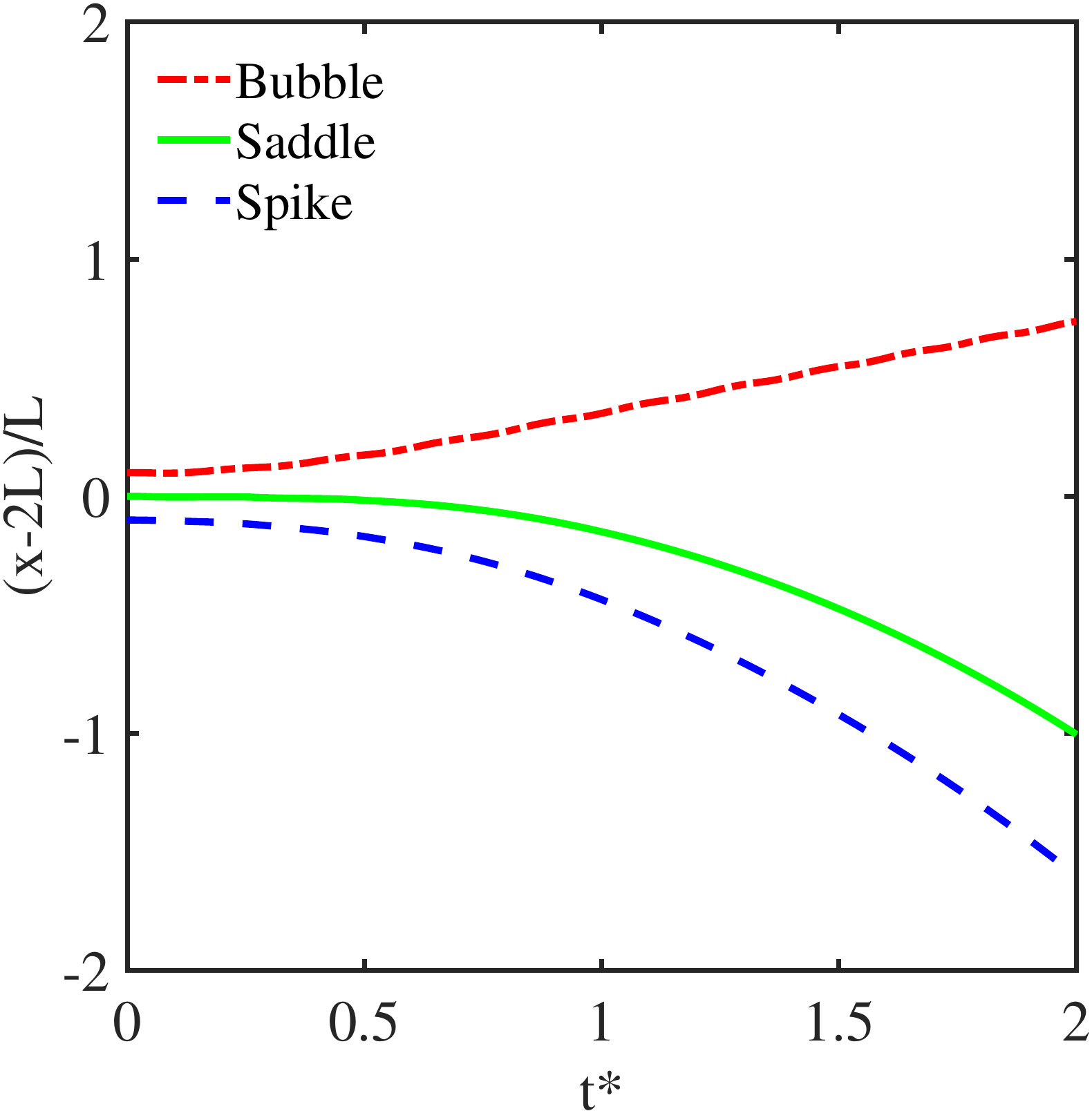}
	\centering
	\caption{Time evolution of the positions of bubble front, saddle point, and spike tip for 3D RTI at $At=0.998$ and $Re=3000$.}
	\label{fig:Timeof3DRT0998}
\end{figure}

\begin{figure}[]
	\centering
	\includegraphics[width=0.18\columnwidth,trim={0cm 0cm 0cm 0cm},clip]{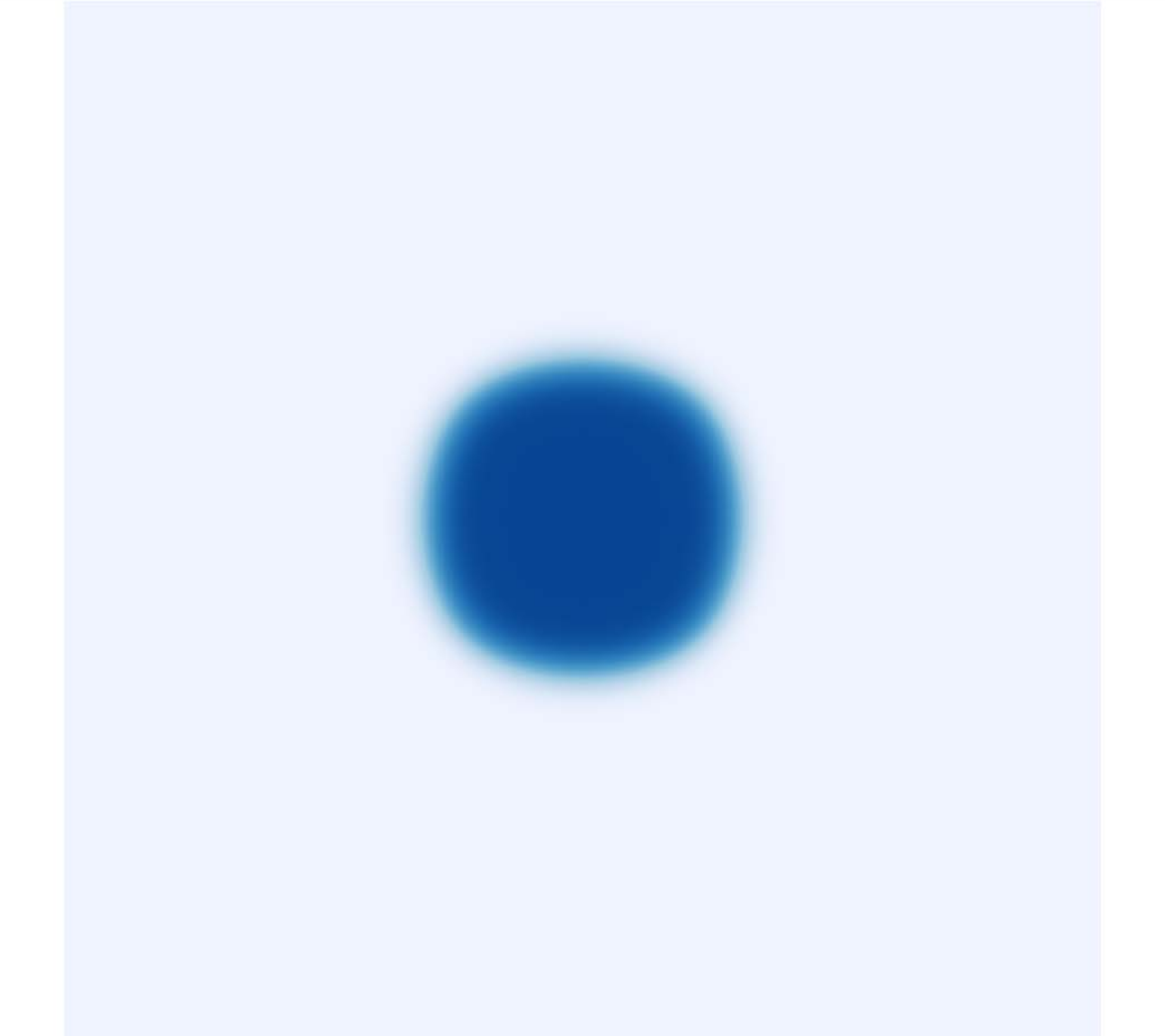}
	\includegraphics[width=0.18\columnwidth,trim={0cm 0cm 0cm 0cm},clip]{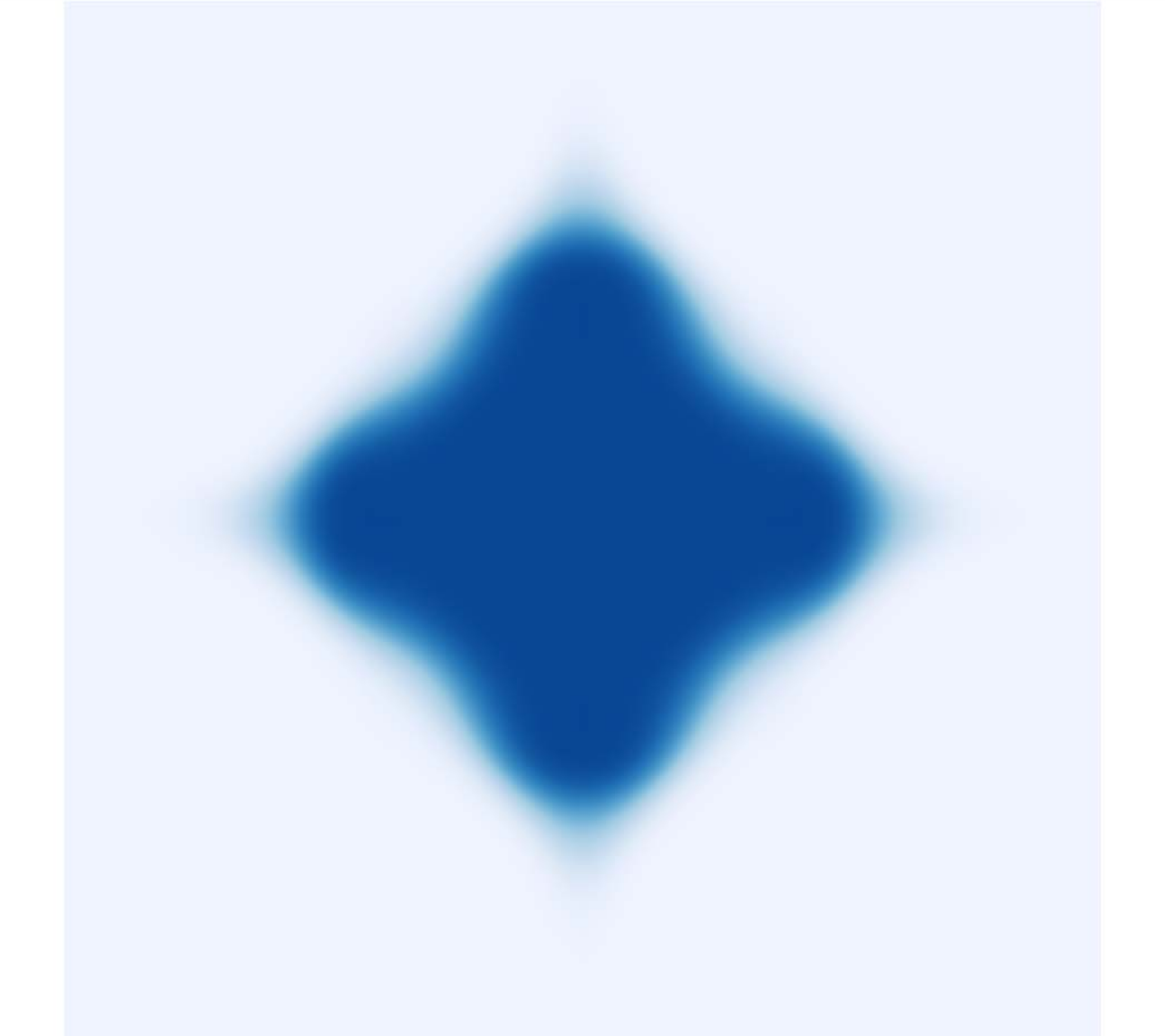}
	\includegraphics[width=0.18\columnwidth,trim={0cm 0cm 0cm 0cm},clip]{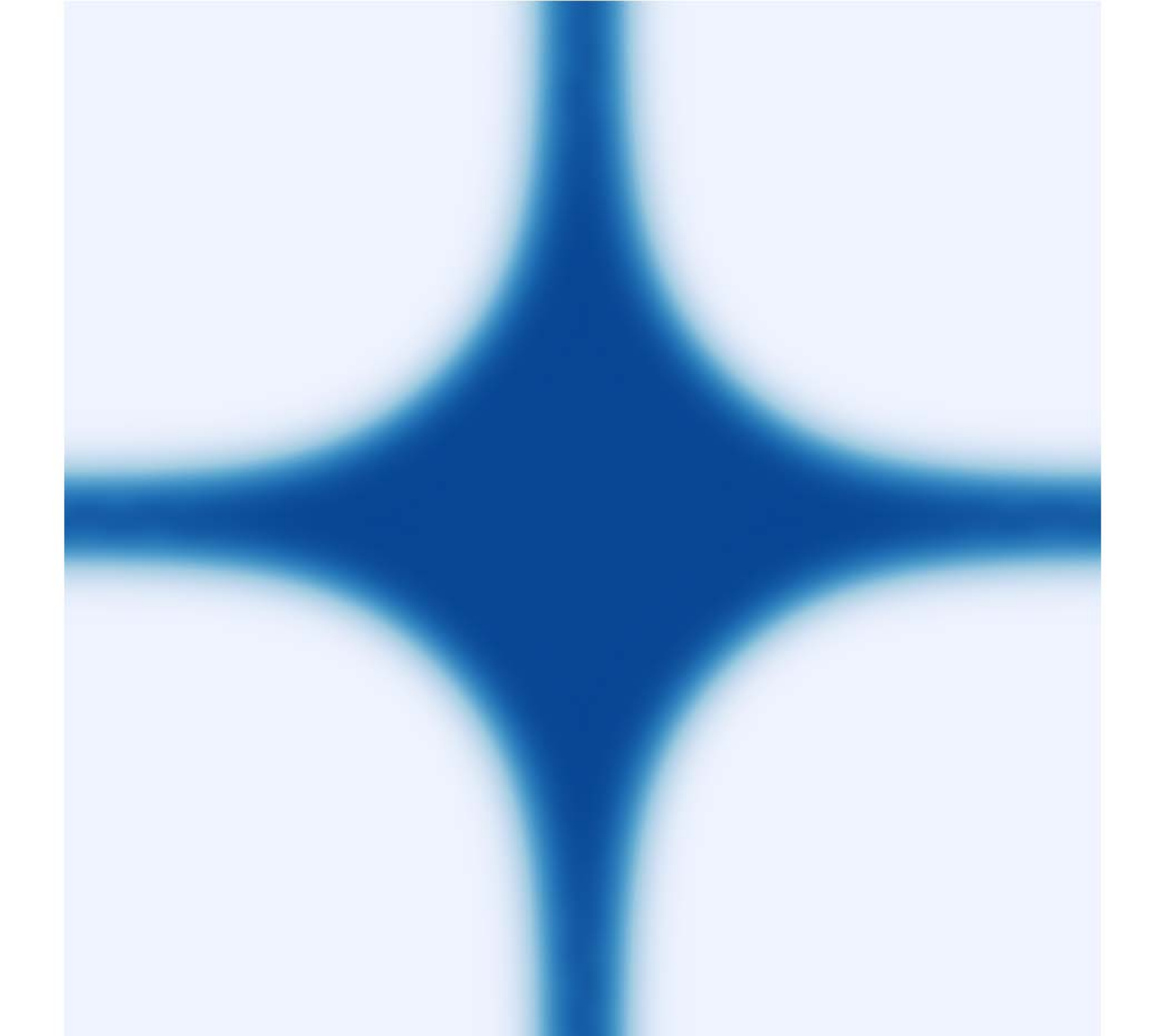}
	\includegraphics[width=0.18\columnwidth,trim={0cm 0cm 0cm 0cm},clip]{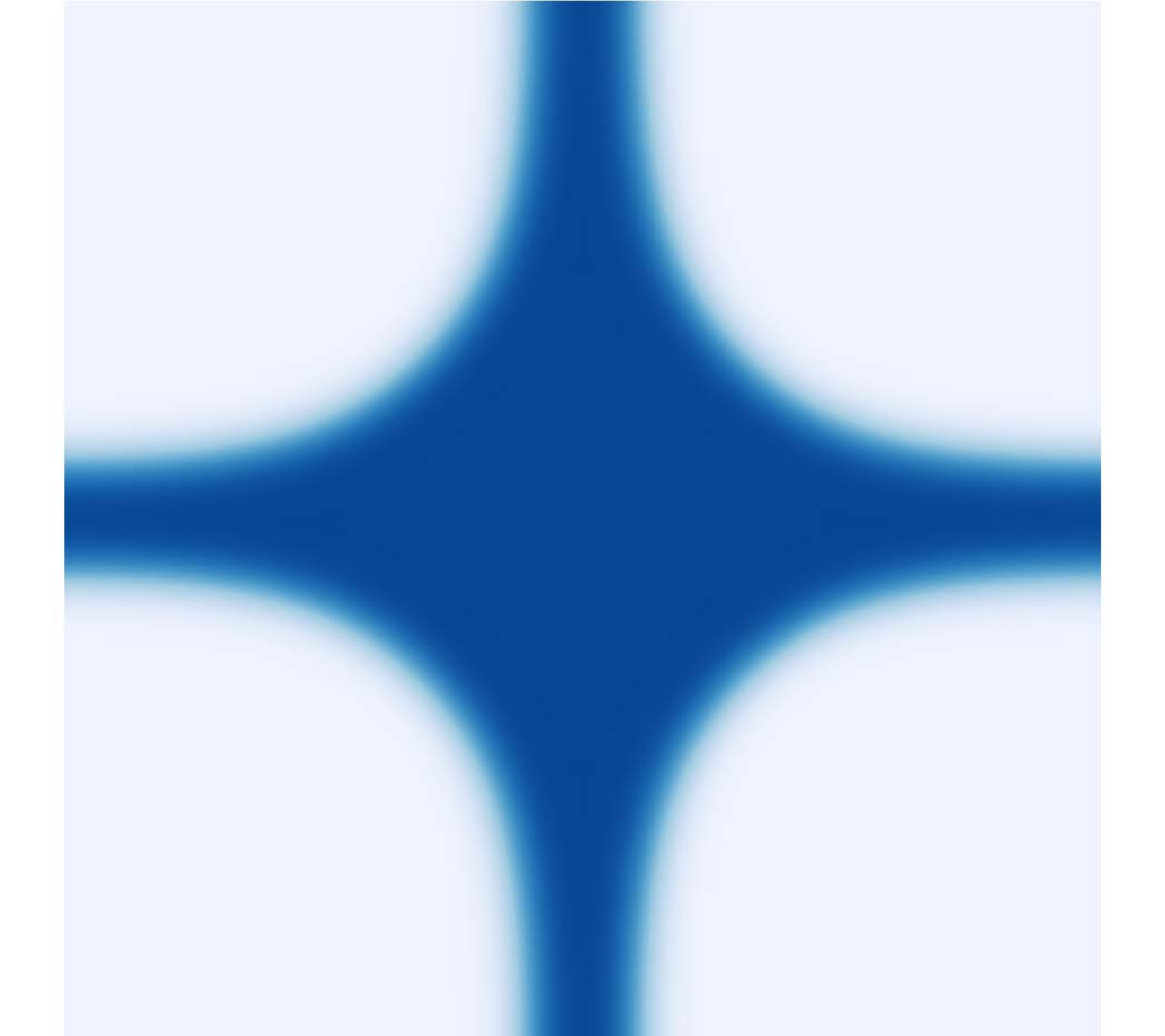}
	\includegraphics[width=0.18\columnwidth,trim={0cm 0cm 0cm 0cm},clip]{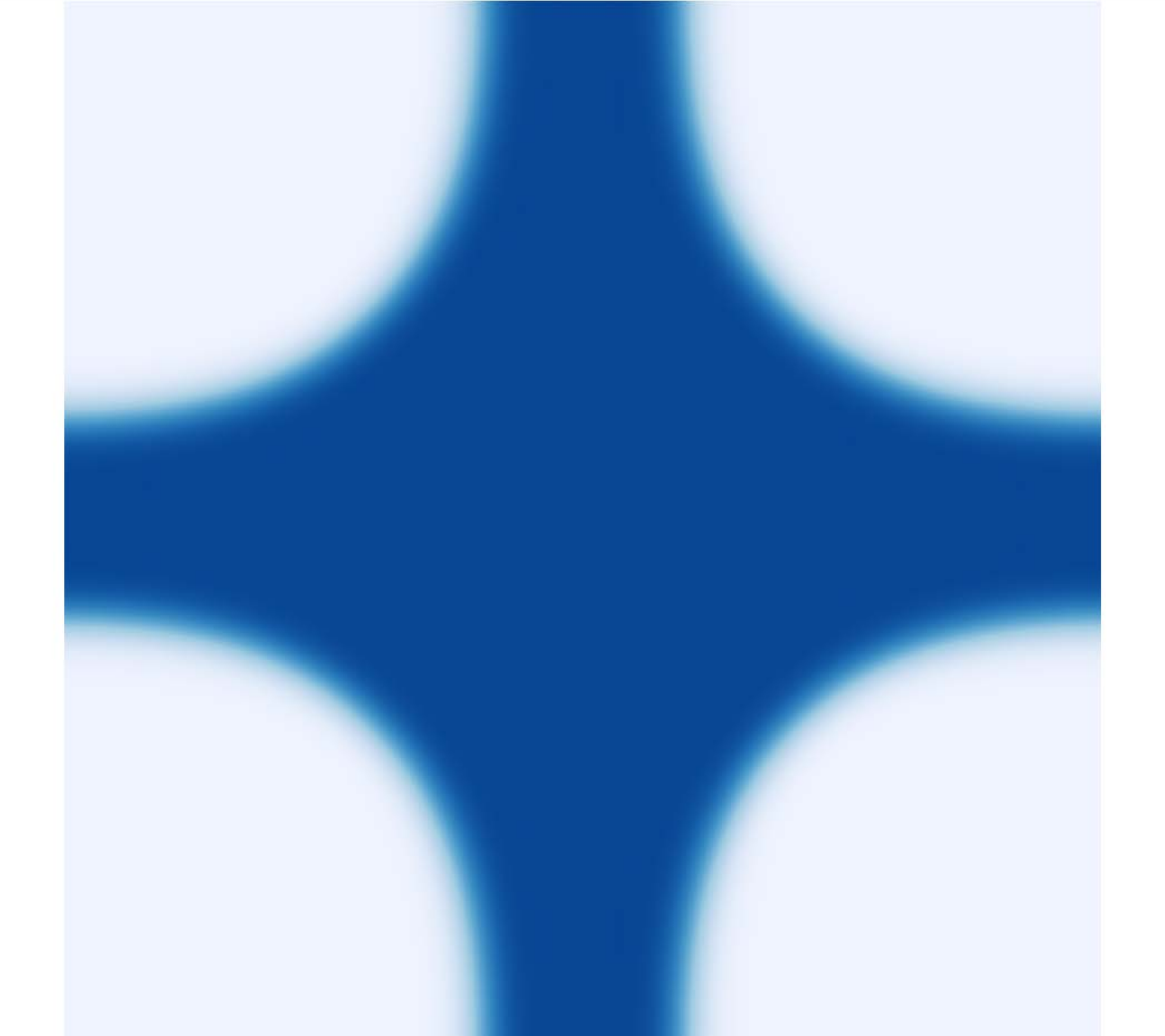}
	\centering
	\caption{Density contours in horizontal $(y,z)$ planes for 3D RTI at $At=0.998$ and $Re=3000$. $t^*=2$.
		$k=60,120,180,240,300$.
		The plane altitude is $x=k/128$.}
	\label{fig:Densitycontours0998}
\end{figure}

To demonstrate the stability of our approach, a large density ratio and high Reynolds number case ($At=0.998$ and $Re=3000$) is simulated.
The other parameters are $U_0=0.02$, $CFL=0.25$, $\mu_B/\mu_A = 2.0$, $W=5.0$, $Ca= 0.44$, $Pe=1000$.
The interface evolution, time evolution of three characteristic positions, and density contours on horizontal planes are shown in Figs.~\ref{fig:RT3Dinterface0998}-\ref{fig:Densitycontours0998}, respectively.
Similar as the 2D large density ratio RTI case in subsection~\ref{subsec: 2DRTI}, the interface profiles are smooth because the light fluid has little effect on the heavy fluid at such a large density ratio.
In this case, we could not find any previous study that simulated the 3D RTI at the same parameters, hence our results can serve as a benchmark for other researchers in the future.


\section{SUMMARY AND CONCLUSIONS} \label{sec: Concl}
In this paper, 
a 3D DUGKS approach is developed, with two essential and logical improvements, in order to simulate the immiscible two-phase flows at large density ratios and high flow Reynolds numbers.
For the two-phase model, the free-energy-based phase-field model is used.
The evolution of order parameter satisfies the conservative Allen-Cahn equation, a second-order partial differential equation.
To better reconstruct the particle
distribution functions at cell interfaces and improve the numerical stability,
a third-order weighted essentially non-oscillatory (WENO) scheme is applied.

Five benchmark problems are simulated to validate the PF(AC)-DUGKS-WENO code.
For all the cases, the density ratio can reach $\rho^*\sim 1000$ without encountering numerical instability.
For a 2D stationary droplet, the density profiles and the Young-Laplace law agree well with the analytical results, for different radii and different surface tensions.
For 2D Rayleigh-Taylor instability, the evolution of the positions of bubble front and spike tip agrees well with the previous studies, and $Re\sim 3000$ is simulated to compare with the previous results when $\rho^*\sim 1000$.
For a 2D droplet impacting on a thin liquid film, both the deposition phenomenon and splashing phenomenon are reproduced, and the evolution of the impact radius fits well with the literature data.

For 3D binary-droplet collision, the evolution process of fluid-fluid interface at $\rho^*\sim 1000$ is very similar 
to the previous experimental result, and a small satellite droplet is reproduced between the two droplets.
This case involves merging of two interfaces as well as breakup at different times.
Furthermore, we studied how the density ratio affect the collision and breakup processes.
When $\rho^*$ becomes smaller, the satellite droplet is found to be also smaller.
When $\rho^*\sim 1$, the two droplets would not break up in the end, due to the strong damping effect of background flow field.

For the case of 3D Rayleigh-Taylor instability, the evolution of the positions of bubble front, saddle point and spike tip at a density ratio of 3 agrees well with the results from previous studies. 
The density contours on the horizontal planes are also similar as those in the previous results.
We then simulated the 3D case with $\rho^* = 1000$ and $Re = 3000$. To our knowledge, this is the first 3D simulation of
RTI at such high levels of density ratio and Reynolds number, and these results could serve as a benchmark for future 3D studies.

Compared to Yang~{\it et al.}~\cite{Yangzeren2019}'s result, PF(AC)-DUGKS coupled with the WENO scheme
 is numerically more stable, which can simulate multiphase flows at larger density ratios and higher Reynolds numbers.
Furthermore, we extend the method to 3D problems, and the results agree well with the experimental results.
Hence our improved DUGKS approach makes the DUGKS scheme more capable in dealing with the realistic
two-phase flow problems.

\begin{acknowledgments}
This work has been supported by 
the National Natural Science Foundation of China (NSFC award numbers 91852205, 91741101 \& 11961131006), NSFC Basic Science Center Program (Award number 11988102), 
the National Numerical Wind Tunnel program, the Taizhou-Shenzhen Innovation Center, 
Guangdong Provincial Key Laboratory of Turbulence Research and
Applications (2019B21203001), Guangdong-Hong Kong-Macao Joint Laboratory for Data-Driven Fluid Mechanics and Engineering Applications (2020B1212030001) and Shenzhen Science and Technology Program (Grant No. KQTD20180411143441009). Computing resources are provided by
the Center for Computational Science and Engineering of Southern University
of Science and Technology. 
\end{acknowledgments}

\section*{DATA AVAILABILITY}
The data that support the findings of this study are available
from the corresponding author upon reasonable request.	
	
\appendix

\section{Integral constraints of the model Boltzmann equations for the ACNS system based on the Chapman-Enskog analysis}\label{ap:InverseDesign}

The purpose of this appendix is to derive the moment-integral constraints for $f^{eq}, g^{eq}, S_{\alpha}^f$, and  $S_{\alpha}^g$  in the model Boltzmann equations, {\it i.e.},
Eqs.~(\ref{Boltzmannf}-\ref{Boltzmanng}),  in order to
recover the ACNS system, Eqs.~(\ref{EqAC}-\ref{EqMo}). 
In particular, Eq.~\eqref{Boltzmannf} is applied to reproduce the conservative AC equation, Eq.~\eqref{EqAC}, while Eq.~\eqref{Boltzmanng} is applied to reproduce the hydrodynamic equations, Eqs.~\eqref{EqMa} and~\eqref{EqMo}.
Here $f$ represents the distribution function for order parameter 
and $g$ the distribution function for velocity/pressure.
The density is obtained from order parameter by a linear relationship.

To reproduce the conservative AC equation, we take the zeroth-order moment of Eq.~\eqref{Boltzmannf} as
\begin{equation}\label{intfBoltzmann_AC}
\int\left[\frac{\partial f}{\partial t}+{\xi}_m\frac{\partial f}{\partial x_m} =-\frac{f-f^{e q}}{\tau_{f}}+S^{f}\right] d \boldsymbol{\xi}.
\end{equation}   
Comparing to Eq.~\eqref{EqAC}, we can define    
\begin{equation}
\phi=\int f d \boldsymbol{\xi}.
\end{equation}
The collision operator in Eq.~\eqref{Boltzmannf} needs to conserve the order parameter, hence the zeroth-order moment of $f^{eq}$ is    
\begin{equation}\label{intfeq0_AC}
\int f^{eq} d \boldsymbol{\xi}=\phi.
\end{equation}    
Furthermore, to obtain the mean advection term, the first-order moment of $f^{eq}$ is chosen to yield
\begin{equation}\label{intfeq1_AC}
\int f^{eq}{\xi}_j d \boldsymbol{\xi}=\phi u_j.
\end{equation}    
Then Eq.~\eqref{intfBoltzmann_AC} becomes
\begin{equation}\label{intBftoAC}
\frac{\partial \phi}{\partial t}+\frac{\partial }{\partial x_m}
\left(\phi u_m \right)  
=\int S^{f} d \boldsymbol{\xi}
-\frac{\partial }{\partial x_m}
\int\left(f-f^{eq} \right){\xi}_m d \boldsymbol{\xi}.
\end{equation} 
Using the Chapman-Enskog expansion, we have
\begin{equation}\label{intfxi_AC}
\begin{aligned}
&\int\left(f-f^{e q}\right) \xi_{j} d \boldsymbol{\xi}
\\=&-\tau_{f}\left[\frac{\partial}{\partial t} \int f^{eq} \xi_{j} d \boldsymbol{\xi}+\frac{\partial}{\partial x_{n}} \int f^{e q} \xi_{n} \xi_{j} d \boldsymbol{\xi}-\int S^{f} \xi_{j} d \boldsymbol{\xi}\right]+{\cal O}\left(\tau_{f}^{2}\right)
\\=&-\tau_{f}\left[\begin{aligned}
\frac{\partial}{\partial t}\left(\phi u_{j}\right)+\frac{\partial}{\partial x_{n}} \int f^{e q} \xi_{n} \xi_{j} d \boldsymbol{\xi}-\int S^{f} \xi_{j} d \boldsymbol{\xi}\end{aligned}\right]+{\cal O}\left(\tau_{f}^{2}\right).
\end{aligned}
\end{equation}
Substituting Eq.~\eqref{intfxi_AC} into Eq.~\eqref{intBftoAC} and compared with Eq.~\eqref{EqAC}, the following integral condition is required,
\begin{equation}\label{intfcondition_AC}
\begin{aligned}
&\int S^{f} d \boldsymbol{\xi}
+\frac{\partial }{\partial x_m}
\left\lbrace 
\tau_{f}\left[\begin{aligned}
\frac{\partial}{\partial t}\left(\phi u_{m}\right)+\frac{\partial}{\partial x_{n}} \int f^{e q} \xi_{n} \xi_{m} d \boldsymbol{\xi}-\int S^{f} \xi_{m} d \boldsymbol{\xi}\end{aligned}\right]
\right\rbrace 
\\ \sim &
\frac{\partial }{\partial x_m}
\left[M_{AC} \left(\frac{\partial \phi}{\partial x_m} -\theta n_m\right) \right] .
\end{aligned}
\end{equation}
In order to match the arrangement of spatial derivatives on the two sides, a convenient choice to accommodate Eq.~\eqref{intfcondition_AC} is that
\begin{subequations}
	\begin{equation}\label{intSf_AC}
	\int S^{f} d \boldsymbol{\xi}=0,
	\end{equation} 
	\begin{equation}\label{convenientcond}
	\frac{\partial}{\partial x_{n}} \int f^{e q} \xi_{n} \xi_{m} d \boldsymbol{\xi}-\int S^{f} \xi_{m} d \boldsymbol{\xi}=\frac{M_{AC}}{\tau_{f}} \left(\frac{\partial \phi}{\partial x_m} -\theta n_m\right)
	-\frac{\partial}{\partial t}\left(\phi u_{m}\right),
	\end{equation} 
\end{subequations}
where $M_{AC}/\tau_{f}=RT=const.$ in this model.
Similarly, one possible choice for Eq.~\eqref{convenientcond} is
\begin{subequations}
	\begin{equation}\label{intfeq2_AC}
	\int f^{e q} \xi_{n} \xi_{m} d \boldsymbol{\xi}=RT\phi \delta_{nm},
	\end{equation} 
	\begin{equation}\label{intSf1_AC}
	\int S^{f} \xi_{m} d \boldsymbol{\xi}=RT \theta n_m
	+\frac{\partial}{\partial t}\left(\phi u_{m}\right).
	\end{equation} 
\end{subequations}
We comment that the time derivative term can be converted to spatial derivative terms via the ${\cal O}(1)$ Euler-like equations, 
but it is not possible here to convert this to a clean form like $\partial A_{mn} / \partial x_n$. If the latter were possible,
then the time derivative term could be merged into the second-order moment of $f^{eq}$ to eliminate the derivative calculation.

To recover the continuity equation for incompressible flow, we take the zeroth-order moment of Eq.~\eqref{Boltzmanng} as
\begin{equation}\label{intgBoltzmann_AC}
\int\left[\frac{\partial g}{\partial t}+{\xi}_m\frac{\partial g}{\partial x_m} =-\frac{g-g^{e q}}{\tau_{g}}+S^{g}\right] d \boldsymbol{\xi}.
\end{equation} 
As usual, the velocity field is obtained by
\begin{equation}\label{intgxi}
\rho u_j=\int g {\xi}_j d \boldsymbol{\xi}.
\end{equation}
Compared to the velocity-divergence-free condition, Eq.~\eqref{EqMa}, and assuming the conservation condition
 of the collision operator, the following integral condition is derived
\begin{equation}
\int S^{g} d \boldsymbol{\xi}-
\frac{\partial}{\partial t}\int g d \boldsymbol{\xi}=u_m
\frac{\partial  \rho }{\partial x_m}.
\end{equation}
A logical choice is then
\begin{subequations}
	\begin{equation}\label{intgeq_AC}
	0=\int g d \boldsymbol{\xi}=\int g^{eq} d \boldsymbol{\xi},
	\end{equation}
	\begin{equation}\label{intSg_AC}
	\int S^{g} d \boldsymbol{\xi}=u_m
	\frac{\partial  \rho }{\partial x_m}.
	\end{equation}
\end{subequations}

Next, the first-order moment of Eq.~\eqref{Boltzmanng} is needed to recover the momentum equation, {\it i.e.},
\begin{equation}\label{intgxiBoltzmann_AC}
\int\left[\frac{\partial g}{\partial t}+{\xi}_m\frac{\partial g}{\partial x_m} =-\frac{g-g^{e q}}{\tau_{g}}+S^{g}\right] {\xi}_j d \boldsymbol{\xi}.
\end{equation} 
Again, considering the conservation of collision operator, then Eq.~\eqref{intgxi} implies that  
\begin{equation}\label{intgeq1_AC}
\int g^{e q} {\xi}_j d \boldsymbol{\xi}=\rho u_j.
\end{equation} 
With the above results, Eq.~\eqref{intgxiBoltzmann_AC} can be written as
\begin{equation}\label{intBgxi}
\begin{aligned}
&\frac{\partial}{\partial t}\left(\rho u_{j}\right)+\frac{\partial}{\partial x_k}\int g^{e q} \xi_{j}\xi_{k} d \boldsymbol{\xi}
\\=&0+\int S^{g}\xi_{j} d \boldsymbol{\xi}
-\frac{\partial}{\partial x_k}\int\left(g-g^{e q}\right) \xi_{j}\xi_{k} d \boldsymbol{\xi}
\\=&\int S^{g}\xi_{j} d \boldsymbol{\xi}
+{\cal O}\left(\tau_{g} \right) .
\end{aligned}
\end{equation}
It is well known that $\mu\sim {\cal O}\left(\tau_{g} \right)$ in the mesoscopic approach. 
Then, to the leading-order, Eq.~\eqref{EqMo} can be written as
\begin{equation}\label{EqMom}
\frac{\partial(\rho \boldsymbol{u})}{\partial t}+\nabla \cdot(\rho \boldsymbol{u} \boldsymbol{u})=-\nabla p+\boldsymbol{F} +{\cal O}\left(\tau_{g} \right).
\end{equation}
Comparing to Eq.~\eqref{EqMom},
a natural choice of Eq.~\eqref{intBgxi} is that the source term provides the external force, while the equilibrium distribution function provides the spatial-derivative terms, {\it i.e.},
\begin{equation}\label{gxixiSxi_AC}
\int g^{eq} {\xi}_{j} {\xi}_{k} d \boldsymbol{\xi}=\rho u_{j} u_{k}+p \delta_{jk}, \quad \int S^{g}\xi_{j} d \boldsymbol{\xi}={F}_{j}.
\end{equation}
The ${\cal O}(\tau_{g})$ term needs to recover the viscous term in Eq.~\eqref{EqMo}, {\it i.e.},
\begin{equation}\label{fxixiviscous}
\begin{aligned}
\int\left(g-g^{e q}\right) \xi_{j}\xi_{k} d \boldsymbol{\xi}
\sim
\left\lbrace -\mu\left(\frac{\partial u_j}{\partial x_{k}}+\frac{\partial u_k}{\partial x_{j}}\right)\right\rbrace .
\end{aligned}
\end{equation}
Using the Chapman-Enskog expansion, we have
\begin{equation}\label{fxixiCE}
\begin{aligned}
&\int\left(g-g^{e q}\right) \xi_{j}\xi_{k} d \boldsymbol{\xi}
\\=&-\tau_{g}\left[\frac{\partial}{\partial t} \int g^{eq} \xi_{j}\xi_{k} d \boldsymbol{\xi}+\frac{\partial}{\partial x_{m}} \int g^{e q} \xi_{m} \xi_{j}\xi_{k} d \boldsymbol{\xi}-\int S^{g} \xi_{j}\xi_{k} d \boldsymbol{\xi}\right]+{\cal O}\left(\tau_{g}^{2}\right)
\\=&-\tau_{g}
\left[ \begin{aligned}  
\frac{\partial}{\partial t}\left(\rho u_{k} u_{j}+p \delta_{k j}\right)
+\frac{\partial}{\partial x_{m}} \int g^{e q} \xi_{m} \xi_{j}\xi_{k} d \boldsymbol{\xi}-\int S^{g} \xi_{j}\xi_{k} d \boldsymbol{\xi}
\end{aligned}\right] +{\cal O}\left(\tau_{g}^{2}\right).
\end{aligned}
\end{equation}
To proceed further, we need to approximate the time derivatives of the macroscopic variables.
Since $M_{AC}\sim {\cal O}\left(\tau_{f} \right)$, Eq.~\eqref{EqMass} can be written as
\begin{equation}\label{EqMass1}
\frac{\partial \rho}{\partial t}+\nabla \cdot(\rho \boldsymbol{u})={\cal O}\left(\tau_{f} \right).
\end{equation}
From Eq.~\eqref{EqMass1}, Eq.~\eqref{EqMom} and Eq.~\eqref{EqMa}, the evolution of density and velocity can be expressed as
\begin{subequations}
	\begin{equation}\label{drhodt}
	\frac{\partial \rho}{\partial t}=-u_{m}\frac{\partial\rho}{\partial x_m}+{\cal O}\left(\tau_{f} \right),
	\end{equation}
	\begin{equation}\label{dudt}
	\frac{\partial u_{j}}{\partial t}=-u_{m} \frac{\partial u_{j}}{\partial x_{m}}-\frac{1}{\rho} \frac{\partial p}{\partial x_{j}}+\frac{F_{j}}{\rho}
	+{\cal O}\left(\tau_{g} \right)
	+{\cal O}\left(\tau_{f}Ma \right).
	\end{equation}
\end{subequations}
Hence the time derivative terms in Eq.~\eqref{fxixiCE} can be converted to
\begin{equation}\label{drhouudt}
\begin{aligned}
&\frac{\partial}{\partial t}\left(\rho u_{k} u_{j}+p \delta_{kj}\right)
\\=&
u_{k} u_{j}\frac{\partial\rho}{\partial t}+
\rho u_{j}\frac{\partial u_{k}}{\partial t}+
\rho u_{k}\frac{\partial u_{j}}{\partial t}+
\frac{\partial p}{\partial t} \delta_{kj}
\\=&
u_{j} F_k+u_{k} F_j
+{\cal O}\left(\tau_{f}Ma^2,\tau_{g}Ma,Ma^2\right).
\end{aligned}
\end{equation}
In the above, the time derivative of the pressure term is assumed to scale as ${\cal O}(Ma^2)$, which applies for
the regions outside the fluid-fluid interfaces. However, within the interfaces, due to
the interface advection, $\partial p /\partial t \sim  u_I \sigma / (W R)$, where $u_I$ is the fluid velocity at the interface. Therefore, it
 follows that a condition $\sigma / W < \rho u^2$ would be required for dynamic interface problems.

Substituting Eq.~\eqref{drhouudt}  into Eq.~\eqref{fxixiCE}, then Eq.~\eqref{fxixiviscous}
yields the following
 moment-integral condition 
\begin{equation}\label{gxixixiSxixi_AC}
\begin{aligned}
&\frac{\partial}{\partial x_{m}} \int g^{eq} \xi_{m} \xi_{j}\xi_{k} d \boldsymbol{\xi}- \int S^{g} \xi_{j}\xi_{k} d \boldsymbol{\xi} 
\\ \sim &
\left\lbrace 
\frac{\mu}{\tau_{g}}\left(\frac{\partial u_j}{\partial x_{k}}+\frac{\partial u_k}{\partial x_{j}}\right) 
-u_{j} F_k-u_{k} F_j
\right\rbrace .
\end{aligned}
\end{equation}
Given the specific truncated form of the equilibrium $g^{eq}$, we have
\begin{equation}
\int g^{eq} \xi_{m} \xi_{j}\xi_{k} d \boldsymbol{\xi} = \rho RT \left( u_k \delta_{mj} + u_j \delta_{mk} + u_m \delta_{jk}  \right),
\end{equation}
then we obtain
\begin{equation}
\int S^{g} \xi_{j}\xi_{k} d \boldsymbol{\xi}  =  u_{j} F_k + u_{k} F_j +  
RT \left( u_j {\partial \rho \over \partial x_k} +  u_k {\partial \rho \over \partial x_j} + \delta_{jk} u_m {\partial \rho \over \partial x_m}  \right).
\end{equation}
Certainly, this is only one of the many possible combinations of the third-order moment of  $g^{eq}$ and the 
second-order moment of $S^{g}$.

To summarize, the moment-integral constraints for the model Boltzmann equations are:
\begin{enumerate}
	\item The conservative AC equation constrains to the zeroth-, first-, and second-order moments of $f^{eq}$, Eqs.~\eqref{intfeq0_AC},~\eqref{intfeq1_AC} and~\eqref{intfeq2_AC}, and the zeroth- and first-order moments for $S^f$, 
	Eqs.~\eqref{intSf_AC} and~\eqref{intSf1_AC};
	\item The continuity equation sets the
	zeroth-order moment for $g^{eq}$, Eq.~(\ref{intgeq_AC}) and the zeroth-order moment for $S^g$, Eq.~(\ref{intSg_AC});
	\item The momentum equation provides the
	first-, second- and third-order moments for $g^{eq}$, first-order and second-order moments for $S^g$, {\it i.e.},
	Eq.~(\ref{intgeq1_AC}), Eq.~(\ref{gxixiSxi_AC}), and Eq.~(\ref{gxixixiSxixi_AC}).
\end{enumerate}

We point out that the above integral conditions are designed as one of the many possibilities. 
Furthermore, even with this design stated by the above integral constraints, there are many ways to specify the
precise forms for $f^{eq}$, $g^{eq}$, $S^f$, and $S^g$.  We can confirm that the specific forms given in Eqs.~(\ref{feqSf}) and~(\ref{geqSg}) 
do satisfy all the requirements stated above. 
We can also introduce other specific forms, 
for example, 
utilizing the Hermite expansion formulae, as done by Chen {\it et al.}.~\cite{2021Inverse}
Eqs.~(\ref{feqSf}) and~(\ref{geqSg}) are used in our large-density-ratio simulations, because their capability have been verified by Yang~{\it et al.},~\cite{Yangzeren2019} and we combine the WENO treatment to improve the numerical stability of the PF(AC)-DUGKS approach.

\bibliography{cas-refs}

\begin{thebibliography}{63}%
\makeatletter
\providecommand \@ifxundefined [1]{%
 \@ifx{#1\undefined}
}%
\providecommand \@ifnum [1]{%
 \ifnum #1\expandafter \@firstoftwo
 \else \expandafter \@secondoftwo
 \fi
}%
\providecommand \@ifx [1]{%
 \ifx #1\expandafter \@firstoftwo
 \else \expandafter \@secondoftwo
 \fi
}%
\providecommand \natexlab [1]{#1}%
\providecommand \enquote  [1]{``#1''}%
\providecommand \bibnamefont  [1]{#1}%
\providecommand \bibfnamefont [1]{#1}%
\providecommand \citenamefont [1]{#1}%
\providecommand \href@noop [0]{\@secondoftwo}%
\providecommand \href [0]{\begingroup \@sanitize@url \@href}%
\providecommand \@href[1]{\@@startlink{#1}\@@href}%
\providecommand \@@href[1]{\endgroup#1\@@endlink}%
\providecommand \@sanitize@url [0]{\catcode `\\12\catcode `\$12\catcode
  `\&12\catcode `\#12\catcode `\^12\catcode `\_12\catcode `\%12\relax}%
\providecommand \@@startlink[1]{}%
\providecommand \@@endlink[0]{}%
\providecommand \url  [0]{\begingroup\@sanitize@url \@url }%
\providecommand \@url [1]{\endgroup\@href {#1}{\urlprefix }}%
\providecommand \urlprefix  [0]{URL }%
\providecommand \Eprint [0]{\href }%
\providecommand \doibase [0]{http://dx.doi.org/}%
\providecommand \selectlanguage [0]{\@gobble}%
\providecommand \bibinfo  [0]{\@secondoftwo}%
\providecommand \bibfield  [0]{\@secondoftwo}%
\providecommand \translation [1]{[#1]}%
\providecommand \BibitemOpen [0]{}%
\providecommand \bibitemStop [0]{}%
\providecommand \bibitemNoStop [0]{.\EOS\space}%
\providecommand \EOS [0]{\spacefactor3000\relax}%
\providecommand \BibitemShut  [1]{\csname bibitem#1\endcsname}%
\let\auto@bib@innerbib\@empty
\bibitem [{\citenamefont {Wang}\ \emph
  {et~al.}(2015{\natexlab{a}})\citenamefont {Wang}, \citenamefont {Shu},
  \citenamefont {Huang},\ and\ \citenamefont {Teo}}]{WANG2015404}%
  \BibitemOpen
  \bibfield  {author} {\bibinfo {author} {\bibfnamefont {Y.}~\bibnamefont
  {Wang}}, \bibinfo {author} {\bibfnamefont {C.}~\bibnamefont {Shu}}, \bibinfo
  {author} {\bibfnamefont {H.}~\bibnamefont {Huang}}, \ and\ \bibinfo {author}
  {\bibfnamefont {C.}~\bibnamefont {Teo}},\ }\bibfield  {title} {\enquote
  {\bibinfo {title} {{Multiphase lattice Boltzmann flux solver for
  incompressible multiphase flows with large density ratio}},}\ }\href
  {\doibase https://doi.org/10.1016/j.jcp.2014.09.035} {\bibfield  {journal}
  {\bibinfo  {journal} {Journal of Computational Physics}\ }\textbf {\bibinfo
  {volume} {280}},\ \bibinfo {pages} {404--423} (\bibinfo {year}
  {2015}{\natexlab{a}})}\BibitemShut {NoStop}%
\bibitem [{\citenamefont {Wang}, \citenamefont {Shu},\ and\ \citenamefont
  {Yang}(2015)}]{WANG201541}%
  \BibitemOpen
  \bibfield  {author} {\bibinfo {author} {\bibfnamefont {Y.}~\bibnamefont
  {Wang}}, \bibinfo {author} {\bibfnamefont {C.}~\bibnamefont {Shu}}, \ and\
  \bibinfo {author} {\bibfnamefont {L.}~\bibnamefont {Yang}},\ }\bibfield
  {title} {\enquote {\bibinfo {title} {{An improved multiphase lattice
  Boltzmann flux solver for three-dimensional flows with large density ratio
  and high Reynolds number}},}\ }\href {\doibase
  https://doi.org/10.1016/j.jcp.2015.08.049} {\bibfield  {journal} {\bibinfo
  {journal} {Journal of Computational Physics}\ }\textbf {\bibinfo {volume}
  {302}},\ \bibinfo {pages} {41--58} (\bibinfo {year} {2015})}\BibitemShut
  {NoStop}%
\bibitem [{\citenamefont {Chen}\ \emph {et~al.}(2018)\citenamefont {Chen},
  \citenamefont {Shu}, \citenamefont {Tan}, \citenamefont {Niu},\ and\
  \citenamefont {Li}}]{PhysRevE.98.063314}%
  \BibitemOpen
  \bibfield  {author} {\bibinfo {author} {\bibfnamefont {Z.}~\bibnamefont
  {Chen}}, \bibinfo {author} {\bibfnamefont {C.}~\bibnamefont {Shu}}, \bibinfo
  {author} {\bibfnamefont {D.}~\bibnamefont {Tan}}, \bibinfo {author}
  {\bibfnamefont {X.~D.}\ \bibnamefont {Niu}}, \ and\ \bibinfo {author}
  {\bibfnamefont {Q.~Z.}\ \bibnamefont {Li}},\ }\bibfield  {title} {\enquote
  {\bibinfo {title} {{Simplified multiphase lattice Boltzmann method for
  simulating multiphase flows with large density ratios and complex
  interfaces}},}\ }\href {\doibase 10.1103/PhysRevE.98.063314} {\bibfield
  {journal} {\bibinfo  {journal} {Phys. Rev. E}\ }\textbf {\bibinfo {volume}
  {98}},\ \bibinfo {pages} {063314} (\bibinfo {year} {2018})}\BibitemShut
  {NoStop}%
\bibitem [{\citenamefont {Liang}\ \emph {et~al.}(2018)\citenamefont {Liang},
  \citenamefont {Xu}, \citenamefont {Chen}, \citenamefont {Wang}, \citenamefont
  {Chai},\ and\ \citenamefont {Shi}}]{PhysRevE.97.033309}%
  \BibitemOpen
  \bibfield  {author} {\bibinfo {author} {\bibfnamefont {H.}~\bibnamefont
  {Liang}}, \bibinfo {author} {\bibfnamefont {J.}~\bibnamefont {Xu}}, \bibinfo
  {author} {\bibfnamefont {J.}~\bibnamefont {Chen}}, \bibinfo {author}
  {\bibfnamefont {H.}~\bibnamefont {Wang}}, \bibinfo {author} {\bibfnamefont
  {Z.}~\bibnamefont {Chai}}, \ and\ \bibinfo {author} {\bibfnamefont
  {B.}~\bibnamefont {Shi}},\ }\bibfield  {title} {\enquote {\bibinfo {title}
  {{Phase-field-based lattice Boltzmann modeling of large-density-ratio
  two-phase flows}},}\ }\href {\doibase 10.1103/PhysRevE.97.033309} {\bibfield
  {journal} {\bibinfo  {journal} {Phys. Rev. E}\ }\textbf {\bibinfo {volume}
  {97}},\ \bibinfo {pages} {033309} (\bibinfo {year} {2018})}\BibitemShut
  {NoStop}%
\bibitem [{\citenamefont {Liang}\ \emph {et~al.}(2019)\citenamefont {Liang},
  \citenamefont {Liu}, \citenamefont {Chai},\ and\ \citenamefont
  {Shi}}]{PhysRevE.99.063306}%
  \BibitemOpen
  \bibfield  {author} {\bibinfo {author} {\bibfnamefont {H.}~\bibnamefont
  {Liang}}, \bibinfo {author} {\bibfnamefont {H.}~\bibnamefont {Liu}}, \bibinfo
  {author} {\bibfnamefont {Z.}~\bibnamefont {Chai}}, \ and\ \bibinfo {author}
  {\bibfnamefont {B.}~\bibnamefont {Shi}},\ }\bibfield  {title} {\enquote
  {\bibinfo {title} {Lattice {Boltzmann} method for contact-line motion of
  binary fluids with high density ratio},}\ }\href {\doibase
  10.1103/PhysRevE.99.063306} {\bibfield  {journal} {\bibinfo  {journal} {Phys.
  Rev. E}\ }\textbf {\bibinfo {volume} {99}},\ \bibinfo {pages} {063306}
  (\bibinfo {year} {2019})}\BibitemShut {NoStop}%
\bibitem [{\citenamefont {Fakhari}\ \emph {et~al.}(2017)\citenamefont
  {Fakhari}, \citenamefont {Mitchell}, \citenamefont {Leonardi},\ and\
  \citenamefont {Bolster}}]{2017Improved}%
  \BibitemOpen
  \bibfield  {author} {\bibinfo {author} {\bibfnamefont {A.}~\bibnamefont
  {Fakhari}}, \bibinfo {author} {\bibfnamefont {T.}~\bibnamefont {Mitchell}},
  \bibinfo {author} {\bibfnamefont {C.}~\bibnamefont {Leonardi}}, \ and\
  \bibinfo {author} {\bibfnamefont {D.}~\bibnamefont {Bolster}},\ }\bibfield
  {title} {\enquote {\bibinfo {title} {{Improved locality of the phase-field
  lattice-Boltzmann model for immiscible fluids at high density ratios}},}\
  }\href {\doibase 10.1103/PhysRevE.96.053301} {\bibfield  {journal} {\bibinfo
  {journal} {Phys. Rev. E}\ }\textbf {\bibinfo {volume} {96}},\ \bibinfo
  {pages} {053301} (\bibinfo {year} {2017})}\BibitemShut {NoStop}%
\bibitem [{\citenamefont {Kumar}, \citenamefont {Sannasiraj},\ and\
  \citenamefont {Sundar}(2019)}]{2019Phase}%
  \BibitemOpen
  \bibfield  {author} {\bibinfo {author} {\bibfnamefont {E.~D.}\ \bibnamefont
  {Kumar}}, \bibinfo {author} {\bibfnamefont {S.~A.}\ \bibnamefont
  {Sannasiraj}}, \ and\ \bibinfo {author} {\bibfnamefont {V.}~\bibnamefont
  {Sundar}},\ }\bibfield  {title} {\enquote {\bibinfo {title} {{Phase field
  lattice Boltzmann model for air-water two phase flows}},}\ }\href {\doibase
  10.1063/1.5100215} {\bibfield  {journal} {\bibinfo  {journal} {Physics of
  Fluids}\ }\textbf {\bibinfo {volume} {31}},\ \bibinfo {pages} {072103}
  (\bibinfo {year} {2019})}\BibitemShut {NoStop}%
\bibitem [{\citenamefont {Yang}, \citenamefont {Zhong},\ and\ \citenamefont
  {Zhuo}(2019)}]{Yangzeren2019}%
  \BibitemOpen
  \bibfield  {author} {\bibinfo {author} {\bibfnamefont {Z.}~\bibnamefont
  {Yang}}, \bibinfo {author} {\bibfnamefont {C.}~\bibnamefont {Zhong}}, \ and\
  \bibinfo {author} {\bibfnamefont {C.}~\bibnamefont {Zhuo}},\ }\bibfield
  {title} {\enquote {\bibinfo {title} {Phase-field method based on discrete
  unified gas-kinetic scheme for large-density-ratio two-phase flows},}\ }\href
  {\doibase 10.1103/PhysRevE.99.043302} {\bibfield  {journal} {\bibinfo
  {journal} {Phys. Rev. E}\ }\textbf {\bibinfo {volume} {99}},\ \bibinfo
  {pages} {043302} (\bibinfo {year} {2019})}\BibitemShut {NoStop}%
\bibitem [{\citenamefont {Yang}\ \emph {et~al.}(2022)\citenamefont {Yang},
  \citenamefont {Liu}, \citenamefont {Zhuo},\ and\ \citenamefont
  {Zhong}}]{doi:10.1063/5.0086723}%
  \BibitemOpen
  \bibfield  {author} {\bibinfo {author} {\bibfnamefont {Z.}~\bibnamefont
  {Yang}}, \bibinfo {author} {\bibfnamefont {S.}~\bibnamefont {Liu}}, \bibinfo
  {author} {\bibfnamefont {C.}~\bibnamefont {Zhuo}}, \ and\ \bibinfo {author}
  {\bibfnamefont {C.}~\bibnamefont {Zhong}},\ }\bibfield  {title} {\enquote
  {\bibinfo {title} {Conservative multilevel discrete unified gas kinetic
  scheme for modeling multiphase flows with large density ratios},}\ }\href
  {\doibase 10.1063/5.0086723} {\bibfield  {journal} {\bibinfo  {journal}
  {Physics of Fluids}\ }\textbf {\bibinfo {volume} {34}},\ \bibinfo {pages}
  {043316} (\bibinfo {year} {2022})}\BibitemShut {NoStop}%
\bibitem [{\citenamefont {Juric}\ and\ \citenamefont
  {Tryggvason}(1996)}]{1996A}%
  \BibitemOpen
  \bibfield  {author} {\bibinfo {author} {\bibfnamefont {D.}~\bibnamefont
  {Juric}}\ and\ \bibinfo {author} {\bibfnamefont {G.}~\bibnamefont
  {Tryggvason}},\ }\bibfield  {title} {\enquote {\bibinfo {title} {{A
  Front-Tracking Method for Dendritic Solidification}},}\ }\href {\doibase
  https://doi.org/10.1006/jcph.1996.0011} {\bibfield  {journal} {\bibinfo
  {journal} {Journal of Computational Physics}\ }\textbf {\bibinfo {volume}
  {123}},\ \bibinfo {pages} {127--148} (\bibinfo {year} {1996})}\BibitemShut
  {NoStop}%
\bibitem [{\citenamefont {Scardovelli}\ and\ \citenamefont
  {Zaleski}(1999)}]{Scardovelli1999DIRECT}%
  \BibitemOpen
  \bibfield  {author} {\bibinfo {author} {\bibfnamefont {R.}~\bibnamefont
  {Scardovelli}}\ and\ \bibinfo {author} {\bibfnamefont {S.}~\bibnamefont
  {Zaleski}},\ }\bibfield  {title} {\enquote {\bibinfo {title} {{DIRECT
  NUMERICAL SIMULATION OF FREE-SURFACE AND INTERFACIAL FLOW}},}\ }\href
  {\doibase 10.1146/annurev.fluid.31.1.567} {\bibfield  {journal} {\bibinfo
  {journal} {Annual Review of Fluid Mechanics}\ }\textbf {\bibinfo {volume}
  {31}},\ \bibinfo {pages} {567--603} (\bibinfo {year} {1999})}\BibitemShut
  {NoStop}%
\bibitem [{\citenamefont {Balcázar}\ \emph {et~al.}(2016)\citenamefont
  {Balcázar}, \citenamefont {Lehmkuhl}, \citenamefont {Jofre}, \citenamefont
  {Rigola},\ and\ \citenamefont {Oliva}}]{2021A}%
  \BibitemOpen
  \bibfield  {author} {\bibinfo {author} {\bibfnamefont {N.}~\bibnamefont
  {Balcázar}}, \bibinfo {author} {\bibfnamefont {O.}~\bibnamefont {Lehmkuhl}},
  \bibinfo {author} {\bibfnamefont {L.}~\bibnamefont {Jofre}}, \bibinfo
  {author} {\bibfnamefont {J.}~\bibnamefont {Rigola}}, \ and\ \bibinfo {author}
  {\bibfnamefont {A.}~\bibnamefont {Oliva}},\ }\bibfield  {title} {\enquote
  {\bibinfo {title} {A coupled volume-of-fluid/level-set method for simulation
  of two-phase flows on unstructured meshes},}\ }\href {\doibase
  https://doi.org/10.1016/j.compfluid.2015.10.005} {\bibfield  {journal}
  {\bibinfo  {journal} {Computers and Fluids}\ }\textbf {\bibinfo {volume}
  {124}},\ \bibinfo {pages} {12--29} (\bibinfo {year} {2016})}\BibitemShut
  {NoStop}%
\bibitem [{\citenamefont {Sussman}\ \emph {et~al.}(1998)\citenamefont
  {Sussman}, \citenamefont {Fatemi}, \citenamefont {Smereka},\ and\
  \citenamefont {Osher}}]{SUSSMAN1998663}%
  \BibitemOpen
  \bibfield  {author} {\bibinfo {author} {\bibfnamefont {M.}~\bibnamefont
  {Sussman}}, \bibinfo {author} {\bibfnamefont {E.}~\bibnamefont {Fatemi}},
  \bibinfo {author} {\bibfnamefont {P.}~\bibnamefont {Smereka}}, \ and\
  \bibinfo {author} {\bibfnamefont {S.}~\bibnamefont {Osher}},\ }\bibfield
  {title} {\enquote {\bibinfo {title} {An improved level set method for
  incompressible two-phase flows},}\ }\href {\doibase
  https://doi.org/10.1016/S0045-7930(97)00053-4} {\bibfield  {journal}
  {\bibinfo  {journal} {Computers and Fluids}\ }\textbf {\bibinfo {volume}
  {27}},\ \bibinfo {pages} {663--680} (\bibinfo {year} {1998})}\BibitemShut
  {NoStop}%
\bibitem [{\citenamefont {Liu}, \citenamefont {Valocchi},\ and\ \citenamefont
  {Kang}(2012)}]{PhysRevE.85.046309}%
  \BibitemOpen
  \bibfield  {author} {\bibinfo {author} {\bibfnamefont {H.}~\bibnamefont
  {Liu}}, \bibinfo {author} {\bibfnamefont {A.~J.}\ \bibnamefont {Valocchi}}, \
  and\ \bibinfo {author} {\bibfnamefont {Q.}~\bibnamefont {Kang}},\ }\bibfield
  {title} {\enquote {\bibinfo {title} {{Three-dimensional lattice Boltzmann
  model for immiscible two-phase flow simulations}},}\ }\href {\doibase
  10.1103/PhysRevE.85.046309} {\bibfield  {journal} {\bibinfo  {journal} {Phys.
  Rev. E}\ }\textbf {\bibinfo {volume} {85}},\ \bibinfo {pages} {046309}
  (\bibinfo {year} {2012})}\BibitemShut {NoStop}%
\bibitem [{\citenamefont {Gunstensen}\ \emph {et~al.}(1991)\citenamefont
  {Gunstensen}, \citenamefont {Rothman}, \citenamefont {Zaleski},\ and\
  \citenamefont {Zanetti}}]{1991Lattice}%
  \BibitemOpen
  \bibfield  {author} {\bibinfo {author} {\bibfnamefont {A.~K.}\ \bibnamefont
  {Gunstensen}}, \bibinfo {author} {\bibfnamefont {D.~H.}\ \bibnamefont
  {Rothman}}, \bibinfo {author} {\bibfnamefont {S.}~\bibnamefont {Zaleski}}, \
  and\ \bibinfo {author} {\bibfnamefont {G.}~\bibnamefont {Zanetti}},\
  }\bibfield  {title} {\enquote {\bibinfo {title} {{Lattice Boltzmann model of
  immiscible fluids}},}\ }\href {\doibase 10.1103/PhysRevA.43.4320} {\bibfield
  {journal} {\bibinfo  {journal} {Phys. Rev. A}\ }\textbf {\bibinfo {volume}
  {43}},\ \bibinfo {pages} {4320--4327} (\bibinfo {year} {1991})}\BibitemShut
  {NoStop}%
\bibitem [{\citenamefont {Shan}\ and\ \citenamefont
  {Chen}(1993)}]{shan1993lattice}%
  \BibitemOpen
  \bibfield  {author} {\bibinfo {author} {\bibfnamefont {X.}~\bibnamefont
  {Shan}}\ and\ \bibinfo {author} {\bibfnamefont {H.}~\bibnamefont {Chen}},\
  }\bibfield  {title} {\enquote {\bibinfo {title} {{Lattice Boltzmann model for
  simulating flows with multiple phases and components}},}\ }\href {\doibase
  10.1103/PhysRevE.47.1815} {\bibfield  {journal} {\bibinfo  {journal} {Phys.
  Rev. E}\ }\textbf {\bibinfo {volume} {47}},\ \bibinfo {pages} {1815--1819}
  (\bibinfo {year} {1993})}\BibitemShut {NoStop}%
\bibitem [{\citenamefont {Shan}\ and\ \citenamefont
  {Chen}(1994)}]{1994Simulation}%
  \BibitemOpen
  \bibfield  {author} {\bibinfo {author} {\bibfnamefont {X.}~\bibnamefont
  {Shan}}\ and\ \bibinfo {author} {\bibfnamefont {H.}~\bibnamefont {Chen}},\
  }\bibfield  {title} {\enquote {\bibinfo {title} {Simulation of nonideal gases
  and liquid-gas phase transitions by the lattice {Boltzmann} equation},}\
  }\href {\doibase 10.1103/PhysRevE.49.2941} {\bibfield  {journal} {\bibinfo
  {journal} {Phys. Rev. E}\ }\textbf {\bibinfo {volume} {49}},\ \bibinfo
  {pages} {2941--2948} (\bibinfo {year} {1994})}\BibitemShut {NoStop}%
\bibitem [{\citenamefont {Swift}\ \emph {et~al.}(1996)\citenamefont {Swift},
  \citenamefont {Orlandini}, \citenamefont {Osborn},\ and\ \citenamefont
  {Yeomans}}]{swift1996lattice}%
  \BibitemOpen
  \bibfield  {author} {\bibinfo {author} {\bibfnamefont {M.~R.}\ \bibnamefont
  {Swift}}, \bibinfo {author} {\bibfnamefont {E.}~\bibnamefont {Orlandini}},
  \bibinfo {author} {\bibfnamefont {W.~R.}\ \bibnamefont {Osborn}}, \ and\
  \bibinfo {author} {\bibfnamefont {J.~M.}\ \bibnamefont {Yeomans}},\
  }\bibfield  {title} {\enquote {\bibinfo {title} {{Lattice Boltzmann
  simulations of liquid-gas and binary fluid systems}},}\ }\href {\doibase
  10.1103/PhysRevE.54.5041} {\bibfield  {journal} {\bibinfo  {journal} {Phys.
  Rev. E}\ }\textbf {\bibinfo {volume} {54}},\ \bibinfo {pages} {5041--5052}
  (\bibinfo {year} {1996})}\BibitemShut {NoStop}%
\bibitem [{\citenamefont {Rothman}\ and\ \citenamefont
  {Keller}(1988)}]{rothman1988immiscible}%
  \BibitemOpen
  \bibfield  {author} {\bibinfo {author} {\bibfnamefont {D.~H.}\ \bibnamefont
  {Rothman}}\ and\ \bibinfo {author} {\bibfnamefont {J.~M.}\ \bibnamefont
  {Keller}},\ }\bibfield  {title} {\enquote {\bibinfo {title} {Immiscible
  cellular-automaton fluids},}\ }\href {\doibase 10.1007/BF01019743} {\bibfield
   {journal} {\bibinfo  {journal} {Journal of Statistical Physics}\ }\textbf
  {\bibinfo {volume} {52}},\ \bibinfo {pages} {1119--1127} (\bibinfo {year}
  {1988})}\BibitemShut {NoStop}%
\bibitem [{\citenamefont {Chen}\ and\ \citenamefont
  {Doolen}(1998)}]{chen1998lattice}%
  \BibitemOpen
  \bibfield  {author} {\bibinfo {author} {\bibfnamefont {S.}~\bibnamefont
  {Chen}}\ and\ \bibinfo {author} {\bibfnamefont {G.~D.}\ \bibnamefont
  {Doolen}},\ }\bibfield  {title} {\enquote {\bibinfo {title} {{LATTICE
  BOLTZMANN METHOD FOR FLUID FLOWS}},}\ }\href {\doibase
  10.1146/annurev.fluid.30.1.329} {\bibfield  {journal} {\bibinfo  {journal}
  {Annual Review of Fluid Mechanics}\ }\textbf {\bibinfo {volume} {30}},\
  \bibinfo {pages} {329--364} (\bibinfo {year} {1998})}\BibitemShut {NoStop}%
\bibitem [{\citenamefont {Ba}\ \emph {et~al.}(2016)\citenamefont {Ba},
  \citenamefont {Liu}, \citenamefont {Li}, \citenamefont {Kang},\ and\
  \citenamefont {Sun}}]{2016Multiple}%
  \BibitemOpen
  \bibfield  {author} {\bibinfo {author} {\bibfnamefont {Y.}~\bibnamefont
  {Ba}}, \bibinfo {author} {\bibfnamefont {H.}~\bibnamefont {Liu}}, \bibinfo
  {author} {\bibfnamefont {Q.}~\bibnamefont {Li}}, \bibinfo {author}
  {\bibfnamefont {Q.}~\bibnamefont {Kang}}, \ and\ \bibinfo {author}
  {\bibfnamefont {J.}~\bibnamefont {Sun}},\ }\bibfield  {title} {\enquote
  {\bibinfo {title} {{Multiple-relaxation-time color-gradient lattice Boltzmann
  model for simulating two-phase flows with high density ratio}},}\ }\href
  {\doibase 10.1103/PhysRevE.94.023310} {\bibfield  {journal} {\bibinfo
  {journal} {Phys. Rev. E}\ }\textbf {\bibinfo {volume} {94}},\ \bibinfo
  {pages} {023310} (\bibinfo {year} {2016})}\BibitemShut {NoStop}%
\bibitem [{\citenamefont {Li}, \citenamefont {Luo},\ and\ \citenamefont
  {Li}(2013)}]{PhysRevE.87.053301}%
  \BibitemOpen
  \bibfield  {author} {\bibinfo {author} {\bibfnamefont {Q.}~\bibnamefont
  {Li}}, \bibinfo {author} {\bibfnamefont {K.~H.}\ \bibnamefont {Luo}}, \ and\
  \bibinfo {author} {\bibfnamefont {X.~J.}\ \bibnamefont {Li}},\ }\bibfield
  {title} {\enquote {\bibinfo {title} {{Lattice Boltzmann modeling of
  multiphase flows at large density ratio with an improved pseudopotential
  model}},}\ }\href {\doibase 10.1103/PhysRevE.87.053301} {\bibfield  {journal}
  {\bibinfo  {journal} {Phys. Rev. E}\ }\textbf {\bibinfo {volume} {87}},\
  \bibinfo {pages} {053301} (\bibinfo {year} {2013})}\BibitemShut {NoStop}%
\bibitem [{\citenamefont {Zhang}, \citenamefont {Guo},\ and\ \citenamefont
  {Li}(2019)}]{ZHANG20191128}%
  \BibitemOpen
  \bibfield  {author} {\bibinfo {author} {\bibfnamefont {C.}~\bibnamefont
  {Zhang}}, \bibinfo {author} {\bibfnamefont {Z.}~\bibnamefont {Guo}}, \ and\
  \bibinfo {author} {\bibfnamefont {Y.}~\bibnamefont {Li}},\ }\bibfield
  {title} {\enquote {\bibinfo {title} {{A fractional step lattice Boltzmann
  model for two-phase flow with large density differences}},}\ }\href {\doibase
  https://doi.org/10.1016/j.ijheatmasstransfer.2019.04.101} {\bibfield
  {journal} {\bibinfo  {journal} {International Journal of Heat and Mass
  Transfer}\ }\textbf {\bibinfo {volume} {138}},\ \bibinfo {pages} {1128--1141}
  (\bibinfo {year} {2019})}\BibitemShut {NoStop}%
\bibitem [{\citenamefont {Qian}, \citenamefont {d'Humi{\`e}res},\ and\
  \citenamefont {Lallemand}(1992)}]{qian1992lattice}%
  \BibitemOpen
  \bibfield  {author} {\bibinfo {author} {\bibfnamefont {Y.-H.}\ \bibnamefont
  {Qian}}, \bibinfo {author} {\bibfnamefont {D.}~\bibnamefont
  {d'Humi{\`e}res}}, \ and\ \bibinfo {author} {\bibfnamefont {P.}~\bibnamefont
  {Lallemand}},\ }\bibfield  {title} {\enquote {\bibinfo {title} {{Lattice BGK
  models for Navier-Stokes equation}},}\ }\href@noop {} {\bibfield  {journal}
  {\bibinfo  {journal} {EPL (Europhysics Letters)}\ }\textbf {\bibinfo {volume}
  {17}},\ \bibinfo {pages} {479} (\bibinfo {year} {1992})}\BibitemShut
  {NoStop}%
\bibitem [{\citenamefont {Allen}\ and\ \citenamefont
  {Cahn}(1976)}]{ALLEN1976425}%
  \BibitemOpen
  \bibfield  {author} {\bibinfo {author} {\bibfnamefont {S.~M.}\ \bibnamefont
  {Allen}}\ and\ \bibinfo {author} {\bibfnamefont {J.~W.}\ \bibnamefont
  {Cahn}},\ }\bibfield  {title} {\enquote {\bibinfo {title} {{Mechanisms of
  phase transformations within the miscibility gap of Fe-rich Fe-Al alloys}},}\
  }\href {\doibase https://doi.org/10.1016/0001-6160(76)90063-8} {\bibfield
  {journal} {\bibinfo  {journal} {Acta Metallurgica}\ }\textbf {\bibinfo
  {volume} {24}},\ \bibinfo {pages} {425--437} (\bibinfo {year}
  {1976})}\BibitemShut {NoStop}%
\bibitem [{\citenamefont {Sun}\ and\ \citenamefont
  {Beckermann}(2007)}]{SUN2007626}%
  \BibitemOpen
  \bibfield  {author} {\bibinfo {author} {\bibfnamefont {Y.}~\bibnamefont
  {Sun}}\ and\ \bibinfo {author} {\bibfnamefont {C.}~\bibnamefont
  {Beckermann}},\ }\bibfield  {title} {\enquote {\bibinfo {title} {Sharp
  interface tracking using the phase-field equation},}\ }\href {\doibase
  https://doi.org/10.1016/j.jcp.2006.05.025} {\bibfield  {journal} {\bibinfo
  {journal} {Journal of Computational Physics}\ }\textbf {\bibinfo {volume}
  {220}},\ \bibinfo {pages} {626--653} (\bibinfo {year} {2007})}\BibitemShut
  {NoStop}%
\bibitem [{\citenamefont {Chiu}\ and\ \citenamefont {Lin}(2011)}]{CHIU2011185}%
  \BibitemOpen
  \bibfield  {author} {\bibinfo {author} {\bibfnamefont {P.-H.}\ \bibnamefont
  {Chiu}}\ and\ \bibinfo {author} {\bibfnamefont {Y.-T.}\ \bibnamefont {Lin}},\
  }\bibfield  {title} {\enquote {\bibinfo {title} {A conservative phase field
  method for solving incompressible two-phase flows},}\ }\href {\doibase
  https://doi.org/10.1016/j.jcp.2010.09.021} {\bibfield  {journal} {\bibinfo
  {journal} {Journal of Computational Physics}\ }\textbf {\bibinfo {volume}
  {230}},\ \bibinfo {pages} {185--204} (\bibinfo {year} {2011})}\BibitemShut
  {NoStop}%
\bibitem [{\citenamefont {Cahn}\ and\ \citenamefont
  {Hilliard}(1958)}]{cahn1958free}%
  \BibitemOpen
  \bibfield  {author} {\bibinfo {author} {\bibfnamefont {J.~W.}\ \bibnamefont
  {Cahn}}\ and\ \bibinfo {author} {\bibfnamefont {J.~E.}\ \bibnamefont
  {Hilliard}},\ }\bibfield  {title} {\enquote {\bibinfo {title} {{Free energy
  of a nonuniform system. I. Interfacial free energy}},}\ }\href@noop {}
  {\bibfield  {journal} {\bibinfo  {journal} {The Journal of chemical physics}\
  }\textbf {\bibinfo {volume} {28}},\ \bibinfo {pages} {258--267} (\bibinfo
  {year} {1958})}\BibitemShut {NoStop}%
\bibitem [{\citenamefont {Cahn}\ and\ \citenamefont
  {Hilliard}(1959)}]{cahn1959free}%
  \BibitemOpen
  \bibfield  {author} {\bibinfo {author} {\bibfnamefont {J.~W.}\ \bibnamefont
  {Cahn}}\ and\ \bibinfo {author} {\bibfnamefont {J.~E.}\ \bibnamefont
  {Hilliard}},\ }\bibfield  {title} {\enquote {\bibinfo {title} {{Free energy
  of a nonuniform system. III. Nucleation in a two-component incompressible
  fluid}},}\ }\href@noop {} {\bibfield  {journal} {\bibinfo  {journal} {The
  Journal of chemical physics}\ }\textbf {\bibinfo {volume} {31}},\ \bibinfo
  {pages} {688--699} (\bibinfo {year} {1959})}\BibitemShut {NoStop}%
\bibitem [{\citenamefont {Wang}\ \emph {et~al.}(2016)\citenamefont {Wang},
  \citenamefont {Chai}, \citenamefont {Shi},\ and\ \citenamefont
  {Liang}}]{H2016Comparative}%
  \BibitemOpen
  \bibfield  {author} {\bibinfo {author} {\bibfnamefont {H.~L.}\ \bibnamefont
  {Wang}}, \bibinfo {author} {\bibfnamefont {Z.~H.}\ \bibnamefont {Chai}},
  \bibinfo {author} {\bibfnamefont {B.~C.}\ \bibnamefont {Shi}}, \ and\
  \bibinfo {author} {\bibfnamefont {H.}~\bibnamefont {Liang}},\ }\bibfield
  {title} {\enquote {\bibinfo {title} {{Comparative study of the lattice
  Boltzmann models for Allen-Cahn and Cahn-Hilliard equations}},}\ }\href
  {\doibase 10.1103/PhysRevE.94.033304} {\bibfield  {journal} {\bibinfo
  {journal} {Phys. Rev. E}\ }\textbf {\bibinfo {volume} {94}},\ \bibinfo
  {pages} {033304} (\bibinfo {year} {2016})}\BibitemShut {NoStop}%
\bibitem [{\citenamefont {Jiang}\ and\ \citenamefont
  {Shu}(1996)}]{jiang1996efficient}%
  \BibitemOpen
  \bibfield  {author} {\bibinfo {author} {\bibfnamefont {G.-S.}\ \bibnamefont
  {Jiang}}\ and\ \bibinfo {author} {\bibfnamefont {C.-W.}\ \bibnamefont
  {Shu}},\ }\bibfield  {title} {\enquote {\bibinfo {title} {{Efficient
  Implementation of Weighted ENO Schemes}},}\ }\href {\doibase
  https://doi.org/10.1006/jcph.1996.0130} {\bibfield  {journal} {\bibinfo
  {journal} {Journal of Computational Physics}\ }\textbf {\bibinfo {volume}
  {126}},\ \bibinfo {pages} {202--228} (\bibinfo {year} {1996})}\BibitemShut
  {NoStop}%
\bibitem [{\citenamefont {Harten}\ \emph {et~al.}(1987)\citenamefont {Harten},
  \citenamefont {Engquist}, \citenamefont {Osher},\ and\ \citenamefont
  {Chakravarthy}}]{HARTEN1987231}%
  \BibitemOpen
  \bibfield  {author} {\bibinfo {author} {\bibfnamefont {A.}~\bibnamefont
  {Harten}}, \bibinfo {author} {\bibfnamefont {B.}~\bibnamefont {Engquist}},
  \bibinfo {author} {\bibfnamefont {S.}~\bibnamefont {Osher}}, \ and\ \bibinfo
  {author} {\bibfnamefont {S.~R.}\ \bibnamefont {Chakravarthy}},\ }\bibfield
  {title} {\enquote {\bibinfo {title} {{Uniformly high order accurate
  essentially non-oscillatory schemes, III}},}\ }\href {\doibase
  https://doi.org/10.1016/0021-9991(87)90031-3} {\bibfield  {journal} {\bibinfo
   {journal} {Journal of Computational Physics}\ }\textbf {\bibinfo {volume}
  {71}},\ \bibinfo {pages} {231--303} (\bibinfo {year} {1987})}\BibitemShut
  {NoStop}%
\bibitem [{\citenamefont {Laniewski-Wollk}\ and\ \citenamefont
  {Rokicki}(2016)}]{LANIEWSKIWOLLK2016833}%
  \BibitemOpen
  \bibfield  {author} {\bibinfo {author} {\bibfnamefont {L.}~\bibnamefont
  {Laniewski-Wollk}}\ and\ \bibinfo {author} {\bibfnamefont {J.}~\bibnamefont
  {Rokicki}},\ }\bibfield  {title} {\enquote {\bibinfo {title} {{Adjoint
  Lattice Boltzmann for topology optimization on multi-GPU architecture}},}\
  }\href {\doibase https://doi.org/10.1016/j.camwa.2015.12.043} {\bibfield
  {journal} {\bibinfo  {journal} {Computers and Mathematics with Applications}\
  }\textbf {\bibinfo {volume} {71}},\ \bibinfo {pages} {833--848} (\bibinfo
  {year} {2016})}\BibitemShut {NoStop}%
\bibitem [{\citenamefont {Guo}, \citenamefont {Xu},\ and\ \citenamefont
  {Wang}(2013)}]{Guo2013}%
  \BibitemOpen
  \bibfield  {author} {\bibinfo {author} {\bibfnamefont {Z.}~\bibnamefont
  {Guo}}, \bibinfo {author} {\bibfnamefont {K.}~\bibnamefont {Xu}}, \ and\
  \bibinfo {author} {\bibfnamefont {R.}~\bibnamefont {Wang}},\ }\bibfield
  {title} {\enquote {\bibinfo {title} {{Discrete unified gas kinetic scheme for
  all Knudsen number flows: Low-speed isothermal case}},}\ }\href {\doibase
  10.1103/PhysRevE.88.033305} {\bibfield  {journal} {\bibinfo  {journal} {Phys.
  Rev. E}\ }\textbf {\bibinfo {volume} {88}},\ \bibinfo {pages} {033305}
  (\bibinfo {year} {2013})}\BibitemShut {NoStop}%
\bibitem [{\citenamefont {Guo}, \citenamefont {Wang},\ and\ \citenamefont
  {Xu}(2015)}]{Guo2015}%
  \BibitemOpen
  \bibfield  {author} {\bibinfo {author} {\bibfnamefont {Z.}~\bibnamefont
  {Guo}}, \bibinfo {author} {\bibfnamefont {R.}~\bibnamefont {Wang}}, \ and\
  \bibinfo {author} {\bibfnamefont {K.}~\bibnamefont {Xu}},\ }\bibfield
  {title} {\enquote {\bibinfo {title} {{Discrete unified gas kinetic scheme for
  all Knudsen number flows. II. Thermal compressible case}},}\ }\href {\doibase
  10.1103/PhysRevE.91.033313} {\bibfield  {journal} {\bibinfo  {journal} {Phys.
  Rev. E}\ }\textbf {\bibinfo {volume} {91}},\ \bibinfo {pages} {033313}
  (\bibinfo {year} {2015})}\BibitemShut {NoStop}%
\bibitem [{\citenamefont {Aidun}\ and\ \citenamefont
  {Clausen}(2010)}]{aidun2010lattice}%
  \BibitemOpen
  \bibfield  {author} {\bibinfo {author} {\bibfnamefont {C.~K.}\ \bibnamefont
  {Aidun}}\ and\ \bibinfo {author} {\bibfnamefont {J.~R.}\ \bibnamefont
  {Clausen}},\ }\bibfield  {title} {\enquote {\bibinfo {title}
  {{Lattice-Boltzmann Method for Complex Flows}},}\ }\href {\doibase
  10.1146/annurev-fluid-121108-145519} {\bibfield  {journal} {\bibinfo
  {journal} {Annual Review of Fluid Mechanics}\ }\textbf {\bibinfo {volume}
  {42}},\ \bibinfo {pages} {439--472} (\bibinfo {year} {2010})}\BibitemShut
  {NoStop}%
\bibitem [{\citenamefont {Xu}\ and\ \citenamefont {Huang}(2010)}]{XU20107747}%
  \BibitemOpen
  \bibfield  {author} {\bibinfo {author} {\bibfnamefont {K.}~\bibnamefont
  {Xu}}\ and\ \bibinfo {author} {\bibfnamefont {J.-C.}\ \bibnamefont {Huang}},\
  }\bibfield  {title} {\enquote {\bibinfo {title} {A unified gas-kinetic scheme
  for continuum and rarefied flows},}\ }\href {\doibase
  https://doi.org/10.1016/j.jcp.2010.06.032} {\bibfield  {journal} {\bibinfo
  {journal} {Journal of Computational Physics}\ }\textbf {\bibinfo {volume}
  {229}},\ \bibinfo {pages} {7747--7764} (\bibinfo {year} {2010})}\BibitemShut
  {NoStop}%
\bibitem [{\citenamefont {Rowlinson}\ and\ \citenamefont
  {Widom}(1989)}]{2002Molecular}%
  \BibitemOpen
  \bibfield  {author} {\bibinfo {author} {\bibfnamefont {J.~S.}\ \bibnamefont
  {Rowlinson}}\ and\ \bibinfo {author} {\bibfnamefont {B.}~\bibnamefont
  {Widom}},\ }\href {\doibase https://doi.org/10.1016/0022-0728(89)80091-9}
  {\emph {\bibinfo {title} {Molecular Theory of Capillarity}}}\ (\bibinfo
  {year} {1989})\BibitemShut {NoStop}%
\bibitem [{\citenamefont {Liu}\ and\ \citenamefont
  {Shen}(2003)}]{liu2003phase}%
  \BibitemOpen
  \bibfield  {author} {\bibinfo {author} {\bibfnamefont {C.}~\bibnamefont
  {Liu}}\ and\ \bibinfo {author} {\bibfnamefont {J.}~\bibnamefont {Shen}},\
  }\bibfield  {title} {\enquote {\bibinfo {title} {{A phase field model for the
  mixture of two incompressible fluids and its approximation by a
  Fourier-spectral method}},}\ }\href {\doibase
  https://doi.org/10.1016/S0167-2789(03)00030-7} {\bibfield  {journal}
  {\bibinfo  {journal} {Physica D: Nonlinear Phenomena}\ }\textbf {\bibinfo
  {volume} {179}},\ \bibinfo {pages} {211--228} (\bibinfo {year}
  {2003})}\BibitemShut {NoStop}%
\bibitem [{\citenamefont {Jacqmin}(1996)}]{jacqmin1996energy}%
  \BibitemOpen
  \bibfield  {author} {\bibinfo {author} {\bibfnamefont {D.}~\bibnamefont
  {Jacqmin}},\ }\bibfield  {title} {\enquote {\bibinfo {title} {An energy
  approach to the continuum surface tension method},}\ }in\ \href@noop {}
  {\emph {\bibinfo {booktitle} {34th Aerospace sciences meeting and exhibit}}}\
  (\bibinfo {year} {1996})\ p.\ \bibinfo {pages} {858}\BibitemShut {NoStop}%
\bibitem [{\citenamefont {YUE}\ \emph {et~al.}(2004)\citenamefont {YUE},
  \citenamefont {FENG}, \citenamefont {LIU},\ and\ \citenamefont
  {SHEN}}]{yue2004diffuse}%
  \BibitemOpen
  \bibfield  {author} {\bibinfo {author} {\bibfnamefont {P.}~\bibnamefont
  {YUE}}, \bibinfo {author} {\bibfnamefont {J.~J.}\ \bibnamefont {FENG}},
  \bibinfo {author} {\bibfnamefont {C.}~\bibnamefont {LIU}}, \ and\ \bibinfo
  {author} {\bibfnamefont {J.}~\bibnamefont {SHEN}},\ }\bibfield  {title}
  {\enquote {\bibinfo {title} {A diffuse-interface method for simulating
  two-phase flows of complex fluids},}\ }\href {\doibase
  10.1017/S0022112004000370} {\bibfield  {journal} {\bibinfo  {journal}
  {Journal of Fluid Mechanics}\ }\textbf {\bibinfo {volume} {515}},\ \bibinfo
  {pages} {293–317} (\bibinfo {year} {2004})}\BibitemShut {NoStop}%
\bibitem [{\citenamefont {Zhang}, \citenamefont {Yang},\ and\ \citenamefont
  {Guo}(2018)}]{zhang2018discrete}%
  \BibitemOpen
  \bibfield  {author} {\bibinfo {author} {\bibfnamefont {C.}~\bibnamefont
  {Zhang}}, \bibinfo {author} {\bibfnamefont {K.}~\bibnamefont {Yang}}, \ and\
  \bibinfo {author} {\bibfnamefont {Z.}~\bibnamefont {Guo}},\ }\bibfield
  {title} {\enquote {\bibinfo {title} {A discrete unified gas-kinetic scheme
  for immiscible two-phase flows},}\ }\href {\doibase
  https://doi.org/10.1016/j.ijheatmasstransfer.2018.06.016} {\bibfield
  {journal} {\bibinfo  {journal} {International Journal of Heat and Mass
  Transfer}\ }\textbf {\bibinfo {volume} {126}},\ \bibinfo {pages} {1326--1336}
  (\bibinfo {year} {2018})}\BibitemShut {NoStop}%
\bibitem [{\citenamefont {Chen}\ \emph {et~al.}(2019)\citenamefont {Chen},
  \citenamefont {Ch{\'e}ron}, \citenamefont {Guo}, \citenamefont {de~Motta},
  \citenamefont {Menard},\ and\ \citenamefont {Wang}}]{chen2019simulation}%
  \BibitemOpen
  \bibfield  {author} {\bibinfo {author} {\bibfnamefont {T.}~\bibnamefont
  {Chen}}, \bibinfo {author} {\bibfnamefont {V.}~\bibnamefont {Ch{\'e}ron}},
  \bibinfo {author} {\bibfnamefont {Z.}~\bibnamefont {Guo}}, \bibinfo {author}
  {\bibfnamefont {J.~C.~B.}\ \bibnamefont {de~Motta}}, \bibinfo {author}
  {\bibfnamefont {T.}~\bibnamefont {Menard}}, \ and\ \bibinfo {author}
  {\bibfnamefont {L.-P.}\ \bibnamefont {Wang}},\ }\bibfield  {title} {\enquote
  {\bibinfo {title} {Simulation of immiscible two-phase flows based on a
  kinetic diffuse interface approach},}\ }in\ \href@noop {} {\emph {\bibinfo
  {booktitle} {International Conference on Multiphase Flow}}}\ (\bibinfo {year}
  {2019})\BibitemShut {NoStop}%
\bibitem [{\citenamefont {Liang}\ \emph {et~al.}(2014)\citenamefont {Liang},
  \citenamefont {Shi}, \citenamefont {Guo},\ and\ \citenamefont
  {Chai}}]{liang2014phase}%
  \BibitemOpen
  \bibfield  {author} {\bibinfo {author} {\bibfnamefont {H.}~\bibnamefont
  {Liang}}, \bibinfo {author} {\bibfnamefont {B.~C.}\ \bibnamefont {Shi}},
  \bibinfo {author} {\bibfnamefont {Z.~L.}\ \bibnamefont {Guo}}, \ and\
  \bibinfo {author} {\bibfnamefont {Z.~H.}\ \bibnamefont {Chai}},\ }\bibfield
  {title} {\enquote {\bibinfo {title} {{Phase-field-based
  multiple-relaxation-time lattice Boltzmann model for incompressible
  multiphase flows}},}\ }\href {\doibase 10.1103/PhysRevE.89.053320} {\bibfield
   {journal} {\bibinfo  {journal} {Phys. Rev. E}\ }\textbf {\bibinfo {volume}
  {89}},\ \bibinfo {pages} {053320} (\bibinfo {year} {2014})}\BibitemShut
  {NoStop}%
\bibitem [{\citenamefont {Geier}, \citenamefont {Fakhari},\ and\ \citenamefont
  {Lee}(2015)}]{PhysRevE.91.063309}%
  \BibitemOpen
  \bibfield  {author} {\bibinfo {author} {\bibfnamefont {M.}~\bibnamefont
  {Geier}}, \bibinfo {author} {\bibfnamefont {A.}~\bibnamefont {Fakhari}}, \
  and\ \bibinfo {author} {\bibfnamefont {T.}~\bibnamefont {Lee}},\ }\bibfield
  {title} {\enquote {\bibinfo {title} {Conservative phase-field lattice
  {Boltzmann} model for interface tracking equation},}\ }\href {\doibase
  10.1103/PhysRevE.91.063309} {\bibfield  {journal} {\bibinfo  {journal} {Phys.
  Rev. E}\ }\textbf {\bibinfo {volume} {91}},\ \bibinfo {pages} {063309}
  (\bibinfo {year} {2015})}\BibitemShut {NoStop}%
\bibitem [{\citenamefont {He}, \citenamefont {Chen},\ and\ \citenamefont
  {Zhang}(1999)}]{HE1999642}%
  \BibitemOpen
  \bibfield  {author} {\bibinfo {author} {\bibfnamefont {X.}~\bibnamefont
  {He}}, \bibinfo {author} {\bibfnamefont {S.}~\bibnamefont {Chen}}, \ and\
  \bibinfo {author} {\bibfnamefont {R.}~\bibnamefont {Zhang}},\ }\bibfield
  {title} {\enquote {\bibinfo {title} {{A Lattice Boltzmann Scheme for
  Incompressible Multiphase Flow and Its Application in Simulation of
  Rayleigh–Taylor Instability}},}\ }\href {\doibase
  https://doi.org/10.1006/jcph.1999.6257} {\bibfield  {journal} {\bibinfo
  {journal} {Journal of Computational Physics}\ }\textbf {\bibinfo {volume}
  {152}},\ \bibinfo {pages} {642--663} (\bibinfo {year} {1999})}\BibitemShut
  {NoStop}%
\bibitem [{\citenamefont {Yang}\ and\ \citenamefont
  {Guo}(2016)}]{PhysRevE.93.043303}%
  \BibitemOpen
  \bibfield  {author} {\bibinfo {author} {\bibfnamefont {K.}~\bibnamefont
  {Yang}}\ and\ \bibinfo {author} {\bibfnamefont {Z.}~\bibnamefont {Guo}},\
  }\bibfield  {title} {\enquote {\bibinfo {title} {{Lattice Boltzmann method
  for binary fluids based on mass-conserving quasi-incompressible phase-field
  theory}},}\ }\href {\doibase 10.1103/PhysRevE.93.043303} {\bibfield
  {journal} {\bibinfo  {journal} {Phys. Rev. E}\ }\textbf {\bibinfo {volume}
  {93}},\ \bibinfo {pages} {043303} (\bibinfo {year} {2016})}\BibitemShut
  {NoStop}%
\bibitem [{\citenamefont {Bhatnagar}, \citenamefont {Gross},\ and\
  \citenamefont {Krook}(1954)}]{PhysRev.94.511}%
  \BibitemOpen
  \bibfield  {author} {\bibinfo {author} {\bibfnamefont {P.~L.}\ \bibnamefont
  {Bhatnagar}}, \bibinfo {author} {\bibfnamefont {E.~P.}\ \bibnamefont
  {Gross}}, \ and\ \bibinfo {author} {\bibfnamefont {M.}~\bibnamefont
  {Krook}},\ }\bibfield  {title} {\enquote {\bibinfo {title} {A model for
  collision processes in gases},}\ }\href@noop {} {\bibfield  {journal}
  {\bibinfo  {journal} {Phys. Rev.}\ }\textbf {\bibinfo {volume} {94}},\
  \bibinfo {pages} {511--525} (\bibinfo {year} {1954})}\BibitemShut {NoStop}%
\bibitem [{\citenamefont {Chapman}\ and\ \citenamefont
  {Cowling}(1970)}]{Chapman1970}%
  \BibitemOpen
  \bibfield  {author} {\bibinfo {author} {\bibfnamefont {S.}~\bibnamefont
  {Chapman}}\ and\ \bibinfo {author} {\bibfnamefont {T.~G.}\ \bibnamefont
  {Cowling}},\ }\href@noop {} {\emph {\bibinfo {title} {The Mathematical Theory
  of Non-Uniform Gases}}}\ (\bibinfo  {publisher} {Cambridge University
  Press},\ \bibinfo {year} {1970})\BibitemShut {NoStop}%
\bibitem [{\citenamefont {Zu}\ and\ \citenamefont {He}(2013)}]{zu2013phase}%
  \BibitemOpen
  \bibfield  {author} {\bibinfo {author} {\bibfnamefont {Y.~Q.}\ \bibnamefont
  {Zu}}\ and\ \bibinfo {author} {\bibfnamefont {S.}~\bibnamefont {He}},\
  }\bibfield  {title} {\enquote {\bibinfo {title} {{Phase-field-based lattice
  Boltzmann model for incompressible binary fluid systems with density and
  viscosity contrasts}},}\ }\href {\doibase 10.1103/PhysRevE.87.043301}
  {\bibfield  {journal} {\bibinfo  {journal} {Phys. Rev. E}\ }\textbf {\bibinfo
  {volume} {87}},\ \bibinfo {pages} {043301} (\bibinfo {year}
  {2013})}\BibitemShut {NoStop}%
\bibitem [{\citenamefont {Chen}\ \emph {et~al.}(2020)\citenamefont {Chen},
  \citenamefont {Wen}, \citenamefont {Wang}, \citenamefont {Guo}, \citenamefont
  {Wang},\ and\ \citenamefont {Chen}}]{2020Simulation}%
  \BibitemOpen
  \bibfield  {author} {\bibinfo {author} {\bibfnamefont {T.}~\bibnamefont
  {Chen}}, \bibinfo {author} {\bibfnamefont {X.}~\bibnamefont {Wen}}, \bibinfo
  {author} {\bibfnamefont {L.-P.}\ \bibnamefont {Wang}}, \bibinfo {author}
  {\bibfnamefont {Z.}~\bibnamefont {Guo}}, \bibinfo {author} {\bibfnamefont
  {J.}~\bibnamefont {Wang}}, \ and\ \bibinfo {author} {\bibfnamefont
  {S.}~\bibnamefont {Chen}},\ }\bibfield  {title} {\enquote {\bibinfo {title}
  {Simulation of three-dimensional compressible decaying isotropic turbulence
  using a redesigned discrete unified gas kinetic scheme},}\ }\href {\doibase
  10.1063/5.0029424} {\bibfield  {journal} {\bibinfo  {journal} {Physics of
  Fluids}\ }\textbf {\bibinfo {volume} {32}},\ \bibinfo {pages} {125104}
  (\bibinfo {year} {2020})}\BibitemShut {NoStop}%
\bibitem [{\citenamefont {Liu}, \citenamefont {Osher},\ and\ \citenamefont
  {Chan}(1994)}]{LIU1994200}%
  \BibitemOpen
  \bibfield  {author} {\bibinfo {author} {\bibfnamefont {X.-D.}\ \bibnamefont
  {Liu}}, \bibinfo {author} {\bibfnamefont {S.}~\bibnamefont {Osher}}, \ and\
  \bibinfo {author} {\bibfnamefont {T.}~\bibnamefont {Chan}},\ }\bibfield
  {title} {\enquote {\bibinfo {title} {{Weighted Essentially Non-oscillatory
  Schemes}},}\ }\href {\doibase https://doi.org/10.1006/jcph.1994.1187}
  {\bibfield  {journal} {\bibinfo  {journal} {Journal of Computational
  Physics}\ }\textbf {\bibinfo {volume} {115}},\ \bibinfo {pages} {200--212}
  (\bibinfo {year} {1994})}\BibitemShut {NoStop}%
\bibitem [{\citenamefont {Lycett-Brown}\ and\ \citenamefont
  {Luo}(2016)}]{lycett2016cascaded}%
  \BibitemOpen
  \bibfield  {author} {\bibinfo {author} {\bibfnamefont {D.}~\bibnamefont
  {Lycett-Brown}}\ and\ \bibinfo {author} {\bibfnamefont {K.~H.}\ \bibnamefont
  {Luo}},\ }\bibfield  {title} {\enquote {\bibinfo {title} {Cascaded lattice
  boltzmann method with improved forcing scheme for large-density-ratio
  multiphase flow at high reynolds and weber numbers},}\ }\href@noop {}
  {\bibfield  {journal} {\bibinfo  {journal} {Physical Review E}\ }\textbf
  {\bibinfo {volume} {94}},\ \bibinfo {pages} {053313} (\bibinfo {year}
  {2016})}\BibitemShut {NoStop}%
\bibitem [{\citenamefont {Hamzehloo}, \citenamefont {Bartholomew},\ and\
  \citenamefont {Laizet}(2021)}]{2021Direct}%
  \BibitemOpen
  \bibfield  {author} {\bibinfo {author} {\bibfnamefont {A.}~\bibnamefont
  {Hamzehloo}}, \bibinfo {author} {\bibfnamefont {P.}~\bibnamefont
  {Bartholomew}}, \ and\ \bibinfo {author} {\bibfnamefont {S.}~\bibnamefont
  {Laizet}},\ }\bibfield  {title} {\enquote {\bibinfo {title} {Direct numerical
  simulations of incompressible rayleigh–taylor instabilities at low and
  medium atwood numbers},}\ }\href {\doibase 10.1063/5.0049867} {\bibfield
  {journal} {\bibinfo  {journal} {Physics of Fluids}\ }\textbf {\bibinfo
  {volume} {33}},\ \bibinfo {pages} {054114} (\bibinfo {year}
  {2021})}\BibitemShut {NoStop}%
\bibitem [{\citenamefont {Ding}, \citenamefont {Spelt},\ and\ \citenamefont
  {Shu}(2007)}]{DING20072078}%
  \BibitemOpen
  \bibfield  {author} {\bibinfo {author} {\bibfnamefont {H.}~\bibnamefont
  {Ding}}, \bibinfo {author} {\bibfnamefont {P.~D.}\ \bibnamefont {Spelt}}, \
  and\ \bibinfo {author} {\bibfnamefont {C.}~\bibnamefont {Shu}},\ }\bibfield
  {title} {\enquote {\bibinfo {title} {Diffuse interface model for
  incompressible two-phase flows with large density ratios},}\ }\href {\doibase
  https://doi.org/10.1016/j.jcp.2007.06.028} {\bibfield  {journal} {\bibinfo
  {journal} {Journal of Computational Physics}\ }\textbf {\bibinfo {volume}
  {226}},\ \bibinfo {pages} {2078--2095} (\bibinfo {year} {2007})}\BibitemShut
  {NoStop}%
\bibitem [{\citenamefont {Li}\ \emph {et~al.}(2012)\citenamefont {Li},
  \citenamefont {Luo}, \citenamefont {Gao},\ and\ \citenamefont
  {He}}]{li2012additional}%
  \BibitemOpen
  \bibfield  {author} {\bibinfo {author} {\bibfnamefont {Q.}~\bibnamefont
  {Li}}, \bibinfo {author} {\bibfnamefont {K.~H.}\ \bibnamefont {Luo}},
  \bibinfo {author} {\bibfnamefont {Y.~J.}\ \bibnamefont {Gao}}, \ and\
  \bibinfo {author} {\bibfnamefont {Y.~L.}\ \bibnamefont {He}},\ }\bibfield
  {title} {\enquote {\bibinfo {title} {Additional interfacial force in lattice
  boltzmann models for incompressible multiphase flows},}\ }\href {\doibase
  10.1103/PhysRevE.85.026704} {\bibfield  {journal} {\bibinfo  {journal} {Phys.
  Rev. E}\ }\textbf {\bibinfo {volume} {85}},\ \bibinfo {pages} {026704}
  (\bibinfo {year} {2012})}\BibitemShut {NoStop}%
\bibitem [{\citenamefont {Ren}\ \emph {et~al.}(2016)\citenamefont {Ren},
  \citenamefont {Song}, \citenamefont {Sukop},\ and\ \citenamefont
  {Hu}}]{ren2016improved}%
  \BibitemOpen
  \bibfield  {author} {\bibinfo {author} {\bibfnamefont {F.}~\bibnamefont
  {Ren}}, \bibinfo {author} {\bibfnamefont {B.}~\bibnamefont {Song}}, \bibinfo
  {author} {\bibfnamefont {M.~C.}\ \bibnamefont {Sukop}}, \ and\ \bibinfo
  {author} {\bibfnamefont {H.}~\bibnamefont {Hu}},\ }\bibfield  {title}
  {\enquote {\bibinfo {title} {{Improved lattice Boltzmann modeling of binary
  flow based on the conservative Allen-Cahn equation}},}\ }\href {\doibase
  10.1103/PhysRevE.94.023311} {\bibfield  {journal} {\bibinfo  {journal} {Phys.
  Rev. E}\ }\textbf {\bibinfo {volume} {94}},\ \bibinfo {pages} {023311}
  (\bibinfo {year} {2016})}\BibitemShut {NoStop}%
\bibitem [{\citenamefont {Josserand}\ and\ \citenamefont
  {Zaleski}(2003)}]{josserand2003droplet}%
  \BibitemOpen
  \bibfield  {author} {\bibinfo {author} {\bibfnamefont {C.}~\bibnamefont
  {Josserand}}\ and\ \bibinfo {author} {\bibfnamefont {S.}~\bibnamefont
  {Zaleski}},\ }\bibfield  {title} {\enquote {\bibinfo {title} {Droplet
  splashing on a thin liquid film},}\ }\href@noop {} {\bibfield  {journal}
  {\bibinfo  {journal} {Physics of fluids}\ }\textbf {\bibinfo {volume} {15}},\
  \bibinfo {pages} {1650--1657} (\bibinfo {year} {2003})}\BibitemShut {NoStop}%
\bibitem [{\citenamefont {Lee}\ and\ \citenamefont
  {Lin}(2005)}]{lee2005stable}%
  \BibitemOpen
  \bibfield  {author} {\bibinfo {author} {\bibfnamefont {T.}~\bibnamefont
  {Lee}}\ and\ \bibinfo {author} {\bibfnamefont {C.-L.}\ \bibnamefont {Lin}},\
  }\bibfield  {title} {\enquote {\bibinfo {title} {A stable discretization of
  the lattice boltzmann equation for simulation of incompressible two-phase
  flows at high density ratio},}\ }\href@noop {} {\bibfield  {journal}
  {\bibinfo  {journal} {Journal of Computational Physics}\ }\textbf {\bibinfo
  {volume} {206}},\ \bibinfo {pages} {16--47} (\bibinfo {year}
  {2005})}\BibitemShut {NoStop}%
\bibitem [{\citenamefont {Wang}\ \emph
  {et~al.}(2015{\natexlab{b}})\citenamefont {Wang}, \citenamefont {Shu},
  \citenamefont {Huang},\ and\ \citenamefont {Teo}}]{wang2015multiphase}%
  \BibitemOpen
  \bibfield  {author} {\bibinfo {author} {\bibfnamefont {Y.}~\bibnamefont
  {Wang}}, \bibinfo {author} {\bibfnamefont {C.}~\bibnamefont {Shu}}, \bibinfo
  {author} {\bibfnamefont {H.}~\bibnamefont {Huang}}, \ and\ \bibinfo {author}
  {\bibfnamefont {C.}~\bibnamefont {Teo}},\ }\bibfield  {title} {\enquote
  {\bibinfo {title} {Multiphase lattice boltzmann flux solver for
  incompressible multiphase flows with large density ratio},}\ }\href@noop {}
  {\bibfield  {journal} {\bibinfo  {journal} {Journal of Computational
  Physics}\ }\textbf {\bibinfo {volume} {280}},\ \bibinfo {pages} {404--423}
  (\bibinfo {year} {2015}{\natexlab{b}})}\BibitemShut {NoStop}%
\bibitem [{\citenamefont {Pan}, \citenamefont {Chou},\ and\ \citenamefont
  {Tseng}(2009)}]{pan2009binary}%
  \BibitemOpen
  \bibfield  {author} {\bibinfo {author} {\bibfnamefont {K.-L.}\ \bibnamefont
  {Pan}}, \bibinfo {author} {\bibfnamefont {P.-C.}\ \bibnamefont {Chou}}, \
  and\ \bibinfo {author} {\bibfnamefont {Y.-J.}\ \bibnamefont {Tseng}},\
  }\bibfield  {title} {\enquote {\bibinfo {title} {Binary droplet collision at
  high weber number},}\ }\href@noop {} {\bibfield  {journal} {\bibinfo
  {journal} {Physical Review E}\ }\textbf {\bibinfo {volume} {80}},\ \bibinfo
  {pages} {036301} (\bibinfo {year} {2009})}\BibitemShut {NoStop}%
\bibitem [{\citenamefont {He}\ \emph {et~al.}(1999)\citenamefont {He},
  \citenamefont {Zhang}, \citenamefont {Chen},\ and\ \citenamefont
  {Doolen}}]{1999On}%
  \BibitemOpen
  \bibfield  {author} {\bibinfo {author} {\bibfnamefont {X.}~\bibnamefont
  {He}}, \bibinfo {author} {\bibfnamefont {R.}~\bibnamefont {Zhang}}, \bibinfo
  {author} {\bibfnamefont {S.}~\bibnamefont {Chen}}, \ and\ \bibinfo {author}
  {\bibfnamefont {G.~D.}\ \bibnamefont {Doolen}},\ }\bibfield  {title}
  {\enquote {\bibinfo {title} {{On the three-dimensional Rayleigh–Taylor
  instability}},}\ }\href {\doibase 10.1063/1.869984} {\bibfield  {journal}
  {\bibinfo  {journal} {Physics of Fluids}\ }\textbf {\bibinfo {volume} {11}},\
  \bibinfo {pages} {1143--1152} (\bibinfo {year} {1999})}\BibitemShut {NoStop}%
\bibitem [{\citenamefont {Chen}\ \emph {et~al.}(2021)\citenamefont {Chen},
  \citenamefont {Wang}, \citenamefont {Lai},\ and\ \citenamefont
  {Chen}}]{2021Inverse}%
  \BibitemOpen
  \bibfield  {author} {\bibinfo {author} {\bibfnamefont {T.}~\bibnamefont
  {Chen}}, \bibinfo {author} {\bibfnamefont {L.-P.}\ \bibnamefont {Wang}},
  \bibinfo {author} {\bibfnamefont {J.}~\bibnamefont {Lai}}, \ and\ \bibinfo
  {author} {\bibfnamefont {S.}~\bibnamefont {Chen}},\ }\bibfield  {title}
  {\enquote {\bibinfo {title} {Inverse design of mesoscopic models for
  compressible flow using the {Chapman-Enskog} analysis},}\ }\href
  {https://doi.org/10.1186/s42774-020-00059-2} {\bibfield  {journal} {\bibinfo
  {journal} {Advances in Aerodynamics}\ }\textbf {\bibinfo {volume} {3}}
  (\bibinfo {year} {2021})}\BibitemShut {NoStop}%
\end{thebibliography}%
\newpage

\end{document}